\documentclass{SciPost}

\hypersetup{
    colorlinks,
    linkcolor={red!50!black},
    citecolor={blue!50!black},
    urlcolor={blue!80!black}
}

\usepackage[bitstream-charter]{mathdesign}
\usepackage{orcidlink} 

\DeclareSymbolFont{usualmathcal}{OMS}{cmsy}{m}{n}
\DeclareSymbolFontAlphabet{\mathcal}{usualmathcal}

\fancypagestyle{SPstyle}{
\fancyhf{}
\rhead{\colorbox{scipostblue}{\bf \color{white} ~Insulating and conducting spin liquids}}

\fancyfoot[C]{\textbf{\thepage}}
}

\usepackage{amsmath,bm} 
\usepackage{graphicx}  
\usepackage{color} 
\usepackage{enumitem}
\usepackage{dsfont}
\usepackage{tcolorbox}
\newcommand{\diff}{d}

\usepackage{braket}
\usepackage{mathtools}
\usepackage[section]{placeins}
\usepackage{multirow}
\def\beq{\begin{equation}}
\def\eeq{\end{equation}}
\def\nn{\nonumber \\}
\newcommand{\mathbfcal}[1]{\boldsymbol{\mathcal{#1}}}
\newcommand{\vi}{{\boldsymbol{i}}}
\newcommand{\vj}{{\boldsymbol{j}}}
\newcommand{\vP}{{\boldsymbol{P}}}
\newcommand{\pdagger}{{\phantom{\dagger}}}
\def\bea{\begin{eqnarray}}
\def\eea{\end{eqnarray}}

\begin{document}

\pagestyle{SPstyle}

\begin{center}{\Large \textbf{\color{scipostdeepblue}{
Lectures on insulating and conducting quantum spin liquids\\
}}}\end{center}

\begin{center}\textbf{
Subir Sachdev\,\orcidlink{0000-0002-2432-7070}
}\end{center}

\begin{center}
 Department of Physics, Harvard University, Cambridge, Massachusetts, 02138, USA.\\
 Center for Computational Quantum Physics, Flatiron Institute,\\ 162 5th Avenue, New York, NY 10010, USA.\\
The Abdus Salam International Centre for Theoretical Physics,\\ Strada Costiera 11, I-34151, Trieste, Italy.
\\[\baselineskip]
\href{mailto:sachdev@g.harvard.edu}{\small sachdev@g.harvard.edu}
\end{center}

\section*{\color{scipostdeepblue}{Abstract}}
\textbf{\boldmath{%
Two of the iconic phases of the hole-doped cuprate materials are the intermediate temperature pseudogap metal and the lower temperature $d$-wave superconductor. Following the prescient suggestion of P.~W.~Anderson, there were numerous early theories of these phases as doped quantum spin liquids. However, these theories have had difficulties with two prominent observations:\\ 
({\it i\/}) angle-dependent magnetoresistance measurements (ADMR) in the pseudogap metal, including observation of the Yamaji effect, present convincing evidence of small hole pockets which can tunnel coherently between square lattice layers (Fang {\it et al.\/}, \href{https://doi.org/10.1038/s41567-022-01514-1}{Nature Physics {\bf 18}, 558 (2022)}; Chan {\it et al.\/}, \href{https://doi.org/10.1038/s41567-025-03032-2}{Nature Physics {\bf 21}, 1753 (2025)}) and\\
 ({\it ii\/}) the velocities of the nodal Bogoliubov quasiparticles in the $d$-wave superconductor are highly anisotropic, with $v_F \gg v_\Delta$
 (Chiao {\it et al.\/}, \href{https://doi.org/10.1103/PhysRevB.62.3554}{Phys. Rev. B {\bf 62}, 3554 (2000)}). \\~\\
 These lecture notes review how the fractionalized Fermi Liquid (FL*) state, which dopes quantum spin liquids with gauge-neutral electron-like quasiparticles, resolves both difficulties. Theories of insulating quantum spin liquids employing fractionalization of the electron spin into bosonic or fermionic partons are discussed. Doping the bosonic parton theory leads to a holon metal theory: while not appropriate for the cuprate pseudogap, this theory is argued to apply to ultracold atom experiments on the Lieb lattice.
Doping the fermionic parton theory leads to a $d$-wave superconductor with nearly isotropic quasiparticle velocities. The construction of the FL* state in a single band Hubbard-type model is described using a quantum dimer model, followed by a more realistic description using the Ancilla Layer Model (ALM), which is then used to obtain the theory of the pseudogap and the $d$-wave superconductor. The ALM also leads to a variational wavefunction for the FL* state of the Hubbard model, and its local correlations agree well with ultracold atom observations.
}}


\begin{center}
 {\tt Advanced School and Conference on Quantum Matter, Dec 1-12, 2025\\
International Centre for Theoretical Physics, Trieste\\}
Lecture Videos: \href{https://www.youtube.com/watch?v=z6uo8wGBOxc}{Lecture I}, \href{https://www.youtube.com/watch?v=THW1QjaOjqs}{Lecture II}.  \\
\href{https://www.youtube.com/watch?v=WdFyyIuXCqk}{Mathematical Physics Webinar Video}, Rutgers University, January 14, 2026.
\end{center}
\begin{itemize}
\item Sections~\ref{sec:sdw}, \ref{sec:spinliquids} are mostly adapted from \href{https://www.cambridge.org/gb/universitypress/subjects/physics/condensed-matter-physics-nanoscience-and-mesoscopic-physics/quantum-phases-matter?format=HB}{\it Quantum Phases of Matter}, by Subir Sachdev, Cambridge University Press (2023).
\item  Sections~\ref{sec:fermions}, \ref{sec:halffilling}, \ref{sec:FLs} are mostly adapted from {\it Fractionalized Fermi liquids and the cuprate phase diagram\/}, Pietro M. Bonetti, Maine Christos, Alexander Nikolaenko, Aavishkar A. Patel, and Subir Sachdev,  \href{https://doi.org/10.1088/1361-6633/ae530d}{Reports on Progress in Physics {\bf 89}, 044501 (2026)}.
\end{itemize}


\newpage
\vspace{10pt}
\noindent\rule{\textwidth}{1pt}
\tableofcontents
\noindent\rule{\textwidth}{1pt}
\vspace{10pt}


\section{Introduction}
\label{sec:intro}

These lectures begin  by presenting the conventional Hartree-Fock theory of antiferromagnetic order in the square lattice Hubbard model in Section~\ref{sec:sdw}. The state with antiferromagnetic order is generically a metal, with electron and/or hole pocket Fermi surfaces. However, precisely at half-filling, and {\it only if\/} the antiferromagnetic order is large enough, the Fermi surfaces disappear, and we obtain an insulating antiferromagnet. 

Sections~\ref{sec:spinliquids} and \ref{sec:fermions} present the theory of insulating quantum spin liquids in two spatial dimensions, involving {\it many-boson ({\it i.e.\/} spin) entanglement\/}.  These are states at half-filling which are insulating without any antiferromagnetic order or other broken symmetry. No such states appear in the theory of Section~\ref{sec:sdw}, or in other conventional approaches found in condensed matter texts published before the 1980s. 
We describe these states by the powerful parton method, which requires the presence of emergent gauge fields. 
Section~\ref{sec:spinliquids} expresses the electron spin operator in terms of bosonic partons, while Section~\ref{sec:fermions} employs fermionic partons.

The bosonic parton theory on triangular lattice leads to the theory of the $\mathbb{Z}_2$ spin liquid \cite{NRSS91,XGW91,SSkagome}. 
This is the simplest quantum spin liquid which can have time-reversal symmetry, and its spectrum of fractionalized `anyon' excitations is the same as that of the toric code \cite{KitaevToric}.

The bosonic parton theory on the square lattice leads to a U(1) gauge theory which maps at low energy to the $\mathbb{CP}^1$ field theory, along with additional Berry phases for Dirac monopole configurations in the U(1) gauge field \cite{NRSS89prl,NRSS90} (Section~\ref{sec:U1}).
This theory describes the transition from the antiferromagnetically ordered N\'eel state to a valence bond solid (VBS), with a gapless spin liquid regime between them.

The fermionic parton theory on the square lattice is discussed in Section~\ref{sec:fermions}. Its low energy theory is a SU(2) gauge theory with 2 flavors ($N_f=2$) of massless Dirac spinons \cite{AM88,Affleck-SU2,Fradkin88,LeeWen96}. Remarkably this SU(2) gauge theory has a duality mapping \cite{Wang17} to the U(1) gauge theory of bosonic partons in Section~\ref{sec:U1}.

The remaining sections involve entangled states with mobile {\it many-fermion entanglement\/}. The greater complexity of many-fermion entanglemet has also been explored in some recent mathematical papers \cite{Lake19,King25}.

Section~\ref{sec:holonmetal} moves the bosonic parton theory of Section~\ref{sec:spinliquids} away from half-filling. This leads a `holon metal' state, which is relevant for the Hubbard model on the Lieb lattice, as realized in ultracold atom experiments \cite{LiebScience,LiebPRB}. But holons cannot tunnel coherently between square lattice layers, and so we argue that ADMR experiments \cite{Ramshaw22,Yamaji24} imply that the holon metal cannot describe the cuprate pseudogap metal.

Section~\ref{sec:halffilling} examines charge fluctuations in the fermionic parton theory of Section~\ref{sec:spinliquids} at half-filling. 
As the charger gap is reduced, we obtain a $d$-wave superconductor and also a variety of competing charge-order states. But the $d$-wave superconductor has nearly isotropic velocities for the Bogoliubov quasiparticles, in contrast to observations \cite{Chiao00}.

The Fractionalized Fermi liquid (FL*) state \cite{TSSSMV03,TSSSMV04} is introduced in Section~\ref{sec:FLs}. The theory of FL* in a single-band Hubbard-type model is developed, and argued to resolve the problems with applications to the cuprates noted above. The FL* state does not obey the conventional Luttinger constraint on the size of the Fermi surface: in Section~\ref{sec:gLutt}, we discuss generalized Luttinger constraints which are obeyed by FL* and other fractionalized states. The half-filled charge fluctuation theory of Section~\ref{sec:halffilling} is extended to non-zero doping in this framework.

The concluding Section~\ref{sec:conc} discusses the implication of the FL* theory of the pseudogap metal for the rest of the cuprate phase diagram.

\section{Spin density wave order in the Hubbard model}
\label{sec:sdw}

We consider the onset of magnetism at a non-zero wavevector in a metal, often called a spin density wave (SDW).
We will focus on the case where the wavevector of the SDW is ${\bm K} = (\pi, \pi)$ on the square lattice, and so the ordering has the same symmetry as the N\'eel state in an insulating antiferromagnet. The main ingredient here will be a bosonic collective mode representing antiferromagnetic spin fluctuations in the metal: this boson is the `paramagnon'. \index{paramagnon}

Near the transition from the Fermi liquid to the antiferromagnetic metal, it is possible to derive a systematic approach to the paramagnon modes of a metal. We begin with an electronic Hubbard model
\beq
H = \sum_{{\bm k}, \alpha} \varepsilon_{\bm k} c_{{\bm k} \alpha}^\dagger c_{{\bm k} \alpha} + U \sum_\vi n_{\vi \uparrow} n_{\vi \downarrow}
\label{dwave1}
\eeq
where $n_{\vi\uparrow} \equiv c_{\vi \uparrow}^\dagger c_{\vi \uparrow}$, and similarly for $n_{\vi \downarrow}$.
Upon using the single-site identity
\begin{equation}
U \left(n_{\vi\uparrow} - \frac{1}{2} \right) \left(n_{\vi\downarrow} - \frac{1}{2} \right) = -\frac{2U}{3} {\bm S}_{\vi}^{2} + \frac{U}{4} \,,
\end{equation}
(which is easily established from the electron commutation relations) it becomes possible to decouple the 4-fermion term in a particle-hole channel. We decouple the interaction term in the Hubbard model in (\ref{dwave1}), by the Hubbard-Stratonovich transformation
\begin{equation} 
\exp \left( \frac{2U}{3} \sum_{\vi} \int d \tau {\bm S}^{2}_{\vi} \right) = \int \mathcal{D} \vP_{\vi} (\tau) \exp
\left( - \sum_{\vi} \int d \tau \left[ \frac{3U}{8} \vP^{2}_{\vi} - U \,\vP_{\vi} \cdot c_{\vi \alpha}^\dagger \frac{{\bm \sigma}_{\alpha\beta}}{2} c_{\vi \beta} \right] \right) 
\label{UPhi}
\end{equation}
We now have a new field $\vP_\vi (\tau)$ which will play the role of the paramagnon field. 

\begin{figure}
\begin{center}
\includegraphics[height=3cm]{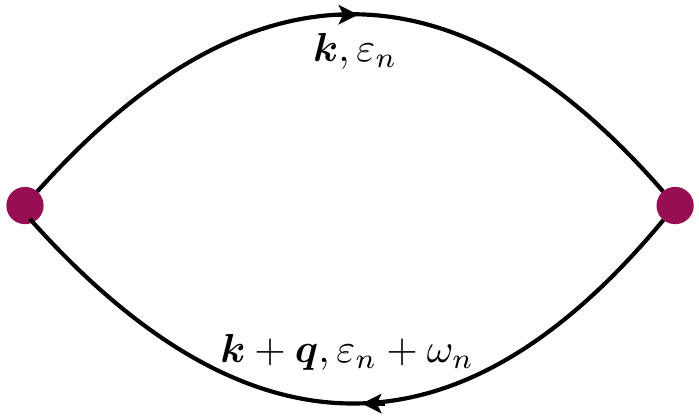}
\end{center}
\caption{Feynman diagram leading to (\ref{spara3}).}
\label{fig:lindhard}
\end{figure}

The path integral of the Hubbard model can now be written exactly as:
\bea
\mathcal{Z} &=& \int \mathcal{D} c_{\vi \alpha} (\tau) \mathcal{D} \vP_{\vi} (\tau) \exp \Biggl( -  \int d \tau \Biggl\{ \sum_{{\bm k},\alpha} c_{{\bm k}\alpha}^\dagger \left[ \frac{\partial}{\partial \tau} + \varepsilon_{{\bm k}} \right] c_{{\bm k} \alpha}   \nonumber \\
&~&~~~~~~~~~+ \sum_{\vi}\left[ \frac{3U}{8} \vP^{2}_{\vi} - U \, \vP_{\vi} \cdot c_{\vi \alpha}^\dagger \frac{{\bm \sigma}_{\alpha\beta}}{2} c_{\vi \beta} \right] \Biggr\} \Biggr)\,. \label{ZHubbard}
\eea
We can now formally integrate out the electrons, and obtain 
\beq
\frac{\mathcal{Z}}{\mathcal{Z}_0} = \int \prod_{\vi} \mathcal{D} \vP_{\vi} (\tau)  \exp \Biggl( - \mathcal{S}_{\rm paramagnon} \left[ \vP_{\vi} (\tau) \right] \Biggr)\,, \label{spara1}
\eeq
where $\mathcal{Z}_0$ is the free electron partition function.
Close to the onset of SDW order (but still on the non-magnetic side), we can expand the action in powers of $\vP$
\beq
\mathcal{S}_{\rm paramagnon} \left[ \vP_{\vi} (\tau) \right] = \frac{T}{2} \sum_{{\bm q}, \omega_n} |\vP({\bm q}, \omega_n) |^2 
U \left[ \frac{3}{4} - \frac{U \chi_0 ({\bm q}, i\omega_n)}{2} \right] + \ldots \label{spara2}
\eeq
where $\chi_0 ({\bm q}, \omega_n)$ is the frequency-dependent Lindhard susceptibility, given by the particle-hole bubble graph shown in Fig.~\ref{fig:lindhard}
\beq
\chi_0 ({\bm q}, i\omega_n) = -\frac{T}{V} \sum_{{\bm p}, \epsilon_n} \frac{1}{( i \epsilon_n - \varepsilon_{{\bm k}})
( i \epsilon_n + i \omega_n - \varepsilon_{{\bm k}+{\bm q}})} \label{spara3}
\eeq
Performing the sum over frequencies by partial fractions, we obtain
\beq
\chi_0 ({\bm q}, i \omega_n) = \frac{1}{V} \sum_{{\bm k}} \frac{f(\varepsilon_{{\bm k} + {\bm q}}) - f(\varepsilon_{{\bm k}})}{i \omega_n + \varepsilon_{{\bm k}}- \varepsilon_{{\bm k} + {\bm q}}}\,, \label{spara4}
\eeq

From the structure of the $\vP$ propagator, it is clear that $\vP$ will first condense at the wavevector ${\bm q}_{\rm max}$ at which $\chi_0 ({\bm q}, i\omega=0)$ is a maximum, and ${\bm q}_{\rm max}$ is then the wavevector of the SDW. In the mean field treatment of (\ref{spara2}), the appearance of the this condensate requires that $U$ is large enough to obey the \index{Stoner criterion} `Stoner criterion':
\beq
\frac{3}{4} - \frac{U \chi_0 ({\bm q}_{\rm max}, i\omega=0)}{2} < 0 \,. \label{stoner}
\eeq
This wavevector is in turn determined by the dispersion $\varepsilon_{\bm k}$ of the underlying fermions. For simplicitly, we will only consider the case of a SDW with wavevector ${\bm K} = (\pi, \pi)$.
The frequency dependence of $\chi_0 ({\bm q}, i\omega)$ also has an important influence on the dynamics of the paramagnon fluctuations. 

\subsection{Fermi surface reconstruction}
\label{sec:fsreconstruction}

\begin{figure}
\centering
\includegraphics[width=2.3in]{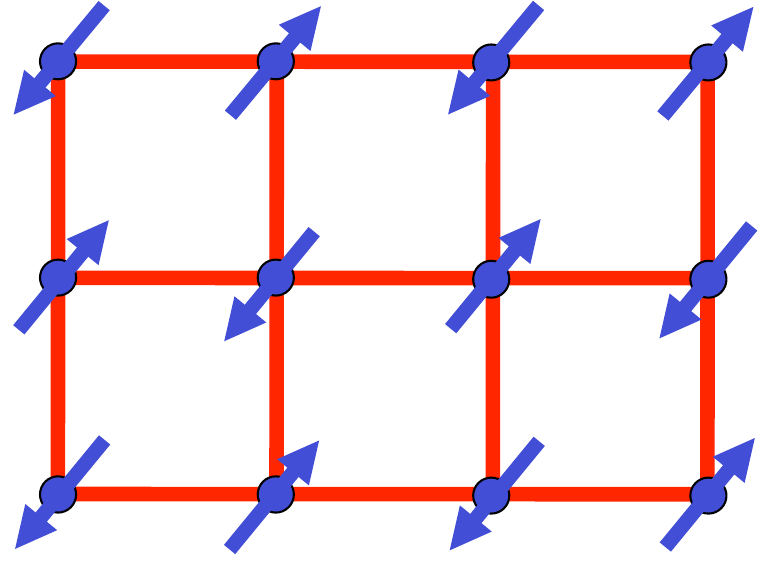}
\caption{Antiferromagnetic (N\'eel) order $\mathcal{N}$ of the spin density wave state. This can be an insulator at $p=0$, provided $\Delta = \mathcal{N}$ is large enough, as shown in Fig.~\ref{fig:sdw0}. Otherwise, it is a metal with electron and/or hole pocket Fermi surfaces, as shown in Fig.~\ref{fig:sdw0}-\ref{fig:sdw2}.}
\label{fig:Neel}
\end{figure}

Let us now move into the antiferromagnetic metal phase, where we assume there is a $\vP$ condensate at wavevector ${\bm K} = (\pi, \pi)$
\beq
\left\langle \vP_{\vi} \right\rangle = \eta_{\vi} \,\mathcal{N} \hat{\bm z}\,, \label{Phicondensate}
\eeq
where the factor 
\beq 
\eta_{\vi} = \pm 1 \label{defeta}
\eeq 
on the two checkerboard sublattices of the square lattice, and $\mathcal{N}$ measuring the strength of the N\'eel ordered moment shown in Fig.~\ref{fig:Neel}. 

We wish to describe the excitations of this state. One class of excitations are spin waves: these can be obtained by considering transverse fluctuations of $\vP$ about the condensate in (\ref{Phicondensate}) using the full action in (\ref{spara1}). However, there are also low energy fermionic excitations in the antiferromagnetic metal, which are gapped in the insulator. We can determine the spectrum of the fermions by inserting (\ref{Phicondensate}) into the Yukawa coupling;
using $\eta_{\vi} = e^{i {\bm K} \cdot {\bm r}_{\vi}}$, with ${\bm K} = (\pi, \pi)$, we can write the fermion Hamiltonian in momentum space
\beq
H_{\rm AFM} = \sum_{{\bm k}} \left[ \varepsilon_{{\bm k}} c_{{\bm k} \alpha}^\dagger c_{{\bm k} \alpha} -
\Delta  \, c_{{\bm k} \alpha}^\dagger \sigma^z_{\alpha\alpha} c_{{\bm k} + {\bm K}, \alpha}
\right] + \mbox{constant}.
\label{HAFM}
\eeq
The is the analog of the BCS Hamiltonian for superconductivity, and the analog of the pairing gap is the energy 
\beq
\Delta = \mathcal{N}\,.
\eeq
But, in general, the spectrum of $H_{\rm AFM}$ does not have a gap, as we will see below.
As in BCS theory, the value of $\mathcal{N}$ has to be determined self-consistently from the mean-field equations. \index{Fermi surface reconstruction}
\index{spin density wave}

\begin{figure}
\begin{center}
\includegraphics[width=3in]{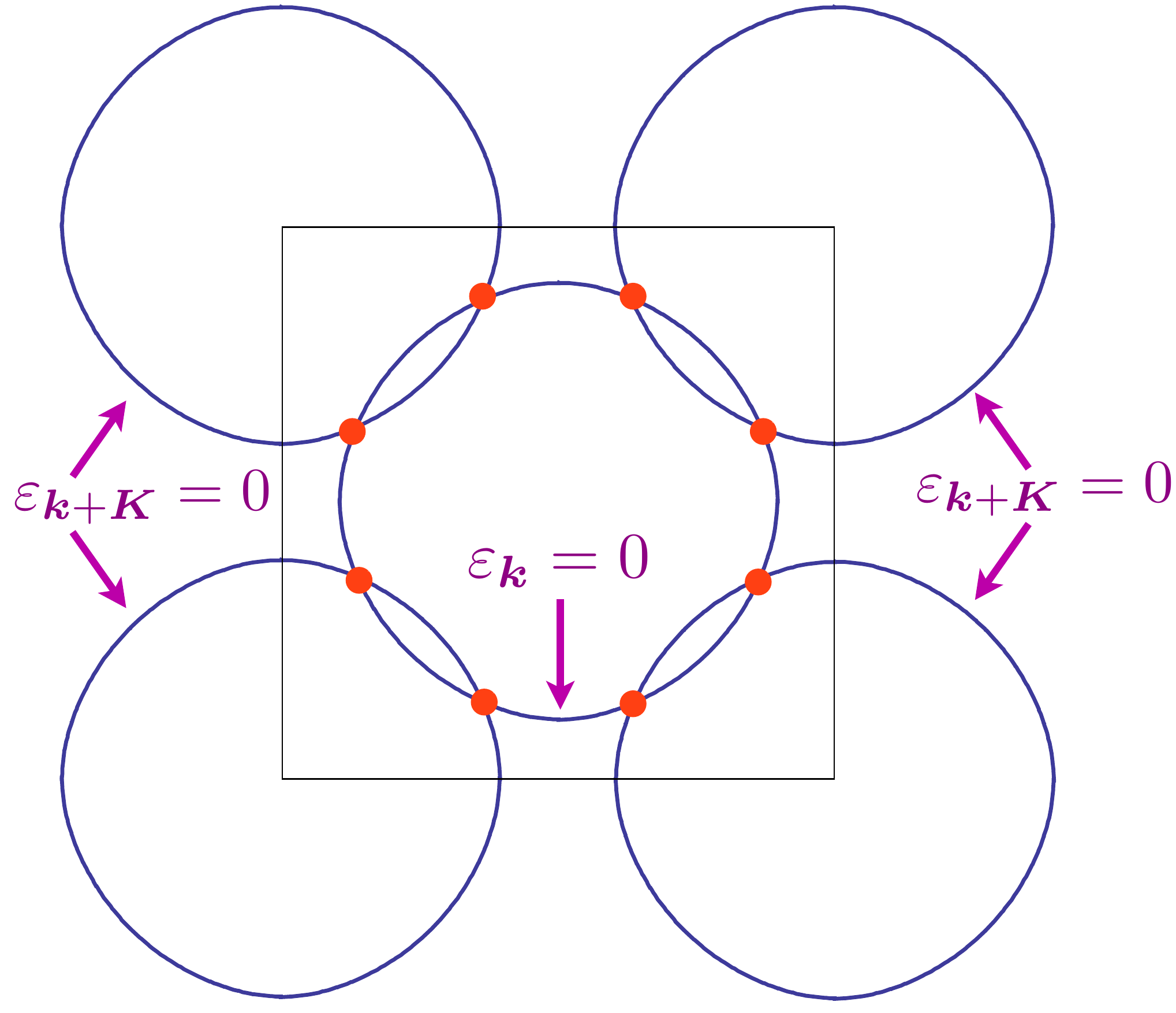}
\end{center}
\caption{Fermi surface of the original dispersion $\varepsilon_{\bm k}$, and those obtained by scattering off the antiferromagnetic order by a momentum ${\bm K}$. The intersections of the Fermi surfaces are the hotspots, where the antiferromagnetism opens a gap. In this figure only, the center is the momentum ${\bm K}$, around which the large hole-like Fermi surface is centered in the cuprates.}
\label{fig:hotspots}
\end{figure}
\begin{figure}
\begin{center}
\includegraphics[width=4.5in]{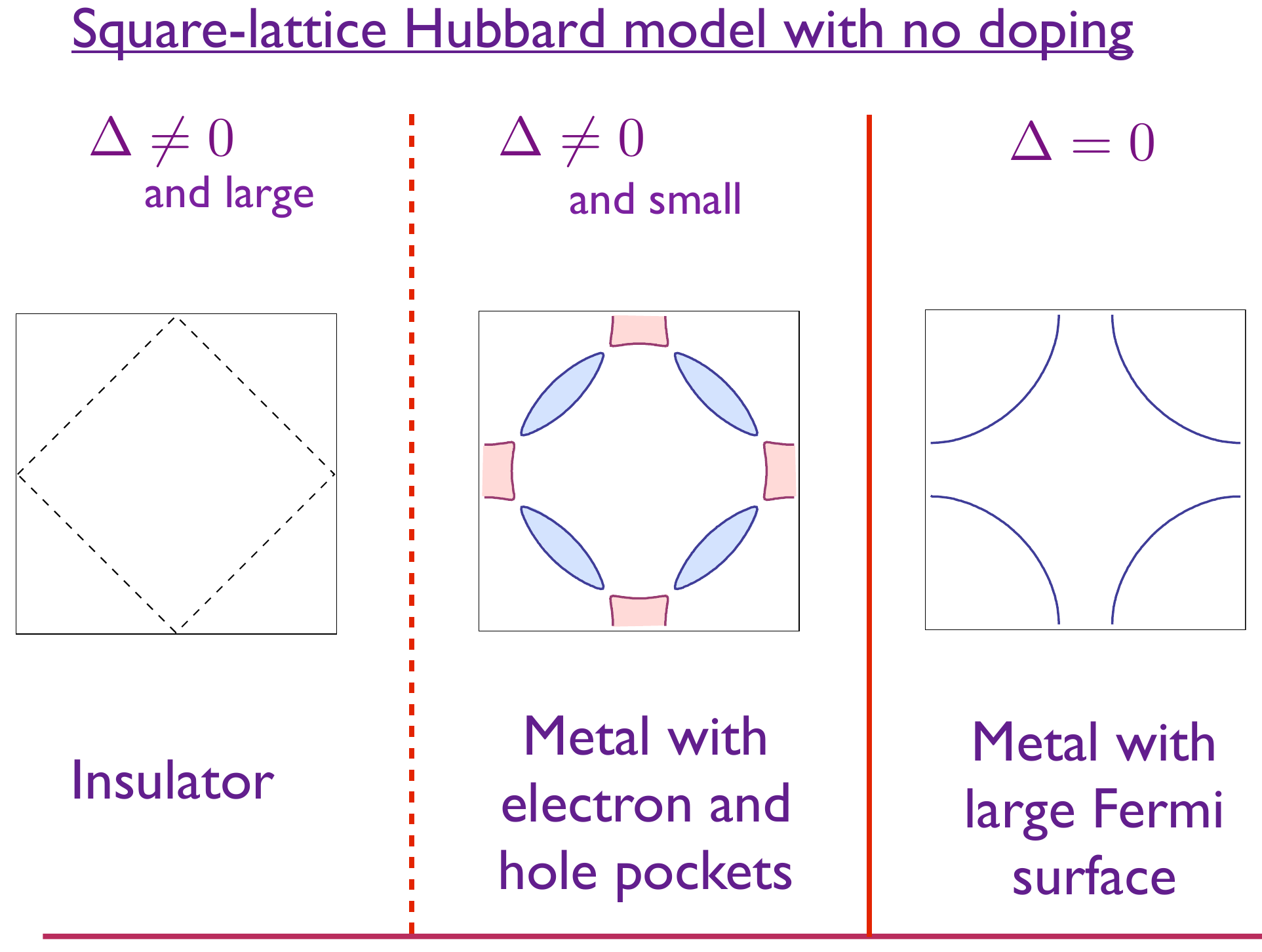}
\end{center}
\caption{Fermi surfaces of the N\'eel state at half-filling, {\it i.e.\/} doping $p=0$. The pockets intersecting the diagonals of the Brillouin zone have both bands in (\ref{bcs52}) empty and so form hole pockets, while the remaining pockets have both bands occupied and form electron pockets. The dashed line in the insulator shows the boundary of the Brillouin zone of the N\'eel state.}
\label{fig:sdw0}
\end{figure}
\begin{figure}
\begin{center}
\includegraphics[width=4.5in]{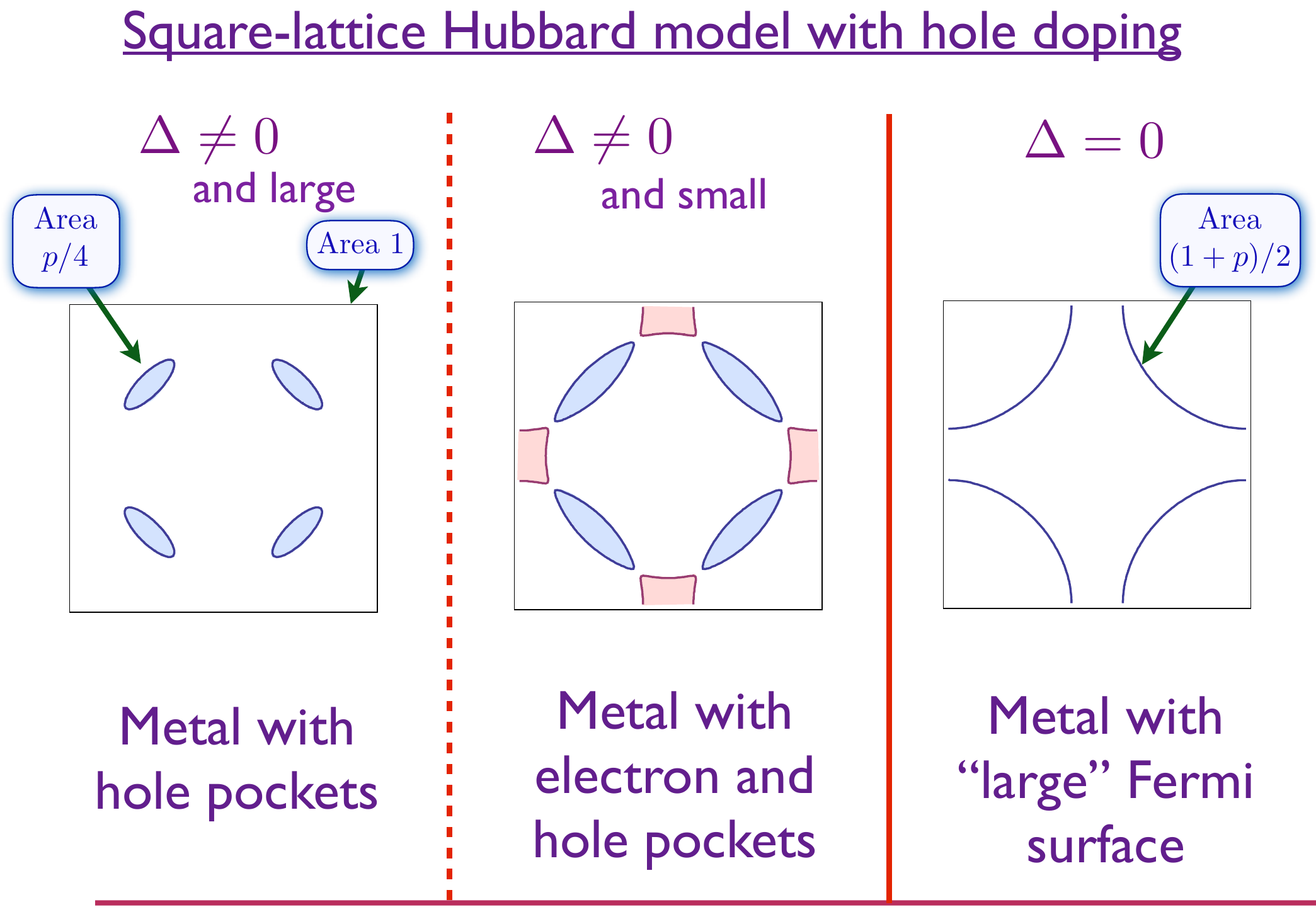}
\end{center}
\caption{Fermi surfaces of the N\'eel state at $p>0$. The pockets are as in Fig.~\ref{fig:sdw0}. From (\ref{eq:sdwLuttinger}), the area of each hole pocket on the left (when $\Delta$ is large and there are no electron pockets) is $p/4$, in units with the square lattice Brillouin zone having unit area. This follows from the existence of 2 independent hole pockets in the magnetic Brillouin zone, each with a spin degeneracy of 2. On the right is the large hole-like Fermi surface centered at $(\pi, \pi)$ with area $(1+p)/2$. }
\label{fig:sdw1}
\end{figure}
\begin{figure}
\begin{center}
\includegraphics[width=4.5in]{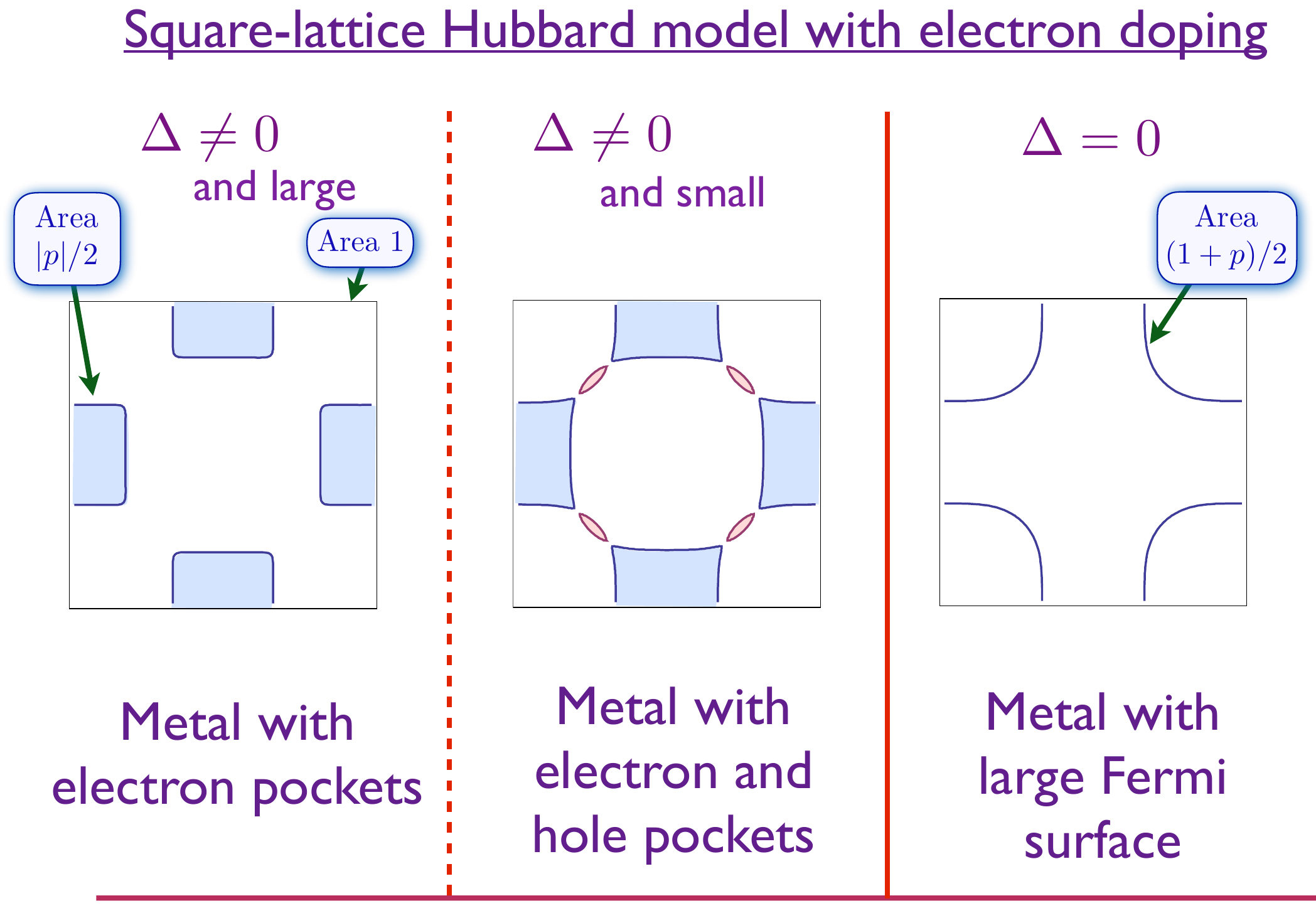}
\end{center}
\caption{Fermi surfaces of the N\'eel state at $p<0$, with pockets as in Figs.~\ref{fig:sdw0}. From (\ref{eq:sdwLuttinger}), the area of each electron pocket on the left (when $\Delta$ is large and there are no hole pockets) is $|p|/2$, in units with the square lattice Brillouin zone having unit area. This follows from the existence of 1 independent electron pocket in the magnetic Brillouin zone with a spin degeneracy of 2.}
\label{fig:sdw2}
\end{figure}

To obtain the fermionic excitation spectrum, we have to perform the analog of the Bogoliubov rotation in BCS theory. This is achieved by writing $H_{\rm AFM}$ in a $2\times 2$ matrix form by using the fact that $2 {\bm K}$ is a reciprocal lattice vector, and so $\varepsilon_{{\bm k} + 2 {\bm K} } = \varepsilon_{{\bm k}}$;
correspondingly, the prime over the summation indicates that it only extends over half the Brillouin zone of the underlying lattice, shown in the left panel of Fig.~\ref{fig:sdw0}, which is the Brillouin zone of the lattice with N\'eel order. 
\beq
H_{\rm AFM} = \sum_{{\bm k}}^{\prime} ( c_{{\bm k} \alpha}^\dagger, c_{{\bm k} + {\bm K},\alpha}^\dagger ) 
\left( \begin{array}{cc}
\varepsilon_{\bm k} & - \Delta \sigma^z_{\alpha\alpha} \\
-\Delta \sigma^z_{\alpha\alpha} & \varepsilon_{{\bm k} + {\bm K}}
\end{array} \right)
\left( \begin{array}{c}
c_{{\bm k} \alpha} \\ c_{{\bm k} + {\bm K},\alpha}
\end{array} \right)\,. \label{dwave51}
\eeq
It is now easy to diagonalize the $2 \times 2$ matrix in (\ref{dwave51}), and we obtain 
\beq
E_{{\bm k}\pm} = \frac{ \varepsilon_{\bm k} + \varepsilon_{{\bm k} + {\bm K}}}{2} \pm \left[ \left(\frac{ \varepsilon_{\bm k} - \varepsilon_{{\bm k} + {\bm K}}}{2} \right)^2 + \Delta^2 \right]^{1/2} \label{bcs52}
\eeq
The spectrum in (\ref{bcs52}) is not gapped, or even positive definite. Rather, it is the spectrum of a metal, in which the negative energy states are filled, and bounded by a Fermi surface. For small $\Delta$, the structure of the Fermi surface can be deduced from the pictorial argument in Fig.~\ref{fig:hotspots}: gaps open at the `hotspots' where both $\varepsilon_{\bm k}$ and $\varepsilon_{{\bm k} + {\bm K}}$ vanish, reconstructing the large Fermi surface into small pockets.
The Fermi surfaces so obtained is shown in Figs.~\ref{fig:sdw0}, \ref{fig:sdw1}, \ref{fig:sdw2} for different values of the electron density $1-p$. Here $p$ is conventionally the hole doping, and electron doping corresponds to $p<0$. The dispersion $\varepsilon_{\bm k}$ has hoppings $t_{1,2,3}$ to first, second, and third neighbors with $t_{1,3}>0$ and $t_2<0$ as appropriate for the cuprates.

We observe that the `large' Fermi surface of the paramagnetic metal has `reconstructed' into small pocket Fermi surfaces in the SDW state.
The excitations of the SDW metal are hole-like quasiparticles on the Fermi surfaces surrounding the hole pockets, and electron-like quasiparticles on the Fermi surfaces surrounding the electron pockets. The spin wave excitations interact rather weakly with the fermionic quasiparticle excitations: this can be see from a somewhat involved computation from the effective action. 

Finally, we discuss the fate of the Luttinger relation in this metal. \index{Luttinger relation} The Luttinger relation 
connects the volume enclosed by the Fermi surface to the density of electrons, modulo 2 electrons per unit cell. It
should be applied in the Brillouin zone of the N\'eel state, which is half the size of the Brillouin zone of the underlying square lattice, as shown in Fig.~\ref{fig:sdw0}. In real space, this corresponds to the fact that the unit cell has doubled, and so the density of electrons per unit cell is $2(1-p)$. For spinful electrons, the Luttinger relations measures electron density modulo 2, and so the density appearing in the Luttinger relation is $-2 p$. This has to be equated to twice the volumes enclosed by the electron and hole pockets within the diamond shaped Brillouin zone in Fig.~\ref{fig:sdw0}. Let $\mathcal{A}_h$ be the area of a single elliptical hole pocket: there are 4 such pockets in the complete Brillouin zone of the square lattice or 2 pockets in the Brillouin zone of the N\'eel state, as is apparent from Figs.~\ref{fig:sdw0}, \ref{fig:sdw1}, \ref{fig:sdw2}. Similarly, let  $\mathcal{A}_e$ be the area of a single elliptical electron pocket: there are 2 such pockets in the complete Brillouin zone of the square lattice or 1 pocket in the Brillouin zone of the N\'eel state. These arguments show that the Luttinger relation becomes
\beq
2 \times \frac{1}{(2 \pi)^2/2} \times \left(-2 \mathcal{A}_h + \mathcal{A}_e \right) = - 2 p \,. \label{eq:sdwLuttinger}
\eeq
On the left hand side, the first factor is the spin degeneracy, and the second factor is the inverse of the volume of the Brillouin zone of the N\'eel state. To reiterate, this is the conventional Luttinger relation applied after accounting for the doubling of the unit cell, and it determines a linear constraint on the areas of the electron and hole pockets. 

\begin{figure}
\begin{center}
\includegraphics[width=6.5in]{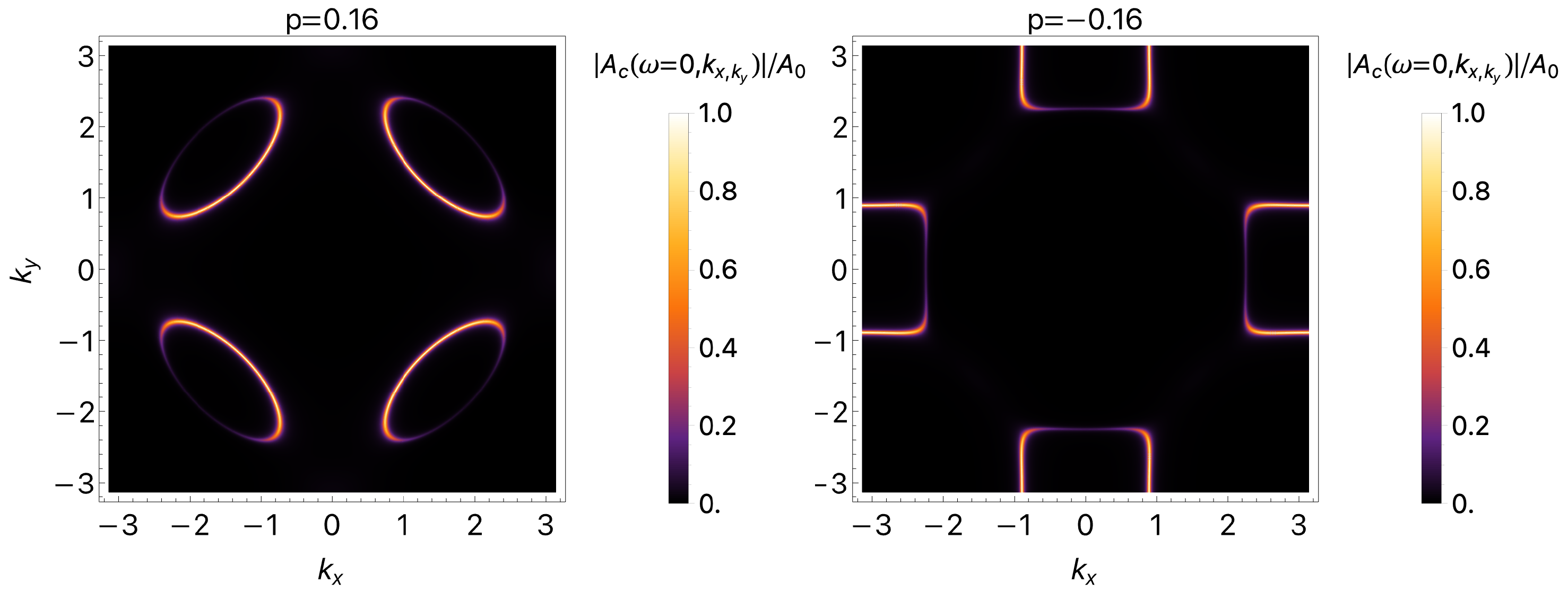}
\end{center}
\caption{Spectral density of hole (left) and electron (right) pockets at $p = 0.16$ and $p=-0.16$ respectively for the SDW state. The fractional area of each hole pocket is $p/4$, and the fractional area of the electron pocket is $|p|/2$. Figure by A. Nikolaenko.}
\label{fig:sdwphoto}
\end{figure}
Fig~\ref{fig:sdwphoto} shows the electron spectral density (as measured by photoemission) at zero frequency for both the electron and hole pockets. This is computed by taking the imaginary part of the Green's function $(\omega - H_{\rm AFM}+ i \eta)^{-1}$ of the matrix Hamiltonian in Eq.~(\ref{dwave51}), where $\eta$ is a positive infinitesimal.


\section{Bosonic spinon theory of quantum spin liquids}
\label{sec:spinliquids}

Section~\ref{sec:sdw} has obtained an insulator at $p=0$ with N\'eel order $\mathcal{N}$. Decreasing $\mathcal{N}$ eventually leads to a metallic state with $\mathcal{N}=0$, even at $p=0$. But there was no insulator at $p=0$ with no broken symmetry, and no such insulator is possible within the conventional Hartree-Fock framework.
In this section we show that an insulator can indeed be obtained at $p=0$ without any broken symmetry, but only after introducing fractionalized spin excitations. 

As long as the charge gap is finite, we can perform a canonical transformation of the Hubbard model in Eq.~(\ref{dwave1}) to spin-only Heisenberg antiferromagnet
\begin{equation}
{\cal H}_J = \sum_{\vi < \vj} J_{\vi\vj} \,{\bm S}_\vi \cdot {\bm S}_\vj\,.
\label{hamil}
\end{equation}
The $J_{\vi\vj}$ are short-ranged antiferromagnetic
exchange interactions between $S=1/2$ spins ${\bm S}_\vi$ on sites $\vi$. We will  consider here the square and triangular lattices with nearest neighbor
interactions, but the methods generalize to a wide class of lattices and interaction ranges. Our analysis will be carried out for the nearest-neighbor Hamiltonian, and we will search for mean-field theories without any broken symmetries. In practice, we know that the nearest neighbor models on both the square and triangular lattices have antiferromagnetic order. But the strategy here is to understand the general structure of the spin liquid states, which can then be realized in more general Hamiltonians.

In this section, we employ a method which fractionalizes the spin operator into bosonic partons. On the square lattice, this leads to the low energy U(1) gauge theory with complex scalars in Eq.~(\ref{cp3}), and  to the phase diagram in Fig.~\ref{fig:Neelvbs}.  This phase diagram is in good agreement with numerical studies \cite{Becca20,Imada21,Gu24} of the square lattice antiferromagnet with first and second neighbor exchange (the $J_1$-$J_2$ model), and also of Sandvik's $J$-$Q$ model \cite{Sandvik24}.

Section~\ref{sec:fermions} fractionalizes the spin operator into fermionic partons. This leads ultimately to a seemingly different low energy theory on the square lattice: a SU(2) gauge theory with massless Dirac fermions in Eq.~(\ref{eq:fermionhop2}), which initially did not agree with numerical studies of square lattice antiferromagnets.
Wang {\it et al.\/} \cite{Wang17} argued that the bosonic and fermionic theories are equivalent, and that the confining phases of the SU(2) gauge theory lead to the same phase diagram as the bosonic partons. This equivalence is powerful, as it yields a toolbox of different approaches to study the square lattice spin liquid. In particular, upon including charge fluctuations, it is the fermionic parton theory that allows study of the confining instability of the square lattice spin liquid to $d$-wave superconductivity that we study in Sections~\ref{sec:halffilling} and \ref{sec:aniso}.

In the bosonic parton approach, we introduce the Schwinger
boson description \cite{AA88}, in terms of elementary $S=1/2$ bosons.
For the group SU(2) the complete set of $(2S+1)$
states on site $\vi$ are represented as follows
\begin{equation}
|S , m \rangle \equiv \frac{1}{\sqrt{(S+m)! (S-m)!}}
(b_{\vi\uparrow}^{\dagger})^{S+m}
(b_{\vi\downarrow}^{\dagger})^{S-m} | 0 \rangle,
\end{equation}
where $m = -S, \ldots S$ is the $z$ component of the spin ($2m$ is an integer).
We have introduced two flavors of Schwinger bosons on each site, \index{Schwinger boson}
created by the canonical operator
$b_{\vi\alpha}^{\dagger}$, with $\alpha = \uparrow, \downarrow$, and
$|0\rangle$ is the vacuum with no Schwinger bosons. The total number of Schwinger bosons, $n_b$,
is the same for all the states; therefore
\begin{equation}
b_{\vi\alpha}^{\dagger}b_{\vi \alpha} = n_b
\label{boseconst}
\end{equation}
with 
\beq
n_b = 2S\,.
\eeq
The above
representation of the states is completely equivalent to the 
operator identity between the spin and Schwinger boson operators
\begin{equation}
{\bm S}_{\vi} =  \frac{1}{2}
b_{\vi\alpha}^{\dagger}\, {\bm \sigma}_{\alpha\beta} \, b_{\vi\beta} \label{sparton}
\end{equation}
where $\ell=x,y,z$ and the ${\bm \sigma}$ are the usual $2\times 2$ Pauli
matrices. 

The spin-states on two sites $\vi,\vj$ can combine to form a singlet in a unique
manner - the wavefunction of the (unnormalized) singlet state is particularly simple
in the boson formulation:
\begin{equation}
\left( \varepsilon_{\alpha\beta} b_{\vi\alpha}^{\dagger}
b_{\vj\beta}^{\dagger} \right)^{2S} |0\rangle
\end{equation}
Also, using the constraint in Eq.~(\ref{boseconst}), the
following Fierz-type identity can be established
\begin{equation}
\left( \varepsilon_{\alpha \beta}
b_{\vi \alpha}^{\dagger} b_{\vj \beta}^{\dagger} \right)
\left( \varepsilon_{\gamma \delta}
b_{\vi\gamma} b_{\vj\delta} \right) =
 - 2 {\bm S}_\vi \cdot {\bm S}_\vj
+ n_b^2 /2 + \delta_{\vi\vj} n_b
\label{su2}
\end{equation}
where $\varepsilon$ is the totally antisymmetric $2\times2$ tensor
\begin{equation}
\varepsilon = \left( \begin{array}{cc}
0 & 1 \\
-1 & 0 \end{array} \right)\,. \label{defepsilon}
\end{equation}
This implies that ${\cal H}_J$ can be rewritten in the form (apart from
an additive constant)
\begin{equation}
{\cal H}_J = - \frac{1}{2} \sum_{\vi < \vj} J_{\vi\vj} \Bigl( \varepsilon_{\alpha \beta}
b_{\vi \alpha}^{\dagger} b_{\vj \beta}^{\dagger} \Bigr)
\Bigl( \varepsilon_{\gamma \delta}
b_{\vi\gamma} b_{\vj\delta} \Bigr)
\label{hafkag}
\end{equation}
This form makes it clear that ${\cal H}_J$ counts the number of singlet
bonds.

\subsection{Mean-field theory}
\label{sec:6A}

We begin by the coherent state path integral of ${\cal H}_J$ in imaginary time $\tau$ at a temperature $\beta = 1/T$
\begin{equation}
\mathcal{Z}_J = \int {\cal D} \mathcal{Q} {\cal D} b  {\cal D} \lambda
\exp \left( - \int_0^{\beta} {\cal L}_J \, d\tau \right) ,
\label{zfunct0}
\end{equation}
where
\begin{displaymath}
{\cal L}_J = \sum_{\vi} \left [
b_{\vi\alpha }^{\dagger}  \left( \frac{d}{d\tau} + i\lambda_\vi \right)
b_{\vi\alpha} - i\lambda_\vi n_b \right ]~~~~~~~~~~~~~~~~~~~~~~
\end{displaymath}
\begin{equation}
~~~~~~~~~~~+  \sum_{<\vi,\vj>} \left [
\frac{J_{\vi\vj} |\mathcal{Q}_{\vi,\vj} |^2}{2}
- \frac{J_{\vi\vj} \mathcal{Q}_{\vi,\vj}^{\ast}}{2} \varepsilon_{\alpha\beta} b_{\vi\alpha}
b_{\vj\beta}
+ H.c. \right] .
\label{zfunct}
\end{equation}
Here the $\lambda_\vi$
fix the boson number of $n_b$ at each site;
$\tau$-dependence of all fields is implicit; $\mathcal{Q}$ was introduced by
a Hubbard-Stratonovich decoupling of ${\cal H}_J$. In contrast to the decoupling in Eq.~(\ref{UPhi}), we have now decoupled in the spin-singlet channel as we are interested in states that preserve spin rotation invariance.

This procedure is similar to that employed in deriving the Landau-Ginzburg theory of superconductivity from electron pairing, with the crucial difference that now the Lagrangian
${\cal L}_J$ has a $U(1)$ gauge invariance associated with the local constraint in Eq.~(\ref{boseconst}):
\begin{eqnarray}
b_{\vi\alpha}^{\dagger} & \rightarrow & b_{\vi\alpha}^{\dagger} 
\exp \left( i\rho_\vi (\tau ) \right ) \nonumber \\
\mathcal{Q}_{\vi,\vj} &\rightarrow & \mathcal{Q}_{\vi,\vj} \exp \left( - i \rho_\vi (\tau ) - i
\rho_\vj (\tau ) \right) \nonumber \\
\lambda_\vi & \rightarrow & \lambda_\vi + \frac{\partial \rho_\vi}{\partial
\tau} (\tau) \,.
\label{gaugetrans}
\end{eqnarray}
The functional integral over ${\cal L}_J$
faithfully represents the partition function, but does require gauge fixing. This gauge invariance leads to emergent gauge field degrees of freedom, as we will see below.

We begin with mean-field saddle point of $\mathcal{Z}_J$ over the path integrals of $\mathcal{Q}$ and $\lambda$. The saddle-point approximation is valid in the limit of a large number of spin flavors, but we do not explore this here.
With the saddle point values $\mathcal{Q}_{\vi \vj} = \bar{\mathcal{Q}}_{\vi\vj}$, $i \lambda_\vi = \bar{\lambda}_\vi$ we obtain a 
mean-field Hamiltonian for the $b_{\vi \alpha}$
\begin{eqnarray}
{\cal H}_{J,MF} = &  \sum_{<\vi,\vj>} \left( \displaystyle
\frac{J_{\vi\vj} |\bar{\mathcal{Q}}_{\vi\vj} |^2}{2}
- \frac{J_{\vi\vj} \bar{\mathcal{Q}}_{\vi\vj}^{\ast}}{2} \varepsilon_{\alpha\beta} b_\vi^{\alpha}
b_\vj^{\beta}
+ H.c. \right) \nonumber \\
&
+ \sum_{\vi} \bar{\lambda}_\vi (
b_{\vi\alpha }^{\dagger} b_{\vi\alpha} -  n_b )\,.
\label{HMF}
\end{eqnarray}
This Hamiltonian is quadratic in the boson operators and all its
eigenvalues can be determined by a Bogoliubov transformation. This
leads in general to an expression of the form
\begin{equation}
{\cal H}_{J,MF} = E_{J,MF}[ \bar{\mathcal{Q}} , \bar{\lambda}] + \sum_{\mu} \omega_{\mu} [\bar{\mathcal{Q}} , \bar{\lambda}]
\gamma_{\mu\alpha}^{\dagger} \gamma_{\mu\alpha}
\end{equation}
The index $\mu$ extends over $1\ldots$number of sites in the system,
$E_{J,MF}$ is the ground state energy and is a functional of $\bar{\mathcal{Q}}$,
$\bar{\lambda}$, $\omega_{\mu}$ is the eigenspectrum of excitation energies
which is also a function of $\bar{\mathcal{Q}}$, $\bar{\lambda}$, and the
$\gamma_{\mu}^{\alpha}$ represent the bosonic eigenoperators. The
excitation spectrum thus consists of non-interacting spinor bosons.
The ground state is determined by minimizing $E_{J,MF}$ with respect to
the $\bar{\mathcal{Q}}_{\vi\vj}$ subject to the constraints
\begin{equation}
\frac{\partial E_{MF}}{\partial \bar{\lambda}_\vi } = 0
\label{sp1}
\end{equation}
The saddle-point value of the $\bar{\mathcal{Q}}$ satisfies
\begin{equation}
\bar{\mathcal{Q}}_{\vi\vj} = \langle \varepsilon_{\alpha\beta} b_{\vi\alpha} b_{\vj\beta}
\rangle
\label{sp2}
\end{equation}
Note that $\bar{\mathcal{Q}}_{\vi\vj} = - \bar{\mathcal{Q}}_{ji}$ indicating that $\bar{\mathcal{Q}}_{\vi\vj}$ is a
directed field - an orientation has to be chosen on every link.

\begin{figure}
\centering
\includegraphics[width=2.5in]{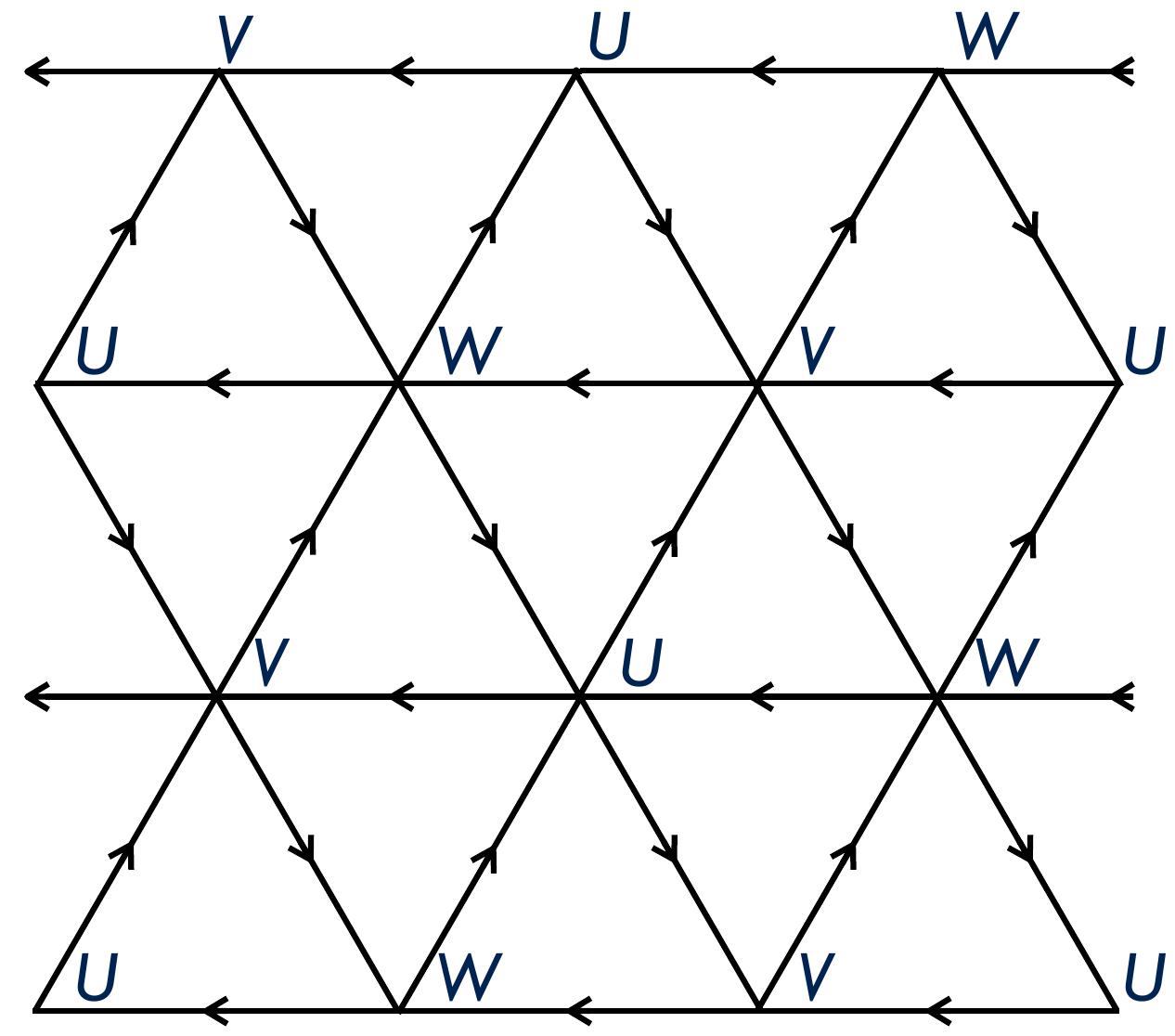}
\caption{Orientation of the nearest neighbor $\bar{\mathcal{Q}}_{\vi\vj}$ on the triangular lattice.
Also shown are the labels of the 3 sublattices.}
\label{fig:triangQ}
\end{figure}

These saddle-point equations have been solved for the square and triangular lattices with nearest neighbor exchange $J$, 
and they lead to stable 
and translationally invariant solutions for $\bar{\lambda}_{\vi}$ and $\bar{\mathcal{Q}}_{\vi\vj}$. The only saddle-point quantity which does not
have the full symmetry of the lattice is the orientation of the $\bar{\mathcal{Q}}_{\vi\vj}$.
Note that although it appears that such a choice of orientation appears to break inversion or reflection symmetries, such symmetries
are actually preserved: the $\bar{\mathcal{Q}}_{\vi\vj}$ are not gauge-invariant, and all gauge-invariant observables do preserve all symmetries of the 
underlying Hamiltonian. For the square lattice, we have $\bar{\lambda}_{\vi} = \bar{\lambda}$, $\bar{\mathcal{Q}}_{\vi,\vi+\hat{x}} = \bar{\mathcal{Q}}_{\vi,\vi+\hat{y}} = \bar{\mathcal{Q}}$.
Similarly, on the triangular lattice we have $\bar{\mathcal{Q}}_{\vi,\vi+\hat{e}_p} = \bar{\mathcal{Q}}$ for $p=1,2,3$, where the unit vectors 
\bea
\hat{e}_1 &=& (1/2, \sqrt{3}/2) \nonumber \\
\hat{e}_2 &=& (1/2, -\sqrt{3}/2) \nonumber \\
\hat{e}_3 &=& (-1, 0) 
\label{defei}
\eea
point between nearest neighbor sites of the triangular lattice. We sketch the orientation of the $\bar{\mathcal{Q}}_{\vi\vj}$ on the triangular lattice in 
Fig.~\ref{fig:triangQ}. 

\begin{figure}
\centering
\includegraphics[width=4in]{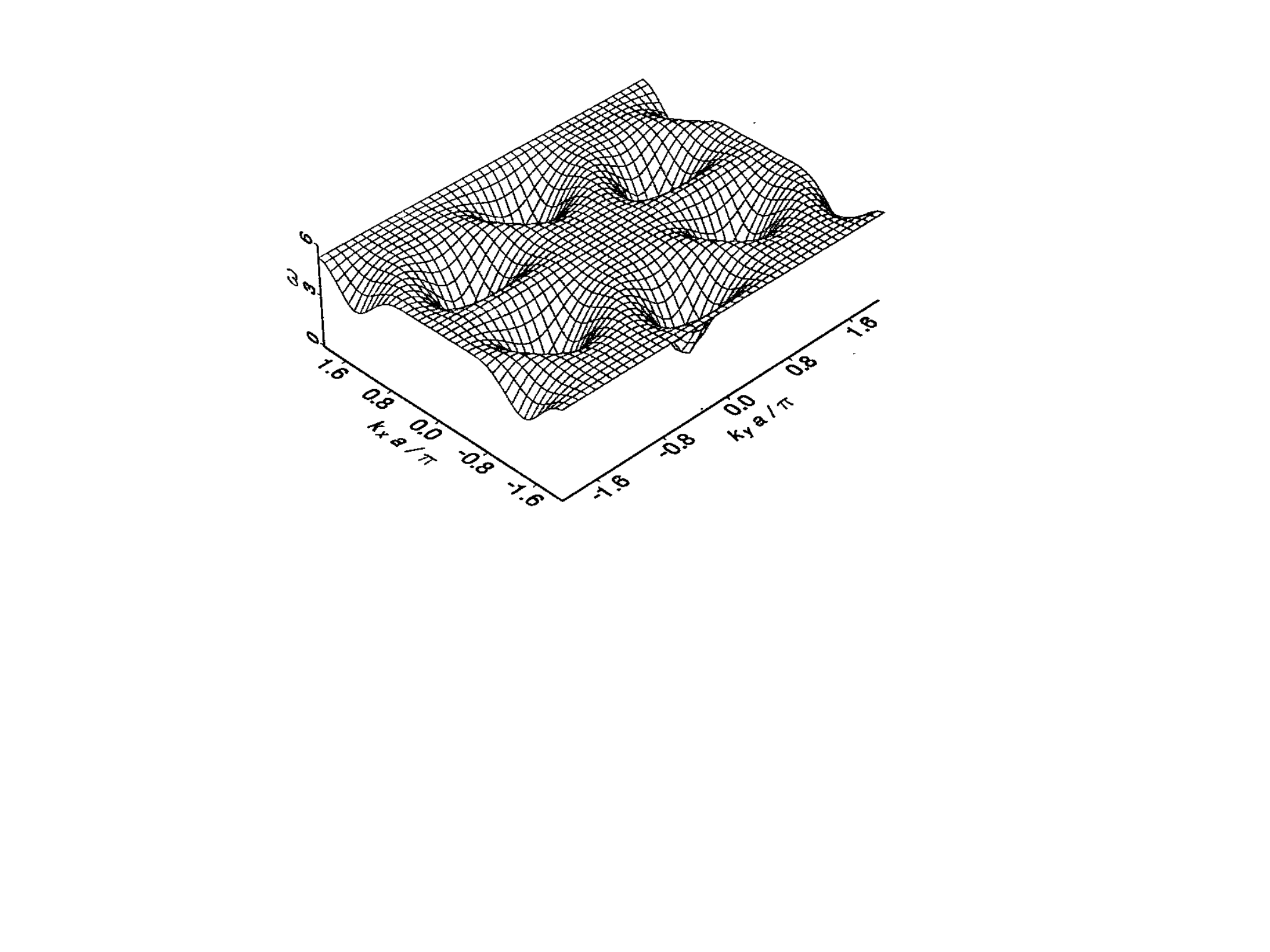}
\caption{Spinon dispersion on the triangular lattice \cite{SSkagome}. Reprinted with permission from APS.}
\label{fig:triangdisp}
\end{figure}

\begin{figure}
\centering
\includegraphics[width=4in]{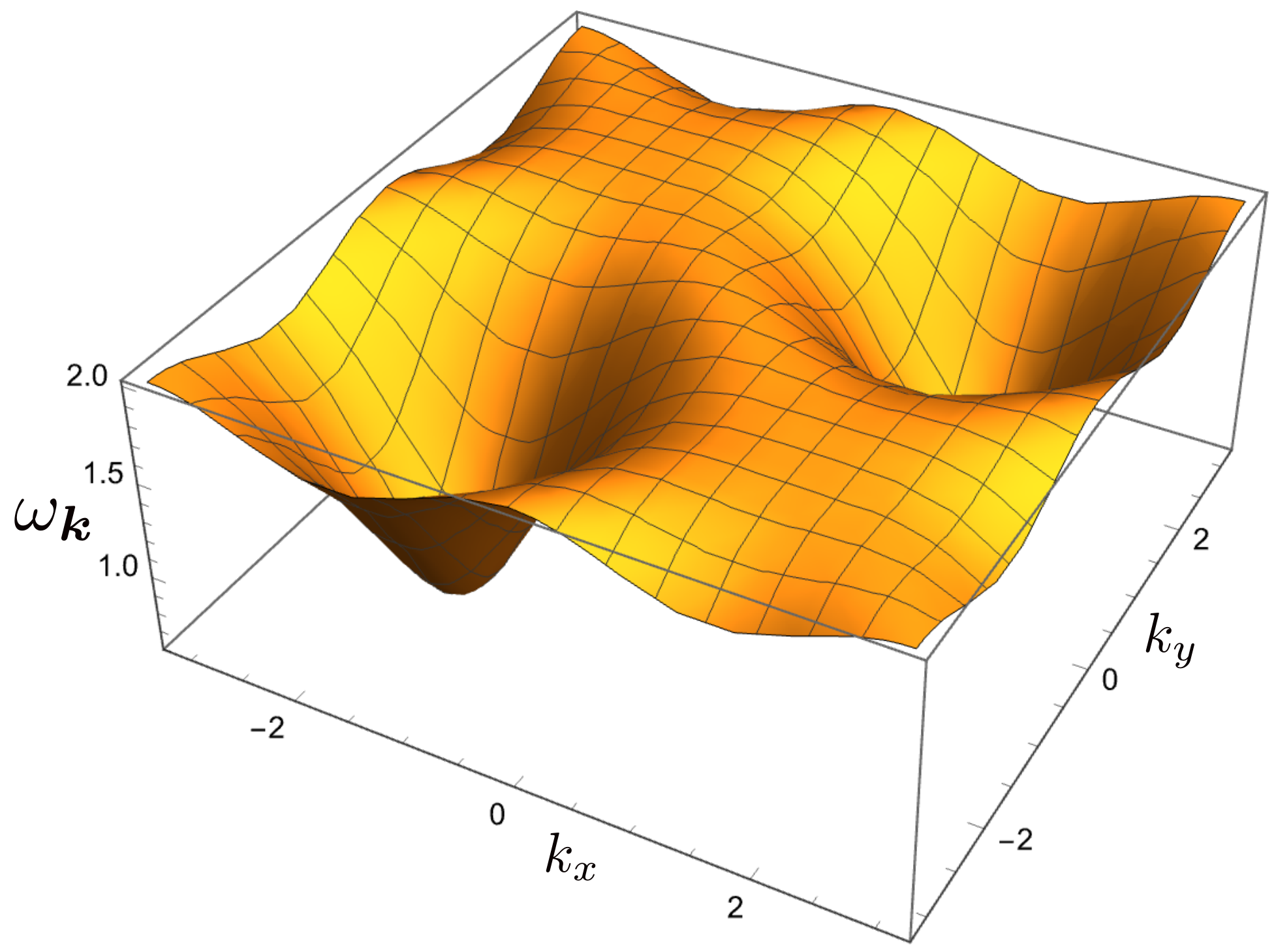}
\caption{Dispersion of bosonic spinons in a square lattice spin liquid, from Eq.~(\ref{eq:bosondisp}).}
\label{fig:bosondisp}
\end{figure}

We can also compute the dispersion $\omega_{\bm k}$ of the $\gamma_{\bm k}$ excitations. These are fractionalized bosonic particles which carry
spin $S=1/2$ and charge 0, named `spinons'. They should be contrasted from the spin-wave excitations of ordered magnets, which carry integer spin and charge 0.
Their dispersion on the square lattice is
\beq
\omega_{\bm k} = \left( \bar{\lambda}^2 - J^2 \bar{\mathcal{Q}}^2 (\sin k_x + \sin k_y)^2 \right)^{1/2} \label{eq:bosondisp}
\eeq
while that on the triangular lattice is \cite{SSkagome}
\beq
\omega_{\bm k} = \left( \bar{\lambda}^2 - J^2 \bar{\mathcal{Q}}^2 (\sin k_1 + \sin k_2 + \sin k_3)^2 \right)^{1/2}
\eeq
with $k_p = {\bm k} \cdot \hat{e}_p$. These are the spinons and the spinon dispersion on the triangular lattice is plotted in Fig.~\ref{fig:triangdisp}, and that on the square lattice in Fig.~\ref{fig:bosondisp}.

Notice that the spinons are gapped, and their minimum gaps are at two degenerate points in the Brillouin zone for both lattices. For the square lattice, the minima
are at ${\bm k} = \pm (\pi/2, \pi/2)$ with an energy gap of $\left( \bar{\lambda}^2 - 4 J^2 \bar{\mathcal{Q}}^2 \right)^{1/2}$. 
For the triangular lattice they are at ${\bm k} = \pm (4\pi/3,0)$ (and at wavevectors separated from 
these by reciprocal lattice vectors). So there are a total of 4 spinon excitations in both cases: 2 associated with the spin degeneracy of $S_z = \pm 1/2$, and 2 associated with the degeneracy in the Brillouin zone spectrum. 

Next, we turn to spin-singlet excitations, {\it i.e.\/} understanding the nature of the spectrum of the $\mathcal{Q}_{\vi\vj}$ and 
$\lambda_i$ fluctuations about the saddle point described above. At the outset, it appears that we can view such fluctuations as composites of 2-spinon
excitations, as both $\mathcal{Q}_{\vi\vj}$ and 
$\lambda_i$ couple to spinon pair operators, and so conclude that such excitations should not be viewed as the `elementary' excitations
of the quantum state found so far. Furthermore, the saddle-point has not broken any global symmetries of the Hamiltonian, and so it
would appear that no such composite excitations has any reason to be low energy without fine-tuning.

However, it does turn out that there are separate elementary excitations in the singlet sector, and these arise from two distinct causes:
({\em i\/}) the gauge invariance in (\ref{gaugetrans}) leads to a gapless ``photon'' excitation; ({\em ii\/}) there are topologically non-trivial
configurations of $\mathcal{Q}_{\vi\vj}$ which lead to excitations which appear as new saddle points.
Excitations in the class ({\em i\/}) arise in the square lattice case, while those in class ({\em ii\/}) appear on the triangular lattice,
and these will be considered separately in the following subsections.

\subsection{Gauge excitations}
\label{sec:gauge}

The gauge transformations in (\ref{gaugetrans}) act on the phases of the $\mathcal{Q}_{\vi\vj}$, and so it is appropriate to just focus
on the fluctuations of the phases of the $\mathcal{Q}_{\vi\vj}$ whose amplitudes are non-zero at the saddle-point. 

\subsubsection{Square lattice}
\label{sec:sqU1}

We define
\bea
\mathcal{Q}_{\vi, \vi + \hat{x}} &=& \bar{\mathcal{Q}} \exp \left(i \Theta_{\vi x} \right) \nonumber \\
\mathcal{Q}_{\vi, \vi + \hat{y}} &=& \bar{\mathcal{Q}} \exp \left(i \Theta_{\vi y} \right)
\label{gaugetrans1a}
\eea
Then, the gauge transformations in (\ref{gaugetrans}) can be written as
\begin{eqnarray}
\Theta_{\vi x} (\tau) &\rightarrow & \Theta_{\vi x} (\tau) - \rho_\vi (\tau ) - 
\rho_{\vi +x} (\tau ) \nonumber \\
\Theta_{\vi y} (\tau) &\rightarrow & \Theta_{\vi y} (\tau) - \rho_\vi (\tau ) - 
\rho_{\vi +y} (\tau ) \nonumber \\
\lambda_\vi & \rightarrow & \lambda_\vi + \frac{\partial \rho_\vi}{\partial
\tau} (\tau) .
\label{gaugetrans2}
\end{eqnarray}
The question before us is whether (\ref{gaugetrans}) imposes on us the presence of a gapless photon in the low energy
and long-wavelength limit. The answer is affirmative, and the needed result is obtained by parameterizing 
such fluctuations as follows
\bea
\Theta_{\vi x} (\tau) &=& \eta_{\vi} a_x ({\bm r}, \tau) \nonumber \\
\Theta_{\vi y} (\tau) &=& \eta_{\vi} a_y ({\bm r}, \tau) \nonumber \\
\lambda_{\vi} &=& - i \bar{\lambda} - \eta_{\vi} a_\tau ({\bm r}, \tau)
\label{gaugetrans3}
\eea
where the $a_\mu$ are assumed to be smooth functions of spacetime parameterized by the continuum spatial co-ordinate $r$,
and imaginary time $\tau$; the factor $\eta_\vi$, defined in Eq.~(\ref{defeta}),
has opposite signs on any pair of nearest-neighbor sites.
Then, taking the continuum limit of (\ref{gaugetrans2}) with $\rho_{\vi} (\tau) = \eta_{\vi} \rho(r, \tau)$, we deduce from
(\ref{gaugetrans3}) that 
\begin{eqnarray}
a_x &\rightarrow& a_x - \partial_x \rho \nn
a_y &\rightarrow& a_y - \partial_y \rho \nn
a_\tau &\rightarrow& a_\tau - \partial_\tau \rho 
\label{gaugetrans4}
\eea
So we reach the very important conclusion that $a_\mu$ transforms just like a continuum U(1) gauge field!
Note that the factor $\eta_{\vi}$ in this U(1) gauge transformation implies from (\ref{gaugetrans}) that the spinons $b_{\vi}$ carry opposite gauge charges on the two sublattices.

As in traditional field-theoretic analyses, (\ref{gaugetrans4}) imposes the requirement that the long-wavelength action of 
the $a_\mu$ fluctuations must have the form \index{photon} \index{emergent!photon}
\beq
\mathcal{S}_b = \int d^3 x \frac{1}{2 K'} (\epsilon_{\mu\nu\lambda} \partial_\nu a_{\lambda})^2,
\label{gaugetrans5}
\eeq
and this describes a gapless $a_\mu$ photon excitation, with a suitable velocity of `light'. 
So, on the square lattice, the spectrum of spin-singlet states includes
a linearly-dispersing photon mode. Such a state is a U(1) spin liquid.
Actually, the gapless photon of this U(1) spin liquid is ultimately not stable because of monopole tunneling events; this  
involves a long and interesting story \cite{NRSS89prl,NRSS90,senthil1,senthil2} which we will discuss briefly in Section~\ref{sec:U1}. 
For now, in the following subsection, we will consider the case of the triangular lattice, where, the U(1) photon is gapped by the Higgs mechanism to yield a $\mathbb{Z}_2$ spin liquid.

\subsubsection{Triangular lattice}
\label{sec:tri}

Now we have to consider 3 separate values of $\mathcal{Q}_{\vi\vj}$ per site, and so we replace (\ref{gaugetrans1a}) by
\begin{equation}
\mathcal{Q}_{\vi ,i+\hat{e}_p} = \bar{\mathcal{Q}}
\exp\left( i \Theta_{p,\vi} \right)
\label{qtheta}
\end{equation}             
where $p=1,2,3$, the vectors $\hat{e}_p$ were defined (\ref{defei}),
$\bar{\mathcal{Q}}$ is the mean-field value, and $\Theta_p$ is a real phase.      
The effective action for the $\Theta_{p,\vi}$ must be invariant under
\begin{equation}
\Theta_{p,\vi} \rightarrow \Theta_{p,\vi} - \rho_{\vi} - \rho_{\vi +\hat{e}_p}.
\end{equation}                                              
Upon performing a Fourier transform, with the link variables $\Theta_p$
placed on the center of the links, the gauge invariance takes the
form
\begin{equation}
\Theta_p ( {\bm k} ) \rightarrow \Theta_p ( {\bm k} ) - 2 \rho ({\bm k} ) \cos( k_p /2)
\label{thetatr}
\end{equation}               
The momentum ${\bm k}$ takes values in the first Brillouin zone of the
triangular lattice. This invariance implies that the effective action
for the $\Theta_p$ can only be a function of the following gauge-invariant
combinations:
\begin{equation}
I_{pq} ({\bm k}) = 2 \cos(k_q /2) \Theta_p ({\bm k} ) -
2 \cos (k_p /2) \Theta_q ({\bm k} )
\label{Ipq}
\end{equation}

We now wish to take the continuum limit at points in the Brillouin zone
where the action involves only gradients of the $\Theta_p$ fields and
thus has the possibility of gapless excitations. The same analysis could have been applied
to the square lattice, in which case there is only one invariant $I_{xy}$. In this case, we choose
${\bm k }  = {\bm g} + {\bm q}$, with ${\bm g} = (\pi, \pi)$ (this corresponds to the choice of $\eta_{\vi}$ above)
and ${\bm q}$ small; then $I_{xy} = q_x \Theta_y - q_y \Theta_x$ which is clearly the U(1) flux
invariant under (\ref{gaugetrans4}).

The situation is more complex for the case of the triangular lattice \cite{SSkagome}. Now there are
3 independent $I_{pq}$ invariants, and it  is not
difficult to see that only two of the three values of $\cos (k_p /2)$
can vanish at any point of the Brillouin zone. One such point is
the wavevector
\begin{equation}
{\bm g} = \frac{2\pi}{\sqrt{3} a} (0, 1)
\label{bmg}
\end{equation}
where
\begin{eqnarray}
{\bm g} \cdot \hat{e}_1 & = & \pi \nonumber \\
{\bm g} \cdot \hat{e}_2 & = & - \pi \nonumber \\
{\bm g} \cdot \hat{e}_3 & = & 0.
\end{eqnarray}  
Taking the continuum limit with the fields varying with momenta with
close to ${\bm g}$ we find that the $I_{pq}$ depend only upon 
gradients of $\Theta_1$ and $\Theta_2$. 
It is also helpful to parametrize the $\Theta_p$ in the following
suggestive manner (analogous to (\ref{gaugetrans3}))
\begin{eqnarray}
\Theta_1 ( {\bm r} ) & = & i a_1 ( {\bm r} ) e^{i {\bm g} \cdot {\bm r} } \nonumber \\
\Theta_2 ( {\bm r} ) & = & - i a_2 ( {\bm r} ) e^{i {\bm g} \cdot {\bm r} } \nonumber \\
\Theta_3 ( {\bm r} ) & = & H ( {\bm r} ) e^{i {\bm g} \cdot {\bm r} } 
\end{eqnarray}                                      
It can be verified that the condition for the reality of $\Theta_p$
is equivalent to demanding that $a_1, a_2, H$ be real.
We will now take the continuum limit with $a_1, a_2, H$ varying
slowly on the scale of the lattice spacing. It is then not difficult to
show that the invariants $I_{pq}$ then reduce to (after a Fourier 
transformation):
\begin{eqnarray}
I_{12} & = & \partial_2 a_1 - 
\partial_1 a_2 \nonumber \\
I_{31} & = & \partial_1 H - 2 a_1 \nonumber \\
I_{32} & = & \partial_2 H - 2 a_2 ,
\end{eqnarray}
where $\partial_{\vi}$ is the spatial gradient along the direction $\hat{e}_{\vi}$.
Thus the $a_1, a_2$ are the components of a U(1) gauge field,
with the components are taken along an `oblique' co-ordinate
system defined by the axes $\hat{e}_1 , \hat{e}_2$; this is just as in the square lattice.
However, in addition to $I_{12}$, we also have the invariants $I_{31}$ and $I_{32}$
in the triangular lattice; we observe that this involves the field $H$ which transforms like 
the phase of charge $\pm 2$ Higgs field under the U(1) gauge invariance.
So the fluctuations of an isotropic triangular lattice will be characterized by an action of the form
\beq
\mathcal{S}_b = \int d^3 x \frac{1}{2 K'} \left[ I_{12}^2 + I_{31}^2 + I_{32}^2 \right],
\label{gaugetrans5a}
\eeq
which replaces (\ref{gaugetrans5}). This is the action expected in the 
{\em Higgs phase} of a U(1) gauge theory. The Higgs condensate gaps out the
U(1) photon, and so there are no gapless singlet excitations on the triangular lattice.
This is a necessary condition for mapping the present state onto a $\mathbb{Z}_2$ 
spin liquid (the reason for this nomenclature will become clearer in the following subsection).

The presentation so far of the gauge fluctuations described by a charge $\pm 2$ Higgs field
coupled to a U(1) gauge field would be appropriate for an anisotropic triangular lattice
in which the couplings along the $\hat{e}_3$ direction are different from those along
$\hat{e}_1$ and $\hat{e}_2$. For an isotropic triangular lattice, all three directions
must be treated equivalently, and then there is no simple way to take the continuum
limit in the gauge sector: we have to work with the action in (\ref{gaugetrans5a}),
but with the invariants specified as in (\ref{Ipq}). Such an action does not
have a gapless photon anywhere in the Brillouin zone, and all gauge excitations remain
gapped. There are other choices for the wavevector ${\bm g}$ in (\ref{bmg}) at which
the other pairs of values of $\cos ( k_p /2) $ vanish; these are the points
\beq
 \frac{2\pi}{\sqrt{3} } 
\left(\frac{\sqrt{3}}{2}, -\frac{1}{2} \right) \quad , \quad
\frac{2\pi}{\sqrt{3} } 
\left(\frac{-\sqrt{3}}{2}, -\frac{1}{2} \right),
\eeq
which are related to the analysis above by the rotational symmetry of the triangular lattice.

\subsection{Topological excitations on the triangular lattice}
\label{sec:topo}

The analysis in Section~\ref{sec:gauge} described small fluctuations in the phases of the $\mathcal{Q}_{\vi\vj}$ about
their saddle-point values $\bar{\mathcal{Q}}$. On the triangular lattice, we found that such fluctuations
led only to gapped excitations, which at higher energies become part of the two-spinon continuum.

Now we consider excitations which involve large deviations from the spatially uniform saddle point values,
and which turn out to be topologically protected. These excitations are closely connected to the vortices in charged superfluids. Consider a scalar field $\Psi$ with charge $q$ coupled to the electromagnetic U(1) gauge field ${\bm A}$. This has stable vortex-like saddle points with flux $n \Phi_0$, with $\Phi_0 = hc/q$, for all integer $n$. We have seen above that the on the triangular lattice, the Schwinger boson state has fluctuations described by a charge 2 Higgs field $H$ coupled to a U(1) gauge field $a_\mu$. In this case, we are normalizing the gauge field so that $\hbar c \Rightarrow 1$, and so we can expect vortex solutions with $a_\mu$ flux $n (2\pi)/2$, for all integer $n$. However, this is not quite correct. A crucial difference between the present theory and the electromagnetic gauge field is that the $a_\mu$ gauge field is `compact': this mean that a gauge field $a_{x,y}$ is identical to $a_{x,y} + 2 \pi$, and tunneling events which change the total flux by $2 \pi$ are allowed (these are `monopoles', which will be considered further in Section~\ref{sec:U1}). This means that all vortex solutions with even $n$ are identical to each other, as are those with odd $n$. The $n=0$ case corresponds to no vortex at all, and so there is only a single non-trivial vortex with $n=1$ and flux $\pi$. This is the sought-after {\it vison\/}. \index{vison} Note that because fluxes $\pi$ and $-\pi$ are identical, the vison saddle point preserves time-reversal 
symmetry. Similar excitations appear in a $\mathbb{Z}_2$ gauge theory, and hence the $\mathbb{Z}_2$ spin liquid nomenclature.

In this section, we will obtain the vison saddle point solution by working with a lattice effective action: this is essential to account for the influence of the monopoles. 
So we look for spatially non-uniform solutions of the saddle-point
equations (\ref{sp1}) and (\ref{sp2}). In general, solving such equations is a demanding numerical task, and so we will be satisfied
with a simplified analysis which is valid when the spin gap is large.  In the large spin gap limit, we can integrate out the Schwinger bosons,
and write the energy as a local functional of the ${\mathcal{Q}}_{\vi\vj}$. This functional is strongly constrained by 
the gauge transformations in (\ref{gaugetrans}): for time-independent
${\mathcal{Q}}_{\vi\vj}$, this functional takes the form
\beq
E[\{{\mathcal{Q}}_{\vi\vj}\}] = - \sum_{\vi <j} \left( \alpha |{\mathcal{Q}}_{\vi\vj}|^2 + \frac{\beta}{2} |{\mathcal{Q}}_{\vi\vj} |^4 \right)  - K \sum_{{\rm even\, loops}} {\mathcal{Q}}_{\vi\vj} {\mathcal{Q}}_{jk}^\ast 
\ldots {\mathcal{Q}}_{\ell i}^\ast
\label{energyfunc}
\eeq
Here $\alpha$, $\beta$, and $K$ are coupling constants determined by the parameters in the Hamiltonian of the antiferromagnet.
We have shown them to be site-independent, 
because we have only displayed terms in which all links/loops are equivalent; they can depend upon links/loops
for longer range couplings 
provided 
the full lattice symmetry is preserved.

We can now search for saddle points of the energy functional in (\ref{energyfunc}). 
Far from the center of the vison, we have $|\mathcal{Q}^{v}_{\vi\vj}| = \bar{\mathcal{Q}}$,  so that the energy differs from the ground
state energy only by a finite amount.
Closer to the center there
are differences in the magnitudes. However, the key difference is in the signs of the  link variables, as illustrated in Fig.~\ref{fig:vison_tri}:
there is a `branch-cut' emerging from the vison core along which $\mbox{sgn}(\mathcal{Q}^{v}_{\vi\vj}) = - \mbox{sgn}(\bar{\mathcal{Q}}_{\vi\vj})$. 
The results of a numerical minimization \cite{HPS11}
of $E[ \{ {\mathcal{Q}}_{\vi\vj} \}]$ on the the triangular lattice are shown in Fig.~\ref{fig:vison_tri}. 
\begin{figure}[ht]
\begin{center}
\includegraphics[width=4.0in]{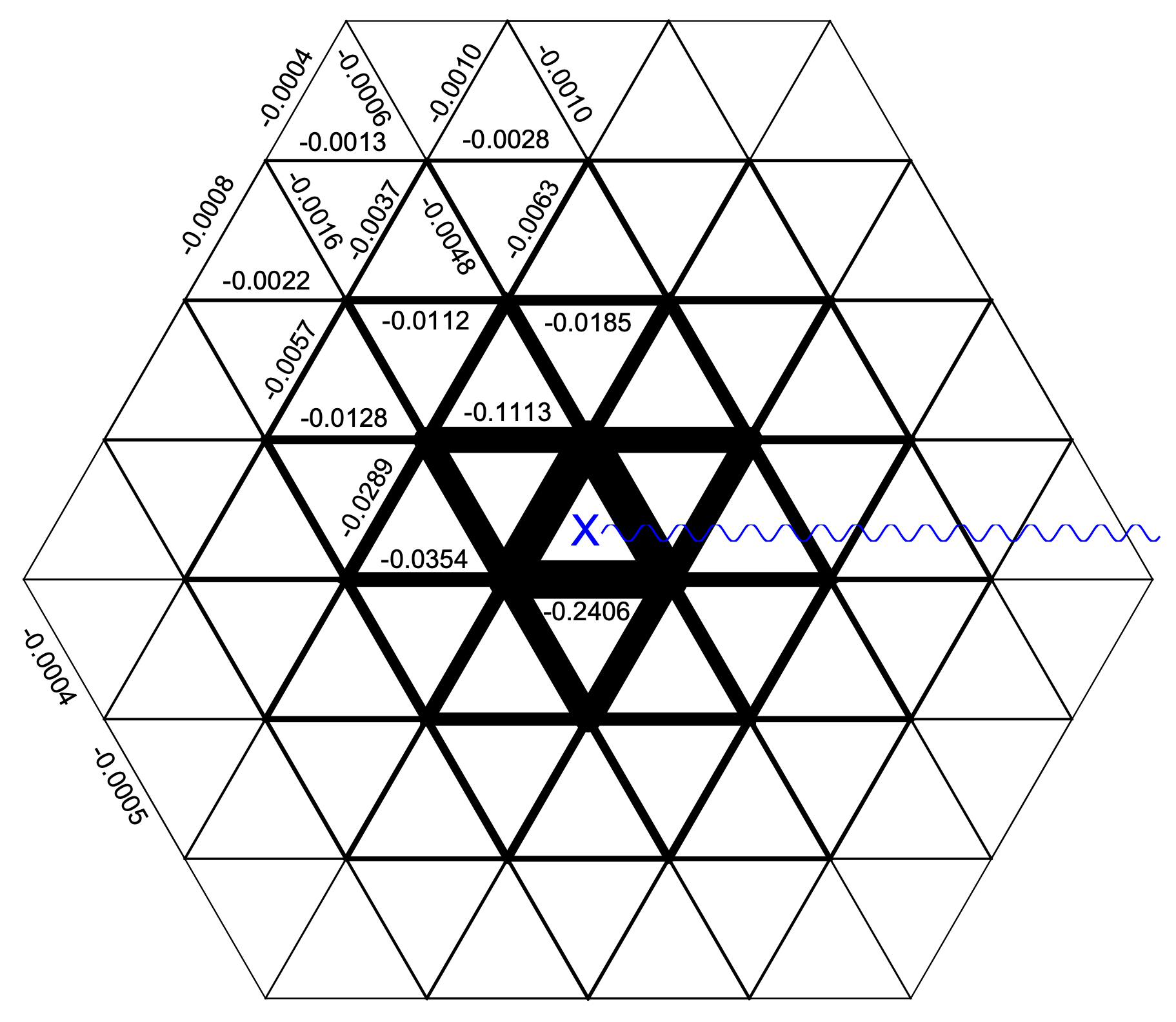}
\end{center}
\caption{A vison on the triangular lattice \cite{HPS11}. The center of the vison is marked by the X. The wavy line is the `branch-cut' where we have $\mbox{sgn}(\mathcal{Q}^{v}_{\vi\vj}) = - \mbox{sgn}(\bar{\mathcal{Q}}_{\vi\vj})$ only on the links crossed by the line. Plotted is the minimization result of $E[ \{ \bar{\mathcal{Q}}_{\vi\vj} \}]$ with $\alpha=1, \beta=-2, K=0.5$. Minimization is done with the cluster embedded in a vison-free lattice with all nearest neighbor links equal to $\bar{\mathcal{Q}}_{\vi\vj}$. The numbers are $(\bar{\mathcal{Q}}_{\vi\vj}-\mathcal{Q}^{v}_{\vi\vj})$ and the thickness of the links are proportional to $(\mathcal{Q}^{v}_{\vi\vj}-\bar{\mathcal{Q}}_{\vi\vj})^{1/2}$. Reprinted with permission from APS.}  
\label{fig:vison_tri}
\end{figure}
The magnitudes of $\mathcal{Q}^v_{\vi\vj}$ are suppressed close to the vison, and converge to $\bar{\mathcal{Q}}_{\vi\vj}$ as we move away from the vison (modulo the sign change associated with the branch cut), analogous to those in Abrikosov vortices. Despite the branch-cut breaking the 3-fold rotation symmetry, 
the gauge-invariant fluxes of $\mathcal{Q}^v_{\vi\vj}$ preserve the rotation symmetry. 

So we have found a stable real-vortex solution which preserves time-reversal, and has a finite excitation energy. We have also
anticipated that this vortex will be identified with the vison particle of the $\mathbb{Z}_2$ spin liquid: more evidence for this identification will appear in Sections~\ref{sec:piflux} and \ref{sec:semions}.

\subsection{Dynamics of excitations on the triangular lattice}

For the case of the triangular lattice, Sections~\ref{sec:6A} and \ref{sec:topo} have identified two types of elementary excitations:
bosonic spinons with a 2-fold spin and a 2-fold lattice degeneracy, and a topological excitation which we have anticipated will
become with vison particle of a $\mathbb{Z}_2$ spin liquid. We will now describe the dynamics of the interactions between these
excitations, and indeed verify that they reproduce the general structure associated with the $\mathbb{Z}_2$ spin liquid. 

A similar analysis can also be carried for the U(1) spin liquid on the square lattice. However, defer consideration of this case to Section~\ref{sec:U1}.

\subsubsection{Motion of spinons}
\label{sec:trispinons}

The general structure of the theory controlling the low energy spectrum becomes clearer upon taking
a suitable continuum limit of the Lagrangian in (\ref{zfunct}), while replacing $\mathcal{Q}_{\vi\vj} = \bar{\mathcal{Q}}_{\vi\vj}$
and $i \lambda_{\vi} = \bar{\lambda}$.
We take the continuum limit after separating 3 sites, $u$, $v$, $w$, in each unit cell (see Fig.~\ref{fig:triangQ}). 
We write the boson operators on these
sites as $b_{u}^{\alpha} = u_\alpha$, $b_{v}^{\alpha} = v_\alpha$ etc. Then to the needed order in spatial gradients,
the Lagrangian density becomes \cite{HFS10}
\begin{eqnarray}\label{eq:lagra}
\mathcal{L} &=& u^\ast_\alpha \frac{\partial u_\alpha}{\partial \tau} + v^\ast_\alpha \frac{\partial v_\alpha}{\partial \tau} + w^\ast_\alpha \frac{\partial w_\alpha}{\partial \tau} + \bar{\lambda} \left( |u_\alpha|^2 + |v_\alpha|^2 + |w_\alpha|^2
\right) \nonumber \\ &~&~~- \frac{3 J \bar{\mathcal{Q}}}{2} \mathcal{J}_{\alpha\beta} \left( u_\alpha v_{\beta} + v_\alpha w_{\beta} 
+ w_\alpha u_{\beta} \right) + \mbox{c.c.}  \nonumber \\ &+&
\frac{3J \bar{\mathcal{Q}}}{8} \mathcal{J}_{\alpha\beta} \left( {\bm \nabla}u_\alpha \cdot {\bm \nabla} v_{\beta} +  
{\bm \nabla} v_\alpha \cdot {\bm \nabla} w_{\beta} + {\bm \nabla}w_\alpha \cdot {\bm \nabla} u_{\beta} \right] + \mbox{c.c.}
\end{eqnarray}
We now perform a unitary transformation to new variables $x_\alpha$, $y_\alpha$, $z_\alpha$. These
are chosen to diagonalize only the non-gradient terms in $\mathcal{L}$. 
\begin{eqnarray}
 \left( \begin{array}{c} u_\alpha \\ v_\alpha \\ w_\alpha  \end{array} \right) &=& \frac{z_\alpha}{\sqrt{6}} \left( \begin{array}{c}
1 \\ \zeta \\ \zeta^2  \end{array} \right)
 + \mathcal{J}_{\alpha\beta}\frac{z_{\beta}^\ast}{\sqrt{6}}  \left( \begin{array}{c}
-i \\ -i \zeta^2 \\  -i \zeta  \end{array} \right) + \frac{y_\alpha}{\sqrt{6}} \left( \begin{array}{c}
1 \\ \zeta \\ \zeta^2 \end{array} \right)
 + \mathcal{J}_{\alpha\beta}\frac{y_{\beta}^\ast}{\sqrt{6}}  \left( \begin{array}{c}
i \\ i \zeta^2 \\  i \zeta  \end{array} \right) \nonumber \\
&~&~~~~~+ \frac{x_\alpha}{\sqrt{3}} \left( \begin{array}{c} 1 \\ 1 \\ 1 \end{array} \right) .
 \label{umat}
\end{eqnarray}
where $\zeta \equiv e^{2 \pi i /3}$. The tensor structure above makes it clear
that this transformation is rotationally invariant, and that $x_\alpha$, $y_\alpha$, $z_\alpha$ 
transform as spinors under SU(2) spin rotations. 
Inserting Eq.~(\ref{umat}) into $\mathcal{L}$ we find
\begin{eqnarray}
\mathcal{L} &=& x_\alpha^\ast \frac{\partial x_\alpha}{\partial \tau} + y_\alpha^\ast \frac{\partial z_\alpha}{\partial \tau} + z_\alpha^\ast \frac{\partial y_\alpha}{\partial \tau} + 
 (\bar\lambda - 3\sqrt{3} J \bar{\mathcal{Q}}/2) |z_\alpha |^2 \label{lz1} \\
&+&  (\bar\lambda + 3\sqrt{3} J \bar{\mathcal{Q}}/2)  |y_\alpha|^2 + 
\bar\lambda |x_\alpha |^2 + \frac{3J \bar{\mathcal{Q}} \sqrt{3}}{8} \left( |\partial_x z_\alpha |^2 + |\partial_y z_\alpha |^2 \right) + \ldots
\nonumber
\end{eqnarray}
The ellipses indicate omitted terms involving 
spatial gradients in the $x_\alpha$ and $y_\alpha$ which we will not keep track of.
This is because the fields $y_\alpha$ and $x_\alpha$ are massive relative to $z_\alpha$, 
and so can be integrated out. This yields the effective Lagrangian \index{spinon}
\begin{eqnarray}
\mathcal{L}_z &=& \frac{1}{(\bar \lambda + 3\sqrt{3}J \bar{\mathcal{Q}}/2) } |\partial_\tau z_\alpha |^2 + 
\frac{3J \bar{\mathcal{Q}} \sqrt{3}}{8} \left( |\partial_x z_\alpha |^2 + |\partial_y z_\alpha |^2 \right) \nonumber \\
&~&~~~~~~
+ (\bar \lambda - 3\sqrt{3}J \bar{\mathcal{Q}}/2) |z_\alpha |^2 + \ldots \label{lhz}
\end{eqnarray}
Note that the omitted spatial gradient terms in $x_\alpha$, $y_\alpha$ do contribute a correction to the
spatial gradient term in (\ref{lhz}), and we have not accounted for this. 

So we reach the important conclusion that the spinons are described by a relativistic complex scalar field $z_\alpha$.
Counting the two values of $\alpha$, and the particle and anti-pariticle excitations, we have a total
of 4 spinons, as expected.

Next, we consider the higher
order terms in (\ref{lhz}), which will arise from including the fluctuations of the gapped fields
$\mathcal{Q}$ and $\lambda$. Rather than computing these from the microscopic Lagrangian, it is more efficient
to deduce their structure from symmetry considerations. The representation in (\ref{umat}), and the 
connection of the $u_\alpha$, $v_\alpha$, $w_\alpha$ to the lattice degrees of freedom, allow us to deduce the
following symmetry transformations of the $x_\alpha$, $y_\alpha$, $z_\alpha$:
\begin{itemize}
\item Under a global spin rotation by the SU(2) matrix $g_{\alpha\beta}$, we have $z_\alpha 
\rightarrow g_{\alpha\beta} z_{\beta}$, and similarly for $x_\alpha$, and $y_\alpha$.
\item Under a 120$^\circ$ lattice rotation, we have $u_\alpha \rightarrow v_\alpha$, 
$v_\alpha \rightarrow w_\alpha$, $w_\alpha \rightarrow u_\alpha$. From (\ref{umat}), we see
that this symmetry is realized by 
\begin{equation}
z_\alpha \rightarrow \zeta z_\alpha~,~y_\alpha \rightarrow \zeta y_\alpha~,~
x_\alpha \rightarrow x_\alpha . \label{trot}
\end{equation}
Note that this is distinct
from the SU(2) rotation because $\mbox{det} (\zeta) \neq 1$.
\item Finally, there is a crucial $\mathbb{Z}_2$ gauge symmetry, which is the remnant of the U(1) gauge symmetry in Eq.~(\ref{gaugetrans}). This is the transformation
\beq
z_{\vi \alpha} \rightarrow \mu_\vi z_{\vi \alpha} \label{z2gauge}
\eeq
where $\mu_\vi = \pm 1$ has an arbitrary site-dependence.
\end{itemize}
It is easy to verify that Eq.~(\ref{lz1}) is invariant under all the symmetry operations above.
These symmetry operators make it clear that the only allowed quartic term for the Heisenberg Hamiltonian
is $\left( \sum_\alpha |z_\alpha|^2 \right)^2$: this quartic term added to $\mathcal{L}_z$ yields a theory with
O(4) symmetry, corresponding to rotations between the 4 real fields that can be extracted from the 2 complex field $z_\alpha$.  
\begin{figure}
\begin{center}
\includegraphics[width=4.5in]{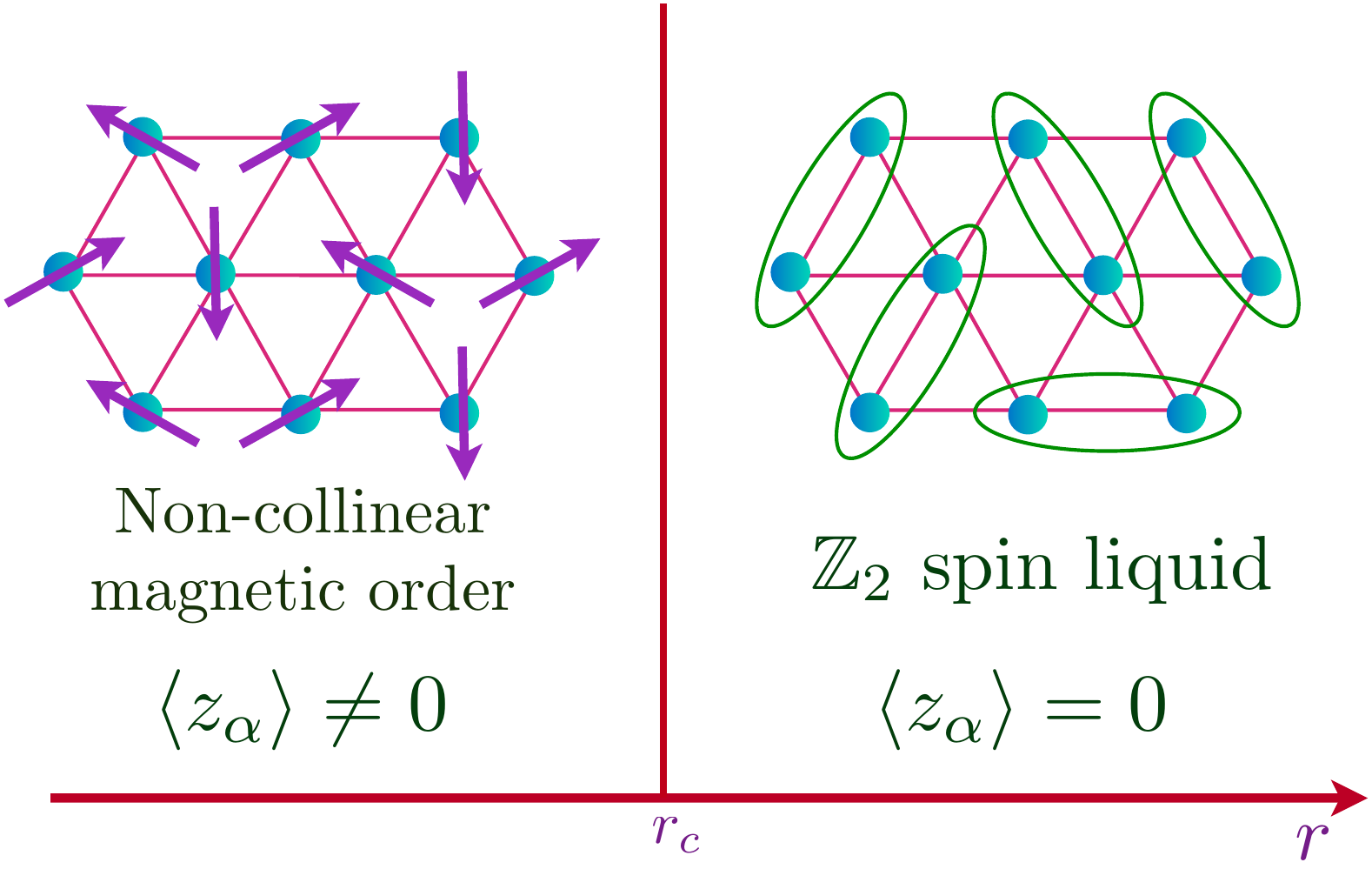}
\end{center}
\caption{Magnetic ordering transition driven by tuning $r$ in (\ref{defr}). Fractionalized anyonic excitations are present only for $r>r_c$, and so there is a `confinement' transition at $r=r_c$. The critical theory is expressed in terms of bosonic spinons $z_\alpha$, and is an example of deconfined critical theory. 
There is evidence for such a transition in KYbSe$_2$ \cite{Batista21,Scheie24}.}  
\label{fig:o4transition}
\end{figure}
We also observe from (\ref{lhz}) that the $z_\alpha$ field will condense when 
\beq
r = (\bar\lambda - 3\sqrt{3}J \bar{\mathcal{Q}}/2)
\label{defr}
\eeq
becomes negative as $\kappa$ is varied across $\kappa_c$. This condensation breaks the spin rotation symmetry, and leads to a quantum phase transition to a phase with coplanar antiferromgnetic long-range order, as illustrated in Fig.~\ref{fig:o4transition}. This order parameter of this coplanar antiferromagnet is related to $\mathbb{Z}_2$ gauge-invariant bilinears of $z_{\alpha}$ by 
\beq
{\bm S}_{\vi} \propto \mbox{Im} \left[ \exp \left( i {\bm Q} \cdot {\bm r} \right) \varepsilon_{\alpha\gamma}
z_\gamma {\bm \sigma}_{\alpha\beta} z_{\beta} \right]\,, \label{defcoplanar}
\eeq
where the wavevector ${\bm Q} = (4 \pi /a) (1/3, 1/\sqrt{3})$.

\subsubsection{Motion of visons}
\label{sec:piflux}

Let us now consider the motion of the vison elementary excitation, which we illustrated earlier in Fig.~\ref{fig:visonberry}.
The vison is located at the center of a triangle, and so can tunnel between neighboring
triangular cells. We are interested here in any possible Berry phases the vison could pick
up upon tunneling around a closed path. 

In Section~\ref{sec:topo}, we characterized the vison by the saddle-point configuration $\mathcal{Q}^v_{\vi\vj}$ of the bond variables
in the Hamiltonian (\ref{HMF}). By diagonalizing this Hamiltonian \cite{SSkagome,HPS11}, we can show that the wavefunction of the vison can
be written as
\beq
\left|\Psi^v \right\rangle = \mathcal{P} \exp \left( \sum_{\vi <j} f_{\vi\vj}^v \, \mathcal{J}_{\alpha\beta} b_{\vi  \alpha}^\dagger b_{j \beta}^\dagger
\right) |0 \rangle, \label{spinl}
\eeq
where $|0 \rangle$ is the boson vaccum, $\mathcal{P}$ is a projection operator which selects only states which obey (\ref{boseconst}), 
and the boson pair wavefunction $f^v_{\vi\vj}=-f^v_{ji}$ is determined from (\ref{HMF}) by a Bogoliubov transformation.

Let us now consider the motion of a single vison \cite{HPS11}. The gauge-invariant Berry phases are those associated with a periodic motion, and so let us consider the motion of a vison along a general
closed loop $\mathcal{C}$. We illustrated the simple case where $\mathcal{C}$ encloses a single site of the triangular lattice 
in Fig.~\ref{fig:visonberry}.
\begin{figure}
\begin{center}
\includegraphics[height=8cm]{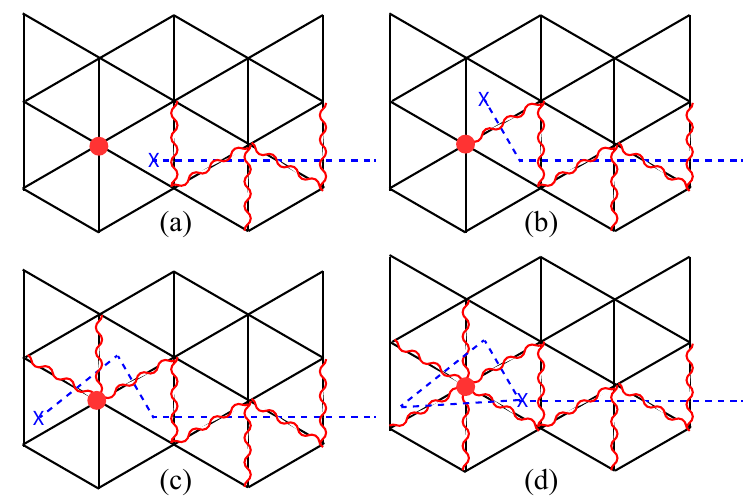}
\end{center}
\caption{Adiabatic motion of a vison (denoted by the X) around a single site of the triangular lattice (denoted by the filled circle). The wavy line is the vison branch cut, as in Fig.~\ref{fig:vison_tri}.
The initial state is in (a), and the final state is in (d), and these differ by a gauge transformation under which 
$b_{\vi \alpha} \rightarrow - b_{\vi \alpha}$
only on the filled circle site.}
\label{fig:visonberry}
\end{figure}
The wavy lines indicate 
$\mbox{sgn}(\mathcal{Q}^{v}_{\vi\vj}) = - \mbox{sgn}(\bar{\mathcal{Q}}_{\vi\vj})$, as in Fig.~\ref{fig:vison_tri}. The last state is gauge-equivalent to the first
state, after the gauge transformation $b_{\vi }^{\alpha} \rightarrow - b_{\vi }^{\alpha}$ only for the site $i$ marked by the filled circle.
As long as the vison wavefunction can be chosen to be purely real, it is clear that no Berry phase is accumulated from the time-evolution
of the wavefunction as the vison tunnels around the path $\mathcal{C}$. However, there can still be a non-zero Berry phase because
a gauge-transformation is required to map the final state to the initial state. The analysis in Fig.~\ref{fig:visonberry} shows that
the required gauge transformation is
\bea
b_{\vi }^{\alpha} \rightarrow - b_{\vi }^{\alpha}, &\quad& \mbox{for $i$ inside $\mathcal{C}$} \nn
b_{\vi }^{\alpha} \rightarrow b_{\vi }^{\alpha}, &\quad& \mbox{for $i$ outside $\mathcal{C}$}. \label{bZ2}
\eea
By Eq.~(\ref{boseconst}), each site has $n_b=2S$ bosons, and so the total Berry phase accumulated by $|\Psi^v \rangle$ is 
\beq
\pi n_b \times \left( \mbox{number of sites enclosed by $\mathcal{C}$} \right). \label{pins}
\eeq
For the important case of $S=1/2$, the vison experiences a flux of $\pi$ for every site of the triangular lattice. 
This phase factor of $\pi $ is related to an `anomaly' associated with the global U(1) boson number symmetry, and translational symmetry \cite{Bonderson16,SenthilElse21,Seiberg23,Meng26}, and was first noted in Refs.~\cite{RJSS91,SSMV99} as a feature of $\mathbb{Z}_2$ spin liquids 
with half-integer spin. In particular, this result implies the RVB state is an {\it odd\/} $\mathbb{Z}_2$ spin liquid. 

A notable feature of (\ref{pins}) is that the quantized integer value of $n_b = 2S$ is important. Memory of this quantization was lost in the mean field theory of Section~\ref{sec:6A}, which was sensible also for non-integer values of $n_b$. So inclusion of the vison fluctuations restores the quantization of spin. A more complete theory for the vison fluctuations appears below in Section~\ref{sec:z2complete}.

\subsubsection{Semions and fermions}
\label{sec:semions}

Proceeding the identification of the present Schwinger boson spin liquid with the $\mathbb{Z}_2$ spin liquid,
we need to establish that the spinons and visons are mutual semions. This is immediately apparent from a glance at Figs.~\ref{fig:vison_tri} and \ref{fig:semion}.
\begin{figure}
\begin{center}
\includegraphics[width=3in]{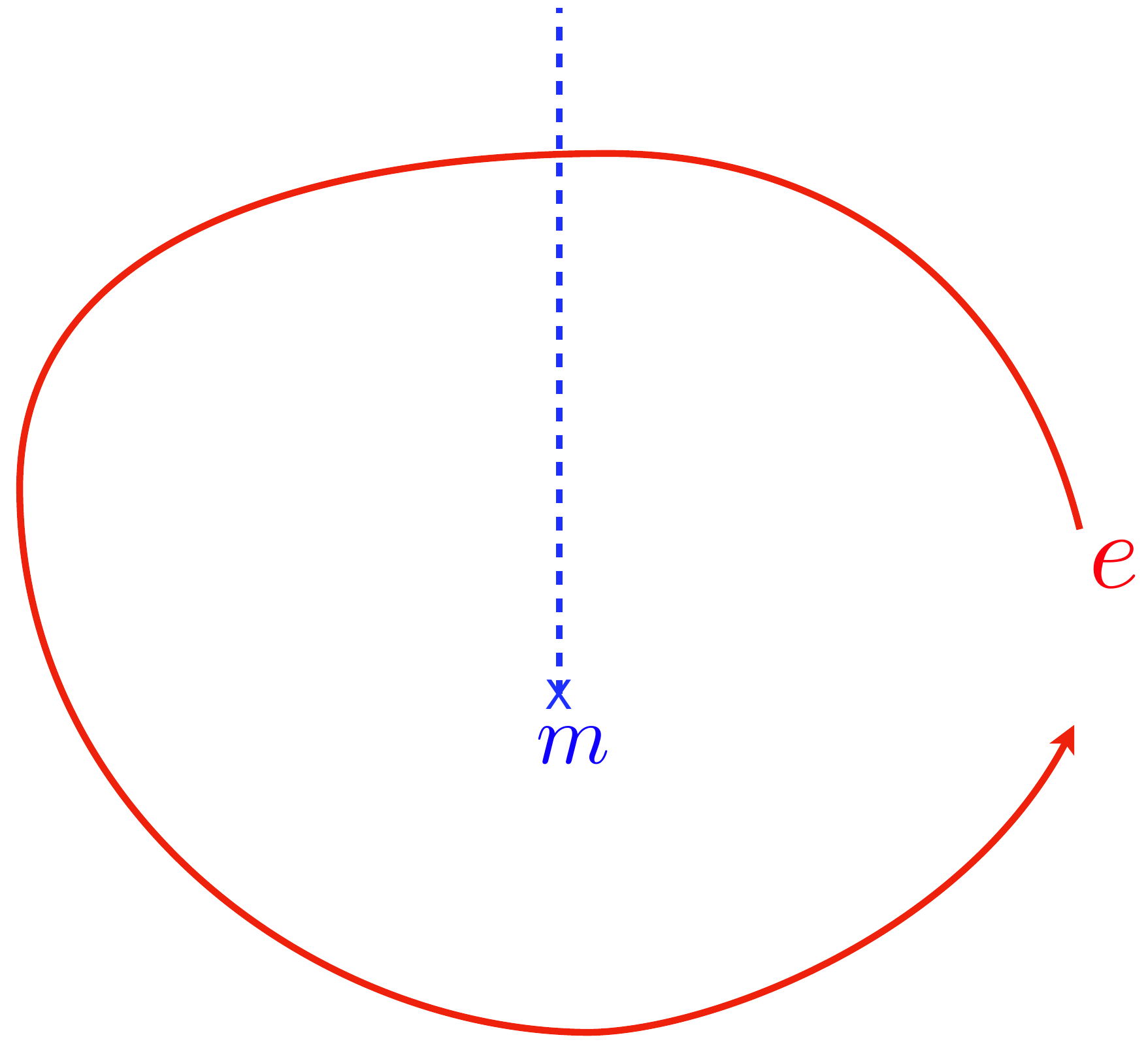}
\end{center}
\caption{Mutual statistics of $e$ and $m$ particles. This process leads to a Berry phase of $-1$, when the $e$ particle crosses the branch cut of the $m$ particle.}  
\label{fig:semion}
\end{figure}

The $\mathcal{Q}^v_{\vi\vj}$ transport the spinons 
from site to site, and for spinon encircling a vison in a large circuit, the only difference between the 
cases with and without the vison is the branch cut. This branch cut yields an additional phase of $\pi$ in the vison amplitude, and provides the
needed phase for mutual semion statistics \cite{NRSS91,XGW91}.

We can now identify the $e$, $\epsilon$, and $m$ anyons, in the abstract topological characterization of the $\mathbb{Z}_2$ spin liquid obtained from the toric code \cite{KitaevToric}. The $e$ anyon is the the Schwinger boson itself, $b_\alpha$. This is mutual semion \index{mutual semion} with respect to the vison, and so we identify the vison with the $m$ particle. Finally, the $\epsilon$ anyon is obtained by the fusion $\epsilon = e \times m$, and so the $\epsilon$ anyon is a bound state of $e$ and $m$. The $\epsilon$ anyon is a {\it fermion\/} as can be deduced by computing the Berry phase associated with exchanging one bound state of $e$ and $m$ with another bound state, as shown in Fig.~\ref{fig:epsilon}.
\begin{figure}
\begin{center}
\includegraphics[width=3in]{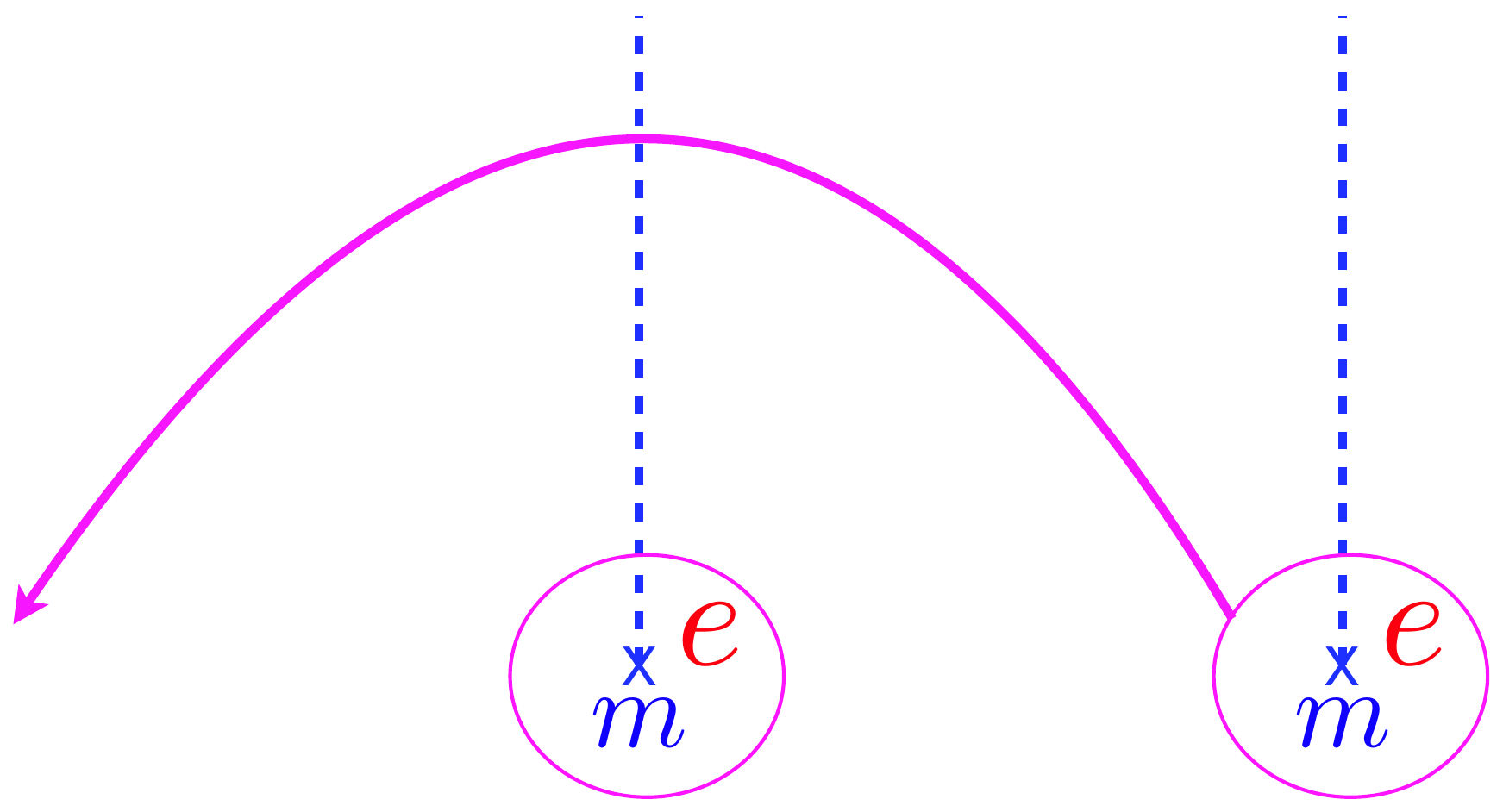}
\end{center}
\caption{Two $\epsilon$ particles undergoing an exchange: after traversing the path shown, a translation returns the $\epsilon$ particles to the original state. Each $\epsilon$ particle is a bound state of a vison (the $m$ particle) and the $s_\alpha$ bosonic spinon (the $e$ particle). This process leads to a Berry phase of $-1$, when the moving $e$ particle crosses the branch cut of the stationary $m$ particle.}  
\label{fig:epsilon}
\end{figure}
(Note that this `long-distance' Berry phase is multiplied by the `short-distance' vison motion phases discussed in Section~\ref{sec:piflux}.)
It is quite remarkable that a microscopic theory of bosonic spins ${\bm S}$, expressed in terms of fractionalized bosons $s_{\alpha}$, yields an excitation which is a fermion: this is one indication of the presence of long-range entanglement and topological order.

An alternative formulation of the $\mathbb{Z}_2$ spin liquid on the triangular lattice proceeds by expressing the spins ${\bm S}$ in terms of Schwinger fermions $f_{\alpha}$. For the $\mathbb{Z}_2$ spin liquid, the $f_\alpha$ spinons would become the $\epsilon$ particles, and the bosonic $e$ would be the bound state of the $f_\alpha$ and the $m$ vison. So ultimately, independent of whether we choose to fractionalize the ${\bm S}$ spins in terms of bosonic or fermionic partons, we obtain the same characterization of the observable excitations in the resulting $\mathbb{Z}_2$ spin liquid. This identity also extends to the symmetry transformations of the anyons, as has been shown in some detail on the triangular \cite{YuanMing}, square \cite{Wang16,ThomsonSS18} and kagome \cite{GilAshvin} lattices. 

\subsubsection{Complete $\mathbb{Z}_2$ gauge theory of spinons and visons}
\label{sec:z2complete}

We now combine the above results for the dynamics of spinons and visons to write down a $\mathbb{Z}_2$ gauge theory for the triangular lattice quantum antiferromagnet which couples the $z_{\vi \alpha}$ to a dynamic $\mathbb{Z}_2$ gauge field $Z_{\vi \vj}$ on the links of a 3-dimensional hexagonal lattice ({\it i.e.\/} stacked triangular lattices) \cite{ParkSS02,ShackletonZ2}: 
\begin{align}
\mathcal{Z}_{\mathbb{Z}_2} & = \sum_{Z_{\vi\vj} = \pm 1}  \prod_i \int  d z_{\vi  \alpha} \delta \left( \sum_\alpha |z_{\vi  \alpha}|^2 - 1 \right) 
\left\{ \prod_i Z_{\vi ,\vi+\hat{\tau}} \right\}^{2S}  \exp \left( - \mathcal{S}_{\mathbb{Z}_2} [  z_\alpha , Z ] \right) \nonumber \\
 \mathcal{S}_{\mathbb{Z}_2} [  z_\alpha , Z ] & = 
- J_2 \sum_{\langle ij \rangle}  Z_{\vi\vj} \left( z_{\vi  \alpha}^\ast z_{\vj \alpha} + \mbox{c.c.} \right) - K \sum_{\triangle, \square} \,\, \prod_{\vi\vj \in \triangle,\square} Z_{\vi\vj}
\,.
\label{xyf22o4}
\end{align}
Here the $z_{\vi  \alpha}$ reside on the sites $\vi$ of the hexagonal lattice, and the $K$ term acts on the 
triangular plaquettes of the spatial plane, and the rectangular plaquettes along the temporal direction. 
Similarly, for quantum antiferromagnets on other two-dimensional lattices, we would work with a three dimensional lattice obtained by stacking the two-dimensional lattice.
The $Z_{\vi \vj}$ descend from the $\mathcal{Q}_{\vi\vj}$ 
by fixing their magnitudes and allowing their signs to be dynamical via
\beq
\mathcal{Q}_{\vi\vj} \Rightarrow \left| \mathcal{Q}_{\vi\vj}  \right| Z_{\vi \vj}.
\eeq
Thus we are approximating the path integral over $\mathcal{Q}_{ij}$ in Eq.~(\ref{zfunct0}) by a sum over $Z_{\vi \vj}$; this is acceptable because the sum is sufficient to allow the vison states to appear.
From Eq.~(\ref{gaugetrans}), 
the $\mathbb{Z}_2$ gauge transformation in Eq.~(\ref{z2gauge}) acts on the $\mathbb{Z}_2$ gauge field via
\beq
Z_{\vi\vj} \rightarrow \mu_\vi Z_{\vi \vj} \mu_\vj\,.
\eeq

The novel term in Eq.~(\ref{xyf22o4}), beyond those found in the lattice gauge theory literature, is 
the prefactor of the exponential in the curly brackets: this is the Berry phase term for half-integer spin $S$. It is a direct consequence of the presence of a background $\mathbb{Z}_2$ gauge charge on each site, as implied by Eqs.~(\ref{boseconst}) and (\ref{bZ2}) \cite{ShackletonZ2}, and so accounts of the motion of visons in a background $\pi$ flux.

The partition function in Eq.~(\ref{xyf22o4}) has been studied by quantum Monte Carlo, and yields the phase diagram in Fig.~\ref{fig:signfree}.
\begin{figure}
\begin{center}
\includegraphics[width=3in]{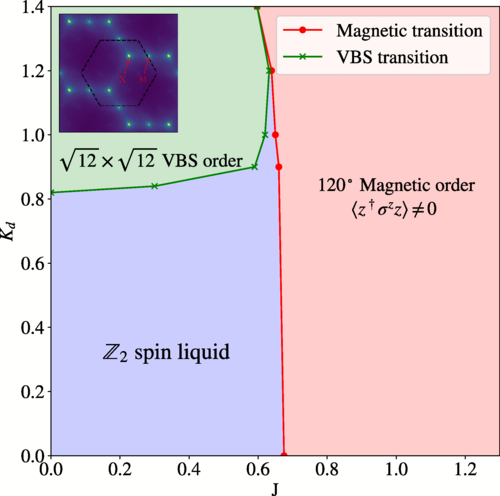}
\end{center}
\caption{Phase diagram of $\mathcal{Z}_{\mathbb{Z}_2}$ in Eq.~(\ref{xyf22o4}) obtained via sign-problem-free quantum Monte Carlo in Ref.~\cite{ShackletonZ2}.}  
\label{fig:signfree}
\end{figure}
Apart from the ordered antiferromagnet and the $\mathbb{Z}_2$ spin liquid in Fig.~\ref{fig:o4transition}, here we obtain a valence bond solid (VBS) state as a consequence of the Berry phase term \cite{RJSS91,SSMV99}. This is gapped state which preserves spin rotation invariance, but breaks lattice translational symmetry.

The transition from the $\mathbb{Z}_2$ spin liquid to the VBS can be studied by integrating out the $z_\alpha$ spinons from Eq.~(\ref{xyf22o4}). This yields a $\mathbb{Z}_2$ gauge theory for the visons in terms of the $Z_{\vi \vj}$ alone. It is convenient to express this $\mathbb{Z}_2$ gauge theory in Hamiltonian form on the triangular lattice, which has exactly the form familiar in the lattice gauge theory literature:
\begin{equation}
 H_{\mathbb{Z}_2} =  - K \sum_{\triangle} \,\, \prod_{\vi\vj \in \triangle} Z_{\vi\vj} - g \sum_{\vi\vj} X_{\vi \vj}\,, \label{wegner}
 \end{equation}
where $X_{\vi\vj}$ are Pauli $X$ operators on the links of the triangular lattice. However, the Berry phase term in Eq.~(\ref{xyf22o4}) does make a crucial difference in selecting the gauge-invariant sector of the Hilbert space of $H_{\mathbb{Z}_2}$. There are infinite number of $\mathbb{Z}_2$ gauge charges which commute with $H_{\mathbb{Z}_2}$, one on each site of the triangular lattice:
\begin{equation}
G_\vi = \prod_{\ell \ni \vi} X_\ell\,, \label{defGi}
\end{equation}
where now $\ell$ labels the links of the triangular lattice.
The Berry phase of Eq.~(\ref{xyf22o4}) implies that we should chose the sector with \cite{RJSS91,SSMV99,SenthilFisher00}
\begin{equation}
G_\vi = (-1)^{2S}\,. \label{mGS}
\end{equation}
For $(-1)^{2S} = -1$, this term is crucial in ensuring that the confinement transition from the $\mathbb{Z}_2$ spin liquid leads to a VBS, and there is no trivial ground state at large $g$.

The transition from the ordered antiferromagnet to the $\mathbb{Z}_2$ spin liquid is not influenced by the Berry phase term (because the visons remains gapped at this transition), and is described by the O(4)
Wilson-Fisher critical theory \cite{CSS93,CSS94}. However, there is an important difference in the structure of the observable order parameter. Note that $z_\alpha$ was obtained from the continuum limit of the spinon $s_\alpha$, and so it is fractionalized degree of freedom, carrying a unit $\mathbb{Z}_2$ charge. Correlators of $z_\alpha$ are therefore not observable, only those of gauge-invariant bilinear combinations. This is denoted by stating the universality class of the transition is actually O(4)$^\ast$. This critical theory has the same exponents as the O(4) theory, but some observables in a finite geometry are different \cite{WhitsittZ2}. As the critical fields of the theory are fractionalized spinons, this is an example of a `deconfined critical point'.

Near the multicritical region where the 3 phases meet in Fig.~\ref{fig:signfree}, there could be additional deconfined criticality \cite{ShackletonZ2} associated with a U(1) gauge theory of fermionic spinons \cite{song2018,song2019,wietek2023,Knolle25} which we do not explore here.

\subsection{Dynamics of excitations on the square lattice}
\label{sec:U1}

We now examine the low energy theory on the square lattice in the regime where the energy gap of the spinon excitations is small. Here, we can take a continuum limit for the spinons, and also account for the fluctuations of $\mathcal{Q}$ and $\lambda$. For the spinons, we introduce the wavevector at the minimum spinon gap
${\bm k}_0 = (\pi/2 , \pi/2 )$ and parameterize on the checkerboard $A$ and $B$ sublattices (with $\vi_x + \vi_y$ even and odd)
\begin{eqnarray}
b_{A\vi\alpha} &=& \psi_{1\alpha} ({\bm r}_\vi ) e^{i {\bm k}_0 \cdot {\bm r}_\vi }
\nonumber \\
b_{B\vi\alpha} &=& -i \varepsilon_{\alpha\beta} \psi_{2\beta} ({\bm r}_\vi )
e^{i {\bm k}_0 \cdot {\bm r}_\vi } \, .
\label{bosepar}
\end{eqnarray}
For $\mathcal{Q}$ and $\lambda$, we use the parameterization already discussed in Section~\ref{sec:sqU1}.

We insert these parameterizations into the spinon action,
perform a gradient expansion, and transform the Lagrangian ${\cal L}_J$ into ($a$ is the lattice spacing)
\begin{eqnarray}
{\cal L}_z &=& \int \frac{d^2 r}{2a^2} \left [
\psi_{1\alpha}^{\ast} \left( \frac{d}{d\tau} + i a_{\tau}
\right)
\psi_{1\alpha} +
\psi_{2\alpha}^{\ast} \left( \frac{d}{d\tau} - i a_{\tau}
\right)
\psi_{2\alpha} \right.
\nonumber \\
&& \qquad\qquad
 + \bar{\lambda} \left( |\psi_{1\alpha} |^2
+ |\psi_{2\alpha} |^2 \right) 
-2 J \bar{\mathcal{Q}} \left ( \psi_{1\alpha}\psi_{2\alpha} +
\psi_{1\alpha}^{\ast}\psi_{2\alpha}^{\ast}
\right )
\nonumber \\
&&\qquad\qquad
+  (J/2) \bar{\mathcal{Q}} a^2 \left [
\left ( {\bm \nabla}  + i  {\bm a} \right ) \psi_{1\alpha}
\left ( {\bm \nabla}  - i {\bm a} \right ) \psi_{2\alpha} \right.
\nonumber \\
&& \qquad\qquad\qquad
+ \left.
\left ( {\bm \nabla} - i {\bm a} \right )
\psi_{1\alpha}^{\ast}
\left ( {\bm \nabla} + i {\bm a} \right )
\psi_{2\alpha}^{\ast} \right ]  \Biggr] \, .
\label{charge}
\end{eqnarray}
We now introduce the fields
\begin{eqnarray*}
z_{\alpha} & = & (\psi_{1\alpha} +
\psi_{2\alpha}^{\ast})/\sqrt{2} \\
\pi_{\alpha} & = & (\psi_{1\alpha} -
\psi_{2\alpha}^{\ast})/\sqrt{2} \, ,
\end{eqnarray*}
to map Eq.~(\ref{charge}) to
\begin{eqnarray}
{\cal L}_z &=& \int \frac{d^2 r}{2a^2} \left [
\pi_{\alpha}^{\ast} \left( \frac{d}{d\tau} + i a_{\tau}
\right)
z_{\alpha} -
\pi_{\alpha} \left( \frac{d}{d\tau} - i a_{\tau}
\right)
z_{\alpha}^\ast \right.
\nonumber \\
&& \qquad\qquad
 + \bar{\lambda} \left( |z_{\alpha} |^2
+ |\pi_{\alpha} |^2 \right) 
-2 J \bar{\mathcal{Q}} \left ( |z_\alpha|^2 - |\pi_\alpha|^2 
\right )
\nonumber \\
&&\qquad\qquad
+  (J/2) \bar{\mathcal{Q}} a^2 \left [
\left|\left ( {\bm \nabla}  + i  {\bm a} \right ) z_{\alpha} \right|^2 - 
\left|\left ( {\bm \nabla}  + i  {\bm a} \right ) \pi_{\alpha} \right|^2   \right ]  \Biggr] \, .
\label{charge2}
\end{eqnarray}
From Eq.~(\ref{charge2}), it is clear that the
the $\pi$ fields have `mass' $\bar{\lambda} + 2J \bar{\mathcal{Q}}$,
while the $z$ fields
have a mass $\bar{\lambda} - 2 J \bar{\mathcal{Q}}$ which vanishes at a quantum phase transition where the $z_\alpha$ condense, leading to N\'eel order. The $\pi$ fields can therefore
be safely integrated out,
and ${\cal L}_z$ yields
the following effective action, valid at distances much
larger than the lattice
spacing~\cite{NRSS89prl,NRSS90}:
\begin{equation}
S_{\rm eff} =
\int \frac{d^2 r}{4\sqrt{2}a} \int
d \tau \left\{
|(\partial_{\mu} - ia_{\mu})z_{\alpha}|^2
+ \frac{\Delta^2}{c^2}
|z^{\alpha} |^2\right\} \,.
\label{sefp}
\end{equation}
Here $\mu$ extends over $x,y,\tau$,
$c = \sqrt{2}J \bar{\mathcal{Q}} a$ is the spin-wave velocity, we have rescaled $\tau \rightarrow \tau/c$, and 
$\Delta = (\bar{\lambda}^2 - 4J^2 \bar{\mathcal{Q}}_1^2 )^{1/2}$ is the gap
towards spinon excitations.
Thus the long-wavelength theory describes a spin liquid with 
of a massive, spin-1/2, relativistic, boson $z_{\alpha}$ (spinon) excitation
coupled to a U(1) gauge field $a_\mu$. 

The continuum theory also makes it easy to determine the fate of the antiferromagnet when the spin energy gap vanishes. We expect that $z_\alpha$ will bose condense, and this will break the spin rotation symmetry; a term quartic in $z_\alpha$ will be needed to stabilize the condensate. But $z_\alpha$ carries a U(1) gauge charge, and so is not directly observable. 
Following the definitions of the underlying spin operators, it is not difficult to show
that the gauge-invariant composite
\begin{equation}
\mathbfcal{N} = z_\alpha^{\ast} {\bm \sigma}_{\alpha\beta} z_\beta \sim \eta_\vi {\bm S}_\vi \label{neelz}
\end{equation}
is just the N\'eel order parameter of Fig.~\ref{fig:Neel}.

However, there is an important ingredient that our low energy theory has not yet considered. These are non-perturbative fluctuations of $a_\mu$ which are Dirac monopoles in 2+1 dimensional spacetime, which carry non-trivial Berry phases \cite{NRSS90}. We will not carry out a full analysis here, and
only note that these monopole Berry phases are closely connected to the vison Berry phases in Section~\ref{sec:piflux} \cite{RJSS91,SSMV99}.
An important result is that the spin liquid noted above is  ultimately not a spin liquid. It is unstable to proliferation of monopoles, and ultimately confines a valence bond solid. But monopoles do not have a significant effect on the N\'eel state.

\subsubsection{Quantum criticality}
\label{sec:dqcp}

On general symmetry grounds, we extend Eq.~(\ref{sefp}) to a theory for the vicinity of the quantum critical point at which
the spinon gap vanishes \cite{SJ90}:
\bea
\mathcal{S}_{U(1)} &=& \int d^3 x \, \left( \mathcal{L}_z + \mathcal{L}_{\rm monopole} \right) + \mathcal{S}_B  \nonumber \\
\mathcal{L}_z &=&
|(\partial_\mu -  i a_\mu) z_\alpha|^2 + g |z_\alpha|^2 + u \left(|z_\alpha|^2\right)^2 + K (\epsilon_{\mu\nu\lambda} \partial_\nu a_\lambda)^2 
\nonumber \\
 \mathcal{L}_{\rm monopole} &=& - y \left( \mathcal{M}_a + \mathcal{M}_a^\dagger \right) \nonumber \\
\mathcal{S}_B &=&  i 2S \sum_\vi \eta_\vi \int d \tau \, a_{\vi \tau} \,. \label{cp11}
\eea
The theory $\mathcal{L}_z$ is also known as the $\mathbb{CP}^1$ model. We have
included monopoles $\mathcal{M}_a$ in the gauge field $a_\mu$, and also the Berry phase of the spinons in the ground state in $\mathcal{S}_B$. 
This Berry phase is closely connected to the vison Berry phase in Section~\ref{sec:piflux} \cite{RJSS91,SSMV99}: note the similarity of $\mathcal{S}_B$ to the Berry phase term in curly brackets of the $\mathbb{Z}_2$ gauge field in Eq.~(\ref{xyf22o4}).
\begin{figure}
\centering
\includegraphics[width=4in]{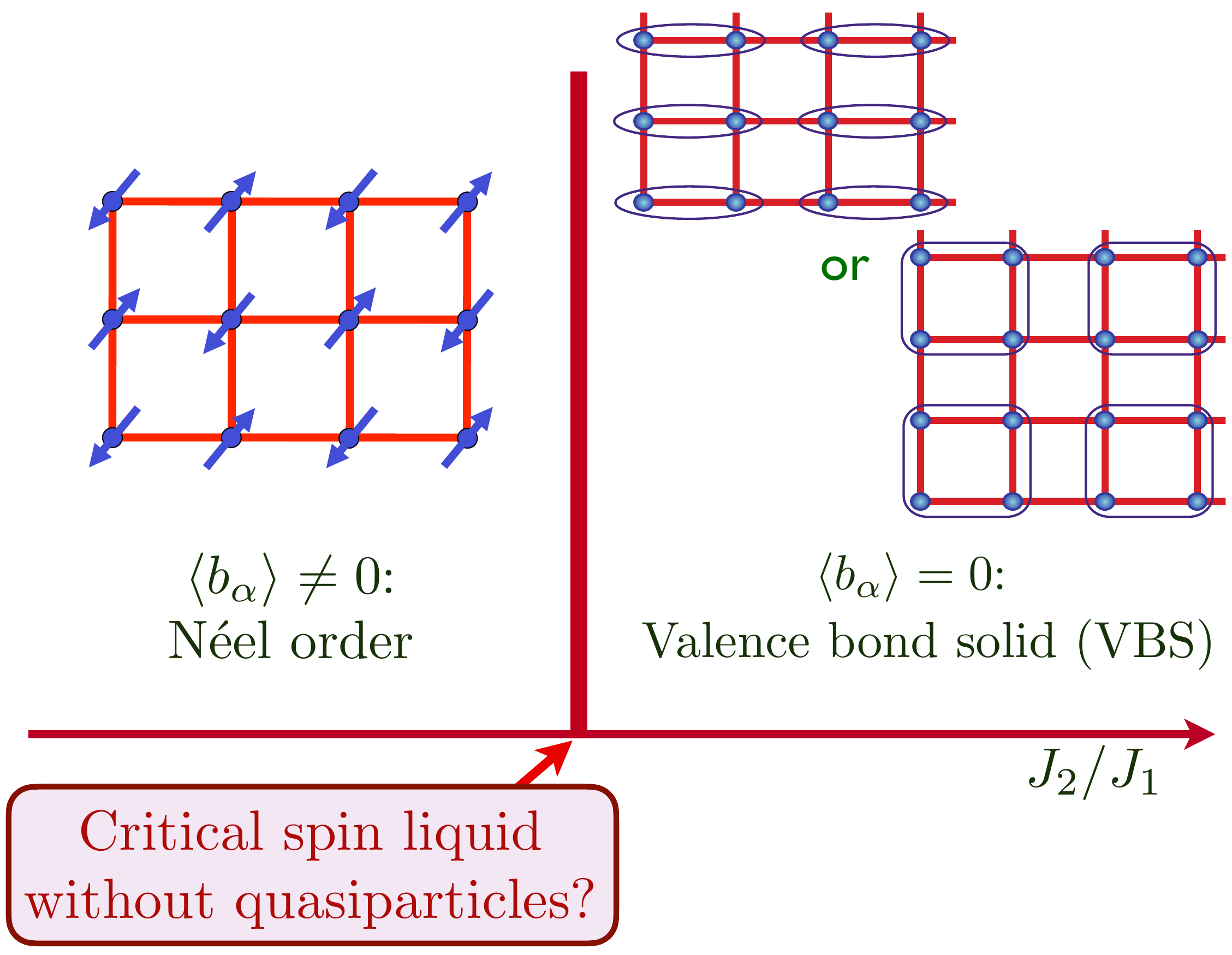}
\caption{Phase diagram of the U(1) gauge theory with bosonic spinons, Eq.~(\ref{cp3}). The N\'eel order appears in a Higgs phase where the bosonic spinons are condensed. The VBS order appears in the confining phase, and is induced by the Berry phases of the confining monopoles. The same phase diagram applies to the fermionic spinon theory in Eq.~(\ref{e105}), and the SO(5) $\sigma$-model with the WZW term in Eq.~(\ref{wzw}).}
\label{fig:Neelvbs}
\end{figure}

As we tune the coupling $g$ in Eq.~(\ref{cp11}), we can expect the 2 phases shown in Fig.~\ref{fig:Neelvbs}:\\
({\it i\/}) N\'eel phase, $g<g_c$: the spinon $z_\alpha$ condenses in a Higgs phase with $\langle z_\alpha \rangle \neq 0$. The $a_\mu$ gauge field is Higgsed, and spin rotation symmetry is broken by opposite polarization of the spins on the two sublattices.\\
({\it ii\/}) Valence bond solid (VBS), $g> g_c$: the spinons are gapped. For half-integer spin $S$, there is broken translational symmetry by the crystallization of valence bonds in the pattern shown in Fig.~\ref{fig:Neelvbs}.

We now obtain a potential gapless spin liquid if there is a continuous quantum phase transition at $g=g_c$. For half-integer spin $S$, the single monopole terms in Eq.~(\ref{cp11}) average to zero at long wavelengths from the Berry phases, and only quadrupoled monopole terms survive. So we can simplify the continuum theory for the vicinity of the quantum critical point to \cite{senthil1,senthil2} 
\beq
\!\!\!\!\!\!\! \!\!\!\!\!\!\! \!\!\!\! 
\mathcal{L}_z =
|(\partial_\mu -  i a_\mu) z_\alpha|^2 + g |z_\alpha|^2 + u \left(|z_\alpha|^2\right)^2 + K (\epsilon_{\mu\nu\lambda} \partial_\nu a_\lambda)^2  - y_4 \left( \mathcal{M}_a^4 + \mathcal{M}_a^{\dagger 4} \right)\,, \label{cp3}
\eeq
where $y_4$ is the quadrupoled monopole fugacity. \index{monopoles} There is ample numerical evidence that $y_4$ is irrelevant near a possible critical point, and so the question reduces to whether the theory $\mathcal{L}_z$ at $y_4=0$ exhibits a critical point which realizes a conformal field theory in 2+1 dimensions. This is a question that has been studied extensively in numerics, and it is clear that a `deconfined critical' description is suitable over a substantial length scale: with fractionalized spinons interacting with a U(1) gauge field in the absence of monopoles \cite{Nahum:2015vka,Meng24,Gu24,Chester24,Sandvik24,Fuzzy24}. 

\section{Fermionic spinon theory of quantum spin liquids}
\label{sec:fermions}

We now present an alternative analysis of the square lattice antiferromagnet in Eq.~(\ref{hamil}), replacing the bosonic partons in Eq.~(\ref{sparton}) by fermionic partons. This will ultimately lead to the same phase diagram as in Fig.~\ref{fig:Neelvbs}, but with a dual description of the phases and the criticality. This dual fermionic description turns out to be the most efficient way to describe the connection between the critical spin liquid and $d$-wave superconductivity in the doped antiferromagnet, as we will see in Section~\ref{sec:halffilling}.

The following Schwinger {\it fermion} representation applies only for $S=1/2$
\begin{equation}
{\bm S}_{\vi} =  \frac{1}{2}
f_{\vi\alpha}^{\dagger}\, {\bm \sigma}_{\alpha\beta} \, f_{\vi \beta}
\label{Schwingerfermion}
\end{equation}
where $f_{\vi \alpha}$ are canonical fermions obeying the constraint
\beq
\sum_{\alpha} f_{\vi \alpha}^\dagger f_{\vi \alpha} = 1 \quad, \quad \mbox{for all $\vi$}. \label{constraintf}
\eeq

While the bosonic parton representation led to the U(1) gauge symmetry in Eq.~(\ref{gaugetrans}), it turns out the Eqs.~(\ref{Schwingerfermion}) and (\ref{constraintf}) have a larger SU(2) gauge invariance, and this will be crucial to our results. The analysis is clearest upon introducing a matrix notation for the fermions 
\beq
\mathcal{F}_\vi \equiv \left(
\begin{array}{cc}
f_{\vi \uparrow} & - f_{\vi \downarrow} \\
f_{\vi \downarrow}^\dagger & f_{\vi \uparrow}^\dagger
\end{array}
\right) \label{Fmatrix}
\eeq
This matrix obeys the `reality' condition
\beq
\mathcal{F}_{\vi}^\dagger = \sigma^y \mathcal{F}_{\vi}^T \sigma^y. \label{fdft}
\eeq
Now we can write Eq.~(\ref{Schwingerfermion}) as
\beq
{\bm S}_{\vi} = -\frac{1}{4} \mbox{Tr} ( \mathcal{F}_{\vi}^{\phantom\dagger} \sigma^z {\bm \sigma}^T \sigma^z \mathcal{F}_{\vi}^{\dagger} ) \,. \label{NambuST}
\eeq
The SU(2) gauge symmetry is now associated with a SU(2) matrix $V_\vi$ under which \cite{Affleck-SU2,Fradkin88,AndreiColeman2}
\beq
\mathcal{F}_\vi \rightarrow V_\vi \, \mathcal{F}_\vi \,, \label{SU2gauge}
\eeq
which is easily seen to leave the spin operator in Eq.~(\ref{NambuST}) invariant. The global spin rotation symmetry is however
\beq
\mathcal{F}_\vi \rightarrow  \mathcal{F}_\vi \, \sigma^z \Omega^T \sigma^z  \,. \label{su212b}
\eeq
where $\Omega$ is the $S=1/2$ spin rotation matrix defined by
\beq
\left( \begin{array}{c}
 f_{\vi \uparrow} \\ f_{\vi \downarrow} \end{array} \right) \rightarrow \Omega  \left( \begin{array}{c}
 f_{\vi \uparrow} \\ f_{\vi \downarrow} \end{array} \right)\,.
 \eeq

Next we insert Eq.~(\ref{NambuST}) into Eq.~(\ref{hamil}), and perform Hubbard-Stratonovich transformation to obtain an effective Hamiltonian for the spinons, following the same procedure as for bosonic spinons. We skip the intermediate steps, and focus directly on the fermion bilinear Hamiltonian on symmetry grounds; we obtain the mean-field nearest-neighbor spin liquid Hamiltonian for the spinons of the $\pi$-flux phase \cite{BZA87,Kotliar87,AM88,Affleck-SU2,Zhang88}:
\bea
    \mathcal{H}_{SLf} & = & \frac{iJ}{2} \sum_{\langle \vi \vj \rangle} e_{\vi \vj} \left[  \mbox{Tr} \left( \mathcal{F}_{\vi}^{\dagger} \mathcal{F}_{\vj }^{\phantom\dagger} \right) -  \mbox{Tr} \left( \mathcal{F}_{\vj }^{\dagger} \mathcal{F}_{\vi }^{\phantom\dagger} \right) \right] \nonumber \\
&=& i J \sum_{\langle \vi\vj\rangle} e_{\vi\vj} \left( \Psi_\vi^\dagger  \Psi_\vj - \Psi_\vj^\dagger  \Psi_\vi \right); \quad
\Psi_\vi \equiv \left( \begin{array}{c} f_{\vi \uparrow} \\ f_{\vi \downarrow}^{\dagger} \end{array} 
\right),
    \label{eq:fermionhop}
\eea
where $e_{\vi \vj} = \pm 1$ represents $\pi$-flux on the fermions as shown in Fig.~\ref{fig:eij}. We choose
$e_{\vi\vj} =- e_{\vj \vi}$ and
\beq
 e_{\vi,\vi+\hat{{\bm x}}}  =  1 \,,\quad
 e_{\vi,\vi+\hat{{\bm y}}}  =  (-1)^{x}      \,, \label{su2ansatz}
\eeq
where $\vi = (x,y)$, $\hat{\bm x} = (1,0)$, $\hat{\bm y} = (0,1)$.
\begin{figure}
\centering
\includegraphics[width=4.5in]{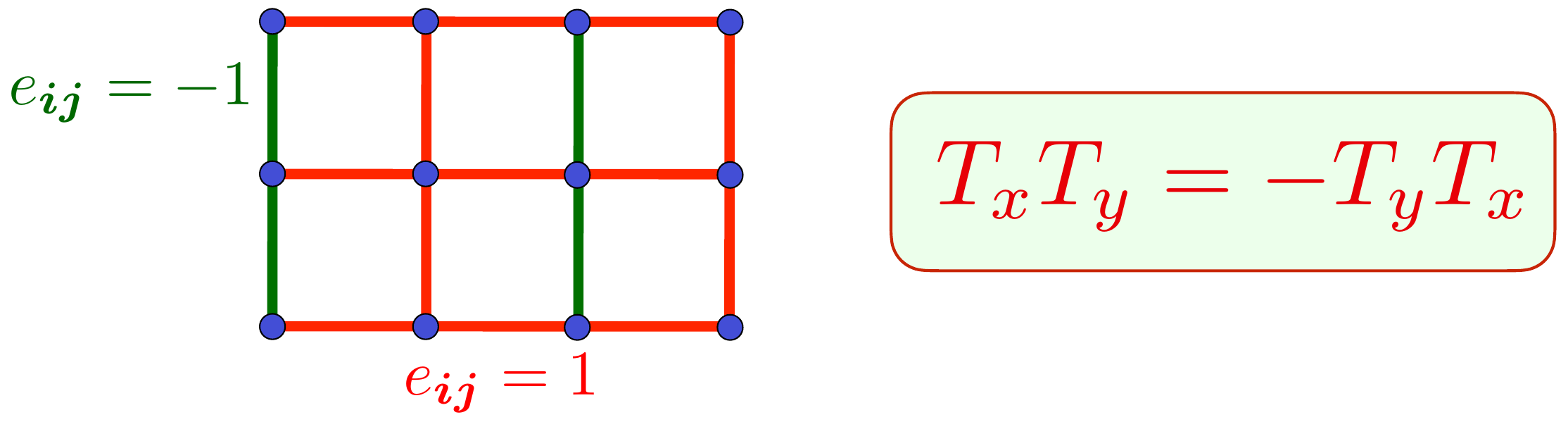}
\caption{Background $\pi$ flux acting on the spinons $f$, and also on the chargons $B$.}
\label{fig:eij}
\end{figure}
In the chosen form, it is evident that $\mathcal{H}_{Slf}$ is invariant under the global spin rotation in Eq.~(\ref{su212b}), and also at least the gobal gauge transformation associate with Eq.~(\ref{SU2gauge}). We have chosen the hopping to be pure imaginary because of the identity
\beq
     \mbox{Tr} \left( \mathcal{F}_{\vi}^\dagger  \mathcal{F}_{\vj }^{\phantom\dagger} \right) = -\mbox{Tr} \left( \mathcal{F}_{\vj }^\dagger  \mathcal{F}_{\vi }^{\phantom\dagger} \right)\,.
\eeq
We can now easily diagonalize the Hamiltonion in Eq.~(\ref{eq:fermionhop}), and obtain the fermionic dispersion spectrum analogous to Eq.~(\ref{eq:bosondisp})
\beq
\omega_{\bm k} = \pm 2J \left( \sin^2 k_x + \sin^2 k_y \right)^{1/2}\,. \label{eq:fermiondisp}
\eeq
We show a plot of this analogous to Fig.~\ref{fig:bosondisp} in Fig.~\ref{fig:fermiondisp}.
\begin{figure}
\centering
\includegraphics[width=4in]{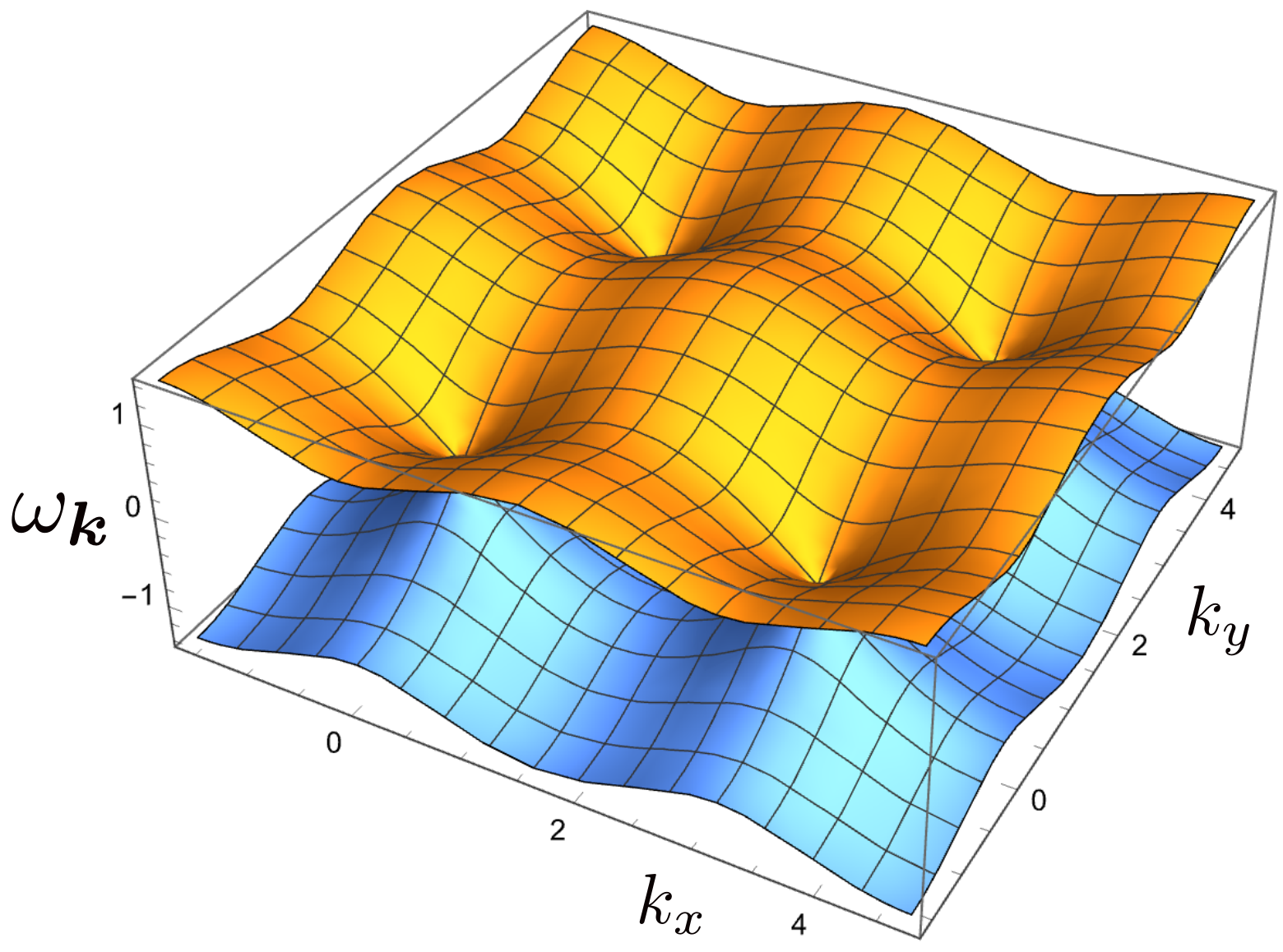}
\caption{Dispersion of fermionic spinons in Eq.~(\ref{eq:fermiondisp}).}
\label{fig:fermiondisp}
\end{figure}
Unlike the bosonic spinons, the energy of the fermionic spinons is allowed to be negative, and the negative energy fermion states are occupied in the ground state. The constraint in Eq.~(\ref{constraintf}) is then automatically satisfied. Notice the two independent nodal Dirac points at ${\bm k}_v$, $v=1,2$ with
\beq
{\bm k}_1 = (0,0) \quad, \quad {\bm k}_2 = (0, \pi)\,.
\eeq
The index $v$ is the `valley'.

\subsection{Low energy SU(2) gauge theory.}
\label{sec:su2theory}

Going beyond mean field theory, while still remaining on the lattice, we extend the mean field Hamiltonian in Eq.~(\ref{eq:fermionhop}) to be invariant under space-dependent SU(2) gauge symmetries. To achieve this, we need to introduce a SU(2) gauge field $U_{\vi\vj} = U_{\vj \vi}^\dagger$ on each lattice link upon which the SU(2) gauge transformation acts as
\beq
U_{\vi\vj} \rightarrow V_\vi^{\phantom\dagger} U_{\vi \vj} V_\vj^{\dagger}\,.
\eeq
Then, by gauge invariance, we extend Eq.~(\ref{eq:fermionhop}) to
\bea
    \mathcal{H}_{SLf} & = & \frac{iJ}{2} \sum_{\langle \vi \vj \rangle} e_{\vi \vj} \left[  \mbox{Tr} \left( \mathcal{F}_{\vi}^{\dagger} U_{\vi \vj}^{\phantom\dagger} \mathcal{F}_{\vj }^{\phantom\dagger} \right) -  \mbox{Tr} \left( \mathcal{F}_{\vj }^{\dagger} U_{\vj \vi}^{\phantom\dagger} \mathcal{F}_{\vi }^{\phantom\dagger} \right) \right] \nonumber \\
&=& i J \sum_{\langle \vi\vj\rangle} e_{\vi\vj} \left( \Psi_\vi^\dagger U_{\vi \vj}^{\phantom\dagger} \Psi_\vj - \Psi_\vj^\dagger U_{\vj \vi}^{\phantom\dagger} \Psi_\vi \right); \quad
\Psi_\vi \equiv \left( \begin{array}{c} f_{\vi \uparrow} \\ f_{\vi \downarrow}^{\dagger} \end{array} 
\right),
    \label{eq:fermionhop2}
\eea
The first form in terms of $\mathcal{F}$ makes both the SU(2) gauge invariance and the SU(2) spin rotation invariance explicit, while in the second form in terms of $\Psi$ only the gauge invariance is explicit.

If we had not used the pure imaginary hopping in Eq.~(\ref{eq:fermionhop}), then the mean-field Hamiltonian would break (`Higgs') the SU(2) gauge symmetry to a smaller symmetry. A `staggered flux' ansatz which breaks the SU(2) down to U(1) has commonly been used in the literature \cite{LeeWenRMP}. However, it is now known that this U(1) spin liquid allows single monopole perturbations \cite{Alicea08,Song1} (unlike the quadrupole monopole perturbations in Eq.~(\ref{cp3})), and such single monopole terms are expected to drive a strong instability to confinement. So we don't consider the staggered-flux U(1) spin liquid.

The continuum formulation of this theory can be obtained by following the same procedure as in Section~\ref{sec:U1}, but we have to carefully account for the SU(2) gauge symmetry. 
First, neglecting gauge fluctuations of $U_{\vi\vj}$ for now, let us write Eq.~(\ref{eq:fermionhop}) in momentum space, in terms of the fermions $\mathcal{F}_s ({\bm k})$, where $s=A,B$ is a sublattice 
index. Now the sublattices refer to sites with $\vi_x$ even and odd, which are the two sites in the unit cell. We obtain
\bea
\mathcal{H}_{SLf} = -J \sum_{\bm k} \mbox{Tr} \left( \mathcal{F}^\dagger ({\bm k}) \left[ \rho^x \sin (k_x) +\rho^z \sin(k_y) \right]  \mathcal{F}^\dagger ({\bm k}) \right)\,, \label{e102}
\eea
where $\rho^\ell$, with $\ell=x,y,z$, are Pauli matrices in sublattice space. Next, analogous to Eq.~(\ref{bosepar}), 
we take the continuum limit near the valley momenta in terms of $\mathcal{X}_{sv} ({\bm r}, \tau)$ 
\bea
\mathcal{F}_{A\vi} &=& \sum_v \mathcal{X}_{Av} ({\bm r}, \tau) e^{i {\bm k}_v \cdot {\bm r}_i} \nonumber \\
\mathcal{F}_{B\vi} &=& \sum_v \mathcal{X}_{Bv} ({\bm r}, \tau) e^{i {\bm k}_v \cdot {\bm r}_i}\label{e101}
\eea
for $\vi$ on the A and B sublattices respectively. This yields the imaginary time Lagrangian density
\beq
\mathcal{L}_{\mathcal{X}} = \frac{1}{2} \mbox{Tr} \left(\mathcal{X}^\dagger \left[ \partial_\tau + 2 J i \rho^x \partial_x + 2 J i \rho^z \mu^z \partial_y \right] 
\mathcal{X} \right)  \,, \label{e103}
\eeq
where $\mu^\ell$ are the Pauli matrices in valley space.
We recall that the fermion $\mathcal{X}_{sv}$ has four-components, and each component is a $2 \times 2$ matrix which obeys the reality condition in Eq.~(\ref{fdft}). We can write this in a relativistic Dirac form 
\beq
\mathcal{L}_{\mathcal{X}} = \frac{i}{2} \mbox{Tr} \left(\bar{\mathcal{X}} \gamma^\mu \partial_\mu \mathcal{X} \right) \,, \label{e104}
\eeq
with the definitions
\beq
\bar{\mathcal{X}} = -i \mathcal{X}^\dagger \gamma^0 \quad, \quad \gamma^0 = \rho^y \mu^z \quad, \quad \gamma^x =  \rho^z \mu^z \quad, \quad \gamma^y = -\rho^x  \,,
\eeq
where we have absorbed factor of $c=2J$ for the velocity of light.
Finally, it is a simple matter to include the SU(2) gauge field by taking the continuum limit by writing
\beq
U_{\vi \vj} = \exp \left(-i A_{\vi\vj}^\ell \sigma^\ell \right)
\label{eq:UA}
\eeq
(where $\sigma^\ell$ are the Pauli matrices in SU(2) gauge space) and expanding the exponential. We then obtain 
\beq
\mathcal{L}_{\mathcal{X}} = \frac{i}{2} \mbox{Tr} \left(\bar{\mathcal{X}} \gamma^\mu \left[\partial_\mu -i A_\mu^\ell \sigma^\ell \right]  \mathcal{X} \right) \,.\label{e105}
\eeq

The theory in Eq.~(\ref{e105}) is the analog of the $\mathcal{L}_z$ in Eq.~(\ref{cp11}) for bosonic spinons. The latter theory was a U(1) gauge theory with two relativistic complex scalars $z_\alpha$. In the present case, we have a SU(2) gauge theory with $N_f =2$ massless Dirac fermions, associated with valley index $v$. The global symmetry of $z_\alpha$ was just spin rotations $z \rightarrow R z$. In contrast, here we have emergent global symmetry which combines spin and valley rotations. A first guess is a SU(4) symmetry generalizing Eq.~(\ref{su212b})
\beq
\mathcal{X} \rightarrow \mathcal{X} L\,,
\eeq
where $L$ acts on spin and valley space with $L^\dagger L = 1$. However, imposition of the reality condition Eq.~(\ref{fdft}) shows that we also need
\beq
L^T = \sigma^y L^\dagger \sigma^y\,,
\eeq
and so the symmetry is only Sp(4)=SO(5)$/\mathbb{Z}_2$ \cite{RanWen06,Wang17}.
In terms of the Hermitian Lie algebra elements $M$, with $L=e^{i M}$, the reality condition is
\beq
M^T = - \sigma^y M \sigma^y\,.
\eeq
Requiring that $M$ commute with the $\gamma^\mu$, we can now write down the 10 elements of the Lie algebra of Sp(4)=SO(5)$/\mathbb{Z}_2$
\beq
M = \{\sigma^x, \sigma^y, \sigma^z, \mu^z \sigma^x, \mu^z \sigma^y, \mu^z \sigma^z, \rho^x \mu^y, \rho^x \mu^x \sigma^x, \rho^x \mu^x \sigma^y,\rho^x \mu^x  \sigma^z\}\,. \label{e200}
\eeq

The remaining 5 SU(4) generators which commute with the $\gamma^\mu$ are ($t = 1\ldots 5$)
\beq
\Gamma^t = \{ \mu^z, \rho^x \mu^x, \rho^x \mu^y \sigma^x, \rho^x \mu^y \sigma^y, \rho^x \mu^y \sigma^z \}\,.
\eeq
The $\Gamma^t$ all anti-commute with each other, and transform as a SO(5) vector under the generators in Eq.~(\ref{e200}).
It is now straightforward to check by working back to the lattice operators from the information above that the vector $i \mbox{Tr} \left( \bar{\mathcal{X}} \Gamma^t \mathcal{X} \right)$ corresponds precisely to the 5 components of the orders parameters shown in Fig.~\ref{fig:Neelvbs}: the first two components are the VBS order, and the last 3 components are the N\'eel order $\mathbfcal{N}$ in Eq.~(\ref{neelz}) \cite{RanWen06,Wang17}.

Wang {\it et al.} \cite{Wang17} have argued that the likely fate of the SU(2) gauge theory upon confinement is a state which the SO(5) symmetry is spontaneously broken with  $\langle i \mbox{Tr} \left( \bar{\mathcal{X}} \Gamma^t \mathcal{X} \right) \rangle \neq 0$. The lattice model does not have exact SO(5) symmetry, and the choice between the N\'eel and VBS components of $\Gamma^t$ is made by additional 4-fermi terms that can be added to Eq.~(\ref{e105}). So the ultimate fate of the theory is essentially identical to the fate of the bosonic spinon theory in Section~\ref{sec:dqcp}, as illustrated in Fig.~\ref{fig:Neelvbs}. This is essentially the reason for the duality between the theories in Section~\ref{sec:dqcp} and \ref{sec:su2theory}, and Wang {\it et al.\/} have provided additional topological arguments.

\subsection{SO(5) non-linear $\sigma$-model}
\label{sec:SO5}

There is a third formulation of the theories in Section~\ref{sec:dqcp} and \ref{sec:su2theory} which is useful for some purposes, as we illustrate in Fig.~\ref{fig:triality}.
\begin{figure}
\centering
\includegraphics[width=5.5in]{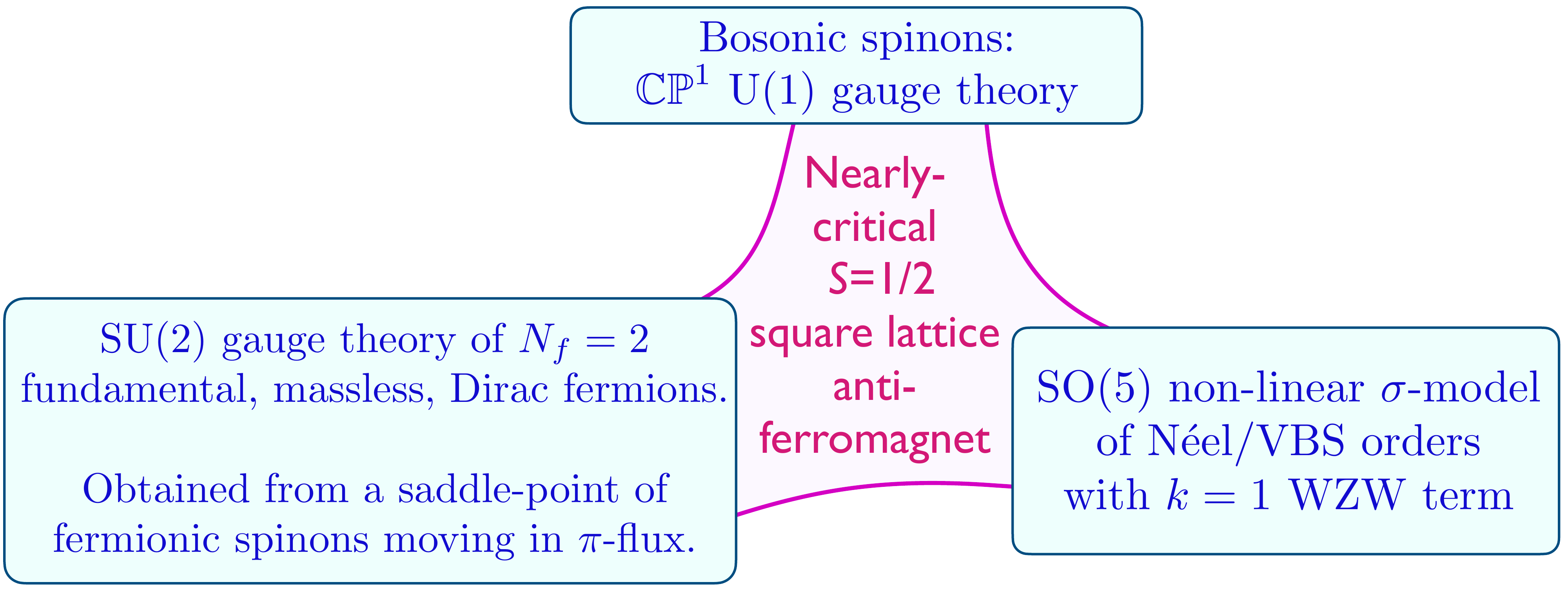}
\caption{Three field-theoretical formulations of the $S=1/2$ square lattice antiferromagnet near the N\'eel-VBS transition in Fig.~\ref{fig:Neelvbs}. All three are valid descriptions \cite{Wang17}, but the SU(2) gauge theory of Dirac fermions is the most convenient starting point to describe the connection to $d$-wave superconductivity.}
\label{fig:triality}
\end{figure}
This is obtained most simply by coupling Eq.~(\ref{e105}) to the SO(5) vector order parameter, and integrating out the fermions. Introducing the SO(5) fundamental unit length field $n_t$, $n_t n_t=1$ to Eq.~(\ref{e105})
\beq
\mathcal{L}_{\mathcal{X}n} = \frac{i}{2} \mbox{Tr} \left(\bar{\mathcal{X}} \gamma^\mu \left[\partial_\mu -i A_\mu^\ell \sigma^\ell \right]  \mathcal{X} \right)  - i n_t \mbox{Tr} \left( \bar{\mathcal{X}} \Gamma^t \mathcal{X} \right) \,, \label{e106}
\eeq
we integrate out the Dirac fermions following the analysis of Ref.~\cite{Abanov:1999qz} and obtain 
\beq
\mathcal{L}_n = \frac{1}{2g} \left( \partial_\mu n_t \right)^2 + 2 \pi i \Gamma[n_t] \label{wzw}
\eeq
The last term is the Wess-Zumino-Witten (WZW) term at level 1: it is a Berry phase associated with spacetime textures of $n_t$, a higher dimensional analog of the Berry phase of a single spin which is proportional to area enclosed by a spherical path \cite{LS15,Wang17}: an explicit expression of $\Gamma[n_t]$ requires 4+1 dimensions with an emergent spatial direction. 
Upon reduction to a O(3) non-linear sigma model for the N\'eel order parameter $\mathbfcal{N}$ in Eq.~(\ref{neelz}), the WZW term reduces \cite{TanakaHu} to the Berry phases of the monopoles noted near Eq.~(\ref{cp3}).

Also, note that while the SO(5) symmetry is explicit in the fermionic spinon theory in Eq.~(\ref{e105}), it is not explicit in the bosonic spinon theory in Eq.~(\ref{cp3}), but expected to be emergent \cite{SenthilFisher06}.

The form in Eq.~(\ref{wzw}) has been exploited in recent numerical work on the fuzzy sphere \cite{Fuzzy24}. 
Their results, and those of a number of other numerical works \cite{Yasir19,Nahum:2015vka,Becca20,Imada21,Meng24,Gu24,Chester24,Sandvik24} show that the critical spin liquid defined by 
Eq.~(\ref{cp3}), Eq.~(\ref{e105}), or (\ref{wzw}) is stable over a substantial intermediate energy and length scales, before ultimately confining into a N\'eel or VBS state. This intermediate range stability is not a bug, but a feature ideal for our purpose in Section~\ref{sec:FLs} of defining a FL* state at intermediate temperatures, which ultimately confines to variety of other states at low temperatures.

\section{Holon metal from a quantum spin liquid with bosonic spinons}
\label{sec:holonmetal}

This section dopes the spin liquid state of Section~\ref{sec:6A} with holes of density $p$. We already discussed the fate of the ordered antiferromagnet upon doping in Section~\ref{sec:sdw}, and here we will discuss the fate of the spin liquid.

For the N\'eel state in Section~\ref{sec:sdw} we obtained a Fermi liquid metal with hole pockets of spin-1/2, charge $e$ quasiparticles. After accounting for the doubling of the unit cell, there are 2 independent pockets in Fig.~\ref{fig:sdw2}, and so the fractional area of each pocket is $p/4$.

In the present section, doping the spin liquid will yield a {\it holon metal\/}, which has Fermi pockets of {\it spinless\/}, charge $e$ quasiparticles, {\it i.e.\/} holons. There is no doubling of the unit cell here, and we will see that there are 4 independent pockets. So the area of each holon pocket is again $p/4$. But such pockets are not directly observable in photoemission, as the electron operator is a composite of a holon and a spinon. However the holon pockets will contribute to magnetotransport observables restricted to motion within each square lattice plane.

We will argue in Section~\ref{sec:FLs} that the holon metal does not provide a satisfactory description of the cuprate pseudogap. However, recent work \cite{LiebScience,LiebPRB} has argued that the holon metal descending from a $\mathbb{Z}_2$ spin liquid is indeed appropriate for the Hubbard model on the Lieb lattice.

A popular way to dope a spin liquid is to introduce a fermionic holon $h$, so that the electron operator can be written as
\beq
c_{\vi \alpha}^{\vphantom\dagger} = h_\vi^\dagger b_{\vi \alpha}^{\vphantom\dagger}\,. \label{eq:chb}
\eeq
However, this approach runs into complications associated with the sublattice assignment of opposite U(1) gauge charges in Eq.~(\ref{bosepar}). We shall instead follow a simpler approach which is ultimately equivalent to Eq.~(\ref{eq:chb}), but is applicable more broadly. The central idea here is to transform the electron operator into a rotating reference frame in spin space \cite{SS80,Schulz90,Dupuis02,Dupuis04,SS09,DCSS15b,DCSS15,CSS17,MSSS18,SSST19,Bonetti22,Bonetti23,LiebPRB}
\beq
\left( \begin{array}{c} c_{\vi, \uparrow} \\ c_{\vi, \downarrow} \end{array} \right) = R_i \left( \begin{array}{c} \psi_{\vi,+} \\ \psi_{\vi,-} \end{array} \right), \quad\quad R_\vi^\dagger R_\vi^{\vphantom\dagger} = R_\vi^{\vphantom\dagger} R_\vi^\dagger = \mathds{1}.
\label{R}
\eeq
The fermions in the rotating reference frame are spinless `chargons' $\psi_{s}$, with $s=\pm$, carrying the electromagnetic charge.
The SU(2) rotation matrix $R_{\vi}$ is represented by two complex numbers
\begin{equation}
R_\vi = \begin{pmatrix}
z_{\vi,\uparrow} & -z_{\vi,\downarrow}^* \\  z_{\vi,\downarrow} & z_{\vi,\uparrow}^*
\end{pmatrix}, 
\qquad |z_{\vi,\uparrow}|^2 + |z_{\vi,\downarrow}|^2 = 1. \label{SpinonInZ}
\end{equation}
Indeed, we identify $z_\alpha$ with the spinons of the spin liquid described by the complex scalars of the $\mathbb{CP}^1$ field theory in Eq.~(\ref{cp11}).
Note that $\sum_{p=\pm} \psi_{\vi, p}^\dagger \psi_{\vi, p} = \sum_{\alpha=\uparrow,\downarrow} c_{\vi,\alpha}^\dagger c_{\vi , \alpha}$, so the density of the chargons is the same as that of the electrons.

A key property of the decomposition in Eq.~(\ref{R}) is that it introduces a SU(2) gauge invariance \cite{SS09}. 
\begin{align}
R_\vi^{\vphantom\dagger} & \rightarrow R_\vi^{\vphantom\dagger} \, W_\vi^\dagger \nonumber \\
 \left( \begin{array}{c} \psi_{\vi,+} \\ \psi_{\vi,-} \end{array} \right)  & \rightarrow W_\vi \left( \begin{array}{c} \psi_{\vi,+} \\ \psi_{\vi,-} \end{array} \right) 
 \,,
 \label{GaugeTrafo}
\end{align}
where $W_\vi$ is the SU(2) matrix generating the gauge transformation.
This gauge invariance is distinct from the SU(2) gauge invariance of the fermionic spinons in Section~\ref{sec:fermions}, which is instead associated with a transformation in pseudospin space. 

We now apply this tranformation to the paramagnon theory in Eq.~(\ref{ZHubbard}). Then, we write the Yukawa coupling between the 
paramagnon $\vP_{\vi}$ and the electrons as
\beq
\vP_{\vi} \cdot c_{\vi \alpha}^\dagger \frac{{\bm \sigma}_{\alpha\beta}}{2} c_{\vi \beta} \Rightarrow {\bm H}_{\vi} \cdot \psi_{\vi, s}^\dagger \frac{{\bm \sigma}_{ss'}}{2} \psi_{\vi , s'}\,.
\eeq
We have now introduced a `Higgs' field ${\bm H}_\vi$ \cite{SS09} given by
\begin{equation}
{\bm \sigma} \cdot {\bm H}_\vi  =  R_\vi^\dagger \, {\bm \sigma} \cdot \vP_{\vi} \, R_\vi \label{H}.
\end{equation}
It is easy to deduce from Eq.~(\ref{H}) that ${\bm H}$ transforms as an adjoint under the SU(2) gauge transformation
\beq
 {\bm \sigma} \cdot {\bm H}_\vi \rightarrow W_\vi \, {\bm \sigma} \cdot {\bm H}_\vi\, W_\vi^\dagger ~
\eeq

The action of the SU(2) gauge transformation $W_i$, should
be distinguished from the action of global SU(2) spin rotations $\Omega$ under which
\begin{align}
R_\vi & \rightarrow \Omega \, R_\vi  \nonumber \\ 
\left( \begin{array}{c} c_{\vi\uparrow} \\ c_{\vi\downarrow} \end{array} \right) & \rightarrow
\Omega \left( \begin{array}{c} c_{\vi\uparrow} \\ c_{\vi\downarrow} \end{array} \right) \nonumber \\
\psi_{i,s} & \rightarrow \psi_{i,s} \nonumber \\
{\bm H}_i & \rightarrow {\bm H}_i\,.
\end{align}
Note that $R$ carries global spin, and so describes spinons. On the other hand ${\bm H}$ and $\psi$ are spinless.

We can now use the SU(2) gauge theory of $\psi$, $R$, and ${\bm H}$ to describe the phases of the Hubbard model. The many possibilities are discussed at some length in Refs.~\cite{SS09,CSS17,MSSS18,SSST19,Bonetti22,Bonetti23,LiebPRB}. We limit ourselves here to the simplest case of the state obtained by rotationally averaging the N\'eel state of Section~\ref{sec:sdw}. The doped spin liquid so obtained, the holon metal, corresponds to condensing the Higgs field as 
\beq
\left\langle {\bm H}_\vi \right\rangle = \eta_{\vi}\, (0,0, H_0) \label{HiggsNeel}
\eeq
where $\eta_\vi$ is the sublattice staggering of Eq.~(\ref{defeta}). Such a condensate preserves spin rotation invariance, but breaks the SU(2) gauge invariance down to U(1). So the low energy theory of the holon metal is a U(1) gauge theory, with a gauge field $a_\mu$ which is the same as that introduced in Eq.~(\ref{gaugetrans3}), and which also appears in the $\mathbb{CP}^1$ theory of $R$ in Eq.~(\ref{cp11}). The effective theory for the chargons $\psi$ can be obtained from Eq.~(\ref{ZHubbard}) after
({\it i\/}) replacing the electrons $c_{\vi \alpha}$ by the chargon $\psi_{\vi,s}$, ({\it ii\/}) replacing the paramagnon $\vP_{\vi}$ by the Higgs condensate in Eq.~(\ref{HiggsNeel}), and ({\it iii\/}) minimally coupling the $\psi$ to the U(1) gauge field $a_\mu$.
So we obtain the effective chargon action
\begin{align}
\mathcal{S}_\psi = \int_0^\beta  \diff\tau\Biggl[&\sum_{\vi,s} \psi^\dagger_{\vi,s}(\partial_\tau - \mu - i a_\tau) \psi^\pdagger_{\vi,s} - \sum_{\vi,\vj,s,s'} t_{\vi\vj} e^{i a_{\vi\vj}} \psi^\dagger_{\vi,s} \psi^\pdagger_{\vj,s} \nonumber \\ &~~~~~~~~+  \sum_{\vi,s,s'} H_0 \cdot \psi^\dagger_{\vi,s} \frac{\sigma^z_{ss'}}{2} \psi^\pdagger_{\vi,s'} \Biggr], \label{ChargonPart0}
\end{align}
where $t_{\vi\vj}$ are the hoppings corresponding to the dispersion $\varepsilon_{\bm k}$ in Eq.~(\ref{ZHubbard}).
At leading order, this yields chargon pockets of spinless fermions which are the same as the pockets of electrons in the ordered N\'eel state presented in Section~\ref{sec:sdw}.
At higher order, the gauge field will renormalize the dispersion so that the pockets lose the reflection symmetry about the magnetic Brillouin zone boundary present in the N\'eel state: the Brillouin zone of the holon metal is the same as that of underlying antiferromagnet without symmetry breaking.

The holon metal in Eq.~(\ref{ChargonPart0}) is associated with a U(1) spin liquid. If we replace Eq.~(\ref{HiggsNeel}) by a non-collinear configuration, then we obtain a $\mathbb{Z}_2$ spin liquid, as is argued to be the case on the Lieb lattice \cite{LiebPRB}.

\section{$d$-wave superconductor and charge orders at half-filling\\ from a quantum spin liquid with fermionic spinons}
\label{sec:halffilling}

We now turn to the fermionic spinon theory of an insulating square lattice spin liquid in Section~\ref{sec:fermions}.
We wish to consider a more general situation in which the gap to charged excitations can vanish \cite{ChristosLuo24}. In the cuprates, gapless charged excitations appear when we dope the antiferromagnet. But we can also consider the case where the charge gap vanishes while the electronic density remains the same as in an insulator {\it i.e.\/} at half-filling. This section will focus on the simpler situation at half-filling, and the full treatment of the doped case is deferred to Section~\ref{sec:FLs}. It turns out that the effective action for the charged degrees of freedom is quite similar to that in the doped case, and so this is a sensible way to proceed.

At half-filling, there can be an emergent particle-hole symmetry, which simplifies considerations of quantum criticality.
We can decrease the charge gap by describing the insulator by an underlying Hubbard model with on-site repulsion $U$, and reducing the value of $U$, or by adding additional off-site interactions. Such models have been considered in numerical studies \cite{AssaadImada,Assaad22,Assaad24,Assaad25,Xu:2020qbj,Scaletter21,Scaletter22,HongYao21,HongYao22,HongYao25,Zhaoyu22,Zhaoyu24}. We will now show that the SU(2) gauge theory of Eq.~(\ref{eq:fermionhop2}) has fates other than those in Fig.~\ref{fig:Neelvbs} once charged excitations are included, the most interesting of which is a $d$-wave superconductor with gapless nodal quasiparticles. 

In terms of adiabatic continuity, this $d$-wave superconductor is the same as the BCS superconductor observed in the cuprates. However, the $d$-wave superconductor obtained in this section has one significant quantitative difference from the observations: it has a Lorentz-invariant form of its dispersion, with the two velocities only the square lattice diagonals, $v_F$ and $v_\Delta$, being equal to each other (see Fig.~\ref{fig:flsdsc}B). The cuprates instead have $v_F \gg v_\Delta$. We will resolve this problem in an interesting manner in Section~\ref{sec:aniso} when we consider the transition from FL* to a $d$-wave superconductor in the doped case.

The only matter field in Section~\ref{sec:fermions} is the fermion $\mathcal{F}$, which has electrical charge 0, spin $1/2$, and is a gauge SU(2) fundamental. As we are allowing for charged fluctuations, we need to define an electron operator, which has charge $-e$, spin $1/2$, and is a gauge SU(2) singlet. This directly leads us to introducing a boson $B$, first introduced in Ref.~\cite{Fradkin88}, which has charge $+e$, spin 0, 
and is a gauge SU(2) fundamental, so that a composite of $\mathcal{F}$ and $B$ will have the same quantum numbers as the electron (see also Ref.~\cite{HermeleHoneycomb} for work on the honeycomb lattice). We now show that this information is sufficient to deduce an effective action for $B$, and to reach our main conclusions. We will give a more microscopic definition of the field $B$ in the doped case later near Eq.~(\ref{Yukawa}).

First, similar to Eq.~(\ref{Fmatrix}), we introduce a matrix notation for the electron $\mathcal{C}$ and the boson $B$:
\bea
& \mathcal{C}_\vi \equiv \left(
\begin{array}{cc}
c_{\vi \uparrow} & - c_{\vi \downarrow} \\
c_{\vi \downarrow}^\dagger & c_{\vi \uparrow}^\dagger
\end{array}
\right) \,, \quad B_\vi \equiv \left( \begin{array}{c} B_{1\vi}  \\ B_{2\vi} \end{array} \right) \,, \quad  \mathcal{B}_\vi \equiv \left( \begin{array}{cc} B_{1\vi} & - B_{2\vi}^\ast \\ B_{2\vi} & B_{1\vi}^\ast \end{array} \right) \,, \label{defB}
\eea
all of which obey the reality condition analagous to Eq.~(\ref{fdft}). Then we write the fact that the electron operator is a composite of the boson $B$ (the `chargon') 
and the spinon $\mathcal{F}$ by
\beq
    \mathcal{C}_{\vi}^{\vphantom\dagger} \sim \mathcal{B}_\vi^\dagger  \,\mathcal{F}_\vi^{\vphantom\dagger} \,. \label{eq:CBF}
\eeq
In terms of the matrix components, we can write Eq.~(\ref{eq:CBF}) as 
\beq
c_{\vi \alpha}^{\dagger} \sim
B_{1\vi}^{\vphantom\dagger} f_{\vi \alpha}^\dagger + B_{2\vi}^{\vphantom\dagger} \varepsilon_{\alpha\beta}^{\vphantom\dagger} f_{\vi \beta}^{\vphantom\dagger}  \,, \label{cBf}
\eeq
where $\varepsilon_{\alpha\beta}$ is the unit antisymmetric tensor for spin SU(2) in Eq.~(\ref{defepsilon}).
Comparing to the transformation to a rotating reference frame in spin space in Eq.~(\ref{R}) in Section~\ref{sec:holonmetal}, we see that Eq.~(\ref{eq:CBF}) corresponds to a transformation to a rotating reference frame in Nambu pseudospin space, with $\mathcal{B}_\vi$ serving as the rotation matrix \cite{MSSS18}.
The parton decomposition in Eqs.~(\ref{eq:CBF},\ref{cBf}) can also be obtained explicitly from a path integral analysis of the Hubbard model \cite{HermeleHoneycomb}.

Now we can use Eq.~(\ref{eq:CBF}) to determine the symmetry transformations of $\mathcal{B}$ from those of $\mathcal{F}$, using the fact the $\mathcal{C}$ is gauge invariant and transforms trivially under all symmetries.
A boson with the same quantum numbers as our $B$ was introduced in earlier work \cite{Fradkin88,AndreiColeman2,LeeWen96}, but with an important difference. In the earlier work, the expectation value of $B^\dagger B$ was set to be equal to the doping density. 
That is not the case here, as we also include second order time derivatives in $B$ (see Eq.~(\ref{Lph})), and the doping density also includes the density of fermionic holes (see Section~\ref{sec:pseudogap}). 

The generalization of the SU(2) gauge transformation in Eq.~(\ref{SU2gauge}) is 
\bea
\mathcal{C}_\vi \rightarrow \mathcal{C}_\vi \quad & , \quad 
\mathcal{F}_\vi \rightarrow V_\vi \, \mathcal{F}_\vi   \nonumber \\
\mathcal{B}_\vi \rightarrow V_\vi \, \mathcal{B}_\vi \quad & , \quad 
U_{\vi\vj} \rightarrow  V_\vi \, U_{\vi\vj} \, 
V_\vj^{\dagger} \,, \label{eq:gauge}
\eea
while the generalization of the global SU(2) spin rotation in Eq.~(\ref{su212b}) is
\bea
\mathcal{C}_\vi \rightarrow \mathcal{C}_\vi \, \sigma^z R^T \sigma^z \quad & , \quad 
\mathcal{F}_\vi \rightarrow \mathcal{F}_\vi \, \sigma^z R^T  \sigma^z \nonumber \\
\mathcal{B}_\vi \rightarrow  \mathcal{B}_\vi \quad & , \quad 
U_{\vi\vj} \rightarrow  U_{\vi\vj}  \,.
\label{eq:spin}
\eea
Finally, the U(1) charge conservation symmetry acts as
\bea
\mathcal{C}_\vi \rightarrow  \Theta \,\mathcal{C}_\vi \, \quad & , \quad 
\mathcal{F}_\vi \rightarrow \mathcal{F}_\vi   \nonumber \\
\mathcal{B}_\vi \rightarrow  \mathcal{B}_\vi \, \Theta^\dagger \quad & , \quad 
U_{\vi\vj} \rightarrow  U_{\vi\vj}  \,,
\label{eq:charge}
\eea
where 
\bea
    \Theta = \left( \begin{array}{cc}
        e^{i\theta} & 0 \\
         0 & e^{-i \theta}
    \end{array}
    \right)\,.
\eea
See also Table~\ref{tab2} later for a summary of these gauge and symmetry transformations.

We now obtain an energy functional for $B$ in a Landau-type expansion \cite{Christos:2023oru}. Such a functional must also involve the gauge field $U_{\vi \vj}$ of Section~\ref{sec:fermions} to maintain gauge invariance. The fermion $f$ experiences a $\pi$ flux with pure imaginary hopping, while the electron $c$ has purely real hopping with zero flux (in the absence of an applied physical magnetic field). From these facts and Eq.~(\ref{eq:CBF}) we reach the important conclusion that the boson $B$ must also have purely imaginary hopping with $\pi$-flux (the $i w$ term in Eq.~(\ref{Bfunctional}) below). So the relation
\beq
T_x T_y = - T_y T_x\,, \label{txty}
\eeq
realizing the $\pi$-flux applies both to the spinons and to $B$.
We can also reach these conclusions, and obtain other constraints, by examining the action of all symmetry operators of $f$, and use Eq.~(\ref{eq:CBF}) to deduce the action of symmetry operations on $B$: the results are summarized in Table~\ref{tab1}.
\begin{table}
    \centering
    \begin{tabular}{|c|c|c|}
\hline
Symmetry & $f_{\alpha}$ & $B_a $  \\
\hline 
\hline
$T_x$ & $(-1)^{y} f_{ \alpha}$ & $(-1)^{y} B_a $ \\
$T_y$ & $ f_{ \alpha}$ & $ B_a $ \\
$P_x$ & $(-1)^{x} f_{ \alpha}$  & $(-1)^{x} B_a $ \\
$P_y$ & $(-1)^{y} f_{ \alpha}$  & $(-1)^{y} B_a $ \\
$P_{xy}$ & $(-1)^{xy} f_{ \alpha}$ & $(-1)^{x y} B_a $ \\
$\mathcal{T}$ & $(-1)^{x+y} \varepsilon_{\alpha\beta} f_{ \beta}$  & $(-1)^{x + y} B_a $\\
\hline
\end{tabular}
    \caption{Projective transformations of the $f$ spinons and $B$ chargons on lattice sites $\vi = (x,y)$ 
    under the symmetries $T_x: (x,y) \rightarrow (x + 1, y)$; $T_y: (x,y) \rightarrow (x,y + 1)$; 
    $P_x: (x,y) \rightarrow (-x, y)$; $P_y: (x,y) \rightarrow (x, -y)$; $P_{xy}: (x,y) \rightarrow (y, x)$; and time-reversal $\mathcal{T}$.
    The indices $\alpha,\beta$ refer to global SU(2) spin, while the index $a=1,2$ refers to gauge SU(2). 
    }
    \label{tab1}
\end{table}
These considerations lead to the energy functional $\mathcal{E}_2 [B, U] +  \mathcal{E}_4 [B, U]$ with terms quadratic and quartic in $B$ respectively.
At half-filling, the methods of Hermele \cite{HermeleHoneycomb} can be employed to obtain an explicit derivation from the Hubbard model, and we will outline a derivation from the ancilla method away from half-filling in Section~\ref{sec:ancilla}. Retaining only short-distance terms, we obtain
\bea
\mathcal{E}_2 [B, U] &=&  (r + 2 \sqrt{2} w) \sum_\vi B^\dagger_\vi B_\vi +  i w \sum_{\langle \vi \vj\rangle} e_{\vi \vj} \left( B_\vi^\dagger U_{\vi \vj} B_\vj - B_\vj^\dagger U_{\vj \vi} B_\vi \right) 
\nonumber \\ &~& + \kappa \sum_{\square} 
\left\{ 1- \frac{1}{2} \mbox{Re} \mbox{Tr} \prod_{\vi\vj \in \square} U_{\vi \vj} \right\} \nonumber \\
\mathcal{E}_4 [B, U] &=& \frac{u}{2} \sum_{\vi} \rho_{\vi}^2   + V_1 \sum_{\vi} \rho_{\vi} \left( \rho_{\vi + \hat{\bm x}} + \rho_{\vi + \hat{\bm y}} \right) 
 + g \sum_{\langle \vi \vj \rangle} \left| \Delta_{\vi\vj} \right|^2 
 + J_1 \sum_{\langle \vi \vj \rangle}  Q_{\vi\vj}^2 \nonumber \\
 &~&+ K_1 \sum_{\langle \vi \vj \rangle}  J_{\vi\vj}^2 + V_{11} \sum_{\vi} \rho_{\vi} \left( \rho_{\vi + \hat{\bm x}+ \hat{\bm y}} + \rho_{\vi +\hat{\bm x}- \hat{\bm y}} \right) \nonumber \\
&~&+ V_{22} \sum_{\vi} \rho_{\vi} \left( \rho_{\vi + 2\hat{\bm x}+ 2\hat{\bm y}} + \rho_{\vi + 2\hat{\bm x}- 2\hat{\bm y}} \right)\,.
\label{Bfunctional}
\eea
The quartic terms are expressed as products of bilinears of $B$ which are associated with various gauge-invariant observables as identified below
\bea
&&\mbox{site charge density:~}\left\langle c_{\vi \alpha}^\dagger c_{\vi \alpha}^{\vphantom\dagger} \right\rangle \sim \rho_{\vi} = B^\dagger_\vi B_\vi^{\vphantom\dagger} \nonumber \\
&&\mbox{bond density:~} \left\langle c_{\vi \alpha}^\dagger c_{\vj \alpha}^{\vphantom\dagger} + c_{\vj \alpha}^\dagger c_{\vi \alpha}^{\vphantom\dagger} \right\rangle~\sim Q_{\vi \vj} = Q_{\vj\vi} = \mbox{Im} \left(B^\dagger_\vi e_{\vi \vj\vphantom\dagger} U_{\vi\vj\vphantom\dagger} B_\vj \right) \nonumber \\
&&\mbox{bond current:~} i\left\langle c_{\vi \alpha}^\dagger c_{\vj \alpha}^{\vphantom\dagger} - c_{\vj \alpha}^\dagger c_{\vi \alpha}^{\vphantom\dagger} \right\rangle \sim J_{\vi \vj} = - J_{\vj\vi} =  \mbox{Re} \left( B^\dagger_\vi e_{\vi \vj\vphantom\dagger} U_{\vi \vj\vphantom\dagger} B_\vj^{\vphantom\dagger} \right) \nonumber \\
&&\mbox{Pairing:~} \left\langle \varepsilon_{\alpha\beta} c_{\vi \alpha} c_{j \beta} \right\rangle \sim \Delta_{\vi \vj} = \Delta_{\vj \vi} = \varepsilon_{ab} B_{a\vi} e_{\vi \vj} U_{\vi \vj} B_{b\vj}\,. 
\label{latticeorders}
\eea
We have retained terms involving nearest neighbor sites, and a few terms with longer-range density-density interactions. 

\begin{figure}
\centering
\includegraphics[width=6in]{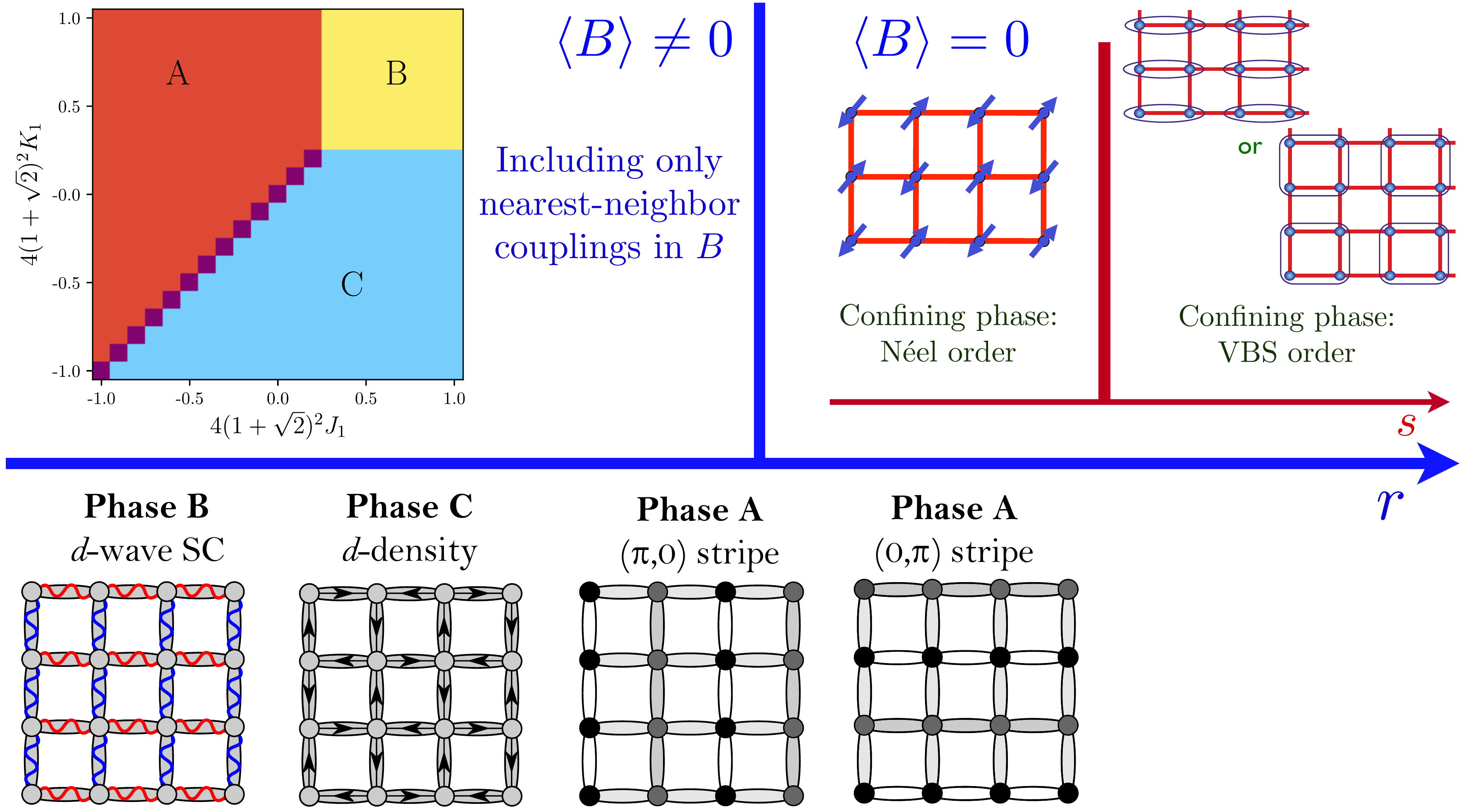}
\caption{Mean field phase diagram obtained by minimization of the Higgs potential of $B$, $\mathcal{E}_2 + \mathcal{E}_4$ (from Ref.~\cite{Christos:2023oru}). For $r>0$, there is no Higgs condensate $\langle B \rangle =0$, and we obtain the same phases as in the insulator from the confinement of the fermionic spinons described by Eq.~(\ref{eq:fermionhop2}). For $r<0$, $\langle B \rangle\neq 0$, and we minimized the Higgs potential with only nearest neighbor interactions by setting $V_{11}=V_{22}=0$.}
\label{fig:pd1}
\end{figure}
A schematic global phase diagram is shown in Fig.~\ref{fig:pd1}, as a function of the tuning parameter $r$, which is the `mass' of $B$. 
\begin{itemize}
\item
When $r$ is large and positive, then we can ignore the $B$ sector, and revert to the spinon only theory of Section~\ref{sec:fermions}. The low energy theory is Eq.~(\ref{e105}), and we expect a confining insulator with either N\'eel or VBS order as the ground state. 
\item 
When $r$ is negative, $B$ condenses, and this has the salutary effect of making the gauge field $A$ massive, as in the Higgs phenomenon. In this case, a mean-field treatment of interactions in the bosonic sector by minimizing $\mathcal{E}_2 [B, U] + \mathcal{E}_4 [B, U]$  is qualitatively valid.
\end{itemize}
For negative $r$, Fig.~\ref{fig:pd1} shows the phases obtained by minimizing the energy functional with nearest-neighbor interactions only, $V_{11} = V_{22}=0$.
Three phases are found:
\begin{enumerate}[label=\Alph*.]
\item This state has charge stripe order with period 2, centered on the sites.
\item A $d$-wave superconductor, with $\Delta_{\vi,\vi+\hat{\bm x}} = - \Delta_{\vi,\vi+\hat{\bm y}}$.
\item A ``$d$-density wave'' state which has a staggered pattern of spontaneous current.
\end{enumerate}
Including further neighbor interactions can induce other phases, including period-4 charge order and pair density waves \cite{Christos:2023oru,BCS24,Sayantan25}), 
as shown in Fig.~\ref{fig:period4}.
\begin{figure}
    \centering
    \includegraphics[width=3.5in]{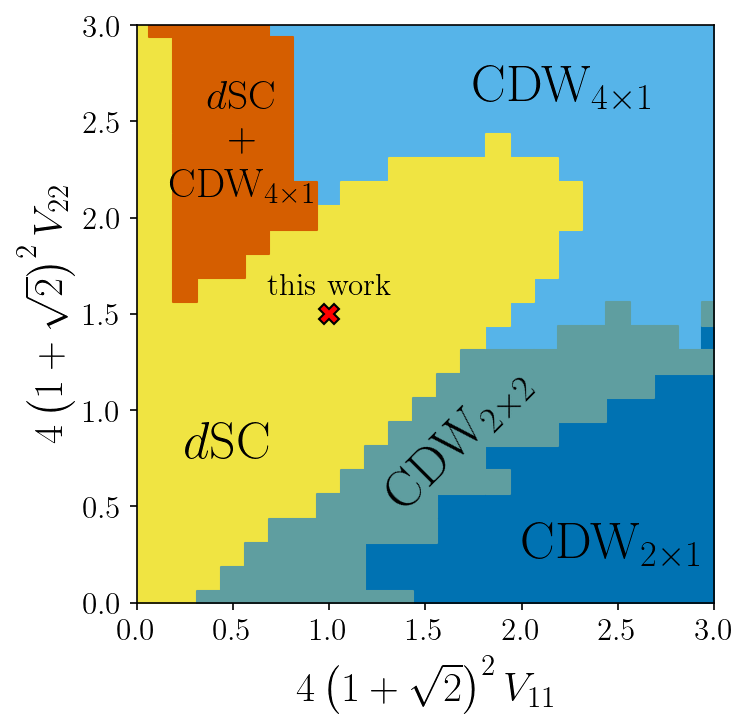}
    \caption{From Ref.~\cite{Sayantan25}. Mean-Field phase diagram of Eq.~(\ref{Bfunctional}) a function of the further neighbor interactions $V_{11}$ and $V_{22}$, extending that in Fig.~\ref{fig:pd1}. The other parameters are
    $r=-0.732$, $w=0.40$, $u=0$, $V_1=0$, $g=0.021446$, $J_1=K_1={2}/({4(1+\sqrt{2})^2})$.
    $d$SC is $d$-wave superconductivity, CDW$_{n\times m}$ is a charge density wave with a supercell with $n\times m$ lattice sites. The red cross marks the parameter values chosen for the Monte Carlo simulations in Ref.~\cite{Sayantan25}.}
    \label{fig:period4}
\end{figure}    

The present theory of multiple orders should be contrasted from more conventional Landau theory approaches \cite{Fradkin10,Hayward:2013jna,Lee14,Nie_15,Fradkin15,Castro17,Pepin23,Fradkin25}. In our case, we employ a fractionalized order parameter $B$, whose gauge-invariant composites can represent multiple ordering patterns and broken symmetries. Upon integrating strong gauge fluctuations of $U$, the fractionalized approach will yield an effective action which is essentially the same as in the Landau theory. But an important advantage of the fractionalized approach is that allows a local description of the fermionic spectrum in the pseudogap metal at non-zero temperatures \cite{Sayantan25}. Moreover, 
an interesting feature of the fractionalized approach is that the gauge-invariant orders are degenerate in the quadratic energy functional $\mathcal{E}_2$, and the degeneracy is broken only at quartic order in $\mathcal{E}_4$. The fact that the leading term is degenerate provides a rationale for nearly-degenerate multiple competing or `intertwined' orders \cite{Christos:2023oru}; Landau theory approaches \cite{Fradkin10,Hayward:2013jna,Lee14,Nie_15,Fradkin15,Pepin23,Fradkin25} do not have any term in which the degeneracy is exact without fine-tuning.

Our primary interest for now is the $d$-wave supercondutor, phase B. Remarkably, the structure of the $\pi$-flux spin liquid, and consequently  the $\pi$-flux on $B$ leads to $d$-wave pairing, and not $s$-wave pairing. Also, once $B$ is condensed, we can identify $c \sim f$ via Eq.~(\ref{eq:CBF}), and so the electron spectral function will inherit nodal Bogoliubov quasiparticles from the massless Dirac spinons \cite{BZA87,Zhang88,KotliarLiu88,LeeWen96,IvanovSenthil,LeeWenRMP}. 
The main phenomenological difficulty is that the Bogoliubov quasiparticles will have isotropic dispersion, as in Eq.~(\ref{eq:fermiondisp}) and Fig.~\ref{fig:fermiondisp}, in contrast to observations \cite{Chiao00}; we will address this difficulty in Section~\ref{sec:FLs}. However, other features of the $d$-wave state obtained from the energy functional in 
Eq.~(\ref{Bfunctional}) do match observations, including vortices with flux $h/(2e)$ (despite the boson $B$ having charge $e$), and competing charge order halos of vortex cores.

Going beyond a classical treatment of $B$ and $U$, we can consider a quantum lattice model to study the interplay of the full array of competing order parameters and deconfined criticality at half-filling. This is obtained by extending the energy functionals in Eq.~(\ref{Bfunctional}) and the lattice fermion theory in Eq.~(\ref{eq:fermionhop2}) to
\beq
\mathcal{L} = \sum_\vi |D_\tau B_\vi |^2 + \mathcal{E}_2 [B, U] + \mathcal{E}_4 [B, U] + \sum_\vi \Psi_\vi^\dagger D_\tau \Psi_\vi + \mathcal{H}_{SLf} + \mathcal{L}_{YM}[U]\,. \label{Lph}
\eeq
Here $D_\tau$ represents a covariant time derivative with the SU(2) gauge field $U$, and $\mathcal{L}_{YM}$ is the lattice Yang-Mills Lagrangian for $U$. We have added only a relativistic time derivative term for $\hat{B}$, which is the allowed term at half-filling with particle-hole symmetry. 
A recent lattice gauge theory simulation \cite{SSMeng26} provides evidence that Eq.~(\ref{Lph}) displays a deconfined quantum critical point separating the insulating N\'eel state from a $d$-wave superconductor with nodal Bogoliubov quasiparticles.
Away from half-filling, we will also have to include the hole pockets to be discussed in Section~\ref{sec:FLs}, and that will likely lead to difficulties with the sign problem.

\section{Fractionalized Fermi liquids (FL*)}
\label{sec:FLs}

Our motivation for studying quantum spin liquids has been to find a basis for a theory of the cuprate phase diagram. Christos {\it et al.\/} proposed \cite{Christos:2023oru}
that the appropriate spin liquid was precisely that observed in various numerical studies of square lattice antiferromagnets: this is the spin liquid described by the $\mathbb{CP}^1$ theory of Section~\ref{sec:dqcp}, which is dual to the SU(2) gauge theory of fermionic spinons in Section~\ref{sec:fermions}. The previous sections have shown that the theories of directly doping such spin liquids run into difficulties:
\begin{itemize}
\item Doping the $\mathbb{CP}^1$ spin liquid yields a holon metal state described in Section~\ref{sec:holonmetal}, which is a candidate for the pseudogap metal. However, as we discuss in Section~\ref{sec:admr}, the holon metal is incompatible with recent angle-dependent magnetoresistance (ADMR) experiments \cite{Ramshaw22,Yamaji24}.
\item The fermionic spin liquid can lead to a $d$-wave superconductor, as described in Section~\ref{sec:halffilling}. However, the velocities of the nodal quasiparticles are nearly isotropic, with $v_F \sim v_\Delta$ \cite{Chiao00}.
\end{itemize}

We now show that replacing the holon metal with another doped spin liquid phase, the fractionalized Fermi liquid (FL*), resolves both (and other) difficulties above. It agrees with the ADMR measurements in the pseudogap, and a confinement transition from FL* yields a $d$-wave superconductor with anisotropic nodal velocities.

We begin by presenting the basic definition of the FL* state. Consider any system of $S=1/2$ fermions of density $\rho$, with spin rotation invariance preserved. In the Fermi liquid (FL) state, Luttinger's argument \cite{Luttinger60} states that the fractional volume of the Brillouin zone enclosed by the Fermi surface should be $\rho/2$, modulo integers to account for filled bands. Oshikawa \cite{MO00} placed Luttinger's argument on a non-perturbative basis by presenting a flux-piercing argument which has a modern interpretation as an `anomaly' associated with translations and charge conversation; see Ref.~\cite{QPMbook} for a review. This is summarized in Fig.~\ref{fig:fldef}.
\begin{figure}
\centering
\includegraphics[width=3.75in]{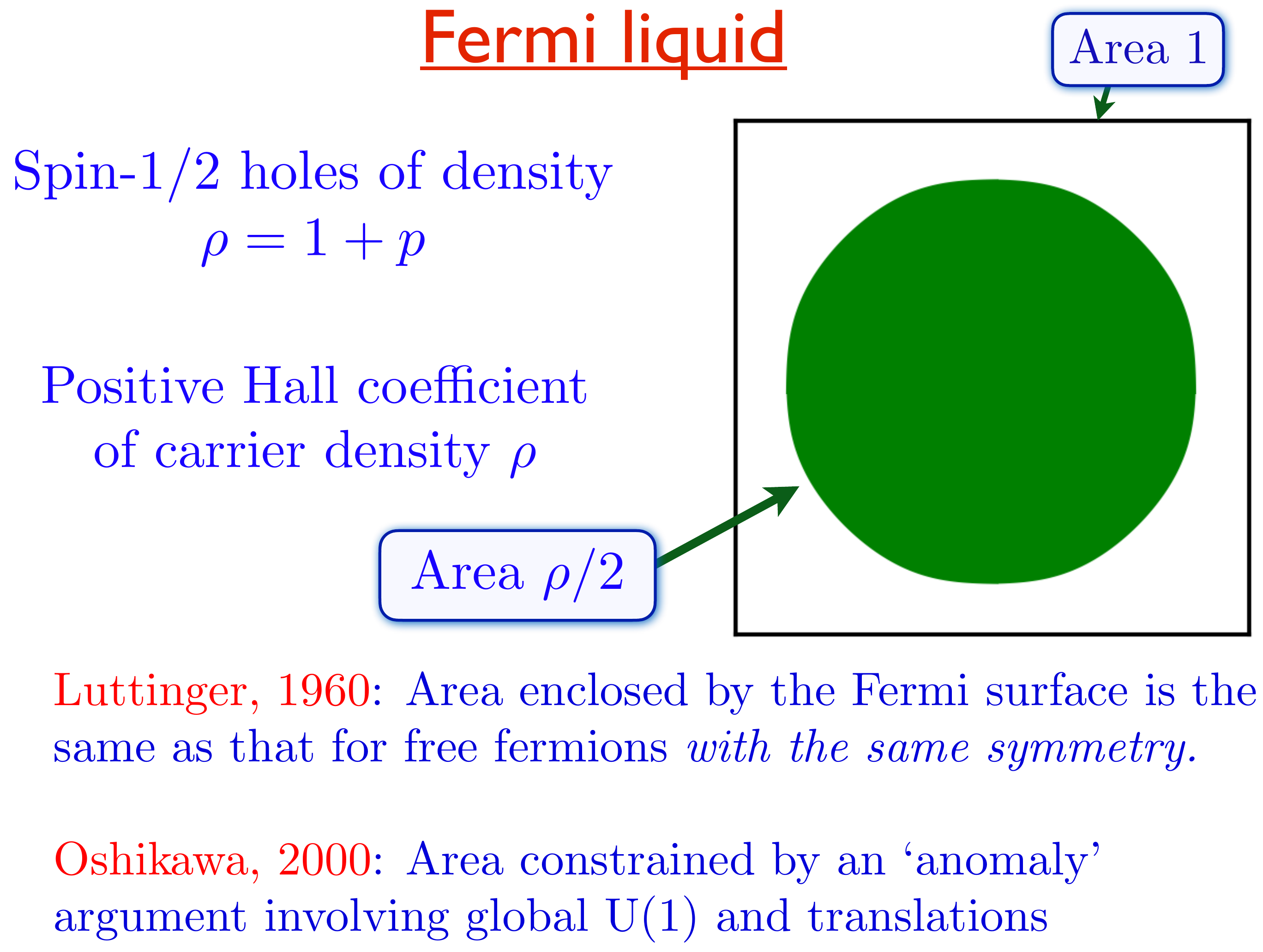}
\caption{Properties of the Fermi liquid (FL) state.}
\label{fig:fldef}
\end{figure}

The idea of a `fractionalized Fermi liquid' (FL*) was introduced in Refs.~\cite{TSSSMV03,TSSSMV04}, as a metallic state in which the Fermi surface did {\it not\/} obey the Luttinger constraint. In the simplest case, the fractional volume enclosed by the Fermi surface in FL* is $(\rho-1)/2$. As stated in Ref.~\cite{TSSSMV03}:
\begin{quote}
``The primary purpose of this paper is to show that there
exist nonmagnetic, metallic states (FL*) in dimensions
$d \geq 2$ with a Fermi surface of ordinary $S=1/2$, charge
$e$, sharp quasiparticles, enclosing a volume
\begin{equation}
\mathcal{V}_{\mathrm{FL}^*} = \mathcal{K}_d \left[ (\rho_a - 1) (\mathrm{mod 2})\right], \nonumber
\end{equation}
over a finite range of parameters.''
\end{quote}
(Here, $\rho_a$ is the total electron density, $\mathcal{K}_d = (2 \pi)^d /(2 v_0)$ is a phase space factor, $v_0$ is the volume of the unit cell, and the (mod 2) accounts for filled bands which we ignore here.)
The central point \cite{TSSSMV03,TSSSMV04,APAV04,Powell05,Coleman05,Qi10,SSMetlitskiPunk12,Bonderson16,SenthilElse21,QPMbook,Seiberg23,Meng26} was that it was possible to satisfy Oshikawa's anomaly-argument by combining the anomaly of a Fermi surface, which contributes an amount equivalent to a density $\rho-1$, with that of a fractionalized spin liquid, which contributes a quantized amount equivalent to a density 1; the results in Eqs.~(\ref{pins}), (\ref{mGS}), (\ref{wzw}) are realizations of such a quantized spin liquid anomaly \cite{Wang17,Metlitski18}.
It is trivially possible to shift $\rho$ by an even integer by adding and removing filled bands, and the novelty is the shift in FL* by an odd integer; note that the shift in density cannot be arbitrary and is quantized by the structure of the spin liquid. The properties of the FL* state are summarized in Fig.~\ref{fig:flsdef}.
\begin{figure}
\centering
\includegraphics[width=3.75in]{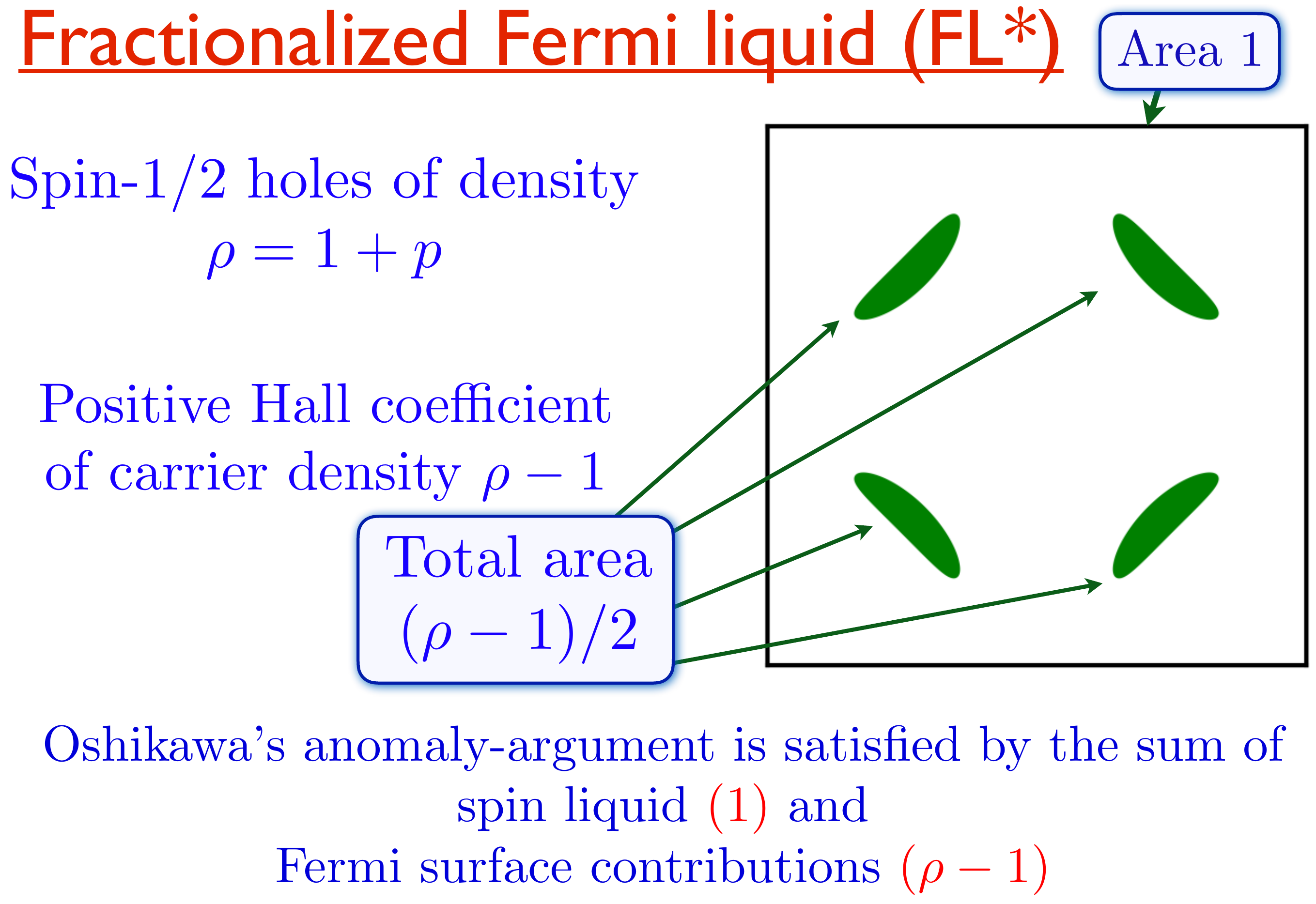}
\caption{Properties of the fractionalized Fermi liquid (FL*) state. The results in Eqs.~(\ref{pins}), (\ref{mGS}), (\ref{wzw}) are realizations of the quantized spin liquid anomaly \cite{Wang17,Metlitski18}.}
\label{fig:flsdef}
\end{figure}

We emphasize that the spin liquid is not required to have a spinon Fermi surface to compensate for the missing Luttinger volume in the FL* state. It is only required that its Oshikawa anomaly be an odd integer.

\FloatBarrier

\subsection{Generalized Luttinger constraints}
\label{sec:gLutt}

The conventional proof of the Luttinger theorem for a Fermi liquid \cite{Luttinger60} rests on an all orders expansion from a free fermion state, and shows that the volume enclosed by the Fermi surface depends only the density of fermions, and is independent of the strength of interactions. The proof involves certain assumptions on the structure of the perturbative expansion, which will be specified below.

In this subsection, we will present Luttinger's analysis in a more general context \cite{Powell05,Coleman05}. Rather than starting from a free fermion state, we will start with a simple saddle-point of a parton theory of a fractionalized metallic state. The perturbative expansion from this saddle point will necessarily involve emergent gauge fields, but this does not preclude application of Luttinger's methods as long the perturbative expansion does not lead to an instability of the saddle point (which we assume). This generalized analysis also leads to constraints on the volume enclosed by the Fermi surface, but the constraints are distinct from those of Luttinger. 

As an example of a parton theory, consider the parton decomposition in Eq.~(\ref{eq:chb}) of an electron into a spinful, neutral boson $b_\alpha$, and 
a spinless fermionic holon, $h$, of charge $e$. Associated with this parton decomposition is an emergent U(1) gauge invariance, and a constraint 
\beq
b_{\vi \alpha}^\dagger b_{\vi \alpha}^{\vphantom\dagger} + h_{\vi}^\dagger h_\vi^{\vphantom\dagger} = 1\,, \label{ll1}
\eeq
which fixes a unit background gauge charge on each site. Note that the value of the r.h.s. of Eq.~(\ref{ll1}) is a {\it rigid\/} property of the spin liquid which does not change upon doping. It is this background gauge charge which is responsible for or the quantized Berry phase in Eq.~(\ref{pins}) of $\pi$ for a vison encircling each site in a $\mathbb{Z}_2$ spin liquid \cite{RJSS91,SSMV99}, and the quantized monopole Berry phase \cite{NRSS90} for the $\mathbb{CP}^1$ spin liquid of Section~\ref{sec:U1}. We will see an explicit example of the constraint Eq.~(\ref{ll1}) being obeyed for the doped case in the quantum dimer model \cite{Punk15} discussed near Fig.~\ref{fig:disp4}.

Distinct from the U(1) gauge invariance, we also have a global U(1) symmetry associated with the conservation of electron density $\rho$. In the parton formulation, an electron density 1 has been absorbed by the background gauge charge of the spin liquid, and the global density constraint is on the density of holons
\beq
\left\langle h_{\vi}^\dagger h_\vi^{\vphantom\dagger} \right\rangle = \rho - 1 \label{ll2}\,.
\eeq
The $-1$ on the r.h.s. of Eq.~(\ref{ll2}) is therefore directly connected to the 1 in r.h.s. of Eq.~(\ref{ll1}).

As a second example of a parton theory, consider the SU(2) spin liquid of Section~\ref{sec:fermions}, which has neutral fermionic spinons $f_\alpha$, and a doublet $(B_1, B_2)$ of spinless, charge $e$ bosons upon doping in Section~\ref{sec:halffilling}. In the parton formulation of Ref.~\cite{LeeWen96}, the analog of the gauge charge constraint in Eq.~(\ref{ll1}) is
\beq
f_{\vi \alpha}^\dagger f_{\vi \alpha}^{\vphantom\dagger} + B_{1\vi}^\dagger B_{1\vi}^{\vphantom\dagger} - B_{2\vi}^\dagger B_{2\vi}^{\vphantom\dagger}  = 1\,, \label{ll3}
\eeq
and there is a similar constraint in the rotor formulation of Ref.~\cite{HermeleHoneycomb}. The r.h.s. of Eq.~(\ref{ll3}) is quantized by the SU(2) gauge symmetry. Now the global electron number symmetry leads to a constraint on the density of bosons
\beq
\left\langle B_{1\vi}^\dagger B_{1\vi}^{\vphantom\dagger} \right\rangle + \left\langle B_{2\vi}^\dagger B_{2\vi}^{\vphantom\dagger} \right\rangle = \rho - 1 \label{ll4}\,,
\eeq
in contrast to the density of fermions in Eq.~(\ref{ll2}).

Next, we follow the analysis of Ref.~\cite{Powell05} to determine the low energy consequences of the microscopic constraints associated with electron number conservation given in Eqs.~(\ref{ll2}) and (\ref{ll4}). Any renormalized theory must have a global U(1) electron number symmetry with a net charge matching that in Eqs.~(\ref{ll2},\ref{ll4}).
We assume that the renormalized theory is well described by a set of fields $F_s$ carrying charges $q_s$ under this global U(1) symmetry, 
so that the theory is invariant under
\beq
F_s \rightarrow F_s e^{i q_s \theta} \label{ll5}\,.
\eeq
The $F_s$ fields can be fermions or bosons, and will generally always include the parton fields of the microscopic theory. However, there can also be additional fields,
associated with `molecular' bound states of the partons. These bound states are associated with a pole in a parton-parton scattering $T$-matrix: such a pole can be replaced by the propagator of 
a canonical fermion or boson field representing the `molecule', along with a coupling allowing decay of the `molecule' into the `atomic' partons: see Fig.~\ref{fig:feshbach}.
\begin{figure}
\centering
\includegraphics[width=5.5in]{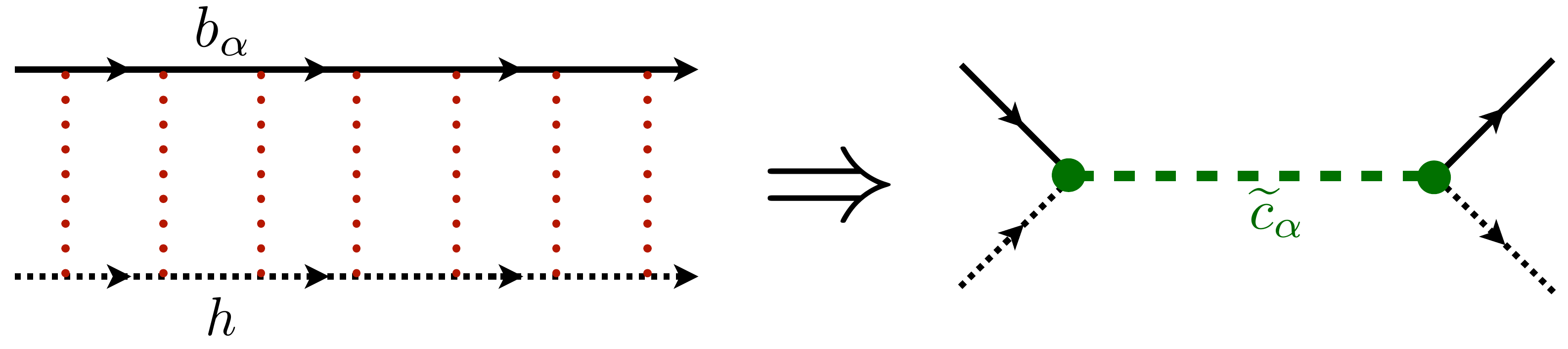}
\caption{$T$-matrix of holon-spinon scattering in the partons of Eq.~(\ref{eq:chb}) realized by an intermediate $\widetilde{c}_\alpha$ bound state. The green dot is the Yukawa coupling, the analog of which in the ancilla model SU(2) gauge theory is in Eq.~(\ref{Yukawa}).}
\label{fig:feshbach}
\end{figure}
With such a representation, it is possible to systematically treat the situation with a finite densities of partons and their bound states \cite{Powell05,NishidaSon1,NishidaSon2,NikolicSS}. This process is similar to the `two channel' approach to the Feshbach resonance \cite{Feshbach58,Feshbach62,Feshbach08}, and effective field theories in nuclear physics \cite{Kaplan97}.
In particular, we include a `molecular' fermionic field which has the same quantum numbers as the electron, $\widetilde{c}_\alpha$, as a renormalized quasiparticle obtained as a bound state of a spinon and a holon \cite{RKK07,RKK08,Qi10}. The complete set of low energy fields is difficult to specify in advance for a particular Hamiltonian, but {\it a posteriori} the requirement is that the Luttinger-type analysis is well behaved, as we will clarify below. We will present an explicit derivation of such a renormalized theory in the discussion on the ancilla model in Section~\ref{sec:ancilla}.

In a perturbative Feynman graph analysis \cite{Potthoff04}, the renormalized theory is characterized by corresponding renormalized Green's functions $G_s ({\bm p}, i \omega) $ and 
self energies $\Sigma_s ({\bm p}, i \omega) $ (assumed diagonal in the field index for convenience) which are related by a Luttinger-Ward functional $\Phi_{LW} [G_s]$ 
\beq
\Sigma_s ({\bm p}, i \omega)= \varsigma_s \frac{ \delta \Phi_{LW}}{\delta G_s ({\bm p}, i \omega)}\,, \label{flt23}
\eeq
where $\varsigma_s = -1 (+1)$ for fermions (bosons).
An important property of the Luttinger-Ward functional, following from the global U(1) symmetry in Eq.~(\ref{ll5}), is its invariance under frequency shifts
\beq
\Phi \left[ G_s({\bm p}, i \omega + i q_s \omega_0) \right] = \Phi \left[ G_s ({\bm p}, i \omega) \right]\,, \label{flt24}
\eeq
for any fixed $\omega_0$. Here, we are regarding $\Phi$ as functional of two sets of distinct functions $f_{s1,s2} (\omega)$, with $f_{s1} (\omega) \equiv G_s ({\bm p}, i \omega + i \omega_0)$ and  $f_{s2} (\omega) = G_s ({\bm p}, i \omega) $, and $\Phi$ evaluates to the same value for these two functions.

In terms of the renormalized theory, the microscopic constraints in Eqs.~(\ref{ll2}) and (\ref{ll4}) can be evaluated by computing the conserved global U(1) density:
\beq
\rho - 1 = -\frac{1}{V} \sum_{\bm p}   \sum_s \varsigma_s q_s d_s \int_{-\infty}^{\infty}\frac{d \omega}{2 \pi} G_s ({\bm p}, i \omega) e^{i \omega 0^+} \,, \label{flt20}
\eeq
where $d_s$ is the spin degeneracy of particle $s$.
Now we can proceed along exactly the lines described by Luttinger \cite{QPMbook}. We insert into Eq.~(\ref{flt20}) the following identity following directly from Dyson's equation
\beq
 G_s ({\bm p}, i \omega) =  i \frac{\partial}{\partial \omega} \ln \left[ G_s ({\bm p}, i \omega) \right]  - i G_s ({\bm p}, i \omega) \frac{\partial}{\partial \omega} \Sigma_s ({\bm p}, i \omega) \,. \label{flt21}
\eeq
Then we can write
\beq
\rho-1 = \mathcal{I}_{1p} + \mathcal{I}_{1z} + \mathcal{I}_2 \label{ll10}
\eeq
where
\bea
\mathcal{I}_{1p} &=&  - \left. \frac{i}{V} \sum_{\bm p} \sum_s \varsigma_s q_s d_s \int_{-\infty}^{\infty}\frac{d \omega}{2 \pi}  \frac{\partial}{\partial \omega} \ln \left[ G_s ({\bm p}, i \omega) \right] e^{i \omega 0^+} \right|_{ \mbox{only from poles of $G_s$}} \nonumber \\
\mathcal{I}_{1z} &=&  - \left. \frac{i}{V} \sum_{\bm p} \sum_s \varsigma_s q_s d_s \int_{-\infty}^{\infty}\frac{d \omega}{2 \pi}  \frac{\partial}{\partial \omega} \ln \left[ G_s ({\bm p}, i \omega) \right] e^{i \omega 0^+} \right|_{ \mbox{only from zeros of $G_s$}} \nonumber \\
\mathcal{I}_2 &=&  \frac{i}{V} \sum_{\bm p} \sum_s \varsigma_s q_s d_s \int_{-\infty}^{\infty}\frac{d \omega}{2 \pi} G_s ({\bm p}, i \omega) \frac{\partial}{\partial \omega} \Sigma_s ({\bm p}, i \omega)\,.
\eea
The $\mathcal{I}_{1p}$ term contains the conventional contribution of the Fermi surface where $G_s^{-1} ({\bm p}, \omega=0) = 0$, while the $\mathcal{I}_{1z}$ is often called the contribution of the ``Luttinger surface'' where $G_s ({\bm p}, \omega=0) = 0$.

Important ingredients in the conventional Luttinger arguments are that $\mathcal{I}_{1z} = 0$ and $\mathcal{I}_2 = 0$. It is assumed that the perturbation theory is stable, there are no poles in the self energy $\Sigma_s$, and consequently no zeros in $G_s$, and so $\mathcal{I}_{1z}$ vanishes. The vanishing of $\mathcal{I}_2$ follows after taking the $\omega_0$ derivative of Eq.~(\ref{flt24}), and integrating by parts in frequency. Then the conventional Luttinger area of $\rho/2$ follows by evaluating $\mathcal{I}_{1p}$ from the electron quasiparticle pole on the Fermi surface, with the l.h.s. in Eq.~(\ref{ll10}) equal to $\rho$ in this case.

For the parton theories being analyzed here, we obtain the FL* state by also assuming $\mathcal{I}_{1z} = \mathcal{I}_{2} =0$ because of the stability of the parton perturbation theory. For $\mathcal{I}_{1p}$, we assume that the only pole contribution is from the emergent electron-like holon-spinon bound state $\widetilde{c}_\alpha$, and there is no contribution to $\mathcal{I}_{1p}$ from any of the spinons and holons. Then we immediately obtain\\
\beq
\rho-1 = 2 \times \mbox{(Area of $\widetilde{c}_\alpha$ Fermi surfaces})
\eeq
{\it i.e.\/}
 the total area of the Fermi surfaces per spin is $(\rho-1)/2$ (in units where the area of the Brillouin zone is 1), illustrated in Fig.~\ref{fig:flsdef}.

There can also be contributions to $\mathcal{I}_{1p}$ from the fermionic holons $h$, and we obtain the holon metal state of Section~\ref{sec:holonmetal} if this is the only non-zero contribution to $\mathcal{I}_{1p}$ (and we have $\mathcal{I}_{1z} = \mathcal{I}_2 = 0$). If there are contributions to $\mathcal{I}_{1p}$ from both $\widetilde{c}_\alpha$ and $h$, then we obtain a hybrid state called the `holon-hole metal' \cite{RKK08}.

We note that bosonic quasiparticles, such as $B_{1,2}$ in Eqs.~(\ref{ll3},\ref{ll4}) cannot have zero frequency poles which contribute to $\mathcal{I}_{1p}$ \cite{Powell05}. Nevertheless Eq.~(\ref{ll4}) can be satisfied in a FL* state even though the l.h.s. has only boson number expectation values on the microscopic lattice scale---the contribution appears in the low energy theory from the appearance of spinon-holon bound states of $B_{1,2}$ and $f_\alpha$.

There are numerous non-perturbative analyses in the literature in which $\mathcal{I}_{1z}$ and/or $\mathcal{I}_2$ are non-zero \cite{Altshuler98,Dz03,Tsvelik06,YRZ,Kotliarzeros,Kane13,Scheurer18,Fabrizio22,Fabrizio22a,Si24,GleisKotliar24,Lucila1,Lucila2,P2_2025,GPS2,GKST,Shang21}.

The non-vanishing of $\mathcal{I}_2$ requires breakdown of integration by parts over frequency, and this is possible if the local density of states is divergent at zero frequency.
This happens for zero-dimensional models like the SYK model \cite{GPS2,GKST,Shang21}. 

The non-vanishing of $\mathcal{I}_{1z}$ is associated with a pole in the self-energy, which indicates the appearance of a new particle in some intermediate channel.
\begin{tcolorbox}
The non-vanishing of $\mathcal{I}_{1z}$ ({\it i.e.} a zero of the Green's function at $\omega=0$) implies a breakdown of perturbation theory, which should be fixed by expanding the set of renormalized fields $F_s$, so that the modified perturbative expansion has $\mathcal{I}_{1z} = 0$.
 \end{tcolorbox} 
 \noindent
This improved theory (which should also have $\mathcal{I}_2 = 0$) will then provide a better handle on the true excitation spectrum of the state under consideration: this is the procedure we follow in Section~\ref{sec:ancilla}.

\FloatBarrier

\subsection{FL* in single band models}
\label{sec:single}

We describe here explicit construction of the FL* phase in a single band model, such as the square lattice Hubbard model (possibly with additional short-range interactions) of interest for the cuprates.
At the level of cartoon pictures, we illustrate the structure of the FL*  and other states in Fig.~\ref{fig:metals} \cite{Punk15}.
This figure describes three distinct metallic phases in a Hubbard time model with electron density $1-p$. Recall that in the absence of a broken symmetry, the Luttinger constraint on hole Fermi surfaces is that they have a fractional area per spin of $(1+p)/2$, relative to the area of the full square lattice Brillouin zone.
\begin{figure}
\centering
\includegraphics[width=5.5in]{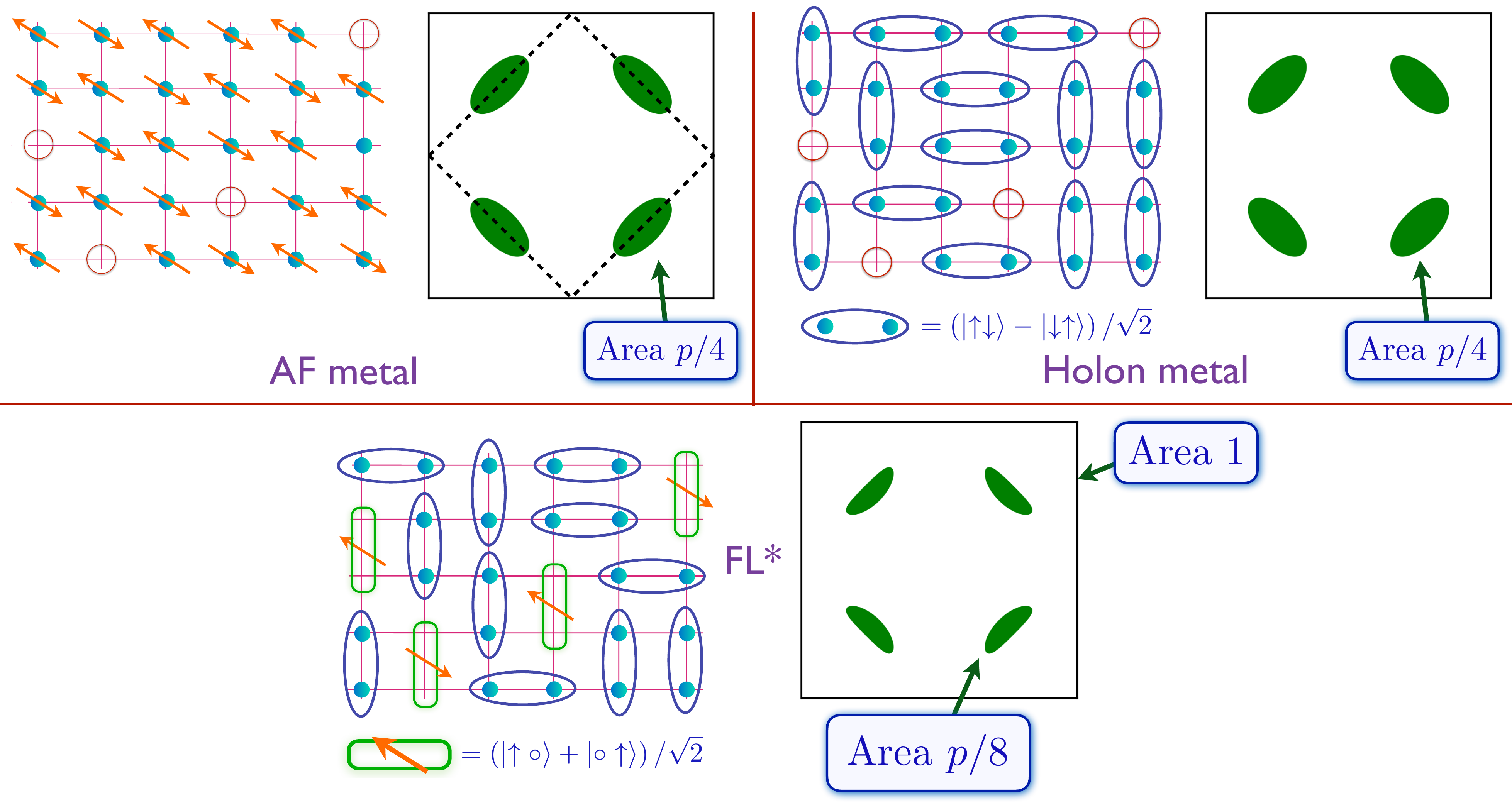}
\caption{Cartoon pictures of different states of doped antiferromagnets; adapted from Ref.~\cite{Punk15}. The areas are those that would be measured by a probe such as quantum oscillation, and are specified in units in which the area of the Brillouin zone is unity. The AF metal has long-range antiferromagnetic order, and the reduced Brillouin zone is shown with the dashed line. The other phases do not break any symmetries. The open circles are holons, and these are assumed fermionic in the holon metal. The green dimers represent bound states of holons and spinons. Note that all 3 states would show a Hall co-efficient of density $p$ positively charged carries.}
\label{fig:metals}
\end{figure}

\begin{itemize}
\item {\it AF Metal.} This is a state with antiferromagnetic long-range order, which we described in Section~\ref{sec:sdw}. We can understand the Fermi surface by considering free electrons moving in a background with the same symmetry {\it i.e.\/} in a background spin-dependent potential which has a modulation at the wavevector $(\pi, \pi)$. This leads to the magnetic Brillouin zone boundary shown by the dashed line, and 4 hole pocket Fermi surfaces. Only two of these pockets are independent within the magnetic Brillouin zone. After accounting for a factor of 2 from spin, we conclude that the fractional area of each pocket is $p/4$. This Fermi surface area obeys the Luttinger constraint. Thermal fluctuations do not move Fermi surfaces, only broaden them, and so we expect that a fluctuating spin density wave state will also have pockets of area $p/4$ \cite{SchmalianPines1,SchmalianPines2,Chubukov23,Chubukov25}.
\item {\it Holon metal.} This is a state with no broken symmetry, in which the electrons have paired up in singlet bonds which resonate with each other. The dopants are realized by spinless mobile vacancies of charge $+e$, known as holons. The density of holons is $p$, and if the holons are fermions, they will form Fermi surfaces corresponding to spinless free fermions of density $p$. If there are four distinct Fermi surfaces in the Brillouin zone (as is the case in many computations), then the fractional area of each pocket will be $p/4$. Although this area is the same as that for the AF metal, the reason is very different. Now there is no broken symmetry, and the fermionic quasiparticles are spinless holons. This Fermi surface area does {\it not\/} obey the usual Luttinger constraint, and this is permitted because of the presence of the spin liquid, as discussed in Section~\ref{sec:gLutt}. A systematic, gauge theoretic treatment of the holon metal state was presented in Section~\ref{sec:holonmetal} \cite{sdw09,DCSS15b,DCSS15,CSS17,WuScheurer1,Scheurer:2017jcp,Sachdev:2018ddg,SSST19,WuScheurer2,Bonetti22,Bonetti23}, and it is relevant to ultracold studies of the Hubbard model on the Lieb lattice \cite{LiebScience,LiebPRB}. The discussion below Eq.~(\ref{ChargonPart0}) explains the equality of pocket areas between the AF metal and the holon metal.
\item {\it FL*.} Finally, we turn to the metallic state of interest. The holon metal state also has spinon excitations, which can be created out of the ground state. Now imagine a situation in which each holons gains energy by binding with a spinon, as in Fig.~\ref{fig:feshbach}, so that the system can pay the price for creating the spinons. In a $t$-$J$ model with electron hopping $t$, and exchange energy $J$, the cost for creating a spinon is order $J$, and the energy gain from the formation of the holon-spinon bound state is of order $t$, and so this bound state formation is very likely in the natural situation with $t \gg J$. Then the ground state will change into one in which the mobile charge carriers are holon-spinon bound states \cite{Spinon-dopon05,RKK07,RKK08,Qi10,Sawatzky11,Moon11,Sawatzky11b,Mei11,Punk12,Punk15,Punk18,Grusdt18,Grusdt19,Grusdt23,Grusdt24,Balents25}. These bound states, the $\widetilde{c}_\alpha$ of Fig.~\ref{fig:feshbach}, are always fermions with charge $+e$, spin $S=1/2$, just like a hole. Treating these holes as free fermions, we conclude that the total area of the Fermi surface should be $p/2$. If there are 4 distinct pockets (as there in the computation below), then each pocket will have the distinctive area of $p/8$. This Fermi surface area also does {\it not\/} obey the usual Luttinger constraint, but does obey the generalized Luttinger constraint discussed in Seçtion~\ref{sec:gLutt}.
\end{itemize}
An important feature of the comparison above is the remarkable distinction between ``thermally disordered'' and ``quantum disordered'' antiferromagnets. The thermally disordered AF Metal has pockets of area $p/4$, while the quantum-disordered FL* state has pockets of area $p/8$. Computations which describe the pseudogap by computing hole motion in an ordered antiferromagnet \cite{Shraiman90,SS94,Grusdt18}, and then rotationally averaging to account for thermal disorder, miss important state-counting constraints of a quantum spin liquid.  

We highlight key features of the single-band FL* state \cite{Punk15} in Fig.~\ref{fig:metals}. We will see that these features will also play a key role in the Ancilla Layer Model of Section~\ref{sec:ancilla}, as we illustrate later in Fig.~\ref{fig:ancilla_dimer}.

\begin{tcolorbox}
  \begin{itemize}
    \item The electron quasiparticle of FL* is a (green) ``dimer'' in Fig.~\ref{fig:metals}, a bound state of a spin and vacancy.
    \item FL* has a background of spinon excitations obtained by breaking singlet bonds (the blue dimers) in Fig.~\ref{fig:metals}.
\end{itemize}  
\end{tcolorbox} 
The FL* picture in Fig.~\ref{fig:metals} was used to motivate an effective quantum dimer model for the FL* phase \cite{Punk15}.
The Hilbert space is defined by the set of dimer packings on the square lattice with 2 species of dimers: blue dimers representing the spin-singlet valence bonds, and green dimers representing the charge $+e$, spin $S=1/2$ bound states, with the blue dimers being bosons, while the green dimers are fermions. 
Note that every configuration of this dimer model obeys the unit gauge charge constraint of Eq.~(\ref{ll1}).
An effective dimer Hamiltonian acts by rearranging dimer configurations on the Hilbert space. Exact diagonalization results on the dispersion spectrum of a single green dimer are shown in Fig.~\ref{fig:disp4}.
\begin{figure}
\begin{center}
\includegraphics[width=3.5in]{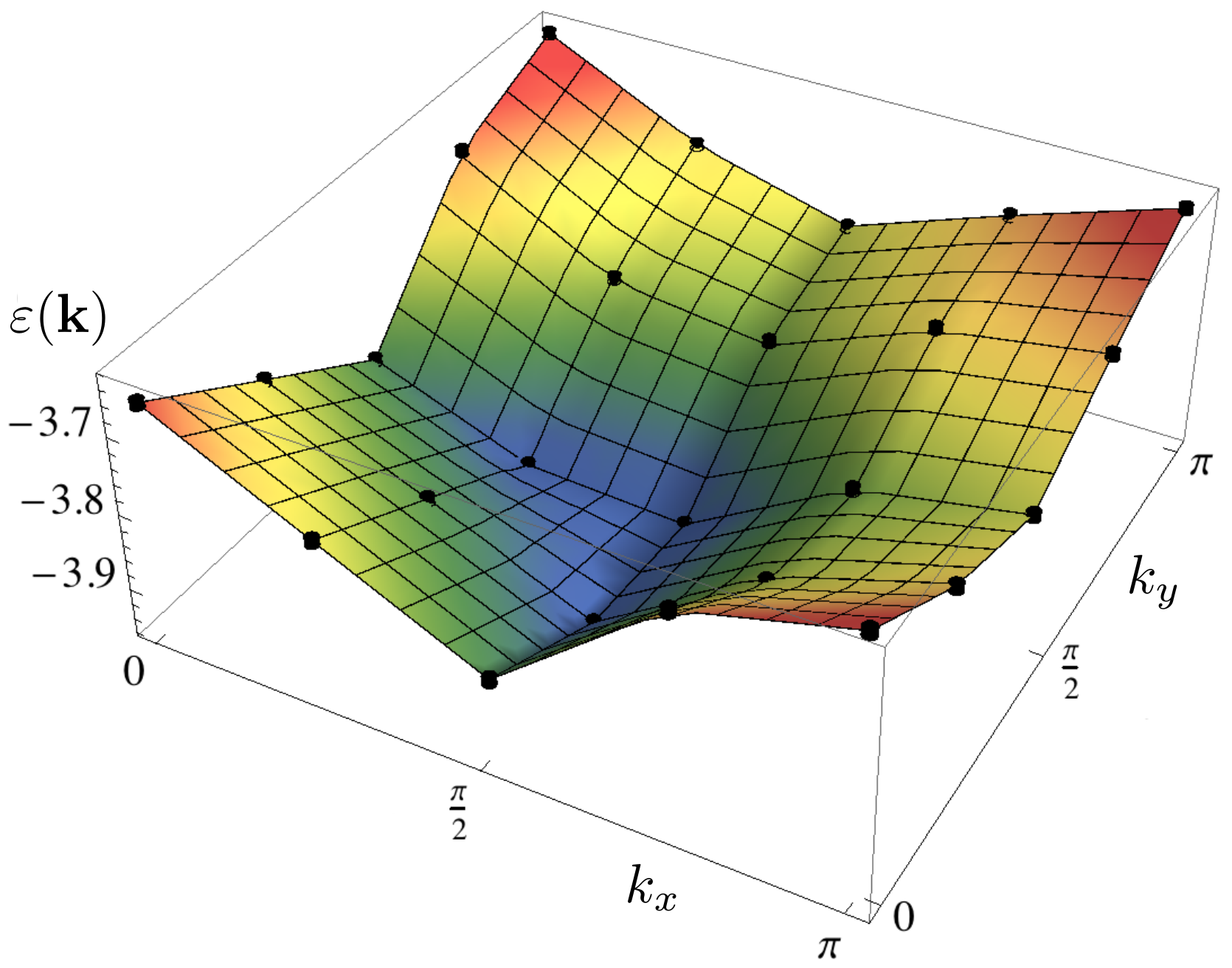}
\end{center}
\caption{From Ref.~\cite{Punk15}. Dispersion of a single fermionic green dimer as a function of momentum ${\bf k}$ obtained by exact diagonalization of an effective quantum dimer model on  
a $8\times 8$ lattice with periodic boundary conditions.
Note that the dispersion is not symmetric about the magnetic Brillouin zone boundary {\it i.e.\/} across the line
connecting $(\pi, 0)$ to $(0, \pi)$.}
\label{fig:disp4}
\end{figure}
This dispersion has minima near $(\pm \pi/2, \pm \pi/2)$, and filling the low-lying states with a density $p$ of green dimers would yield the area $p/8$ pockets of FL* sketched in Fig.~\ref{fig:metals}.

\subsubsection{ADMR}
\label{sec:admr}

Recent ADMR observations  \cite{Ramshaw22,Yamaji24} in the pseudogap metal are well modeled by hole pockets containing quasiparticles which can tunnel coherently between layers. This is not possible for holon metals, as the holons carry charges of distinct emergent gauge fields within each layer. In the AF metal, coherent tunneling requires significant interlayer spin correlations, which are not observed \cite{Greven14,Greven25}, but could perhaps be induced by the applied magnetic field. The FL* quasiparticles are gauge neutral and can indeed tunnel coherently between layers. So even at the outset, before any quantitative comparisons, these observations strongly favor a FL* description of the pseudogap phase of the cuprates \cite{Zhao_Yamaji_25}.

Moreover, the observations \cite{Yamaji24} also provide a quantitative correspondence with the FL* theory.
Their observation
of the Yamaji effect in the cuprate HgBa$_2$CuO$_{4+\delta}$ show pockets of a fractional area of `approximately $1.3\%$' at a doping $p=0.1$, close to the value $p/8 = 1.25\%$ predicted for FL* \cite{TSSSMV03,TSSSMV04,RKK07,Qi10,Joshi23,Zhao_Yamaji_25,FuChun25}.

\subsection{Layer construction with ancilla qubits}
\label{sec:ancilla}

While much insight can be gained from the methods above, they fall short of providing a mean-field theory for the FL* phase, which could be used to study quantum phase transitions out of it. A suitable mean field theory can also lead to a trial wavefunction for the FL* state, which can be used for variational numerical computations and compared to cold atom observations \cite{Koepsell21,Iqbal24,HenryShiwei24,Bloch24}, as discussed in Section~\ref{sec:wavefunction}.
To these ends, we now describe the Ancilla Layer Model (ALM) \cite{YaHui-ancilla1,YaHui-ancilla2}, which is designed to automatically ensure consistency with anomaly-type arguments, which are equivalent to the generalized Luttinger constraints discussed in Section~\ref{sec:gLutt}.

We wish to have a mean-field theory which changes a large hole-like Fermi surface of area $(1+p)/2$ to small hole-like Fermi surfaces of total area $p/2$. The simplest way to achieve this in mean-field theory is to hybridize the large Fermi surface with another band at half-filling (as in the Kondo lattice, and was done by Yang-Rice-Zhang (YRZ) \cite{YRZ,FuChun25})---this leads to a Fermi surface of the needed area $(1+p+1)/2$ (mod 1) $=p/2$. But we are {\it not allowed\/} to add a single band at half-filling because it is not `trivial' {\it i.e.} it is not anomaly-free, and its excitations cannot be integrated out because the extra band can only acquire a gap with a broken symmetry. On the other hand, we {\it are allowed\/} to add two bands at half filling each (or any even number of bands), because they can form a trivial insulator with a gap. After the first added band hybridizes with the physical electrons to yield the small Fermi surfaces, the second added band is left decoupled, and it can form the spin liquid needed to satisfy the Oshikawa anomaly-argument and the constraints in Section~\ref{sec:gLutt}. This is the essence of the ALM.

We begin with a constructive derivation of the ALM starting from the familiar Hubbard model in Eq.~(\ref{dwave1}), and returning to the paramagnon representation in Eq.~(\ref{ZHubbard}).
After some renormalization of the high energy states we give the paramagnon $\vP$ some independent dynamics with a non-zero mass $m_{\vP}$, and obtain its effective Lagrangian 
    \beq
    \mathcal{L}[ \vP] = \sum_\vi \left[ \frac{m_{\vP}}{2} 
    \left( \partial_\tau \vP_\vi \right)^2 + \frac{3U}{8}\vP_\vi^2 \right] \,,
    \eeq
    where the $\vP^2$ term is as in Eq.~(\ref{ZHubbard}).
    This is a paramagnon theory with 3 local harmonic oscillators of frequency $(3 U/(4 m_{\vP}))^{1/2}$ on each site, which are coupled to the electrons with a Yukawa coupling of strength $U$ in Eq.~(\ref{ZHubbard}) . Now we take steps different from conventional paramagnon approaches.
    The ground state of the paramagnon theory is obtained when all three oscillators are in the $n=0$ state: $\left| 0,0,0 \right\rangle$. There is a triplet of first excited states:
    \begin{displaymath}
    \left| 1, 0, 0 \right\rangle  \sim \vP_{\vi x} \left| 0,0,0 \right\rangle ; \left| 0, 1, 0 \right\rangle  \sim \vP_{\vi y} \left| 0,0,0 \right\rangle ; \left| 0, 0, 1 \right\rangle  \sim \vP_{\vi z} \left| 0,0,0 \right\rangle.
    \end{displaymath}
We can map this low energy spectrum to that of a pair of $S=1/2$ ancilla spins with a mutual interaction $J_\perp {\bm S}_{1\vi} \cdot {\bm S}_{2\vi}$. By comparing the above matrix elements to those that couple the singlet and triplet spin states states, we obtain the operator identification
    \beq
    \vP_\vi \sim {\bm S}_{1\vi} - {\bm S}_{2\vi}\,. \label{PS1S2}
    \eeq 
Inserting Eq.~(\ref{PS1S2}) into Eq.~(\ref{ZHubbard}), we obtain the Hamiltonian of the ALM, $\mathcal{H}_{\rm ALM}$.
This has a Kondo lattice Hamiltonian $\mathcal{H}_{\rm KL}$ with couples the electrons $c_\alpha$ to the ${\bm S}_1$ spins
by an antiferromagnetic Kondo coupling $J_K \sim U$, and a second layer of ${\bm S}_2$ spins which have a rung coupling $J_\perp = 
(3 U/(4 m_{\vP}))^{1/2}$ to the ${\bm S}_1$ spins:
\begin{align}
\mathcal{H}_{\rm ALM} 
& = \mathcal{H}_{\rm KL}   + J_\perp \sum_{\vi} {\bm S}_{1\vi} \cdot {\bm S}_{2\vi}  + \sum_{\vi < \vj} J_{\vi\vj} \,{\bm S}_{2\vi} \cdot {\bm S}_{2\vj} \nonumber \\
\mathcal{H}_{\rm KL} & = \sum_{\vi < \vj} J_{1,\vi\vj} \,{\bm S}_{1\vi} \cdot {\bm S}_{1\vj} -\sum_{\vi,\vj} t_{\vi\vj} c^\dagger_{\vi\alpha} c_{\vj\alpha} 
+ \sum_{\vi} \frac{J_K}{2} {\bm S}_{1i} \cdot c_{\vi \alpha}^\dagger {\bm \sigma}_{\alpha\beta} c_{\vi \beta}  \,.
\label{eq:Hancilla}
\end{align}
This Hamiltonian is illustrated in the bottom half of Fig.~\ref{fig:ancilla}.
\begin{figure}
\centering
\includegraphics[width=5in]{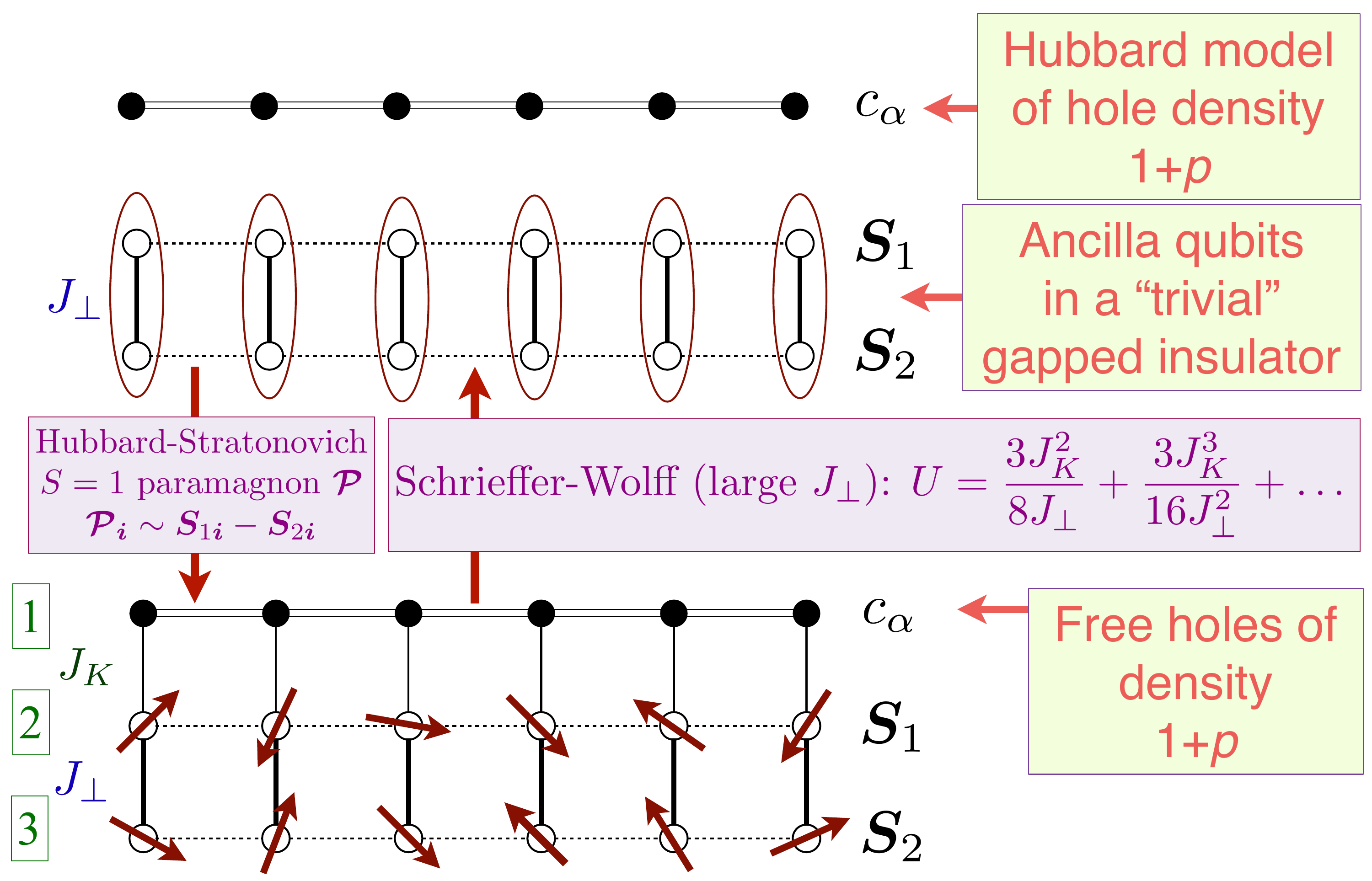}
\caption{Illustration of the mapping from a single-band Hubbard model with decoupled ancilla qubits, to a single band with free fermions coupled to a bilayer antiferromagnet. The Schrieffer-Wolff transformation is derived in Ref.~\cite{Nikolaenko:2021vlw}, and the Hubbard-Stratonovich transformation is derived in Ref.~\cite{Mascot22}. The layer numbers of the layer construction with ancilla qubits are indicated in the bottom picture.}
\label{fig:ancilla}
\end{figure}

So we see that the ancilla spins are simply the fractionalization of the familiar $S=1$ paramagnon into a pair of $S=1/2$ spins \cite{Mascot22}.
The above derivation of the mapping to the ALM yields
 an additional ferromagnetic Kondo interaction between the electrons $c_{\vi\alpha}$ and  spins ${\bm S}_{2 \vi}$. Ferromagnetic Kondo couplings are expected to be irrelevant, and we will note this coupling again in Eq.~(\ref{Yukawa}).  Note that it is Eq.~(\ref{PS1S2}) which breaks the symmetry between ${\bm S}_1$ and ${\bm S}_2$, and we {\it choose\/} the spins in the middle layer to be the ones which have the antiferromagnetic Kondo coupling with the electrons.
We have also explicitly included a direct exchange interaction between the ${\bm S}_{2\vi}$ spins, as it will be important for our purposes below. Note also that there is no Hubbard interaction $U$ in $\mathcal{H}_{\rm ALM}$, and it has been replaced by the Kondo interaction $J_K$ in $\mathcal{H}_{\rm ALM}$.
 
We summarize the phase diagram of $\mathcal{H}_{\rm ALM}$ in Fig.~\ref{fig:oneband}.
\begin{figure}
\centering
\includegraphics[width=5.75in]{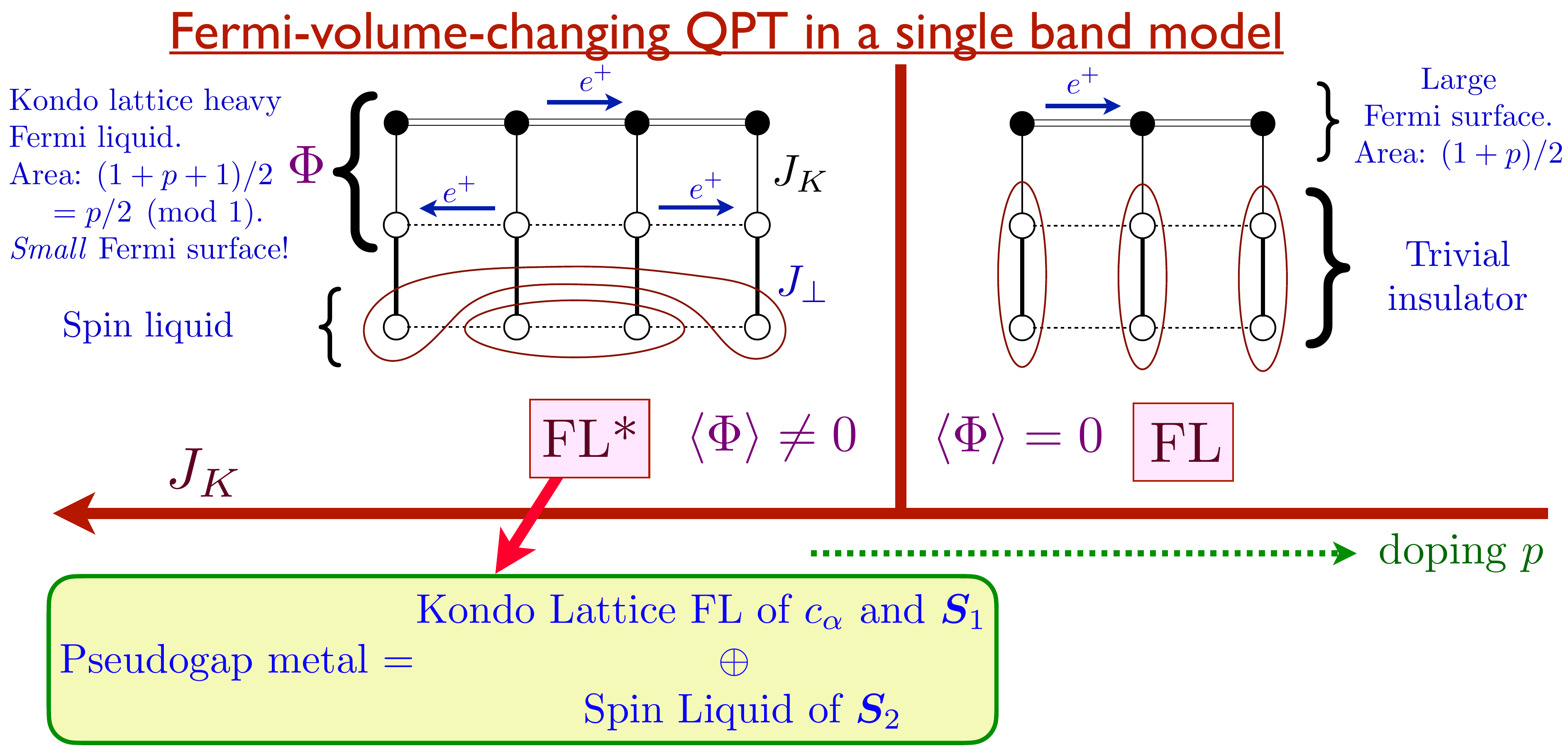}
\caption{Phases of the Ancilla Layer Model. The noted areas are fractions of the square lattice Brillouin zone area.
The phases are distinguished by the condensation of the boson $\Phi$, which hybridizes the conduction electrons in the top layer with the fermionic spinons in the middle layer.}
\label{fig:oneband}
\end{figure}

At small $J_K$, the ancilla spins decouple into rung singlets, and we are back to a $c_{\alpha}$ state adiabatically connected to free electrons, which is the conventional Fermi liquid with a large hole pocket of area $(1+p)/2$.

At large $J_K$, we assume that the $c_\alpha$ electrons and ${\bm S}_1$ spins realize the commonly observed heavy Fermi liquid (FL) phase of $\mathcal{H}_{\rm KL}$. With the density of $c_\alpha$ equal to $1+p$, the total density of the `large' Fermi surface is $2+p$. As a trivial filled band with density 2 can always be removed, we obtain `small' Fermi surface associated with density $p$ in the FL phase of $\mathcal{H}_{\rm KL}$. So the total area of hole pockets is $p/2$.
While this is the Luttinger value for $\mathcal{H}_{\rm KL}$, it is {\it not\/} for $\mathcal{H}_{\rm ALM}$. The ${\bm S}_{2\vi}$ have an effective ferromagnetic Kondo coupling to the conduction electrons $c_\alpha$ (mediated by the antiferromagnetic $J_\perp$ and the antiferromagnetic $J_K$), and so we can expect them to decouple from the top two layers.
Indeed, we obtain a FL* phase for $\mathcal{H}_{\rm ALM}$ if the ${\bm S}_{2\vi}$ layer forms a spin liquid, and the effects of $J_\perp$ in Eq.~(\ref{eq:Hancilla}) can be treated perturbatively. This leads the identification in Fig.~\ref{fig:oneband} of the FL* pseudogap metal with the combination of a Kondo lattice FL state of $c_{\alpha}$ and ${\bm S}_1$, and a spin liquid of ${\bm S}_2$. 

We can now summarize the FL* construction of the Hubbard model in the following simple terms. We express the Hubbard model as a theory of antiferromagnetic paramagnons, and then fractionalize the $S=1$ paramagnon into two $S=1/2$ spins, ${\bm S}_1$ and ${\bm S}_2$. The ${\bm S}_1$ spins form a Kondo lattice heavy Fermi liquid with the electrons $c_\alpha$ to yield the hole pockets. The ${\bm S}_2$ spins form the spin liquid needed to satisfy the constraints in Section~\ref{sec:gLutt}.

\subsection{Mean field theory of the cuprate pseudogap.}
\label{sec:sFL*}

We can now obtain a mean-field theory of the pseudogap by extending the Kondo lattice mean-field theory to the Hamiltonian $\mathcal{H}_{\rm ALM}$ in Eq.~(\ref{eq:Hancilla}).

We begin with the top two layers of electrons $c_\alpha$ and spins ${\bm S}_1$ in  the ALM in Fig.~\ref{fig:ancilla}.
We proceed with parton decomposition of ${\bm S}_1$ in terms of the fermionic spinons $f_{1\alpha}$ as in Eq.~(\ref{Schwingerfermion})
\bea
{\bm S}_{1\vi} = \frac{1}{2} f_{1\vi \alpha}^\dagger {\bm \sigma}_{\alpha\beta}^{\vphantom\dagger} f_{1\vi \beta}^{\vphantom\dagger}\,,
\label{eq:S1f}
\eea
and obtain the mean-field fermion Kondo lattice Hamiltonian for the $c_\alpha$ and $f_{1\alpha}$:
\beq
\mathcal{H}_{\rm KLmf} = \sum_{\vi,\vj} \left[- t_{\vi\vj} c^\dagger_{\vi\alpha} c_{\vj\alpha} -   t_{1,\vi\vj} f^\dagger_{1\vi\alpha} f_{1\vj\alpha} \right] -  \sum_{\vi}(\Phi \,  c^\dagger_{\vi\alpha} f_{1\vi\alpha}+ \Phi^\ast \, f^\dagger_{1\vi\alpha} c_{\vi\alpha}) \,. \label{HKLmf}
\eeq
Here $\Phi$ is decoupling field of the $J_K$ exchange interaction in Eq.~(\ref{eq:Hancilla}), and is illustrated in Fig.~\ref{fig:PhiB}.
At the mean-field level, we can assume the ${\bm S}_2$ spin liquid remains decoupled, and important effects of this spin liquid will be discussed in Section~\ref{sec:pseudogap}.

When considered as a theory of the Kondo lattice model $\mathcal{H}_{\rm KL}$, the FL state corresponds to the condensation of the decoupling field $\Phi$. On the other hand, on the Hilbert space of the full $\mathcal{H}_{\rm ALM}$, this same $\Phi$ condensed phase is the FL* state. This interesting inversion is highlighted in Fig.~\ref{fig:inverted}: the single band model has an `inverted' Kondo lattice transition.
\begin{figure}
\centering
\includegraphics[width=4in]{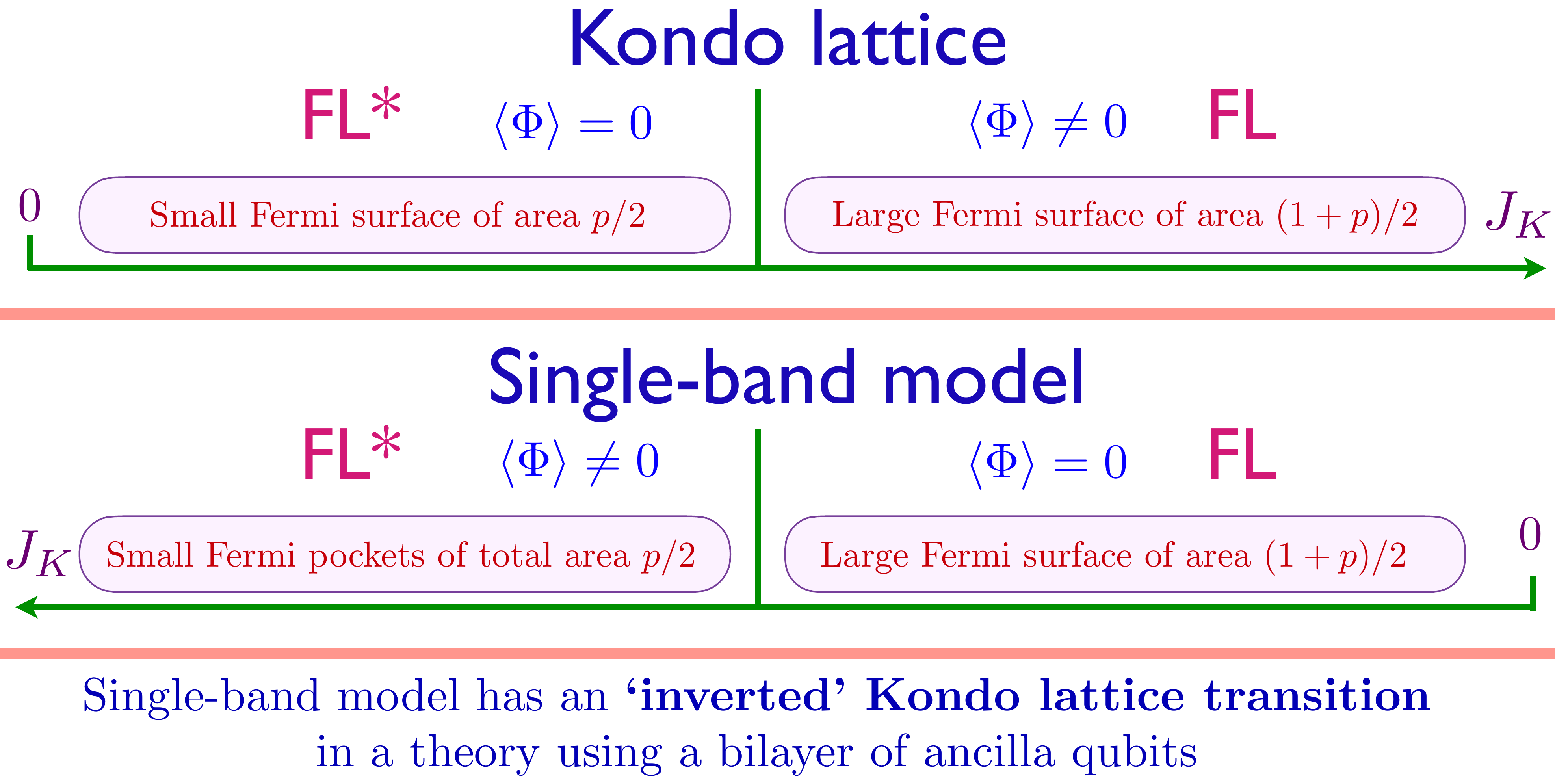}
\caption{Comparison of the phases of the Kondo lattice model and the Ancilla Layer Model of the single band Hubbard model in Fig.~\ref{fig:oneband}. There is an inversion in the phase in which $\Phi$ is condensed.}
\label{fig:inverted}
\end{figure}
The magnitude of $\Phi$ determines the value of the electronic gap in the anti-nodal region of the Brillouin zone (near momenta $(\pi, 0)$, $(0, \pi)$). 

\begin{figure}
\begin{center}
\includegraphics[width=6in]{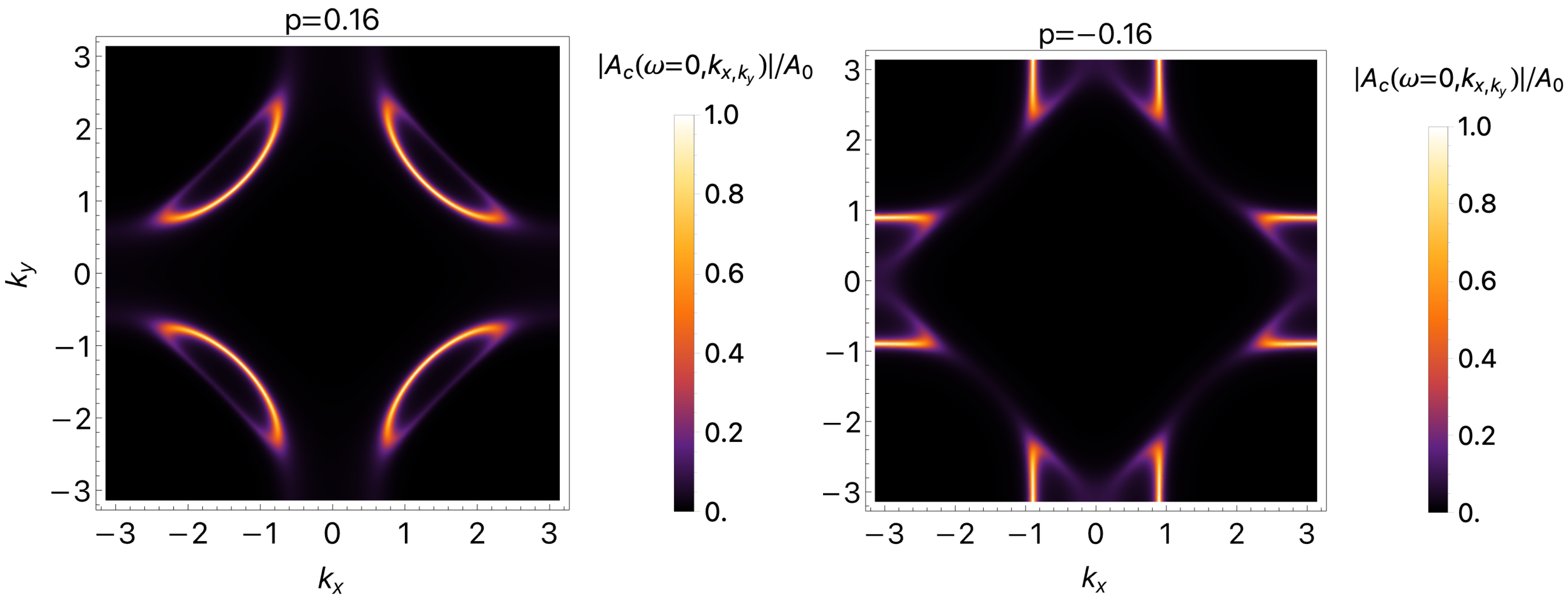}
\end{center}
\caption{Spectral density of hole (left) and electron (right) pockets at $p = 0.16$ and $p=-0.16$ respectively in the FL* state. The fractional area of each hole pocket is $p/8$, and the fractional area of the electron pocket is $|p|/4$. Compared to Fig.~\ref{fig:sdwphoto} for the SDW state, the pocket areas have been halved in the FL* state. Figure by A. Nikolaenko.}
\label{fig:flsphoto}
\end{figure}
Fig~\ref{fig:flsphoto} shows the electron spectral density (as measured by photoemission) at zero frequency for both the electron and hole pockets, computed by diagonalizing Eq.~(\ref{HKLmf}).

We can now present in Fig.~\ref{fig:ancilla_dimer} the close analogy between the FL* state obtained in the ALM and the state presented in Fig.~\ref{fig:metals} as highlighted in the box before the beginning of Section~\ref{sec:ancilla}
\begin{tcolorbox}
    \begin{itemize}
        \item 
        The electron quasiparticle of FL* is the hybridization of $c_\alpha$ and $f_{1\alpha}$ induced by $\Phi$, which creates the bound state $\sim c_{\vi\alpha}^{\vphantom\dagger} f_{1 \vi \alpha}^\dagger$ between a vacancy and a spin, analogous to the green dimers in Fig.~\ref{fig:metals}. 
        \item
        The background of spinons in FL* are obtained from the spin liquid of ${\bm S}_2$ spinons in the bottom layer, analogous the spinons obtained by breaking the blue dimers in Fig.~\ref{fig:metals}. We will later represent these spinons by $f_\alpha$ via 
        Eq.~(\ref{eq:S2f}).
    \end{itemize}
\end{tcolorbox}
\begin{figure}
\centering
\includegraphics[width=6in]{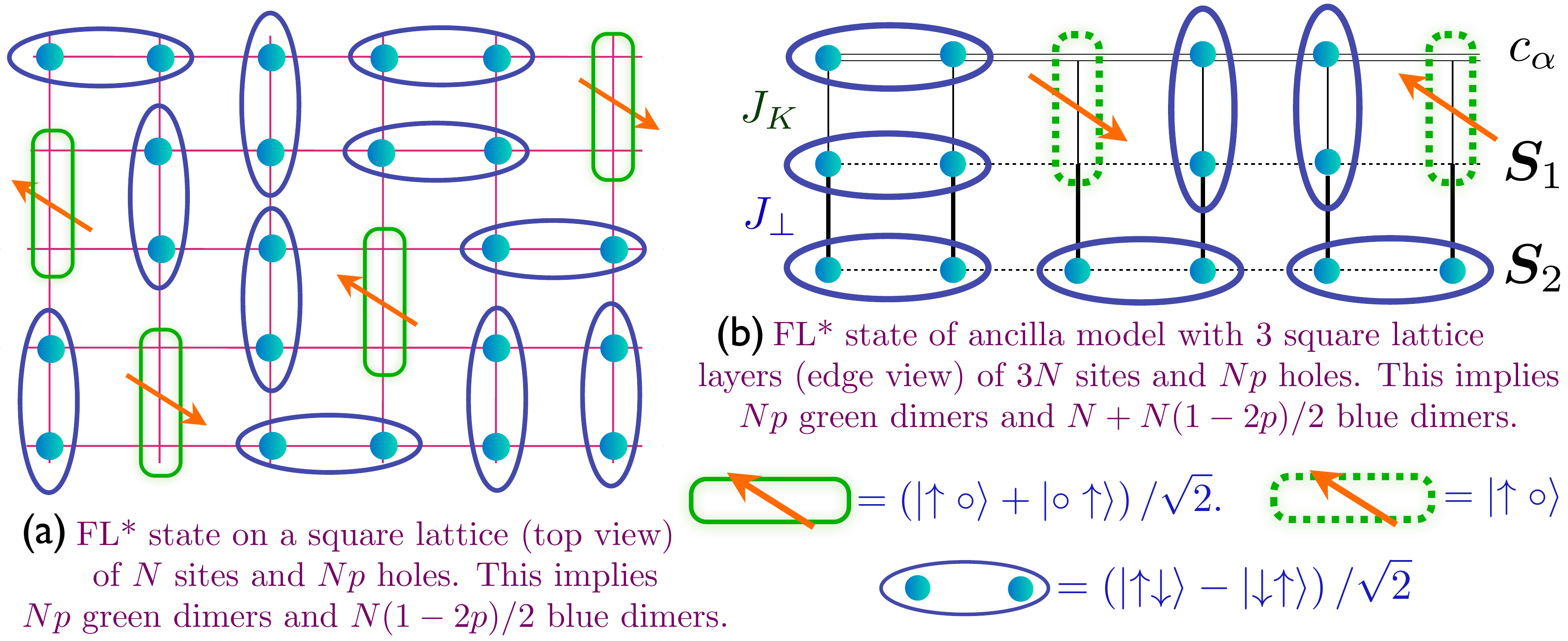}
\caption{Comparison between (a) the quantum dimer theory of the square lattice FL* state \cite{Punk15} and the (b) ALM. The dashed green dimer is the state $c_{\vi\alpha}^{\vphantom\dagger} f_{1 \vi \alpha}^\dagger$ created by the hybridization $\Phi$ in the ALM. Note that the ALM has exactly $N$ blue dimers more than the dimer model, and these extra dimers are the number in a trivial rung-singlet bilayer antiferromagnet.}
\label{fig:ancilla_dimer}
\end{figure}
A key observation is that the number of spin-singlet blue dimers in the ALM is exactly $N$ more than in the dimer model of Fig.~\ref{fig:metals}, while the number of green dimers is the same. This is acceptable because $N$ spin singlets is precisely the number that can be accommodated in a trivial rung-singlet state of a bilayer antiferromagnet.

The connection in Fig.~\ref{fig:ancilla_dimer} shows that the ALM is, in a sense, `minimal'. For a FL* mean field theory with both a small Fermi surface and spinons, we need the middle and bottom ancilla layers to provide the vacancy-spin and singlet valence bonds respectively. And a bonus, not available in the quantum dimer analysis of Ref.~\cite{Punk15,Punk18} or other approaches \cite{Spinon-dopon05,Mei11,Punk12}, is that it also provides a mean-field theory which captures the FL large Fermi when the bottom two layers are in a rung-singlet state. 

\subsection{Wavefunction for FL* and cold atom observations.}
\label{sec:wavefunction}

A separate mean-field approach is to work with variational wavefunctions. The ancilla layer method was the first to provide a variational wavefunction for the FL* phase of the single-band Hubbard model. 
\begin{tcolorbox}
This approach couples all three layers together in
the proposed wavefunction \cite{YaHui-ancilla1}:
\begin{align}
\left|{\rm FL*} \right\rangle_{\rm Hubbard} &=  \left[\mbox{Projection onto rung singlets of ${\bm S}_1, {\bm S}_2$} \right] \nonumber \\
&~~~~~~~\bowtie \left| \mbox{Slater determinant of $(c,f_1)$} \right\rangle \nonumber \\
&~~~~~~~~~~~~~~~\otimes \left| \mbox{Spin liquid of ${\bm S}_2$} \right\rangle \,. \label{wavefunction}
\end{align}
\end{tcolorbox}
Here $f_1$ is obtained from the parton decomposition of ${\bm S}_1$ in Eq.~(\ref{eq:S1f}), and the Slater determinant of $(c,f_1)$ is the ground state of $\mathcal{H}_{\rm KLmf}$ in Eq.~(\ref{HKLmf}).
Note that this wavefunction depends only upon the co-ordinates of the $c$ electrons alone, as the co-ordinates of the $f,f_1$ fermions have been projected out. This wavefunction replaces the `vanilla' Gutzwiller projected wavefunctions of the $c$ electrons alone \cite{vanilla} in the underdoped region. We see rather explicitly in Eq.~(\ref{wavefunction}) the expressivity of the layered construction \cite{Hinton86}, allowing us to include both a small Fermi surface and a spin liquid in the same state.

A summary of this wavefunction, and its connection to the cuprate phase diagram appear in Figs.~\ref{fig:pdancilla1} and~\ref{fig:pdancilla2}.
\begin{figure}
\centering
\includegraphics[width=5.75in]{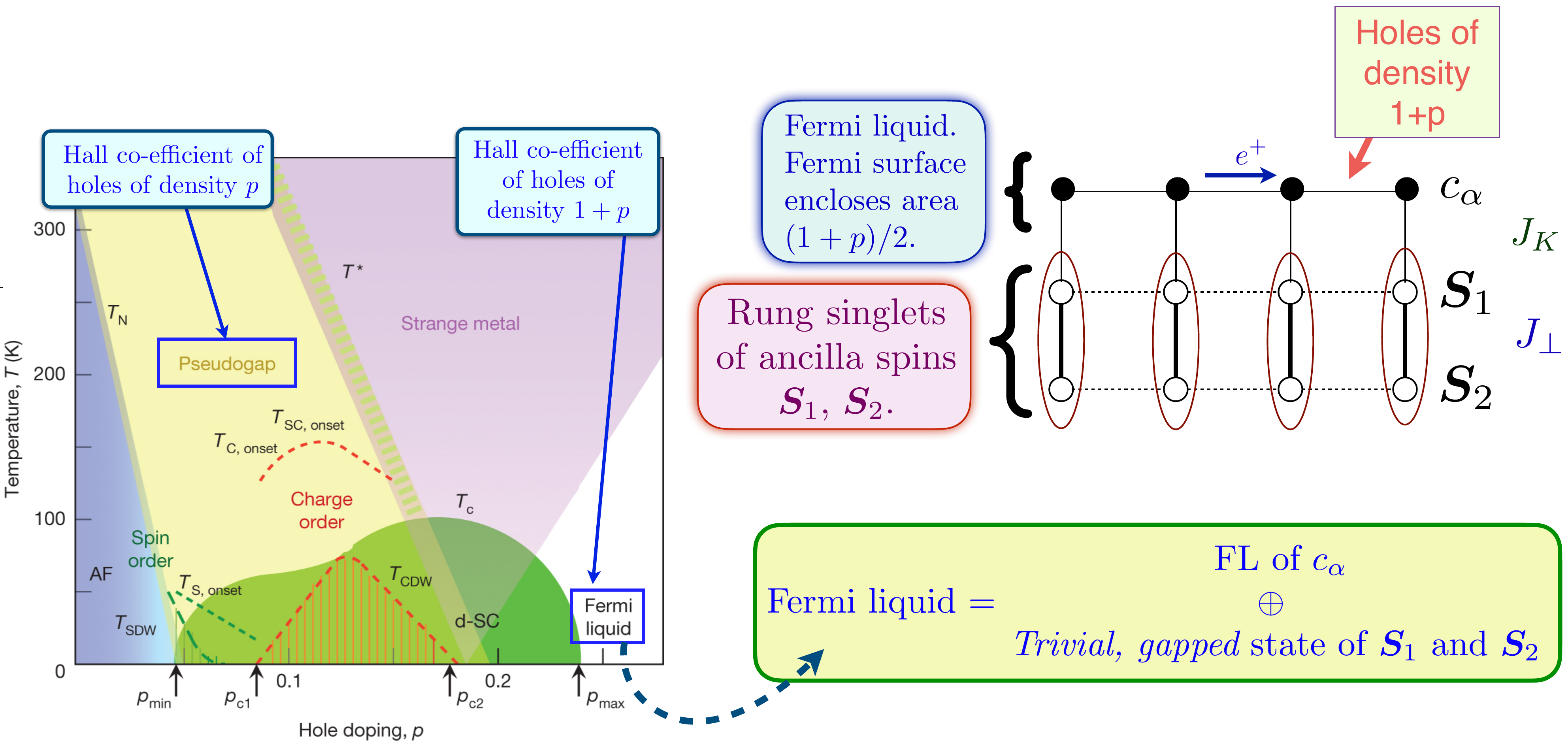}
\caption{Summary of the FL phase of the Hubbard model obtained from the ALM, and its connection to the cuprate phase diagram from Ref.~\cite{phase_diag}.}
\label{fig:pdancilla1}
\end{figure}
\begin{figure}
\centering
\includegraphics[width=5.75in]{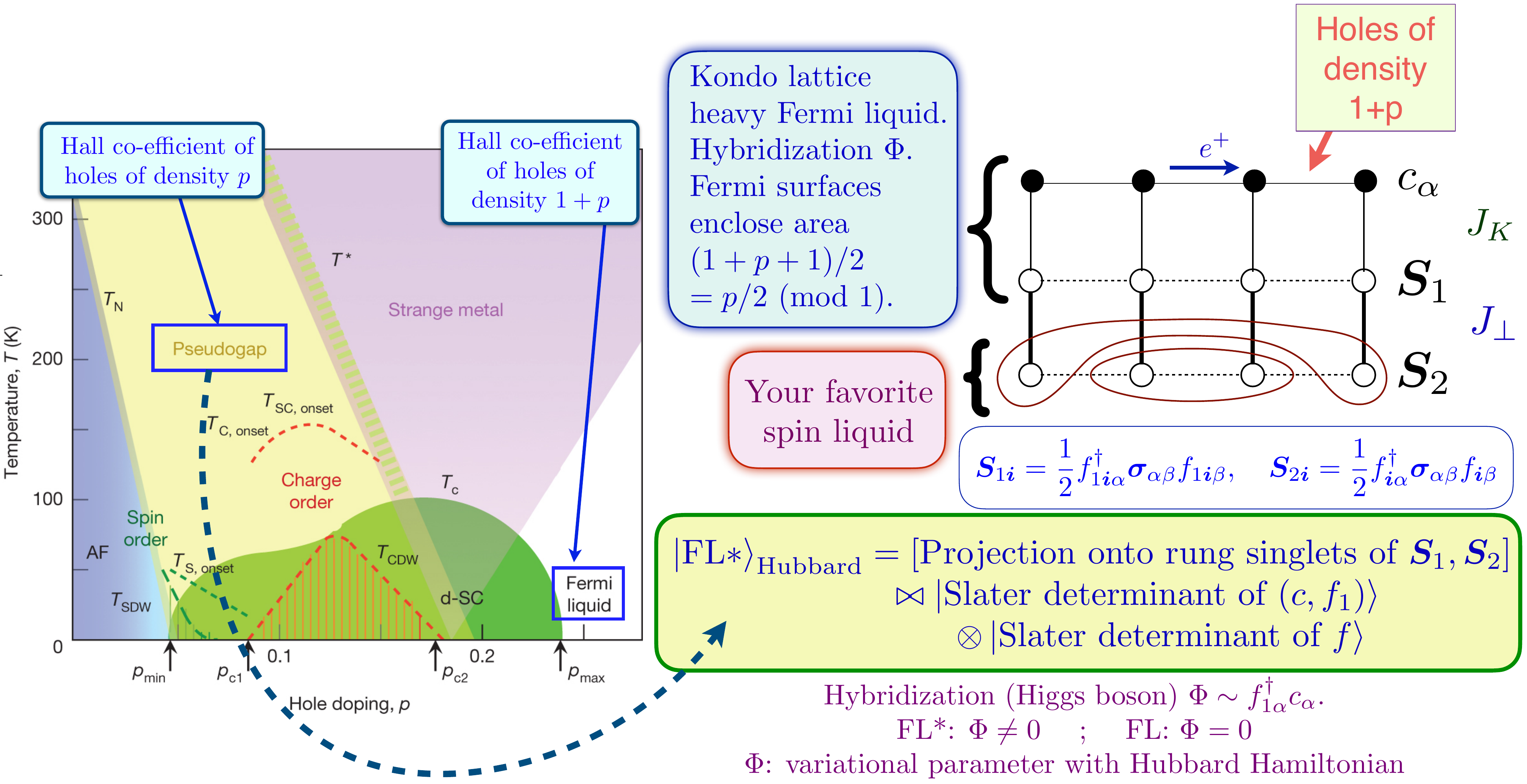}
\caption{Wavefunction of the FL* phase of the Hubbard model obtained from the ALM, and its connection to the cuprate phase diagram from Ref.~\cite{phase_diag}. The hybridization $\Phi$ is that appearing in Eq.~(\ref{HKLmf}).}
\label{fig:pdancilla2}
\end{figure}

Ref.~\cite{YaHui24} has given an appealing quantum information theoretic interpretation of the wavefunction in Eq.~(\ref{wavefunction}). 
We can view the projection into singlet pairs as a set of Bell-pair measurements of the ancilla spins. This measurement teleports the entanglement of the ${\bm S}_2$ spins to the $c$ electrons.

The wavefunction Eq.~(\ref{wavefunction}) has been studied numerically in Refs.~\cite{Iqbal24,HenryShiwei24,YaHui24}, with the couplings in $\mathcal{H}_{\rm KLmf}$ treated as variational parameters. One of the variational parameters is the hybridization/Higgs field $\Phi$, which is non-zero in the FL* phase and decreases with increasing doping until it vanishes at a FL*-FL transition.
The results of Refs.~\cite{Iqbal24,HenryShiwei24} successfully capture the evolution of 
local, multi-point spin and charge correlations with doping as measured in cold atom experiments on the square lattice fermionic Hubbard model \cite{Koepsell21,Bloch24,Kendrick:2025ujy}. The results of Ref.~\cite{YaHui24} compare well with the exact diagonalization of Hubbard models in one and two dimensions.

\subsection{SU(2) gauge theory of the onset of nodal $d$-wave superconductivity from FL*}
\label{sec:pseudogap}

This section addresses the fate of the FL* pseudogap as the temperature is lowered, upon including the coupling to the ${\bm S}_2$ spin liquid in $\mathcal{H}_{\rm pseudogap}$ in Eq.~(\ref{eq:WS}).
This Hamiltonian specifically choses the $\pi$-flux spin liquid of Section~\ref{sec:fermions} for the ${\bm S}_2$ layer.
We will show that for this spin liquid there is a transition to a conventional BCS-type $d$-wave superconductor, with anisotropic nodal velocities for the Boboliubov quasiparticles, and $h/(2e)$ vortices. Nevertheless the transition itself is not of the BCS type with a Cooper-pairing instability of a Fermi surface. Instead, the transition is driven by the confinement of the fractionalized excitations of the ${\bm S}_2$ spin liquid. We also find nearby instabilities to charge ordering, consistent with observations---see the review in Ref.~\cite{ROPP25} for more information.

We note here recent numerics which support the idea of the $d$-wave superconductor emerging from the doping-induced confinement of the $\pi$-flux spin liquid. As we have discussed in Section~\ref{sec:fermions}, the $\pi$-flux spin liquid is one description of the quantum-criticality between the N\'eel and VBS states. The numerical studies of Refs.~\cite{Jiang21,Jiang23} examined the $J_1$-$J_2$ square lattice antiferromagnet near the N\'eel-VBS transition, and indeed found $d$-wave superconductivity upon doping.

Now we return to the analysis of the pseudogap using the ALM in Eq.~(\ref{eq:Hancilla}). In Section~\ref{sec:sFL*}, we presented a mean field analysis in terms of decoupled Kondo lattice and spin liquid models. This section will couple them using the methods developed in Sections~\ref{sec:spinliquids} and \ref{sec:halffilling}.

On the spin liquid layer of ${\bm S}_2$ spins we write a parton decomposition which parallels that in Eq.~(\ref{Schwingerfermion}) and (\ref{eq:S1f})
\bea
{\bm S}_{2\vi} = \frac{1}{2} f_{\vi \alpha}^\dagger {\bm \sigma}_{\alpha\beta}^{\vphantom\dagger} f_{\vi \beta}^{\vphantom\dagger}\,.
\label{eq:S2f}
\eea
Then the analysis of the exchange interactions within the ${\bm S}_2$ layer is precisely that in Section~\ref{sec:fermions}.
W decouple the $J_{ \vi\vj}$ term in Eq.~(\ref{eq:Hancilla}) to $\mathcal{H}_{SLf}$ in Eq.~(\ref{eq:fermionhop2}) realizing a $\pi$-flux state of the ${\bm S}_2$ spins with 
a SU(2) gauge field. 

To couple this spin liquid to the Kondo lattice, 
we have to decouple the 
$J_\perp$ term coupling the $f_{1}$ and $f$ spinons in Eq.~(\ref{eq:Hancilla}). Given the SU(2) gauge structure of the ${\bm S}_2$ layer, it pays to decouple the $J_\perp$ term in a manner which keeps the SU(2) gauge invariance explicit. In fact, the needed decoupling field is precisely the boson $\mathcal{B}_\vi$ introduced in 
Eq.~(\ref{defB}) by different methods for the half-filled case. We also introduce a matrix fermion operator $\mathcal{F}_{1\vi}$
\beq
\mathcal{F}_{1\vi} \equiv \left(
\begin{array}{cc}
f_{1\vi \uparrow} & - f_{1\vi \downarrow} \\
f_{1\vi \downarrow}^\dagger & f_{1\vi \uparrow}^\dagger
\end{array}
\right), \label{F1matrix}
\eeq
whose transformations under the symmetries in Eqs.~(\ref{eq:gauge},\ref{eq:spin},\ref{eq:charge}) are the same as those of $\mathcal{C}_\vi$.
We summarize the gauge and symmetry properties in Table~\ref{tab2}.
\begin{table}
    \centering
    \begin{tabular}{|c|c||c||c|c|}
\hline
\multirow{2}{*}{Field} & \multirow{2}{*}{Layer} & Gauge & \multicolumn{2}{c|}{Global} \\
\cline{3-5}
 & & SU(2) & SU(2) & U(1) \\ \hline
 $c$ or $\mathcal{C}$ & 1 & ${\bm 1}$ & ${\bm 2}_R$ & -1 \\
 \hline
 $f_1$ or $\mathcal{F}_1$ & 2 & ${\bm 1}$ & ${\bm 2}_R$ & -1 \\ \hline
 $f$ or $\mathcal{F}$ & 3 & ${\bm 2}_L$ & ${\bm 2}_R$ & 0 \\ \hline
 $B$ or $\mathcal{B}$ & $2 \leftrightarrow 3$ & ${\bm 2}_L$ & ${\bm 1}$ & 1 \\ \hline
\end{tabular}
    \caption{Summary of gauge and global symmetry transformations for the fields of the ALM in the FL* phase. The representations of the SU(2) are indicated by their dimension; the subscripts $L$/$R$ indicate whether the SU(2) acts by left/right multiplication in the matrix form of the field. The representations of the global U(1) is the electrical charge in units of $e$. For the fermions, the layer column indicates the layer number is Fig.~\ref{fig:ancilla}. For the bosons,  the layer column indicates the layers between which there is a Yukawa coupling to the fermions.}
    \label{tab2}
\end{table}

Then, from the $J_\perp$ term, symmetry considerations are sufficient to constrain the structure of the Yukawa term between ${B}$ and the fermions \cite{Christos:2023oru}, which follows from Eq.~(\ref{eq:CBF}) and is illustrated in Fig.~\ref{fig:PhiB}:
\begin{figure}
\centering
\includegraphics[width=3in]{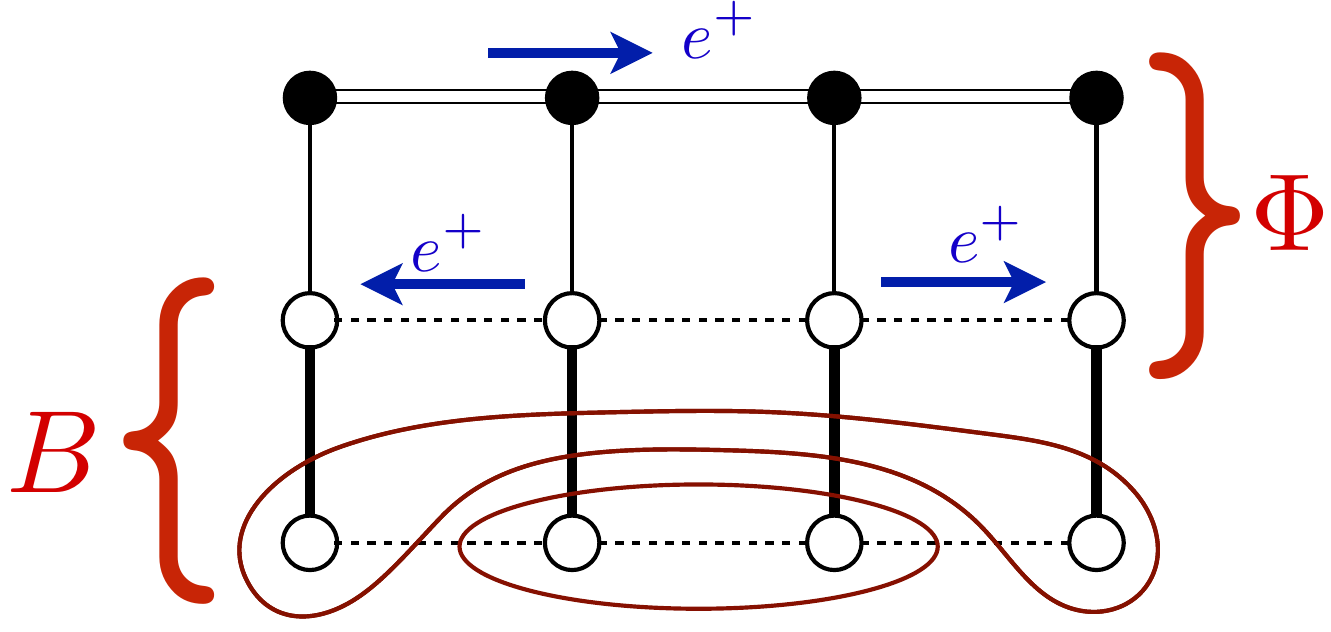}
\caption{The two distinct Higgs fields in the ancilla layer theory of the single band Hubbard model. $\Phi$ hybridizes conduction electrons in the top layer with spinons in the middle layer. $B$ couples the spinons of the bottom layer to the upper layers.}
\label{fig:PhiB}
\end{figure}
\bea
   \mathcal{H}_Y  && = -\frac{1}{2} \sum_\vi \left[ i \, \mbox{Tr} \left(\mathcal{F}_{1\vi}^\dagger \mathcal{B}_{\vi}^\dagger \mathcal{F}_{i}^{\vphantom\dagger} \right) +  i g \, \mbox{Tr} \left(\mathcal{C}_{\vi}^\dagger \mathcal{B}_{\vi}^\dagger \mathcal{F}_{i}^{\vphantom\dagger}\right) +\mbox{H.c.} \right] \nonumber \\
   && =  \sum_\vi \left[  i \left( B_{1\vi}^{\vphantom\dagger} f_{\vi\alpha}^\dagger f_{1\vi \alpha}^{\vphantom\dagger} - B_{2 \vi}^{\vphantom\dagger} \varepsilon_{\alpha\beta}^{\vphantom\dagger} f_{\vi \alpha}^{\vphantom\dagger} f_{1\vi \beta}^{\vphantom\dagger} \right)
   + \mbox{H.c.}\right. \nonumber \\
   && ~~~~~~~~~~~~~~~ \left.   + i g \left( B_{1\vi}^{\vphantom\dagger} f_{\vi\alpha}^\dagger c_{\vi \alpha}^{\vphantom\dagger} - B_{2 \vi}^{\vphantom\dagger} \varepsilon_{\alpha\beta}^{\vphantom\dagger} f_{\vi \alpha}^{\vphantom\dagger} c_{\vi \beta}^{\vphantom\dagger} \right)
   + \mbox{H.c.}\right] \,,
   \label{Yukawa}
\eea
We have also included a Yukawa coupling to $c_\alpha$ from an allowed term $\sim {\bm S}_{2\vi} \cdot c_{\vi \alpha}^\dagger {\bm \sigma}_{\alpha\beta} c_{\vi \beta}$, which descends from the Kondo coupling noted below Eq.~(\ref{PS1S2}). These Yukawa couplings are analogs of the green dot in Fig.~\ref{fig:feshbach}.

We can now collect all terms to write down the complete Hamiltonian needed for our analysis of the pseudogap metal, and its low temperature instabilities.
\beq
\mathcal{H}_{\rm pseudogap} = \mathcal{H}_{\rm KLmf} + \mathcal{H}_{SLf} + \mathcal{H}_Y + \mathcal{E}_2 [B,U] + \mathcal{E}_4 [B,U] \label{eq:WS}
\eeq
specified in Eqs.~(\ref{HKLmf}), (\ref{eq:fermionhop2}), (\ref{Yukawa}), (\ref{Bfunctional}). This Hamiltonian has 3 fermions $c_\alpha$, $f_{1 \alpha}$, $f_\alpha$ whose transformations under SU(2) gauge, spin rotation, and electromagnetic charge symmetries are in Eqs.~(\ref{eq:gauge}), (\ref{eq:spin}), (\ref{eq:charge}), with $f_{1\alpha}$ transforming just like $c_\alpha$. As illustrated in Fig.~\ref{fig:PhiB}, the Higgs boson $\Phi$ controls the hybridization between the electrons $c$ and the spinons $f_1$ in the Kondo lattice formed by the top two layers, while the   
the Higgs boson $B$ couples the Kondo lattice to the spin liquid on the bottom layer, with SU(2) gauge field $U_{\vi \vj}$. The energy symmetry considerations on $B$ are the same as at half-filling, and so the energy functionals $\mathcal{E}_{2,4}$ are the same as those at half-filling. The only difference from half-filling is the absence of particle-hole symmetry, and this allows linear time-derivative terms in the $B$ effective action, which will be generated from $\mathcal{H}_Y$ \cite{Christos:2023oru}.

It would be interesting to extend the variational computations, using the wavefunction in Eq.~(\ref{wavefunction}), of the metallic states  \cite{Iqbal24,HenryShiwei24,YaHui24} to the superconducting and charge-ordered states now permitted by the introduction of $B$. Such a computation would start with the ground state of the fermions $(c,f_1,f)$ for the Hamiltonian $\mathcal{H}_{\rm KLmf} + \mathcal{H}_{SLf} + \mathcal{H}_Y$ (at $U_{ij} = 1$), project out the 
spin-singlets of the ancilla layers as in Eq.~(\ref{wavefunction}), and then use this projected wavefunction to optimize the energy of a Hubbard-type model. In addition to the variational parameters in $\mathcal{H}_{\rm KLmf}$, such a computation would also optimize over an arbitrary spatial dependence of $B$.

Finally, we note in passing that there is a remarkable similarity between Eq.~(\ref{eq:WS}), and the Weinberg-Salam theory of weak interactions \cite{Christos:2023oru}. Although the dispersions of the fermions and bosons have a lattice structure, the SU(2)$\times$ U(1) gauge structure (we treat the electromagnetic U(1) as global), and the Yukawa couplings between the Higgs and the fermions are similar, with the spinons mapping to neutrinos, and the electrons mapping to  electrons.

\subsection{Anisotropic velocities in the $d$-wave superconductor}
\label{sec:aniso}

We now show how the problem of isotropic quasiparticle velocities (in contrast to observations in Ref.~\cite{Chiao00}), noted in Section~\ref{sec:halffilling}, is resolved by the presence of the pocket Fermi surfaces described by $\mathcal{H}_{\rm KLmf}$ in Eq.~(\ref{HKLmf}). The discussion below is based on the detailed computations presented in Refs.~\cite{Chatterjee16,CS23}.

Given the pocket Fermi surfaces and spinons in the FL* normal state, we imagine imposing a BCS type pairing on the Fermi surface excitations. 
If the pairing is $d$-wave, it would lead to 8 nodal Bogoliubov points as shown in Fig.~\ref{fig:flsdsc}A. 
However this state also has the 4 nodal quasiparticles of the ${\bm S}_2$ spin liquid, associated with the dispersion in Fig.~\ref{fig:fermiondisp}. So strictly speaking, this state remains fractionalized, and is {\it not\/} a conventional $d$-wave superconductor. It would be appropriate to call it d-SC*.

\begin{figure}
\centering
\includegraphics[width=4.5in]{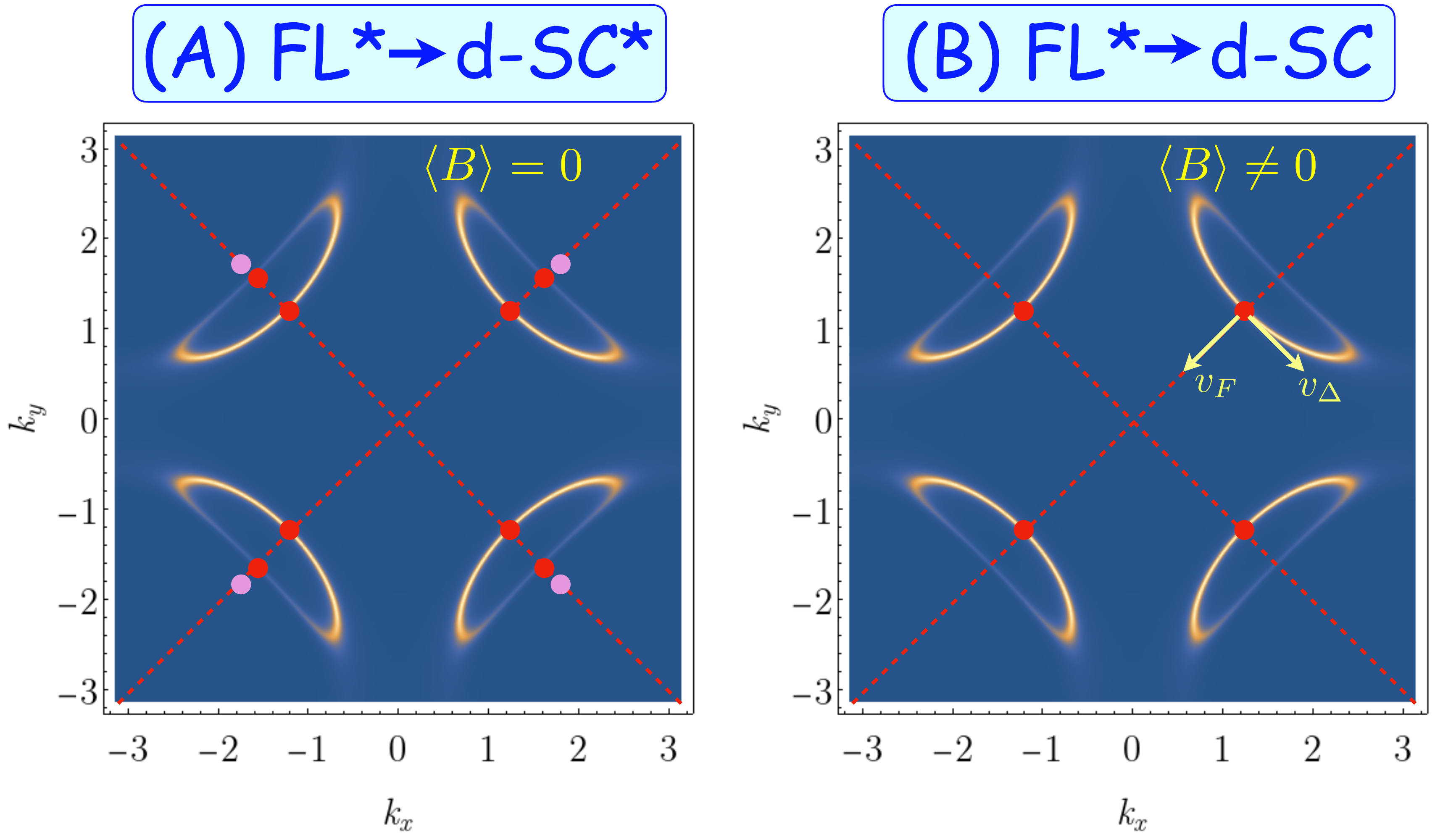}
\caption{(A) Cooper pairing the Fermi surface quasiparticles in FL* leads to d-SC* state, with 8 nodal Bogoliubov quasiparticles (red), and 4 nodal spinons (pink).
(B) Upon condensing $B$, the spinons mutually annihilate 4 of the Bogoliubov quasiparticles, leaving 4 Bogoliubov quasiparticles with $v_F \gg v_\Delta$. }
\label{fig:flsdsc}
\end{figure}
However, if we induce the pairing by the $B$ condensate in Eq.~(\ref{Yukawa}), the SU(2) gauge field is higgsed. Morever, the Yukawa coupling allows the nodal quasiparticles of the ${\bm S}_2$ spin liquid to hybridize with the Bogoliubov quasiparticles of the pocket Fermi surfaces.  The net result, sketched in Fig.~\ref{fig:flsdsc}B, is that the $B$ condensate can enable the nodal points on the `backsides' of the pocket Fermi surfaces to mutually annihilate with the spinons of the ${\bm S}_2$ spin liquid. We are then left with the 4 nodal quasiparticles on front sides of the pocket Fermi surfaces. The number of nodal points are the same as those obtained in conventional BCS theory from $d$-wave pairing of a Fermi liquid. These remaining nodal points are associated with pairing on the pocket Fermi surfaces, and these is no reason for their velocity to be isotropic (unlike the spinons).

The annihilation of the extra nodal points occurs via hybridization between the electrons and spinons within a mean-field band structure of $\mathcal{H}_{\rm KLmf} + \mathcal{H}_{SLf} + \mathcal{H}_Y$ in Eqs.~(\ref{HKLmf}), (\ref{eq:fermionhop}), and (\ref{Yukawa}) .
Furthermore, computations which diagonalize this Hamiltonian with $B$ fixed and $U=1$ do indeed yield anisotropic velocities similar to those observed. Unlike the situation in Section~\ref{sec:halffilling}, the spinons do {\it not\/} become the Bogoliubov quasiparticles in the doped case; instead the spinons are needed to annihilate the extraneous Bogoliubov quasiparticles.

An interesting prediction can be made in the particular case of the electron-doped cuprates. In these materials, photoemission experiments have observed a normal state to superconductivity which is gapped near $(\pi/2,\pi/2)$ and has spectral weight only near electron pockets at the antinode \cite{PhysRevLett.88.257001}. If $d$-wave superconductivity were to emerge as a BCS instability from such a normal state, the resulting superconductor should be gapped. However, the FL$^*$ theory yields a different prediction. Similar to how the spinon degrees of freedom emerge to annhilate with the backside pocket when $B$ Higgs condenses in the hole-doped case, the Dirac node of the spin liquid will appear with a finite spectral weight near $(\pi/2,\pi/2)$ when $B$ condenses, leading to a nodal superconductor. Such a prediction can be explored in future photoemission experiments in the electron doped cuprates \cite{Xu_2023} and could serve as an experimental test of the FL$^*$ theory.

\section{The FL*-FL transition}
\label{sec:conc}

There are two broad classes of theories for the hole-doped cuprate pseudogap in the literature. In the first class are theories in the `fluctuating order' class, with thermal fluctuations between $d$-wave superconductivity, charge and other orders  \cite{Fradkin10,Hayward:2013jna,Lee14,Nie_15,Fradkin15,Castro17,Pepin23,Fradkin25}. The present FL* approach falls into the second class, with a novel, intrinsincally quantum ground state which builds in {\it fermionic quantum entanglement\/} of mobile electrons (to be contrasted with the {\it bosonic quantum entanglement\/} of insulating spin liquids).
The existence of the FL* state sets up the possibility of a quantum phase transition from FL* to the conventional Fermi liquid (FL): a transition between two metals {\it without\/} any broken symmetry order parameter. Such an underlying FL*-FL transition has been argued \cite{SSMetlitskiPunk12} to be the key to understanding the cuprate phase diagram across optimal doping, as is reviewed in more detail elsewhere \cite{ROPP25}. 

Fig.~\ref{fig:phasediag} (a similar figure is in Ref.~\cite{SSMetlitskiPunk12}) shows the metallic phases as a function of the doping $p$. Below a low $p=p_{\rm sdw}$ we have the AF Metal, which is distinguished from FL* by the presence of long-range antiferromagnetic order. Even so, the transition from the AF metal to FL* cannot be described in the Landau theory framework: the presence of a background spin liquid requires deconfined criticality with a fractionalized order parameter \cite{RKK08b,Joshi23}.
\begin{figure}
\centering
\includegraphics[width=5in]{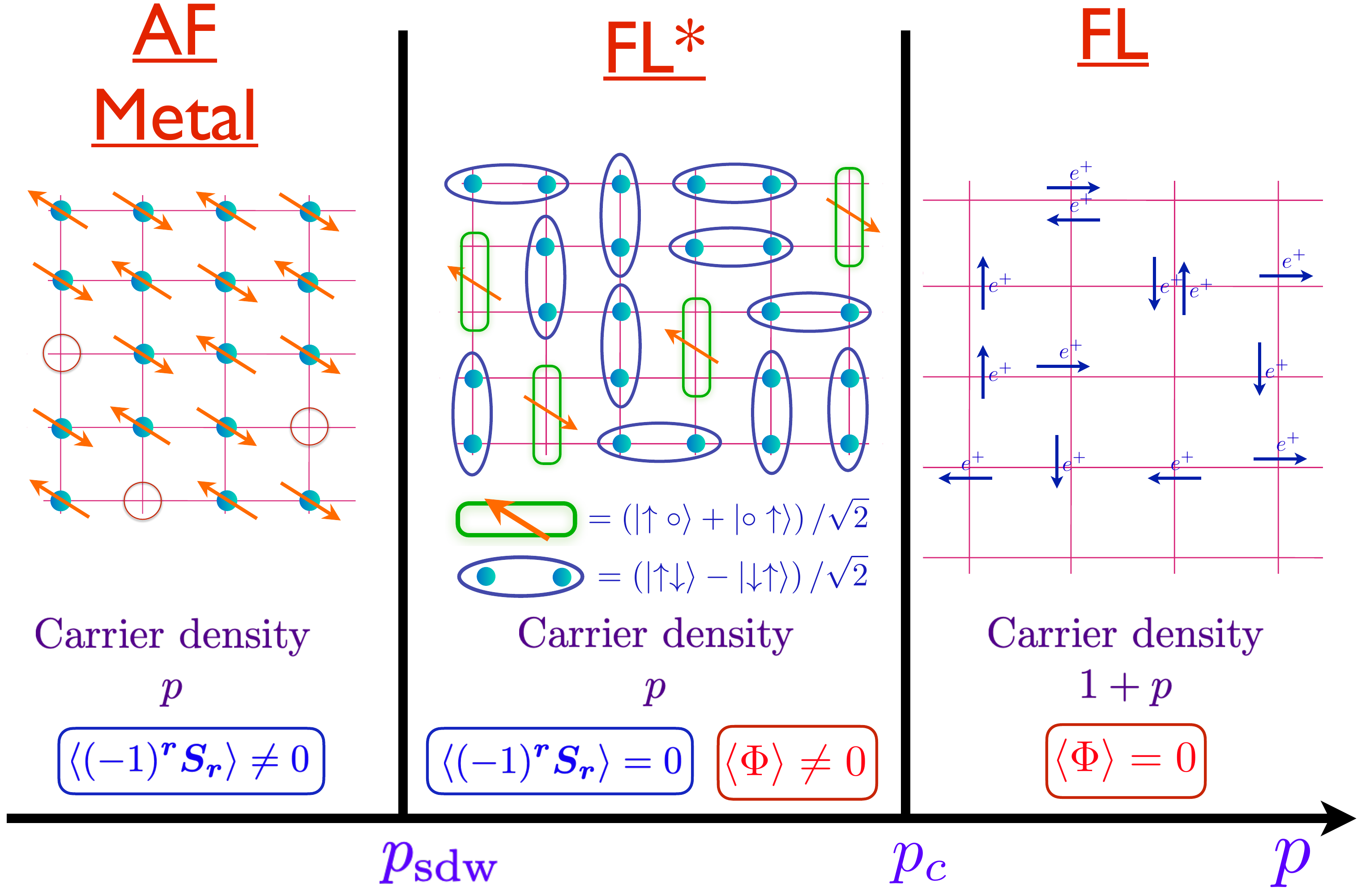}
\caption{Metallic phases for the hole-doped cuprates as a function of doping $p$, as in Ref.~\cite{SSMetlitskiPunk12}.}
\label{fig:phasediag}
\end{figure}

At a larger $p=p_c$, we have the proposed transition between FL* and FL. A theory for this has been discussed elsewhere, and it involves a gauge theory with the Higgs field $\Phi$ \cite{YaHui-ancilla1,LZDC20,YaHui-ancilla2}. This Higgs field $\Phi$ also appears as a variational parameter in the wavefunction for the FL* phase and its transition to the FL phase discussed in Section~\ref{sec:wavefunction}.
Note that $\Phi$ is distinct from the Higgs field $B$ in Sections~\ref{sec:halffilling} and~\ref{sec:pseudogap}, as is illustrated in Fig.~\ref{fig:PhiB}.

A key constraint satisfied by the theory is that the FL and FL* states have a transition at low temperatures to the {\it same\/} $d$-wave superconductor, {\it i.e.\/} there is no transition within the superconducting state between over- and under-doping. In the over-doped regime, the transition from FL to the superconductor is described by the conventional BCS framework. In the under-doped regime, the transition from FL* to the same $d$-wave superconductor is described by a confinement transition of a SU(2) gauge theory \cite{Christos:2023oru,CS23,ChristosLuo24,BCS24,Sayantan25}, employing the SU(2) gauge field of Section~\ref{sec:fermions}. But an important point is that the underlying FL*-FL transition is visible within the superconducting state in the differences in the vortex core structure \cite{ZhangSS24,Sayantan25}. In the over-doped superconductor, the vortices exhibit \cite{Renner21} the Wang-MacDonald peak \cite{WM95} obtained from a conventional BCS theory of a $d$-wave superconductor; in the under-doped superconductor the underlying FL* normal state leads to  \cite{ZhangSS24,Sayantan25} modulations in the local density of states in the halo of the vortex core \cite{Hoffman02}.

The underlying FL*-FL transition is also useful in developing a theory of the strange metal phase at higher temperatures, appearing in Figs.~\ref{fig:pdancilla1} and \ref{fig:pdancilla2}. In the presence of impurities, the strange metal has been argued \cite{Patel2,Li:2024kxr,PLA24,LPS24,Davide25} to be described by a two-dimensional extension of the Sachdev-Ye-Kitaev model \cite{SY,kitaev_talk}. 

\subsection*{Acknowledgements}

I thank Pietro Bonetti, Shubhayu Chatterjee, Maine Christos, Markus Greiner, Yasir Iqbal, Anant Kale, Martin Lebrat, Zhu-Xi Luo, Eric Mascot, Dirk Morr, Tobias M\"uller, Alexander Nikolaenko, Harshit Pandey, Matthias Scheurer, Leyna Shackleton, Ravi Shanker, Sayantan Sharma, Ronny Thomale, Maria Tikhanovskaya, Ya-Hui Zhang, and Jing-Yu Zhao for recent collaborations \cite{YaHui-ancilla1,Mascot22,Nikolaenko:2021vlw,CS23,Christos:2023oru,ShackletonZ2,ChristosLuo24,Iqbal24,LiebPRB,Sayantan25,Zhao_Yamaji_25} on the work reviewed here. I thank my co-authors of the review in Ref.~\cite{ROPP25}, from which some of the text is drawn, and A. Nikolaenko for assistance with Figs.~\ref{fig:sdwphoto} and~\ref{fig:flsphoto}. I thank Henning Schl\"omer for valuable comments on the manuscript.
This research was supported by NSF Grant DMR-2245246 and by the Simons Collaboration on Ultra-Quantum Matter which is a grant from the Simons Foundation (651440, S. S.). 

\bibliography{refs}

\begin{thebibliography}{100}
\providecommand{\url}[1]{\texttt{#1}}
\providecommand{\urlprefix}{URL }
\expandafter\ifx\csname urlstyle\endcsname\relax
  \providecommand{\doi}[1]{doi:\discretionary{}{}{}#1}\else
  \providecommand{\doi}{doi:\discretionary{}{}{}\begingroup
  \urlstyle{rm}\Url}\fi
\providecommand{\eprint}[2][]{\url{#2}}

\bibitem{NRSS91}
N.~Read and S.~Sachdev,
\newblock \emph{{Large $N$ expansion for frustrated quantum antiferromagnets}},
\newblock Phys. Rev. Lett. \textbf{66}, 1773 (1991),
\newblock \doi{10.1103/PhysRevLett.66.1773}.

\bibitem{XGW91}
X.~G. Wen,
\newblock \emph{Mean-field theory of spin-liquid states with finite energy gap
  and topological orders},
\newblock Phys. Rev. B \textbf{44}, 2664 (1991),
\newblock \doi{10.1103/PhysRevB.44.2664}.

\bibitem{SSkagome}
S.~Sachdev,
\newblock \emph{{Kagome and triangular-lattice Heisenberg antiferromagnets:
  Ordering from quantum fluctuations and quantum-disordered ground states with
  unconfined bosonic spinons}},
\newblock Phys. Rev. B \textbf{45}, 12377 (1992),
\newblock \doi{10.1103/PhysRevB.45.12377}.

\bibitem{KitaevToric}
A.~Y. {Kitaev},
\newblock \emph{{Fault-tolerant quantum computation by anyons}},
\newblock Annals of Physics \textbf{303}(1), 2 (2003),
\newblock \doi{10.1016/S0003-4916(02)00018-0},
\newblock \eprint{quant-ph/9707021}.

\bibitem{NRSS89prl}
N.~Read and S.~Sachdev,
\newblock \emph{{Valence-bond and spin-Peierls ground states of low-dimensional
  quantum antiferromagnets}},
\newblock Phys. Rev. Lett. \textbf{62}, 1694 (1989),
\newblock \doi{10.1103/PhysRevLett.62.1694}.

\bibitem{NRSS90}
N.~Read and S.~Sachdev,
\newblock \emph{{Spin-Peierls, valence-bond solid, and N\'eel ground states of
  low-dimensional quantum antiferromagnets}},
\newblock Phys. Rev. B \textbf{42}, 4568 (1990),
\newblock \doi{10.1103/PhysRevB.42.4568}.

\bibitem{AM88}
I.~Affleck and J.~B. Marston,
\newblock \emph{{Large-$n$ limit of the Heisenberg-Hubbard model: Implications
  for high-${T}_{c}$ superconductors}},
\newblock Phys. Rev. B \textbf{37}, 3774 (1988),
\newblock \doi{10.1103/PhysRevB.37.3774}.

\bibitem{Affleck-SU2}
I.~Affleck, Z.~Zou, T.~Hsu and P.~W. Anderson,
\newblock \emph{{SU(2) gauge symmetry of the large-$U$ limit of the Hubbard
  model}},
\newblock Phys. Rev. B \textbf{38}, 745 (1988),
\newblock \doi{10.1103/PhysRevB.38.745}.

\bibitem{Fradkin88}
E.~Dagotto, E.~Fradkin and A.~Moreo,
\newblock \emph{{SU(2) gauge invariance and order parameters in strongly
  coupled electronic systems}},
\newblock Phys. Rev. B \textbf{38}, 2926 (1988),
\newblock \doi{10.1103/PhysRevB.38.2926}.

\bibitem{LeeWen96}
X.-G. {Wen} and P.~A. {Lee},
\newblock \emph{{Theory of Underdoped Cuprates}},
\newblock Phys. Rev. Lett. \textbf{76}(3), 503 (1996),
\newblock \doi{10.1103/PhysRevLett.76.503},
\newblock \eprint{cond-mat/9506065}.

\bibitem{Wang17}
C.~Wang, A.~Nahum, M.~A. Metlitski, C.~Xu and T.~Senthil,
\newblock \emph{{Deconfined quantum critical points: symmetries and
  dualities}},
\newblock Phys. Rev. X \textbf{7}(3), 031051 (2017),
\newblock \doi{10.1103/PhysRevX.7.031051},
\newblock \eprint{1703.02426}.

\bibitem{Lake19}
D.~{Aasen}, E.~{Lake} and K.~{Walker},
\newblock \emph{{Fermion condensation and super pivotal categories}},
\newblock Journal of Mathematical Physics \textbf{60}(12), 121901 (2019),
\newblock \doi{10.1063/1.5045669},
\newblock \eprint{1709.01941}.

\bibitem{King25}
E.~R. {Anschuetz}, C.-F. {Chen}, B.~T. {Kiani} and R.~{King},
\newblock \emph{{Strongly Interacting Fermions Are Nontrivial yet Nonglassy}},
\newblock Phys. Rev. Lett. \textbf{135}(3), 030602 (2025),
\newblock \doi{10.1103/cbqf-d24r},
\newblock \eprint{2408.15699}.

\bibitem{LiebScience}
M.~{Lebrat}, A.~{Kale}, L.~H. {Kendrick}, M.~{Xu}, Y.~{Gang}, A.~{Nikolaenko},
  P.~M. {Bonetti}, S.~{Sachdev} and M.~{Greiner},
\newblock \emph{{Ferrimagnetism of ultracold fermions in a multiband Hubbard
  system}},
\newblock Science \textbf{392}(6798), 612 (2026),
\newblock \doi{10.1126/science.adq2411},
\newblock \eprint{2404.17555}.

\bibitem{LiebPRB}
A.~{Nikolaenko}, P.~M. {Bonetti}, A.~{Kale}, M.~{Lebrat}, M.~{Greiner} and
  S.~{Sachdev},
\newblock \emph{{Canted magnetism and $Z_2$ fractionalization in metallic
  states of the Lieb lattice Hubbard model near quarter filling}},
\newblock Phys. Rev. B \textbf{112}(4), 045129 (2025),
\newblock \doi{10.1103/l9xn-h95s},
\newblock \eprint{2502.12235}.

\bibitem{Ramshaw22}
Y.~Fang, G.~Grissonnanche, A.~Legros, S.~Verret, F.~Lalibert{\'e},
  C.~Collignon, A.~Ataei, M.~Dion, J.~Zhou, D.~Graf, M.~J. Lawler, P.~A.
  Goddard \emph{et~al.},
\newblock \emph{Fermi surface transformation at the pseudogap critical point of
  a cuprate superconductor},
\newblock Nature Physics \textbf{18}(5), 558 (2022),
\newblock \doi{10.1038/s41567-022-01514-1},
\newblock \eprint{2004.01725}.

\bibitem{Yamaji24}
M.~K. {Chan}, K.~A. {Schreiber}, O.~E. {Ayala-Valenzuela}, E.~D. {Bauer},
  A.~{Shekhter} and N.~{Harrison},
\newblock \emph{{Observation of the Yamaji effect in a cuprate
  superconductor}},
\newblock Nature Physics \textbf{21}(11), 1753 (2025),
\newblock \doi{10.1038/s41567-025-03032-2},
\newblock \eprint{2411.10631}.

\bibitem{Chiao00}
M.~{Chiao}, R.~W. {Hill}, C.~{Lupien}, L.~{Taillefer}, P.~{Lambert},
  R.~{Gagnon} and P.~{Fournier},
\newblock \emph{{Low-energy quasiparticles in cuprate superconductors: A
  quantitative analysis}},
\newblock Phys. Rev. B \textbf{62}(5), 3554 (2000),
\newblock \doi{10.1103/PhysRevB.62.3554},
\newblock \eprint{cond-mat/9910367}.

\bibitem{TSSSMV03}
T.~{Senthil}, S.~{Sachdev} and M.~{Vojta},
\newblock \emph{{Fractionalized Fermi Liquids}},
\newblock Phys. Rev. Lett. \textbf{90}(21), 216403 (2003),
\newblock \doi{10.1103/PhysRevLett.90.216403},
\newblock \eprint{cond-mat/0209144}.

\bibitem{TSSSMV04}
T.~{Senthil}, M.~{Vojta} and S.~{Sachdev},
\newblock \emph{{Weak magnetism and non-Fermi liquids near heavy-fermion
  critical points}},
\newblock Phys. Rev. B \textbf{69}(3), 035111 (2004),
\newblock \doi{10.1103/PhysRevB.69.035111},
\newblock \eprint{cond-mat/0305193}.

\bibitem{Becca20}
F.~{Ferrari} and F.~{Becca},
\newblock \emph{{Gapless spin liquid and valence-bond solid in the
  J$_{1}$-J$_{2}$ Heisenberg model on the square lattice: Insights from singlet
  and triplet excitations}},
\newblock Phys. Rev. B \textbf{102}(1), 014417 (2020),
\newblock \doi{10.1103/PhysRevB.102.014417},
\newblock \eprint{2005.12941}.

\bibitem{Imada21}
Y.~{Nomura} and M.~{Imada},
\newblock \emph{{Dirac-Type Nodal Spin Liquid Revealed by Refined Quantum
  Many-Body Solver Using Neural-Network Wave Function, Correlation Ratio, and
  Level Spectroscopy}},
\newblock Physical Review X \textbf{11}(3), 031034 (2021),
\newblock \doi{10.1103/PhysRevX.11.031034},
\newblock \eprint{2005.14142}.

\bibitem{Gu24}
W.-Y. {Liu}, D.~{Poilblanc}, S.-S. {Gong}, W.-Q. {Chen} and Z.-C. {Gu},
\newblock \emph{{Tensor network study of the spin-1/2 square-lattice
  J$_{1}$-J$_{2}$-J$_{3}$ model: Incommensurate spiral order, mixed
  valence-bond solids, and multicritical points}},
\newblock Phys. Rev. B \textbf{109}(23), 235116 (2024),
\newblock \doi{10.1103/PhysRevB.109.235116},
\newblock \eprint{2309.13301}.

\bibitem{Sandvik24}
J.~{Takahashi}, H.~{Shao}, B.~{Zhao}, W.~{Guo} and A.~W. {Sandvik},
\newblock \emph{{SO(5) multicriticality in two-dimensional quantum magnets}},
\newblock arXiv e-prints arXiv:2405.06607 (2024),
\newblock \doi{10.48550/arXiv.2405.06607},
\newblock \eprint{2405.06607}.

\bibitem{AA88}
D.~P. Arovas and A.~Auerbach,
\newblock \emph{{Functional integral theories of low-dimensional quantum
  Heisenberg models}},
\newblock Phys. Rev. B \textbf{38}, 316 (1988),
\newblock \doi{10.1103/PhysRevB.38.316}.

\bibitem{senthil1}
T.~{Senthil}, A.~{Vishwanath}, L.~{Balents}, S.~{Sachdev} and M.~P.~A.
  {Fisher},
\newblock \emph{{Deconfined Quantum Critical Points}},
\newblock Science \textbf{303}, 1490 (2004),
\newblock \doi{10.1126/science.1091806},
\newblock \eprint{cond-mat/0311326}.

\bibitem{senthil2}
T.~{Senthil}, L.~{Balents}, S.~{Sachdev}, A.~{Vishwanath} and M.~P.~A.
  {Fisher},
\newblock \emph{{Quantum criticality beyond the Landau-Ginzburg-Wilson
  paradigm}},
\newblock Phys. Rev. B \textbf{70}(14), 144407 (2004),
\newblock \doi{10.1103/PhysRevB.70.144407},
\newblock \eprint{cond-mat/0312617}.

\bibitem{HPS11}
Y.~{Huh}, M.~{Punk} and S.~{Sachdev},
\newblock \emph{{Vison states and confinement transitions of $\mathbb{Z}_{2}$
  spin liquids on the kagome lattice}},
\newblock Phys. Rev. B \textbf{84}(9), 094419 (2011),
\newblock \doi{10.1103/PhysRevB.84.094419},
\newblock \eprint{1106.3330}.

\bibitem{HFS10}
Y.~{Huh}, L.~{Fritz} and S.~{Sachdev},
\newblock \emph{{Quantum criticality of the kagome antiferromagnet with
  Dzyaloshinskii-Moriya interactions}},
\newblock Phys. Rev. B \textbf{81}(14), 144432 (2010),
\newblock \doi{10.1103/PhysRevB.81.144432},
\newblock \eprint{1003.0891}.

\bibitem{Batista21}
A.~O. {Scheie}, E.~A. {Ghioldi}, J.~{Xing}, J.~A.~M. {Paddison}, N.~E.
  {Sherman}, M.~{Dupont}, L.~D. {Sanjeewa}, S.~{Lee}, A.~J. {Woods},
  D.~{Abernathy}, D.~M. {Pajerowski}, T.~J. {Williams} \emph{et~al.},
\newblock \emph{{Proximate spin liquid and fractionalization in the triangular
  antiferromagnet KYbSe$_{2}$}},
\newblock Nature Physics \textbf{20}(1), 74 (2024),
\newblock \doi{10.1038/s41567-023-02259-1},
\newblock \eprint{2109.11527}.

\bibitem{Scheie24}
A.~O. {Scheie}, M.~{Lee}, K.~{Wang}, P.~{Laurell}, E.~S. {Choi},
  D.~{Pajerowski}, Q.~{Zhang}, J.~{Ma}, H.~D. {Zhou}, S.~{Lee}, C.~{Huan},
  S.~M. {Thomas} \emph{et~al.},
\newblock \emph{{Spectrum and low-temperature bulk properties of triangular
  quantum spin liquid candidate NaYbSe$_2$}},
\newblock arXiv e-prints arXiv:2406.17773 (2024),
\newblock \doi{10.48550/arXiv.2406.17773},
\newblock \eprint{2406.17773}.

\bibitem{Bonderson16}
P.~{Bonderson}, M.~{Cheng}, K.~{Patel} and E.~{Plamadeala},
\newblock \emph{{Topological Enrichment of Luttinger's Theorem}},
\newblock arXiv e-prints arXiv:1601.07902 (2016),
\newblock \doi{10.48550/arXiv.1601.07902},
\newblock \eprint{1601.07902}.

\bibitem{SenthilElse21}
D.~V. {Else}, R.~{Thorngren} and T.~{Senthil},
\newblock \emph{{Non-Fermi Liquids as Ersatz Fermi Liquids: General Constraints
  on Compressible Metals}},
\newblock Physical Review X \textbf{11}(2), 021005 (2021),
\newblock \doi{10.1103/PhysRevX.11.021005},
\newblock \eprint{2007.07896}.

\bibitem{Seiberg23}
M.~{Cheng} and N.~{Seiberg},
\newblock \emph{{Lieb-Schultz-Mattis, Luttinger, and 't Hooft - anomaly
  matching in lattice systems}},
\newblock SciPost Physics \textbf{15}(2), 051 (2023),
\newblock \doi{10.21468/SciPostPhys.15.2.051},
\newblock \eprint{2211.12543}.

\bibitem{Meng26}
L.~{Zou} and M.~{Cheng},
\newblock \emph{{Lieb-Schultz-Mattis Anomalies and Anomaly Matching}},
\newblock arXiv e-prints arXiv:2604.00347 (2026),
\newblock \doi{10.48550/arXiv.2604.00347},
\newblock \eprint{2604.00347}.

\bibitem{RJSS91}
R.~A. Jalabert and S.~Sachdev,
\newblock \emph{{Spontaneous alignment of frustrated bonds in an anisotropic,
  three-dimensional Ising model}},
\newblock Phys. Rev. B \textbf{44}, 686 (1991),
\newblock \doi{10.1103/PhysRevB.44.686}.

\bibitem{SSMV99}
S.~{Sachdev} and M.~{Vojta},
\newblock \emph{{Translational symmetry breaking in two-dimensional
  antiferromagnets and superconductors}},
\newblock J. Phys. Soc. Jpn {\bf 69}, Supp. B, 1  (1999),
\newblock \doi{10.48550/arXiv.cond-mat/9910231},
\newblock \eprint{cond-mat/9910231}.

\bibitem{YuanMing}
Y.-M. {Lu},
\newblock \emph{{Symmetric Z$_{2}$ spin liquids and their neighboring phases on
  triangular lattice}},
\newblock Phys. Rev. B \textbf{93}(16), 165113 (2016),
\newblock \doi{10.1103/PhysRevB.93.165113},
\newblock \eprint{1505.06495}.

\bibitem{Wang16}
X.~{Yang} and F.~{Wang},
\newblock \emph{{Schwinger boson spin-liquid states on square lattice}},
\newblock Phys. Rev. B \textbf{94}(3), 035160 (2016),
\newblock \doi{10.1103/PhysRevB.94.035160},
\newblock \eprint{1507.07621}.

\bibitem{ThomsonSS18}
A.~{Thomson} and S.~{Sachdev},
\newblock \emph{{Fermionic Spinon Theory of Square Lattice Spin Liquids near
  the N{\'e}el State}},
\newblock Physical Review X \textbf{8}(1), 011012 (2018),
\newblock \doi{10.1103/PhysRevX.8.011012},
\newblock \eprint{1708.04626}.

\bibitem{GilAshvin}
Y.-M. Lu, G.~Y. Cho and A.~Vishwanath,
\newblock \emph{Unification of bosonic and fermionic theories of spin liquids
  on the kagome lattice},
\newblock Phys. Rev. B \textbf{96}, 205150 (2017),
\newblock \doi{10.1103/PhysRevB.96.205150},
\newblock \eprint{1403.0575}.

\bibitem{ParkSS02}
K.~{Park} and S.~{Sachdev},
\newblock \emph{{Bond and N{\'e}el order and fractionalization in ground states
  of easy-plane antiferromagnets in two dimensions}},
\newblock Phys. Rev. B \textbf{65}(22), 220405 (2002),
\newblock \doi{10.1103/PhysRevB.65.220405},
\newblock \eprint{cond-mat/0112003}.

\bibitem{ShackletonZ2}
L.~{Shackleton} and S.~{Sachdev},
\newblock \emph{{Sign-problem-free effective models of triangular lattice
  quantum antiferromagnets}},
\newblock Phys. Rev. B \textbf{111}(7), 075101 (2025),
\newblock \doi{10.1103/PhysRevB.111.075101},
\newblock \eprint{2311.01572}.

\bibitem{SenthilFisher00}
T.~{Senthil} and M.~P.~A. {Fisher},
\newblock \emph{{Z$_{2}$ gauge theory of electron fractionalization in strongly
  correlated systems}},
\newblock Phys. Rev. B \textbf{62}(12), 7850 (2000),
\newblock \doi{10.1103/PhysRevB.62.7850},
\newblock \eprint{cond-mat/9910224}.

\bibitem{CSS93}
A.~V. {Chubukov}, T.~{Senthil} and S.~{Sachdev},
\newblock \emph{{Universal magnetic properties of frustrated quantum
  antiferromagnets in two dimensions}},
\newblock Phys. Rev. Lett. \textbf{72}, 2089 (1994),
\newblock \doi{10.1103/PhysRevLett.72.2089},
\newblock \eprint{cond-mat/9311045}.

\bibitem{CSS94}
A.~V. {Chubukov}, S.~{Sachdev} and T.~{Senthil},
\newblock \emph{{Quantum phase transitions in frustrated quantum
  antiferromagnets}},
\newblock Nucl. Phys. B \textbf{426}, 601 (1994),
\newblock \doi{10.1016/0550-3213(94)90023-X},
\newblock \eprint{cond-mat/9402006}.

\bibitem{WhitsittZ2}
S.~{Whitsitt} and S.~{Sachdev},
\newblock \emph{{Transition from the Z$_{2}$ spin liquid to antiferromagnetic
  order: Spectrum on the torus}},
\newblock Phys. Rev. B \textbf{94}(8), 085134 (2016),
\newblock \doi{10.1103/PhysRevB.94.085134},
\newblock \eprint{1603.05652}.

\bibitem{song2018}
X.-Y. Song, Y.-C. He, A.~Vishwanath and C.~Wang,
\newblock \emph{{From spinon band topology to the symmetry quantum numbers of
  monopoles in Dirac spin liquids}},
\newblock Phys. Rev. X \textbf{10}(1), 011033 (2020),
\newblock \doi{10.1103/PhysRevX.10.011033},
\newblock \eprint{1811.11182}.

\bibitem{song2019}
X.-Y. {Song}, C.~{Wang}, A.~{Vishwanath} and Y.-C. {He},
\newblock \emph{{Unifying description of competing orders in two-dimensional
  quantum magnets}},
\newblock Nature Communications \textbf{10}, 4254 (2019),
\newblock \doi{10.1038/s41467-019-11727-3},
\newblock \eprint{1811.11186}.

\bibitem{wietek2023}
A.~{Wietek}, S.~{Capponi} and A.~M. {L{\"a}uchli},
\newblock \emph{{Quantum Electrodynamics in 2 +1 Dimensions as the Organizing
  Principle of a Triangular Lattice Antiferromagnet}},
\newblock Physical Review X \textbf{14}(2), 021010 (2024),
\newblock \doi{10.1103/PhysRevX.14.021010},
\newblock \eprint{2303.01585}.

\bibitem{Knolle25}
J.~{Willsher} and J.~{Knolle},
\newblock \emph{{Dynamics and stability of U(1) spin liquids beyond mean-field
  theory: Triangular-lattice $J_1$-$J_2$ Heisenberg model}},
\newblock arXiv e-prints arXiv:2503.13831 (2025),
\newblock \doi{10.48550/arXiv.2503.13831},
\newblock \eprint{2503.13831}.

\bibitem{SJ90}
S.~Sachdev and R.~Jalabert,
\newblock \emph{Effective lattice models for two-dimensional antiferromagnets},
\newblock Modern Physics Letters B \textbf{04}(16), 1043 (1990),
\newblock \doi{10.1142/S0217984990001318}.

\bibitem{Nahum:2015vka}
A.~Nahum, P.~Serna, J.~T. Chalker, M.~Ortu\~no and A.~M. Somoza,
\newblock \emph{{Emergent SO(5) Symmetry at the N\'eel to Valence-Bond-Solid
  Transition}},
\newblock Phys. Rev. Lett. \textbf{115}(26), 267203 (2015),
\newblock \doi{10.1103/PhysRevLett.115.267203},
\newblock \eprint{1508.06668}.

\bibitem{Meng24}
B.-B. {Chen}, X.~{Zhang}, Y.~{Wang}, K.~{Sun} and Z.~Y. {Meng},
\newblock \emph{{Phases of (2+1)$D$ SO(5) Nonlinear Sigma Model with a
  Topological Term on a Sphere: Multicritical Point and Disorder Phase}},
\newblock Phys. Rev. Lett. \textbf{132}(24), 246503 (2024),
\newblock \doi{10.1103/PhysRevLett.132.246503},
\newblock \eprint{2307.05307}.

\bibitem{Chester24}
S.~M. Chester and N.~Su,
\newblock \emph{{Bootstrapping Deconfined Quantum Tricriticality}},
\newblock Phys. Rev. Lett. \textbf{132}(11), 111601 (2024),
\newblock \doi{10.1103/PhysRevLett.132.111601},
\newblock \eprint{2310.08343}.

\bibitem{Fuzzy24}
Z.~{Zhou}, L.~{Hu}, W.~{Zhu} and Y.-C. {He},
\newblock \emph{{SO(5) Deconfined Phase Transition under the Fuzzy-Sphere
  Microscope: Approximate Conformal Symmetry, Pseudo-Criticality, and Operator
  Spectrum}},
\newblock Physical Review X \textbf{14}(2), 021044 (2024),
\newblock \doi{10.1103/PhysRevX.14.021044},
\newblock \eprint{2306.16435}.

\bibitem{AndreiColeman2}
P.~Coleman and N.~Andrei,
\newblock \emph{Kondo-stabilised spin liquids and heavy fermion
  superconductivity},
\newblock Journal of Physics: Condensed Matter \textbf{1}(26), 4057 (1989),
\newblock \doi{10.1088/0953-8984/1/26/003}.

\bibitem{BZA87}
G.~Baskaran, Z.~Zou and P.~Anderson,
\newblock \emph{{The resonating valence bond state and high-$T_c$
  superconductivity — A mean field theory}},
\newblock Solid State Communications \textbf{63}(11), 973 (1987),
\newblock \doi{10.1016/0038-1098(87)90642-9}.

\bibitem{Kotliar87}
G.~Kotliar,
\newblock \emph{{Resonating valence bonds and $d$-wave superconductivity}},
\newblock Phys. Rev. B \textbf{37}, 3664 (1988),
\newblock \doi{10.1103/PhysRevB.37.3664}.

\bibitem{Zhang88}
F.~C. {Zhang}, C.~{Gros}, T.~M. {Rice} and H.~{Shiba},
\newblock \emph{{A renormalised Hamiltonian approach for a resonant valence
  bond wavefunction}},
\newblock Superconductor Science and Technology \textbf{1}, 36 (1988),
\newblock \doi{10.1088/0953-2048/1/1/009},
\newblock \eprint{cond-mat/0311604}.

\bibitem{LeeWenRMP}
P.~A. Lee, N.~Nagaosa and X.-G. Wen,
\newblock \emph{{Doping a Mott insulator: Physics of high-temperature
  superconductivity}},
\newblock Rev. Mod. Phys. \textbf{78}, 17 (2006),
\newblock \doi{10.1103/RevModPhys.78.17},
\newblock \eprint{cond-mat/0410445}.

\bibitem{Alicea08}
J.~{Alicea},
\newblock \emph{{Monopole quantum numbers in the staggered flux spin liquid}},
\newblock Phys. Rev. B \textbf{78}(3), 035126 (2008),
\newblock \doi{10.1103/PhysRevB.78.035126},
\newblock \eprint{0804.0786}.

\bibitem{Song1}
X.-Y. Song, Y.-C. He, A.~Vishwanath and C.~Wang,
\newblock \emph{{From spinon band topology to the symmetry quantum numbers of
  monopoles in Dirac spin liquids}},
\newblock Phys. Rev. X \textbf{10}(1), 011033 (2020),
\newblock \doi{10.1103/PhysRevX.10.011033},
\newblock \eprint{1811.11182}.

\bibitem{RanWen06}
Y.~{Ran} and X.-g. {Wen},
\newblock \emph{{Continuous quantum phase transitions beyond Landau's paradigm
  in a large-N spin model}},
\newblock arXiv e-prints cond-mat/0609620 (2006),
\newblock \doi{10.48550/arXiv.cond-mat/0609620},
\newblock \eprint{cond-mat/0609620}.

\bibitem{Abanov:1999qz}
A.~G. Abanov and P.~B. Wiegmann,
\newblock \emph{{Theta terms in nonlinear sigma models}},
\newblock Nucl. Phys. B \textbf{570}, 685 (2000),
\newblock \doi{10.1016/S0550-3213(99)00820-2},
\newblock \eprint{hep-th/9911025}.

\bibitem{LS15}
J.~{Lee} and S.~{Sachdev},
\newblock \emph{{Wess-Zumino-Witten Terms in Graphene Landau Levels}},
\newblock Phys. Rev. Lett. \textbf{114}(22), 226801 (2015),
\newblock \doi{10.1103/PhysRevLett.114.226801},
\newblock \eprint{1411.5684}.

\bibitem{TanakaHu}
A.~{Tanaka} and X.~{Hu},
\newblock \emph{{Many-Body Spin Berry Phases Emerging from the
  {\ensuremath{\pi}}-Flux State: Competition between Antiferromagnetism and the
  Valence-Bond-Solid State}},
\newblock Phys. Rev. Lett. \textbf{95}(3), 036402 (2005),
\newblock \doi{10.1103/PhysRevLett.95.036402},
\newblock \eprint{cond-mat/0501365}.

\bibitem{SenthilFisher06}
T.~{Senthil} and M.~P.~A. {Fisher},
\newblock \emph{{Competing orders, nonlinear sigma models, and topological
  terms in quantum magnets}},
\newblock Phys. Rev. B \textbf{74}(6), 064405 (2006),
\newblock \doi{10.1103/PhysRevB.74.064405},
\newblock \eprint{cond-mat/0510459}.

\bibitem{Yasir19}
M.~{Hering}, J.~{Sonnenschein}, Y.~{Iqbal} and J.~{Reuther},
\newblock \emph{{Characterization of quantum spin liquids and their spinon band
  structures via functional renormalization}},
\newblock Phys. Rev. B \textbf{99}(10), 100405 (2019),
\newblock \doi{10.1103/PhysRevB.99.100405},
\newblock \eprint{1806.05021}.

\bibitem{SS80}
B.~I. Shraiman and E.~D. Siggia,
\newblock \emph{{Mobile Vacancies in a Quantum Heisenberg Antiferromagnet}},
\newblock Phys. Rev. Lett. \textbf{61}, 467 (1988),
\newblock \doi{10.1103/PhysRevLett.61.467}.

\bibitem{Schulz90}
H.~J. Schulz,
\newblock \emph{Effective action for strongly correlated fermions from
  functional integrals},
\newblock Phys. Rev. Lett. \textbf{65}, 2462 (1990),
\newblock \doi{10.1103/PhysRevLett.65.2462}.

\bibitem{Dupuis02}
N.~{Dupuis},
\newblock \emph{{Spin fluctuations and pseudogap in the two-dimensional
  half-filled Hubbard model at weak coupling}},
\newblock Phys. Rev. B \textbf{65}(24), 245118 (2002),
\newblock \doi{10.1103/PhysRevB.65.245118},
\newblock \eprint{cond-mat/0110138}.

\bibitem{Dupuis04}
K.~{Borejsza} and N.~{Dupuis},
\newblock \emph{{Antiferromagnetism and single-particle properties in the
  two-dimensional half-filled Hubbard model: A nonlinear sigma model
  approach}},
\newblock Phys. Rev. B \textbf{69}(8), 085119 (2004),
\newblock \doi{10.1103/PhysRevB.69.085119},
\newblock \eprint{cond-mat/0307238}.

\bibitem{SS09}
S.~{Sachdev}, M.~A. {Metlitski}, Y.~{Qi} and C.~{Xu},
\newblock \emph{{Fluctuating spin density waves in metals}},
\newblock Phys. Rev. B \textbf{80}(15), 155129 (2009),
\newblock \doi{10.1103/PhysRevB.80.155129},
\newblock \eprint{0907.3732}.

\bibitem{DCSS15b}
D.~{Chowdhury} and S.~{Sachdev},
\newblock \emph{{Higgs criticality in a two-dimensional metal}},
\newblock Phys. Rev. B \textbf{91}(11), 115123 (2015),
\newblock \doi{10.1103/PhysRevB.91.115123},
\newblock \eprint{1412.1086}.

\bibitem{DCSS15}
D.~{Chowdhury} and S.~{Sachdev},
\newblock \emph{{The Enigma of the Pseudogap Phase of the Cuprate
  Superconductors}},
\newblock In J.~{Jedrzejewski}, ed., \emph{Quantum criticality in condensed
  matter}, 50th Karpacz Winter School of Theoretical Physics, pp. 1--43. {World
  Scientific},
\newblock \doi{10.1142/9789814704090_0001} (2015), \eprint{1501.00002}.

\bibitem{CSS17}
S.~Chatterjee, S.~Sachdev and M.~Scheurer,
\newblock \emph{{Intertwining topological order and broken symmetry in a theory
  of fluctuating spin density waves}},
\newblock Phys. Rev. Lett. \textbf{119}(22), 227002 (2017),
\newblock \doi{10.1103/PhysRevLett.119.227002},
\newblock \eprint{1705.06289}.

\bibitem{MSSS18}
M.~S. {Scheurer} and S.~{Sachdev},
\newblock \emph{{Orbital currents in insulating and doped antiferromagnets}},
\newblock Phys. Rev. B \textbf{98}(23), 235126 (2018),
\newblock \doi{10.1103/PhysRevB.98.235126},
\newblock \eprint{1808.04826}.

\bibitem{SSST19}
S.~Sachdev, H.~D. Scammell, M.~S. Scheurer and G.~Tarnopolsky,
\newblock \emph{{Gauge theory for the cuprates near optimal doping}},
\newblock Phys. Rev. B \textbf{99}(5), 054516 (2019),
\newblock \doi{10.1103/PhysRevB.99.054516},
\newblock \eprint{1811.04930}.

\bibitem{Bonetti22}
P.~M. {Bonetti} and W.~{Metzner},
\newblock \emph{{SU(2) gauge theory of the pseudogap phase in the
  two-dimensional Hubbard model}},
\newblock Phys. Rev. B \textbf{106}(20), 205152 (2022),
\newblock \doi{10.1103/PhysRevB.106.205152},
\newblock \eprint{2207.00829}.

\bibitem{Bonetti23}
R.~{Scholle}, P.~M. {Bonetti}, D.~{Vilardi} and W.~{Metzner},
\newblock \emph{{Comprehensive mean-field analysis of magnetic and charge
  orders in the two-dimensional Hubbard model}},
\newblock Phys. Rev. B \textbf{108}(3), 035139 (2023),
\newblock \doi{10.1103/PhysRevB.108.035139},
\newblock \eprint{2303.15358}.

\bibitem{ChristosLuo24}
M.~{Christos}, L.~{Shackleton}, S.~{Sachdev} and Z.-X. {Luo},
\newblock \emph{{Deconfined quantum criticality of nodal d -wave
  superconductivity, N{\'e}el order, and charge order on the square lattice at
  half-filling}},
\newblock Physical Review Research \textbf{6}(3), 033018 (2024),
\newblock \doi{10.1103/PhysRevResearch.6.033018},
\newblock \eprint{2402.09502}.

\bibitem{AssaadImada}
F.~F. {Assaad}, M.~{Imada} and D.~J. {Scalapino},
\newblock \emph{{Quantum Transition between an Antiferromagnetic Mott Insulator
  and $d_{x^2-y^2}$ Superconductor in Two Dimensions}},
\newblock Phys. Rev. Lett. \textbf{77}(22), 4592 (1996),
\newblock \doi{10.1103/PhysRevLett.77.4592},
\newblock \eprint{cond-mat/9609034}.

\bibitem{Assaad22}
A.~{G{\"o}tz}, S.~{Beyl}, M.~{Hohenadler} and F.~F. {Assaad},
\newblock \emph{{Valence-bond solid to antiferromagnet transition in the
  two-dimensional Su-Schrieffer-Heeger model by Langevin dynamics}},
\newblock Phys. Rev. B \textbf{105}(8), 085151 (2022),
\newblock \doi{10.1103/PhysRevB.105.085151},
\newblock \eprint{2102.08899}.

\bibitem{Assaad24}
A.~{G{\"o}tz}, M.~{Hohenadler} and F.~F. {Assaad},
\newblock \emph{{Phases and exotic phase transitions of a two-dimensional
  Su-Schrieffer-Heeger model}},
\newblock Phys. Rev. B \textbf{109}(19), 195154 (2024),
\newblock \doi{10.1103/PhysRevB.109.195154},
\newblock \eprint{2307.07613}.

\bibitem{Assaad25}
A.~{G{\"o}tz}, F.~F. {Assaad} and N.~C. {Costa},
\newblock \emph{{Tuning the order of a deconfined quantum critical point}},
\newblock arXiv e-prints arXiv:2412.17215 (2024),
\newblock \doi{10.48550/arXiv.2412.17215},
\newblock \eprint{2412.17215}.

\bibitem{Xu:2020qbj}
X.~Y. Xu and T.~Grover,
\newblock \emph{{Competing Nodal $d$-Wave Superconductivity and
  Antiferromagnetism}},
\newblock Phys. Rev. Lett. \textbf{126}(21), 217002 (2021),
\newblock \doi{10.1103/PhysRevLett.126.217002},
\newblock \eprint{2009.06644}.

\bibitem{Scaletter21}
B.~Xing, W.~Chiu, D.~Poletti, R.~T. Scalettar and G.~G. Batrouni,
\newblock \emph{{Quantum Monte Carlo Simulations of the 2D
  Su‑Schrieffer‑Heeger Model}},
\newblock Physical Review Letters \textbf{126}(1), 017601 (2021),
\newblock \doi{10.1103/PhysRevLett.126.017601},
\newblock \eprint{2005.09673}.

\bibitem{Scaletter22}
C.~Feng, B.~Xing, D.~Poletti, R.~T. Scalettar and G.~G. Batrouni,
\newblock \emph{{Phase Diagram of the Su–Schrieffer–Heeger–Hubbard model
  on a square lattice}},
\newblock Physical Review B \textbf{106}(8), L081114 (2022),
\newblock \doi{10.1103/PhysRevB.106.L081114},
\newblock \eprint{2109.09206}.

\bibitem{HongYao21}
X.~{Cai}, Z.-X. {Li} and H.~{Yao},
\newblock \emph{{Antiferromagnetism Induced by Bond Su-Schrieffer-Heeger
  Electron-Phonon Coupling: A Quantum Monte Carlo Study}},
\newblock Phys. Rev. Lett. \textbf{127}(24), 247203 (2021),
\newblock \doi{10.1103/PhysRevLett.127.247203},
\newblock \eprint{2102.05060}.

\bibitem{HongYao22}
X.~{Cai}, Z.-X. {Li} and H.~{Yao},
\newblock \emph{{Robustness of antiferromagnetism in the Su-Schrieffer-Heeger
  Hubbard model}},
\newblock Phys. Rev. B \textbf{106}(8), L081115 (2022),
\newblock \doi{10.1103/PhysRevB.106.L081115},
\newblock \eprint{2112.14744}.

\bibitem{HongYao25}
H.-X. Wang, Y.-F. Jiang and H.~Yao,
\newblock \emph{{Robust $d$-wave superconductivity from the
  Su-Schrieffer-Heeger-Hubbard model: Possible route to high-temperature
  superconductivity}},
\newblock Science Bulletin \textbf{70}(14), 2260 (2025),
\newblock \doi{https://doi.org/10.1016/j.scib.2025.04.055},
\newblock \eprint{2211.09143}.

\bibitem{Zhaoyu22}
Z.~{Han} and S.~A. {Kivelson},
\newblock \emph{{Resonating Valence Bond States in an Electron-Phonon System}},
\newblock Phys. Rev. Lett. \textbf{130}(18), 186404 (2023),
\newblock \doi{10.1103/PhysRevLett.130.186404},
\newblock \eprint{2210.16321}.

\bibitem{Zhaoyu24}
X.~{Cai}, Z.~{Han}, Z.-X. {Li}, S.~A. {Kivelson} and H.~{Yao},
\newblock \emph{{Quantum spin liquid from electron-phonon coupling}},
\newblock Proceedings of the National Academy of Science \textbf{122}(33),
  e2426111122 (2025),
\newblock \doi{10.1073/pnas.2426111122},
\newblock \eprint{2408.04002}.

\bibitem{HermeleHoneycomb}
M.~{Hermele},
\newblock \emph{{SU(2) gauge theory of the Hubbard model and application to the
  honeycomb lattice}},
\newblock Phys. Rev. B \textbf{76}(3), 035125 (2007),
\newblock \doi{10.1103/PhysRevB.76.035125},
\newblock \eprint{cond-mat/0701134}.

\bibitem{Christos:2023oru}
M.~Christos, Z.-X. Luo, L.~Shackleton, Y.-H. Zhang, M.~S. Scheurer and
  S.~Sachdev,
\newblock \emph{{A model of $d$-wave superconductivity, antiferromagnetism, and
  charge order on the square lattice}},
\newblock Proc. Nat. Acad. Sci. \textbf{120}(21), e2302701120 (2023),
\newblock \doi{10.1073/pnas.2302701120},
\newblock \eprint{2302.07885}.

\bibitem{BCS24}
P.~M. {Bonetti}, M.~{Christos} and S.~{Sachdev},
\newblock \emph{{Quantum oscillations in the hole-doped cuprates and the
  confinement of spinons}},
\newblock Proceedings of the National Academy of Sciences \textbf{121},
  e2418633121 (2024),
\newblock \doi{10.1073/pnas.2418633121},
\newblock \eprint{2405.08817}.

\bibitem{Sayantan25}
H.~{Pandey}, M.~{Christos}, P.~M. {Bonetti}, R.~{Shanker}, A.~{Nikolaenko},
  S.~{Sharma} and S.~{Sachdev},
\newblock \emph{{Thermal SU(2) lattice gauge theory of the cuprate pseudogap:
  reconciling Fermi arcs and hole pockets}},
\newblock Proceedings of the National Academy of Sciences \textbf{123},
  e2606117123 (2026),
\newblock \doi{10.1073/pnas.2606117123},
\newblock \eprint{2507.05336}.

\bibitem{Fradkin10}
A.~{Jaefari}, S.~{Lal} and E.~{Fradkin},
\newblock \emph{{Charge-density wave and superconductor competition in stripe
  phases of high-temperature superconductors}},
\newblock Phys. Rev. B \textbf{82}(14), 144531 (2010),
\newblock \doi{10.1103/PhysRevB.82.144531},
\newblock \eprint{1007.2187}.

\bibitem{Hayward:2013jna}
L.~E. Hayward, D.~G. Hawthorn, R.~G. Melko and S.~Sachdev,
\newblock \emph{{Angular fluctuations of a multi-component order describe the
  pseudogap regime of the cuprate superconductors}},
\newblock Science \textbf{343}, 1336 (2014),
\newblock \doi{10.1126/science.1246310},
\newblock \eprint{1309.6639}.

\bibitem{Lee14}
P.~A. {Lee},
\newblock \emph{{Amperean Pairing and the Pseudogap Phase of Cuprate
  Superconductors}},
\newblock Physical Review X \textbf{4}(3), 031017 (2014),
\newblock \doi{10.1103/PhysRevX.4.031017},
\newblock \eprint{1401.0519}.

\bibitem{Nie_15}
L.~{Nie}, L.~E.~H. {Sierens}, R.~G. {Melko}, S.~{Sachdev} and S.~A. {Kivelson},
\newblock \emph{{Fluctuating orders and quenched randomness in the cuprates}},
\newblock Phys. Rev. B \textbf{92}(17), 174505 (2015),
\newblock \doi{10.1103/PhysRevB.92.174505},
\newblock \eprint{1505.06206}.

\bibitem{Fradkin15}
E.~{Fradkin}, S.~A. {Kivelson} and J.~M. {Tranquada},
\newblock \emph{{Colloquium: Theory of intertwined orders in high temperature
  superconductors}},
\newblock Reviews of Modern Physics \textbf{87}(2), 457 (2015),
\newblock \doi{10.1103/RevModPhys.87.457},
\newblock \eprint{1407.4480}.

\bibitem{Castro17}
S.~Caprara, M.~Grilli, C.~Di~Castro and G.~Seibold,
\newblock \emph{Pseudogap and (an)isotropic scattering in the fluctuating
  charge-density wave phase of cuprates},
\newblock Journal of Superconductivity and Novel Magnetism \textbf{30}(1), 25
  (2017),
\newblock \doi{10.1007/s10948-016-3775-9},
\newblock \eprint{1610.05037}.

\bibitem{Pepin23}
C.~{P{\'e}pin} and H.~{Freire},
\newblock \emph{{Charge order and emergent symmetries in cuprate
  superconductors}},
\newblock Annals of Physics \textbf{456}, 169233 (2023),
\newblock \doi{10.1016/j.aop.2023.169233},
\newblock \eprint{2210.04046}.

\bibitem{Fradkin25}
E.~{Fradkin},
\newblock \emph{{Intertwined Orders and the Physics of High Temperature
  Superconductors}},
\newblock Particles \textbf{8}(3) (2025),
\newblock \doi{10.3390/particles8030070},
\newblock \eprint{2506.21673}.

\bibitem{KotliarLiu88}
G.~Kotliar and J.~Liu,
\newblock \emph{{Superexchange mechanism and $d$-wave superconductivity}},
\newblock Phys. Rev. B \textbf{38}, 5142 (1988),
\newblock \doi{10.1103/PhysRevB.38.5142}.

\bibitem{IvanovSenthil}
D.~A. {Ivanov} and T.~{Senthil},
\newblock \emph{{Projected wave functions for fractionalized phases of quantum
  spin systems}},
\newblock Phys. Rev. B \textbf{66}(11), 115111 (2002),
\newblock \doi{10.1103/PhysRevB.66.115111},
\newblock \eprint{cond-mat/0204043}.

\bibitem{SSMeng26}
C.~{Chen}, S.~{Sachdev} and Z.~Y. {Meng},
\newblock \emph{{Deconfined criticality between an antiferromagnetic insulator
  and a nodal d-wave superconductor: a quantum Monte Carlo study}},
\newblock arXiv e-prints arXiv:2607.00762 (2026),
\newblock \doi{10.48550/arXiv.2607.00762},
\newblock \eprint{2607.00762}.

\bibitem{Luttinger60}
J.~M. Luttinger,
\newblock \emph{{Fermi Surface and Some Simple Equilibrium Properties of a
  System of Interacting Fermions}},
\newblock Phys. Rev. \textbf{119}, 1153 (1960),
\newblock \doi{10.1103/PhysRev.119.1153}.

\bibitem{MO00}
M.~{Oshikawa},
\newblock \emph{{Topological Approach to Luttinger's Theorem and the Fermi
  Surface of a Kondo Lattice}},
\newblock Phys. Rev. Lett. \textbf{84}, 3370 (2000),
\newblock \doi{10.1103/PhysRevLett.84.3370},
\newblock \eprint{cond-mat/0002392}.

\bibitem{QPMbook}
S.~Sachdev,
\newblock \emph{{Quantum Phases of Matter}},
\newblock Cambridge University Press, Cambridge, UK, 1 edn. (2023).

\bibitem{APAV04}
A.~{Paramekanti} and A.~{Vishwanath},
\newblock \emph{{Extending Luttinger's theorem to $\mathbb{Z}_{2}$
  fractionalized phases of matter}},
\newblock Phys. Rev. B \textbf{70}(24), 245118 (2004),
\newblock \doi{10.1103/PhysRevB.70.245118},
\newblock \eprint{cond-mat/0406619}.

\bibitem{Powell05}
S.~{Powell}, S.~{Sachdev} and H.~P. {B{\"u}chler},
\newblock \emph{{Depletion of the Bose-Einstein condensate in Bose-Fermi
  mixtures}},
\newblock Phys. Rev. B \textbf{72}(2), 024534 (2005),
\newblock \doi{10.1103/PhysRevB.72.024534},
\newblock \eprint{cond-mat/0502299}.

\bibitem{Coleman05}
P.~{Coleman}, I.~{Paul} and J.~{Rech},
\newblock \emph{{Sum rules and Ward identities in the Kondo lattice}},
\newblock Phys. Rev. B \textbf{72}(9), 094430 (2005),
\newblock \doi{10.1103/PhysRevB.72.094430},
\newblock \eprint{cond-mat/0503001}.

\bibitem{Qi10}
Y.~{Qi} and S.~{Sachdev},
\newblock \emph{{Effective theory of Fermi pockets in fluctuating
  antiferromagnets}},
\newblock Phys. Rev. B \textbf{81}(11), 115129 (2010),
\newblock \doi{10.1103/PhysRevB.81.115129},
\newblock \eprint{0912.0943}.

\bibitem{SSMetlitskiPunk12}
S.~{Sachdev}, M.~A. {Metlitski} and M.~{Punk},
\newblock \emph{{Antiferromagnetism in metals: from the cuprate superconductors
  to the heavy fermion materials}},
\newblock Journal of Physics Condensed Matter \textbf{24}(29), 294205 (2012),
\newblock \doi{10.1088/0953-8984/24/29/294205},
\newblock \eprint{1202.4760}.

\bibitem{Metlitski18}
M.~A. {Metlitski} and R.~{Thorngren},
\newblock \emph{{Intrinsic and emergent anomalies at deconfined critical
  points}},
\newblock Phys. Rev. B \textbf{98}(8), 085140 (2018),
\newblock \doi{10.1103/PhysRevB.98.085140},
\newblock \eprint{1707.07686}.

\bibitem{Punk15}
M.~Punk, A.~Allais and S.~Sachdev,
\newblock \emph{{A quantum dimer model for the pseudogap metal}},
\newblock Proc. Nat. Acad. Sci. \textbf{112}, 9552 (2015),
\newblock \doi{10.1073/pnas.1512206112},
\newblock \eprint{1501.00978}.

\bibitem{NishidaSon1}
Y.~{Nishida} and D.~T. {Son},
\newblock \emph{{$\epsilon$ Expansion for a Fermi Gas at Infinite Scattering
  Length}},
\newblock Phys. Rev. Lett. \textbf{97}(5), 050403 (2006),
\newblock \doi{10.1103/PhysRevLett.97.050403},
\newblock \eprint{cond-mat/0604500}.

\bibitem{NishidaSon2}
Y.~{Nishida} and D.~T. {Son},
\newblock \emph{{Fermi gas near unitarity around four and two spatial
  dimensions}},
\newblock Phys. Rev. A \textbf{75}(6), 063617 (2007),
\newblock \doi{10.1103/PhysRevA.75.063617},
\newblock \eprint{cond-mat/0607835}.

\bibitem{NikolicSS}
P.~{Nikoli{\'c}} and S.~{Sachdev},
\newblock \emph{{Renormalization-group fixed points, universal phase diagram,
  and 1/N expansion for quantum liquids with interactions near the unitarity
  limit}},
\newblock Phys. Rev. A \textbf{75}(3), 033608 (2007),
\newblock \doi{10.1103/PhysRevA.75.033608},
\newblock \eprint{cond-mat/0609106}.

\bibitem{Feshbach58}
H.~Feshbach,
\newblock \emph{Unified theory of nuclear reactions},
\newblock Annals of Physics \textbf{5}(4), 357 (1958),
\newblock \doi{10.1016/0003-4916(58)90007-1}.

\bibitem{Feshbach62}
H.~Feshbach,
\newblock \emph{{A unified theory of nuclear reactions. II}},
\newblock Annals of Physics \textbf{19}(2), 287 (1962),
\newblock \doi{10.1016/0003-4916(62)90221-X}.

\bibitem{Feshbach08}
C.~{Chin}, R.~{Grimm}, P.~{Julienne} and E.~{Tiesinga},
\newblock \emph{{Feshbach Resonances in Ultracold Gases}},
\newblock Rev. Mod. Phys. \textbf{82}, 1225 (2010),
\newblock \doi{10.1103/RevModPhys.82.1225},
\newblock \eprint{0812.1496}.

\bibitem{Kaplan97}
D.~B. {Kaplan},
\newblock \emph{{More effective field theory for non-relativistic scattering}},
\newblock Nuclear Physics B \textbf{494}(1), 471 (1997),
\newblock \doi{10.1016/S0550-3213(97)00178-8},
\newblock \eprint{nucl-th/9610052}.

\bibitem{RKK07}
R.~K. {Kaul}, A.~{Kolezhuk}, M.~{Levin}, S.~{Sachdev} and T.~{Senthil},
\newblock \emph{{Hole dynamics in an antiferromagnet across a deconfined
  quantum critical point}},
\newblock Phys. Rev. B \textbf{75}(23), 235122 (2007),
\newblock \doi{10.1103/PhysRevB.75.235122},
\newblock \eprint{cond-mat/0702119}.

\bibitem{RKK08}
R.~K. {Kaul}, Y.~B. {Kim}, S.~{Sachdev} and T.~{Senthil},
\newblock \emph{{Algebraic charge liquids}},
\newblock Nature Physics \textbf{4}, 28 (2008),
\newblock \doi{10.1038/nphys790},
\newblock \eprint{0706.2187}.

\bibitem{Potthoff04}
M.~{Potthoff},
\newblock \emph{{Non-perturbative construction of the Luttinger-Ward
  functional}},
\newblock Condens. Mat. Phys \textbf{9}, 557 (2006),
\newblock \doi{10.5488/CMP.9.3.557},
\newblock \eprint{cond-mat/0406671}.

\bibitem{Altshuler98}
B.~L. {Altshuler}, A.~V. {Chubukov}, A.~{Dashevskii}, A.~M. {Finkel'stein} and
  D.~K. {Morr},
\newblock \emph{{Luttinger theorem for a spin-density-wave state}},
\newblock EPL (Europhysics Letters) \textbf{41}(4), 401 (1998),
\newblock \doi{10.1209/epl/i1998-00164-y},
\newblock \eprint{cond-mat/9703120}.

\bibitem{Dz03}
I.~Dzyaloshinskii,
\newblock \emph{{Some consequences of the Luttinger theorem: The Luttinger
  surfaces in non-Fermi liquids and Mott insulators}},
\newblock Phys. Rev. B \textbf{68}, 085113 (2003),
\newblock \doi{10.1103/PhysRevB.68.085113}.

\bibitem{Tsvelik06}
R.~M. {Konik}, T.~M. {Rice} and A.~M. {Tsvelik},
\newblock \emph{{Doped Spin Liquid: Luttinger Sum Rule and Low Temperature
  Order}},
\newblock Phys. Rev. Lett. \textbf{96}(8), 086407 (2006),
\newblock \doi{10.1103/PhysRevLett.96.086407},
\newblock \eprint{cond-mat/0511268}.

\bibitem{YRZ}
K.-Y. {Yang}, T.~M. {Rice} and F.-C. {Zhang},
\newblock \emph{{Phenomenological theory of the pseudogap state}},
\newblock Phys. Rev. B \textbf{73}(17), 174501 (2006),
\newblock \doi{10.1103/PhysRevB.73.174501},
\newblock \eprint{cond-mat/0602164}.

\bibitem{Kotliarzeros}
T.~D. {Stanescu} and G.~{Kotliar},
\newblock \emph{{Fermi arcs and hidden zeros of the Green function in the
  pseudogap state}},
\newblock Phys. Rev. B \textbf{74}(12), 125110 (2006),
\newblock \doi{10.1103/PhysRevB.74.125110},
\newblock \eprint{cond-mat/0508302}.

\bibitem{Kane13}
K.~B. {Dave}, P.~W. {Phillips} and C.~L. {Kane},
\newblock \emph{{Absence of Luttinger's Theorem due to Zeros in the
  Single-Particle Green Function}},
\newblock Phys. Rev. Lett. \textbf{110}(9), 090403 (2013),
\newblock \doi{10.1103/PhysRevLett.110.090403},
\newblock \eprint{1207.4201}.

\bibitem{Scheurer18}
M.~S. {Scheurer}, S.~{Chatterjee}, W.~{Wu}, M.~{Ferrero}, A.~{Georges} and
  S.~{Sachdev},
\newblock \emph{{Topological order in the pseudogap metal}},
\newblock Proceedings of the National Academy of Science \textbf{115}(16),
  E3665 (2018),
\newblock \doi{10.1073/pnas.1720580115},
\newblock \eprint{1711.09925}.

\bibitem{Fabrizio22}
M.~Fabrizio,
\newblock \emph{{Emergent quasiparticles at Luttinger surfaces}},
\newblock Nature Communications \textbf{13}(1), 1561 (2022),
\newblock \doi{10.1038/s41467-022-29190-y}.

\bibitem{Fabrizio22a}
J.~{Skolimowski} and M.~{Fabrizio},
\newblock \emph{{Luttinger's theorem in the presence of Luttinger surfaces}},
\newblock Phys. Rev. B \textbf{106}(4), 045109 (2022),
\newblock \doi{10.1103/PhysRevB.106.045109},
\newblock \eprint{2202.00426}.

\bibitem{Si24}
C.~{Setty}, F.~{Xie}, S.~{Sur}, L.~{Chen}, M.~G. {Vergniory} and Q.~{Si},
\newblock \emph{{Electronic properties, correlated topology, and Green's
  function zeros}},
\newblock Physical Review Research \textbf{6}(3), 033235 (2024),
\newblock \doi{10.1103/PhysRevResearch.6.033235},
\newblock \eprint{2309.14340}.

\bibitem{GleisKotliar24}
A.~{Gleis}, S.-S.~B. {Lee}, G.~{Kotliar} and J.~{von Delft},
\newblock \emph{{Emergent Properties of the Periodic Anderson Model: A
  High-Resolution, Real-Frequency Study of Heavy-Fermion Quantum Criticality}},
\newblock Physical Review X \textbf{14}(4), 041036 (2024),
\newblock \doi{10.1103/PhysRevX.14.041036},
\newblock \eprint{2310.12672}.

\bibitem{Lucila1}
L.~{Peralta Gavensky}, S.~{Sachdev} and N.~{Goldman},
\newblock \emph{{Connecting the Many-Body Chern Number to Luttinger's Theorem
  through St{\v{r}}eda's Formula}},
\newblock Phys. Rev. Lett. \textbf{131}(23), 236601 (2023),
\newblock \doi{10.1103/PhysRevLett.131.236601},
\newblock \eprint{2309.02483}.

\bibitem{Lucila2}
A.~A. {Markov}, A.~M. {Nikishin}, N.~R. {Cooper}, N.~{Goldman} and L.~{Peralta
  Gavensky},
\newblock \emph{{Luttinger's Theorem Violation and Green's Function Topological
  Invariants in a Fractional Chern Insulator}},
\newblock arXiv e-prints arXiv:2603.17006 (2026),
\newblock \doi{10.48550/arXiv.2603.17006},
\newblock \eprint{2603.17006}.

\bibitem{P2_2025}
G.~{La Nave}, J.~{Zhao} and P.~W. {Phillips},
\newblock \emph{{Luttinger Count is the Homotopy Not the Physical Charge:
  Generalized Anomalies Characterize Non-Fermi Liquids}},
\newblock Phys. Rev. Lett. \textbf{135}(24), 246501 (2025),
\newblock \doi{10.1103/lfxt-fspg},
\newblock \eprint{2506.04342}.

\bibitem{GPS2}
A.~{Georges}, O.~{Parcollet} and S.~{Sachdev},
\newblock \emph{{Quantum fluctuations of a nearly critical Heisenberg spin
  glass}},
\newblock Phys. Rev. B \textbf{63}(13), 134406 (2001),
\newblock \doi{10.1103/PhysRevB.63.134406},
\newblock \eprint{cond-mat/0009388}.

\bibitem{GKST}
Y.~Gu, A.~Kitaev, S.~Sachdev and G.~Tarnopolsky,
\newblock \emph{{Notes on the complex Sachdev-Ye-Kitaev model}},
\newblock Journal of High Energy Physics \textbf{02}, 157 (2020),
\newblock \doi{10.1007/JHEP02(2020)157},
\newblock \eprint{1910.14099}.

\bibitem{Shang21}
S.~{Liu}, H.~{Shapourian}, A.~{Vishwanath} and M.~A. {Metlitski},
\newblock \emph{{Magnetic impurities at quantum critical points: Large-N
  expansion and connections to symmetry-protected topological states}},
\newblock Phys. Rev. B \textbf{104}(10), 104201 (2021),
\newblock \doi{10.1103/PhysRevB.104.104201},
\newblock \eprint{2104.15026}.

\bibitem{SchmalianPines1}
J.~{Schmalian}, D.~{Pines} and B.~{Stojkovi{\'c}},
\newblock \emph{{Weak Pseudogap Behavior in the Underdoped Cuprate
  Superconductors}},
\newblock Phys. Rev. Lett. \textbf{80}(17), 3839 (1998),
\newblock \doi{10.1103/PhysRevLett.80.3839},
\newblock \eprint{cond-mat/9708238}.

\bibitem{SchmalianPines2}
J.~{Schmalian}, D.~{Pines} and B.~{Stojkovi{\'c}},
\newblock \emph{{Microscopic theory of weak pseudogap behavior in the
  underdoped cuprate superconductors: General theory and quasiparticle
  properties}},
\newblock Phys. Rev. B \textbf{60}(1), 667 (1999),
\newblock \doi{10.1103/PhysRevB.60.667},
\newblock \eprint{cond-mat/9804129}.

\bibitem{Chubukov23}
M.~{Ye} and A.~V. {Chubukov},
\newblock \emph{{Crucial role of thermal fluctuations and vertex corrections
  for the magnetic pseudogap}},
\newblock Phys. Rev. B \textbf{108}(8), L081118 (2023),
\newblock \doi{10.1103/PhysRevB.108.L081118},
\newblock \eprint{2306.05489}.

\bibitem{Chubukov25}
E.~K. Kokkinis and A.~V. Chubukov,
\newblock \emph{Pseudogap in electron-doped cuprates as a thermal precursor to
  magnetism},
\newblock Nature Communications \textbf{17}(1), 1075 (2025),
\newblock \doi{10.1038/s41467-025-67835-w},
\newblock \eprint{2505.11727}.

\bibitem{sdw09}
S.~Sachdev, M.~A. Metlitski, Y.~Qi and C.~Xu,
\newblock \emph{{Fluctuating spin density waves in metals}},
\newblock Phys. Rev. B \textbf{80}, 155129 (2009),
\newblock \doi{10.1103/PhysRevB.80.155129},
\newblock \eprint{0907.3732}.

\bibitem{WuScheurer1}
W.~{Wu}, M.~S. {Scheurer}, S.~{Chatterjee}, S.~{Sachdev}, A.~{Georges} and
  M.~{Ferrero},
\newblock \emph{{Pseudogap and Fermi-Surface Topology in the Two-Dimensional
  Hubbard Model}},
\newblock Phys. Rev. X \textbf{8}(2), 021048 (2018),
\newblock \doi{10.1103/PhysRevX.8.021048},
\newblock \eprint{1707.06602}.

\bibitem{Scheurer:2017jcp}
M.~S. Scheurer, S.~Chatterjee, W.~Wu, M.~Ferrero, A.~Georges and S.~Sachdev,
\newblock \emph{{Topological order in the pseudogap metal}},
\newblock Proc. Nat. Acad. Sci. \textbf{115}(16), E3665 (2018),
\newblock \doi{10.1073/pnas.1720580115},
\newblock \eprint{1711.09925}.

\bibitem{Sachdev:2018ddg}
S.~Sachdev,
\newblock \emph{{Topological order, emergent gauge fields, and Fermi surface
  reconstruction}},
\newblock Rept. Prog. Phys. \textbf{82}(1), 014001 (2019),
\newblock \doi{10.1088/1361-6633/aae110},
\newblock \eprint{1801.01125}.

\bibitem{WuScheurer2}
W.~Wu, M.~S. Scheurer, M.~Ferrero and A.~Georges,
\newblock \emph{{Effect of van Hove singularities in the onset of pseudogap
  states in Mott insulators}},
\newblock Phys. Rev. Res. \textbf{2}, 033067 (2020),
\newblock \doi{10.1103/PhysRevResearch.2.033067},
\newblock \eprint{2001.00019}.

\bibitem{Spinon-dopon05}
T.~C. {Ribeiro} and X.-G. {Wen},
\newblock \emph{{New Mean-Field Theory of the tt$^{'}$t$^{''}$J Model Applied
  to High-T$_{c}$ Superconductors}},
\newblock Phys. Rev. Lett. \textbf{95}(5), 057001 (2005),
\newblock \doi{10.1103/PhysRevLett.95.057001},
\newblock \eprint{cond-mat/0410750}.

\bibitem{Sawatzky11}
B.~{Lau}, M.~{Berciu} and G.~A. {Sawatzky},
\newblock \emph{{High-Spin Polaron in Lightly Doped CuO$_{2}$ Planes}},
\newblock Phys. Rev. Lett. \textbf{106}(3), 036401 (2011),
\newblock \doi{10.1103/PhysRevLett.106.036401},
\newblock \eprint{1010.1867}.

\bibitem{Moon11}
E.~G. {Moon} and S.~{Sachdev},
\newblock \emph{{Underdoped cuprates as fractionalized Fermi liquids:
  Transition to superconductivity}},
\newblock Phys. Rev. B \textbf{83}(22), 224508 (2011),
\newblock \doi{10.1103/PhysRevB.83.224508},
\newblock \eprint{1010.4567}.

\bibitem{Sawatzky11b}
B.~{Lau}, M.~{Berciu} and G.~A. {Sawatzky},
\newblock \emph{{Computational approach to a doped antiferromagnet:
  Correlations between two spin polarons in the lightly doped CuO$_{2}$
  plane}},
\newblock Phys. Rev. B \textbf{84}(16), 165102 (2011),
\newblock \doi{10.1103/PhysRevB.84.165102},
\newblock \eprint{1107.4141}.

\bibitem{Mei11}
J.-W. {Mei}, S.~{Kawasaki}, G.-Q. {Zheng}, Z.-Y. {Weng} and X.-G. {Wen},
\newblock \emph{{Luttinger-volume violating Fermi liquid in the pseudogap phase
  of the cuprate superconductors}},
\newblock Phys. Rev. B \textbf{85}(13), 134519 (2012),
\newblock \doi{10.1103/PhysRevB.85.134519},
\newblock \eprint{1109.0406}.

\bibitem{Punk12}
M.~{Punk} and S.~{Sachdev},
\newblock \emph{{Fermi surface reconstruction in hole-doped $t$-$J$ models
  without long-range antiferromagnetic order}},
\newblock Phys. Rev. B \textbf{85}(19), 195123 (2012),
\newblock \doi{10.1103/PhysRevB.85.195123},
\newblock \eprint{1202.4023}.

\bibitem{Punk18}
J.~Feldmeier, S.~Huber and M.~Punk,
\newblock \emph{{Exact Solution of a Two-Species Quantum Dimer Model for
  Pseudogap Metals}},
\newblock Phys. Rev. Lett. \textbf{120}, 187001 (2018),
\newblock \doi{10.1103/PhysRevLett.120.187001},
\newblock \eprint{1712.01854}.

\bibitem{Grusdt18}
F.~{Grusdt}, M.~{K{\'a}nasz-Nagy}, A.~{Bohrdt}, C.~S. {Chiu}, G.~{Ji},
  M.~{Greiner}, D.~{Greif} and E.~{Demler},
\newblock \emph{{Parton Theory of Magnetic Polarons: Mesonic Resonances and
  Signatures in Dynamics}},
\newblock Physical Review X \textbf{8}(1), 011046 (2018),
\newblock \doi{10.1103/PhysRevX.8.011046},
\newblock \eprint{1712.01874}.

\bibitem{Grusdt19}
C.~S. {Chiu}, G.~{Ji}, A.~{Bohrdt}, M.~{Xu}, M.~{Knap}, E.~{Demler},
  F.~{Grusdt}, M.~{Greiner} and D.~{Greif},
\newblock \emph{{String patterns in the doped Hubbard model}},
\newblock Science \textbf{365}(6450), 251 (2019),
\newblock \doi{10.1126/science.aav3587},
\newblock \eprint{1810.03584}.

\bibitem{Grusdt23}
F.~{Grusdt}, E.~{Demler} and A.~{Bohrdt},
\newblock \emph{{Pairing of holes by confining strings in antiferromagnets}},
\newblock SciPost Physics \textbf{14}(5), 090 (2023),
\newblock \doi{10.21468/SciPostPhys.14.5.090},
\newblock \eprint{2210.02321}.

\bibitem{Grusdt24}
H.~{Schl{\"o}mer}, U.~{Schollw{\"o}ck}, A.~{Bohrdt} and F.~{Grusdt},
\newblock \emph{{Kinetic-to-magnetic frustration crossover and linear
  confinement in the doped triangular t {\ensuremath{-}}J model}},
\newblock Phys. Rev. B \textbf{110}(4), L041117 (2024),
\newblock \doi{10.1103/PhysRevB.110.L041117},
\newblock \eprint{2305.02342}.

\bibitem{Balents25}
J.~H. {Nyhegn}, K.~K. {Nielsen}, L.~{Balents} and G.~M. {Bruun},
\newblock \emph{{Spin-Charge Bound States and Emerging Fermions in a Quantum
  Spin Liquid}},
\newblock PRX Quantum \textbf{6}(4), 040347 (2025),
\newblock \doi{10.1103/w23h-dhrk},
\newblock \eprint{2507.02508}.

\bibitem{Shraiman90}
B.~I. Shraiman and E.~D. Siggia,
\newblock \emph{Mobile vacancy in a quantum antiferromagnet: Effective
  hamiltonian},
\newblock Phys. Rev. B \textbf{42}, 2485 (1990),
\newblock \doi{10.1103/PhysRevB.42.2485}.

\bibitem{SS94}
S.~{Sachdev},
\newblock \emph{{Quantum phases of the Shraiman-Siggia model}},
\newblock Phys. Rev. B \textbf{49}(10), 6770 (1994),
\newblock \doi{10.1103/PhysRevB.49.6770},
\newblock \eprint{cond-mat/9311037}.

\bibitem{Greven14}
W.~{Tabis}, Y.~{Li}, M.~L. {Tacon}, L.~{Braicovich}, A.~{Kreyssig},
  M.~{Minola}, G.~{Dellea}, E.~{Weschke}, M.~J. {Veit}, M.~{Ramazanoglu}, A.~I.
  {Goldman}, T.~{Schmitt} \emph{et~al.},
\newblock \emph{{Charge order and its connection with Fermi-liquid charge
  transport in a pristine high-T$_{c}$ cuprate}},
\newblock Nature Communications \textbf{5}, 5875 (2014),
\newblock \doi{10.1038/ncomms6875},
\newblock \eprint{1404.7658}.

\bibitem{Greven25}
Z.~W. Anderson, Y.~Tang, V.~Nagarajan, M.~K. Chan, C.~J. Dorow, G.~Yu, D.~L.
  Abernathy, A.~D. Christianson, L.~Mangin-Thro, P.~Steffens, T.~Sterling,
  D.~Reznik \emph{et~al.},
\newblock \emph{Gapped commensurate antiferromagnetic response in a strongly
  underdoped model cuprate superconductor},
\newblock npj Quantum Materials \textbf{10}(1), 93 (2025),
\newblock \doi{10.1038/s41535-025-00804-0},
\newblock \eprint{2412.03524}.

\bibitem{Zhao_Yamaji_25}
J.-Y. {Zhao}, S.~{Chatterjee}, S.~{Sachdev} and Y.-H. {Zhang},
\newblock \emph{{Yamaji effect in models of underdoped cuprates}},
\newblock Physical Review B \textbf{113}, 245150 (2026),
\newblock \doi{10.1103/wlyk-v88c},
\newblock \eprint{2510.13943}.

\bibitem{Joshi23}
A.~{Nikolaenko}, J.~{von Milczewski}, D.~G. {Joshi} and S.~{Sachdev},
\newblock \emph{{Spin density wave, Fermi liquid, and fractionalized phases in
  a theory of antiferromagnetic metals using paramagnons and bosonic spinons}},
\newblock Phys. Rev. B \textbf{108}(4), 045123 (2023),
\newblock \doi{10.1103/PhysRevB.108.045123},
\newblock \eprint{2211.10452}.

\bibitem{FuChun25}
Y.~{Zhong}, F.-C. {Zhang} and K.~{Jiang},
\newblock \emph{{Yamaji effect and quantum oscillation in Yang-Rice-Zhang model
  of underdoped cuprates}},
\newblock arXiv e-prints arXiv:2512.10475 (2025),
\newblock \doi{10.48550/arXiv.2512.10475},
\newblock \eprint{2512.10475}.

\bibitem{Koepsell21}
J.~{Koepsell}, D.~{Bourgund}, P.~{Sompet}, S.~{Hirthe}, A.~{Bohrdt}, Y.~{Wang},
  F.~{Grusdt}, E.~{Demler}, G.~{Salomon}, C.~{Gross} and I.~{Bloch},
\newblock \emph{{Microscopic evolution of doped Mott insulators from polaronic
  metal to Fermi liquid}},
\newblock Science \textbf{374}(6563), 82 (2021),
\newblock \doi{10.1126/science.abe7165},
\newblock \eprint[arXiv]{2009.04440}.

\bibitem{Iqbal24}
T.~{M{\"u}ller}, R.~{Thomale}, S.~{Sachdev} and Y.~{Iqbal},
\newblock \emph{{Polaronic correlations from optimized ancilla wave functions
  for the Fermi-Hubbard model}},
\newblock Proceedings of the National Academy of Science \textbf{122}(20),
  e2504261122 (2025),
\newblock \doi{10.1073/pnas.2504261122},
\newblock \eprint{2408.01492}.

\bibitem{HenryShiwei24}
L.~{Shackleton} and S.~{Zhang},
\newblock \emph{{Emergent polaronic correlations in doped spin liquids}},
\newblock arXiv e-prints arXiv:2408.02190 (2024),
\newblock \doi{10.48550/arXiv.2408.02190},
\newblock \eprint{2408.02190}.

\bibitem{Bloch24}
T.~{Chalopin}, P.~{Bojovi{\'c}}, S.~{Wang}, T.~{Franz}, A.~{Sinha}, Z.~{Wang},
  D.~{Bourgund}, J.~{Obermeyer}, F.~{Grusdt}, A.~{Bohrdt}, L.~{Pollet},
  A.~{Wietek} \emph{et~al.},
\newblock \emph{{Observation of emergent scaling of spin-charge correlations at
  the onset of the pseudogap}},
\newblock Proceedings of the National Academy of Sciences \textbf{123}(4),
  e2525539123 (2026),
\newblock \doi{10.1073/pnas.2525539123},
\newblock \eprint{2412.17801}.

\bibitem{YaHui-ancilla1}
Y.-H. {Zhang} and S.~{Sachdev},
\newblock \emph{{From the pseudogap metal to the Fermi liquid using ancilla
  qubits}},
\newblock Physical Review Research \textbf{2}(2), 023172 (2020),
\newblock \doi{10.1103/PhysRevResearch.2.023172},
\newblock \eprint{2001.09159}.

\bibitem{YaHui-ancilla2}
Y.-H. {Zhang} and S.~{Sachdev},
\newblock \emph{{Deconfined criticality and ghost Fermi surfaces at the onset
  of antiferromagnetism in a metal}},
\newblock Phys. Rev. B \textbf{102}(15), 155124 (2020),
\newblock \doi{10.1103/PhysRevB.102.155124},
\newblock \eprint{2006.01140}.

\bibitem{Nikolaenko:2021vlw}
A.~Nikolaenko, M.~Tikhanovskaya, S.~Sachdev and Y.-H. Zhang,
\newblock \emph{{Small to large Fermi surface transition in a single band
  model, using randomly coupled ancillas}},
\newblock Phys. Rev. B \textbf{103}(23), 235138 (2021),
\newblock \doi{10.1103/PhysRevB.103.235138},
\newblock \eprint{2103.05009}.

\bibitem{Mascot22}
E.~{Mascot}, A.~{Nikolaenko}, M.~{Tikhanovskaya}, Y.-H. {Zhang}, D.~K. {Morr}
  and S.~{Sachdev},
\newblock \emph{{Electronic spectra with paramagnon fractionalization in the
  single-band Hubbard model}},
\newblock Phys. Rev. B \textbf{105}(7), 075146 (2022),
\newblock \doi{10.1103/PhysRevB.105.075146},
\newblock \eprint{2111.13703}.

\bibitem{vanilla}
P.~W. {Anderson}, P.~A. {Lee}, M.~{Randeria}, T.~M. {Rice}, N.~{Trivedi} and
  F.~C. {Zhang},
\newblock \emph{{The physics behind high-temperature superconducting cuprates:
  the 'plain vanilla' version of RVB}},
\newblock Journal of Physics Condensed Matter \textbf{16}(24), R755 (2004),
\newblock \doi{10.1088/0953-8984/16/24/R02},
\newblock \eprint{cond-mat/0311467}.

\bibitem{Hinton86}
D.~E. Rumelhart, G.~E. Hinton and R.~J. Williams,
\newblock \emph{Learning representations by back-propagating errors},
\newblock Nature \textbf{323}(6088), 533 (1986),
\newblock \doi{10.1038/323533a0}.

\bibitem{phase_diag}
B.~Keimer, S.~A. Kivelson, M.~R. Norman, S.~Uchida and J.~Zaanen,
\newblock \emph{From quantum matter to high-temperature superconductivity in
  copper oxides},
\newblock Nature \textbf{518}(7538), 179 (2015),
\newblock \doi{10.1038/nature14165}.

\bibitem{YaHui24}
B.~{Zhou}, H.-K. {Jin} and Y.-H. {Zhang},
\newblock \emph{{Variational wavefunction for a Mott insulator at finite U
  using ancilla qubits}},
\newblock Phys. Rev. B \textbf{112}(11), 115159 (2025),
\newblock \doi{10.1103/8knn-mr5x},
\newblock \eprint{2409.07512}.

\bibitem{Kendrick:2025ujy}
L.~{Haldar Kendrick}, A.~{Kale}, Y.~{Gang}, A.~{Dennisovich Deters},
  M.~{Lebrat}, A.~W. {Young} and M.~{Greiner},
\newblock \emph{{Pseudogap in a Fermi-Hubbard quantum simulator}},
\newblock arXiv e-prints arXiv:2509.18075 (2025),
\newblock \doi{10.48550/arXiv.2509.18075},
\newblock \eprint{2509.18075}.

\bibitem{ROPP25}
P.~M. {Bonetti}, M.~{Christos}, A.~{Nikolaenko}, A.~A. {Patel} and
  S.~{Sachdev},
\newblock \emph{{Fractionalized Fermi liquids and the cuprate phase diagram}},
\newblock Reports on Progress in Physics \textbf{89}, 044501 (2026),
\newblock \doi{10.1088/1361-6633/ae530d},
\newblock \eprint{2508.20164}.

\bibitem{Jiang21}
H.-C. {Jiang} and S.~A. {Kivelson},
\newblock \emph{{High Temperature Superconductivity in a Lightly Doped Quantum
  Spin Liquid}},
\newblock Phys. Rev. Lett. \textbf{127}(9), 097002 (2021),
\newblock \doi{10.1103/PhysRevLett.127.097002},
\newblock \eprint{2104.01485}.

\bibitem{Jiang23}
H.-C. {Jiang}, S.~A. {Kivelson} and D.-H. {Lee},
\newblock \emph{{Superconducting valence bond fluid in lightly doped eight-leg
  $t$-$J$ cylinders}},
\newblock Phys. Rev. B \textbf{108}(5), 054505 (2023),
\newblock \doi{10.1103/PhysRevB.108.054505},
\newblock \eprint{2302.11633}.

\bibitem{Chatterjee16}
S.~{Chatterjee} and S.~{Sachdev},
\newblock \emph{{Fractionalized Fermi liquid with bosonic chargons as a
  candidate for the pseudogap metal}},
\newblock Phys. Rev. B \textbf{94}(20), 205117 (2016),
\newblock \doi{10.1103/PhysRevB.94.205117},
\newblock \eprint{1607.05727}.

\bibitem{CS23}
M.~{Christos} and S.~{Sachdev},
\newblock \emph{{Emergence of nodal Bogoliubov quasiparticles across the
  transition from the pseudogap metal to the d-wave superconductor}},
\newblock npj Quantum Materials \textbf{9}(1), 4 (2024),
\newblock \doi{10.1038/s41535-023-00608-0},
\newblock \eprint{2308.03835}.

\bibitem{PhysRevLett.88.257001}
N.~P. Armitage, F.~Ronning, D.~H. Lu, C.~Kim, A.~Damascelli, K.~M. Shen, D.~L.
  Feng, H.~Eisaki, Z.-X. Shen, P.~K. Mang, N.~Kaneko, M.~Greven \emph{et~al.},
\newblock \emph{{Doping Dependence of an $\mathit{n}$-Type Cuprate
  Superconductor Investigated by Angle-Resolved Photoemission Spectroscopy}},
\newblock Phys. Rev. Lett. \textbf{88}, 257001 (2002),
\newblock \doi{10.1103/PhysRevLett.88.257001}.

\bibitem{Xu_2023}
K.-J. Xu, Q.~Guo, M.~Hashimoto, Z.-X. Li, S.-D. Chen, J.~He, Y.~He, C.~Li,
  M.~H. Berntsen, C.~R. Rotundu, Y.~S. Lee, T.~P. Devereaux \emph{et~al.},
\newblock \emph{{Bogoliubov quasiparticle on the gossamer Fermi surface in
  electron-doped cuprates}},
\newblock Nature Physics \textbf{19}(12), 1834–1840 (2023),
\newblock \doi{10.1038/s41567-023-02209-x}.

\bibitem{RKK08b}
R.~K. {Kaul}, M.~A. {Metlitski}, S.~{Sachdev} and C.~{Xu},
\newblock \emph{{Destruction of N{\'e}el order in the cuprates by electron
  doping}},
\newblock Phys. Rev. B \textbf{78}(4), 045110 (2008),
\newblock \doi{10.1103/PhysRevB.78.045110},
\newblock \eprint{0804.1794}.

\bibitem{LZDC20}
L.~{Zou} and D.~{Chowdhury},
\newblock \emph{{Deconfined metallic quantum criticality: A U (2)
  gauge-theoretic approach}},
\newblock Physical Review Research \textbf{2}(2), 023344 (2020),
\newblock \doi{10.1103/PhysRevResearch.2.023344},
\newblock \eprint{2002.02972}.

\bibitem{ZhangSS24}
J.-X. {Zhang} and S.~{Sachdev},
\newblock \emph{{Vortex structure in a $d$-wave superconductor obtained by a
  confinement transition from the pseudogap metal}},
\newblock Phys. Rev. B \textbf{110}(23), 235120 (2024),
\newblock \doi{10.1103/PhysRevB.110.235120},
\newblock \eprint{2406.12964}.

\bibitem{Renner21}
T.~{Gazdi{\'c}}, I.~{Maggio-Aprile}, G.~{Gu} and C.~{Renner},
\newblock \emph{{Wang-MacDonald $d$-Wave Vortex Cores Observed in Heavily
  Overdoped Bi$_{2}$Sr$_{2}$CaCu$_{2}$O$_{8 +{\ensuremath{\delta}}}$}},
\newblock Physical Review X \textbf{11}(3), 031040 (2021),
\newblock \doi{10.1103/PhysRevX.11.031040},
\newblock \eprint{2103.05994}.

\bibitem{WM95}
Y.~{Wang} and A.~H. {MacDonald},
\newblock \emph{{Mixed-state quasiparticle spectrum for d-wave
  superconductors}},
\newblock Phys. Rev. B \textbf{52}(6), R3876 (1995),
\newblock \doi{10.1103/PhysRevB.52.R3876},
\newblock \eprint{cond-mat/9505142}.

\bibitem{Hoffman02}
J.~E. {Hoffman}, E.~W. {Hudson}, K.~M. {Lang}, V.~{Madhavan}, H.~{Eisaki},
  S.~{Uchida} and J.~C. {Davis},
\newblock \emph{{A Four Unit Cell Periodic Pattern of Quasi-Particle States
  Surrounding Vortex Cores in
  Bi$_{2}$Sr$_{2}$CaCu$_{2}$O$_{8+{\ensuremath{\delta}}}$}},
\newblock Science \textbf{295}(5554), 466 (2002),
\newblock \doi{10.1126/science.1066974},
\newblock \eprint{cond-mat/0201348}.

\bibitem{Patel2}
A.~A. Patel, H.~Guo, I.~Esterlis and S.~Sachdev,
\newblock \emph{{Universal theory of strange metals from spatially random
  interactions}},
\newblock Science \textbf{381}, 790 (2023),
\newblock \doi{10.1126/science.abq6011},
\newblock \eprint{2203.04990}.

\bibitem{Li:2024kxr}
C.~Li, D.~Valentinis, A.~A. Patel, H.~Guo, J.~Schmalian, S.~Sachdev and
  I.~Esterlis,
\newblock \emph{{Strange Metal and Superconductor in the Two-Dimensional
  Yukawa-Sachdev-Ye-Kitaev Model}},
\newblock Phys. Rev. Lett. \textbf{133}(18), 186502 (2024),
\newblock \doi{10.1103/PhysRevLett.133.186502},
\newblock \eprint{2406.07608}.

\bibitem{PLA24}
A.~A. {Patel}, P.~{Lunts} and M.~S. {Albergo},
\newblock \emph{{Strange metals and planckian transport in a gapless phase from
  spatially random interactions}},
\newblock Physical Review X \textbf{15}, 031064 (2025),
\newblock \doi{10.1103/611k-yxb9},
\newblock \eprint{2410.05365}.

\bibitem{LPS24}
P.~{Lunts}, A.~A. {Patel} and S.~{Sachdev},
\newblock \emph{{Thermopower across Fermi-volume-changing quantum phase
  transitions without translational symmetry breaking}},
\newblock Phys. Rev. B \textbf{111}(24), 245151 (2025),
\newblock \doi{10.1103/3639-byq1},
\newblock \eprint{2412.15330}.

\bibitem{Davide25}
D.~{Valentinis}, J.~{Schmalian}, S.~{Sachdev} and A.~A. {Patel},
\newblock \emph{{Superlinear Hall angle and carrier mobility from non-Boltzmann
  magnetotransport in the spatially disordered Yukawa-Sachdev-Ye-Kitaev model
  on a square lattice}},
\newblock Physical Review Research \textbf{8}(1), 013299 (2026),
\newblock \doi{10.1103/32ts-qh8d},
\newblock \eprint{2511.01030}.

\bibitem{SY}
S.~{Sachdev} and J.~{Ye},
\newblock \emph{{Gapless spin-fluid ground state in a random quantum Heisenberg
  magnet}},
\newblock Phys. Rev. Lett. \textbf{70}(21), 3339 (1993),
\newblock \doi{10.1103/PhysRevLett.70.3339},
\newblock \eprint{cond-mat/9212030}.

\bibitem{kitaev_talk}
A.~Kitaev,
\newblock \emph{{A simple model of quantum holography, talk given at KITP
  program: entanglement in Strongly-Correlated Quantum Matter}},
\newblock University of California, Santa Barbara  (2015).

\end{thebibliography}

\end{document}